# Solar System Elemental Abundances from the Solar Photosphere and CI-Chondrites

K. Lodders[1*], M. Bergemann[2*], and H. Palme[3]

[1]Dept of Earth, Environmental, & Planetary Sciences and McDonnell Center for Space Sciences, Washington Univ., St. Louis, MO 63130, USA. lodders@wustl.edu. [2]Max Planck Institute for Astronomy, Heidelberg, Germany. bergemann@mpia-hd.mpg.de. [3]Senckenberg Forschungsinstitut und Naturmuseum, Frankfurt, Germany

[*]corresponding authors

Keywords: elemental abundances, sun, solar photosphere, meteorites, chondrites, CI-chondrites

## Abstract

Solar photospheric abundances and CI-chondrite compositions are reviewed and updated to obtain representative solar system abundances of the elements and their isotopes. The new photospheric abundances obtained here lead to higher solar metallicity. Full 3D NLTE photospheric analyses are only available for 11 elements. A quality index for analyses is introduced. For several elements, uncertainties remain large. Protosolar mass fractions are H (X = 0.7060), He (Y = 0.2753), and for metals Li to U (Z = 0.0187). The protosolar (C+N)/H agrees within 13% with the ratio for the solar core from the Borexino experiment. Elemental abundances in CI-chondrites were screened by analytical methods, sample sizes, and evaluated using concentration frequency distributions. Aqueously mobile elements (e.g., alkalis, alkaline earths, etc.) often deviate from normal distributions indicating mobilization and/or sequestration into carbonates, phosphates, and sulfates. Revised CI-chondrite abundances of non-volatile elements are similar to earlier estimates. The moderately volatile elements F and Sb are higher than before, as are C, Br and I, whereas the CI-abundances of Hg and N are now significantly lower. The solar system nuclide distribution curves of s-process elements agree within 4% with s-process predictions of Galactic chemical evolution models. P-process nuclide distributions are assessed. No obvious correlation of CI-chondritic to solar elemental abundance ratios with condensation temperatures is observed, nor is there one for ratios of CI-chondrites/solar wind abundances.

## Introduction

The solar system or proto-solar elemental abundances are a widely used reference set in astronomy, astrophysics, cosmochemistry, geosciences/Earth sciences and planetary sciences. Figure 1 illustrates the wide-ranging impacts of solar abundances.

The Sun's chemical composition - like that of other stars - is revealed through absorption spectra. Gustav Kirchhoff (1824 - 1887) was the first to recognize that the dark lines in the sunlight spectrum are characteristic of chemical elements in the outer cooler layers of the Sun absorbing radiation from the hotter underlying parts. First attempts to quantify the information contained in stellar and solar absorption lines were made in the early 20th century. Henry N. Russell (1929) published the first comprehensive list for photospheric abundances of 56 elements. Improvements in instrumentation and in interpretation of absorption spectra using increased knowledge of atomic properties and better modeling of the solar photosphere produced increasingly accurate compositional data. Today solar atmospheric abundances can be determined within ±10% to 20% for many elements.

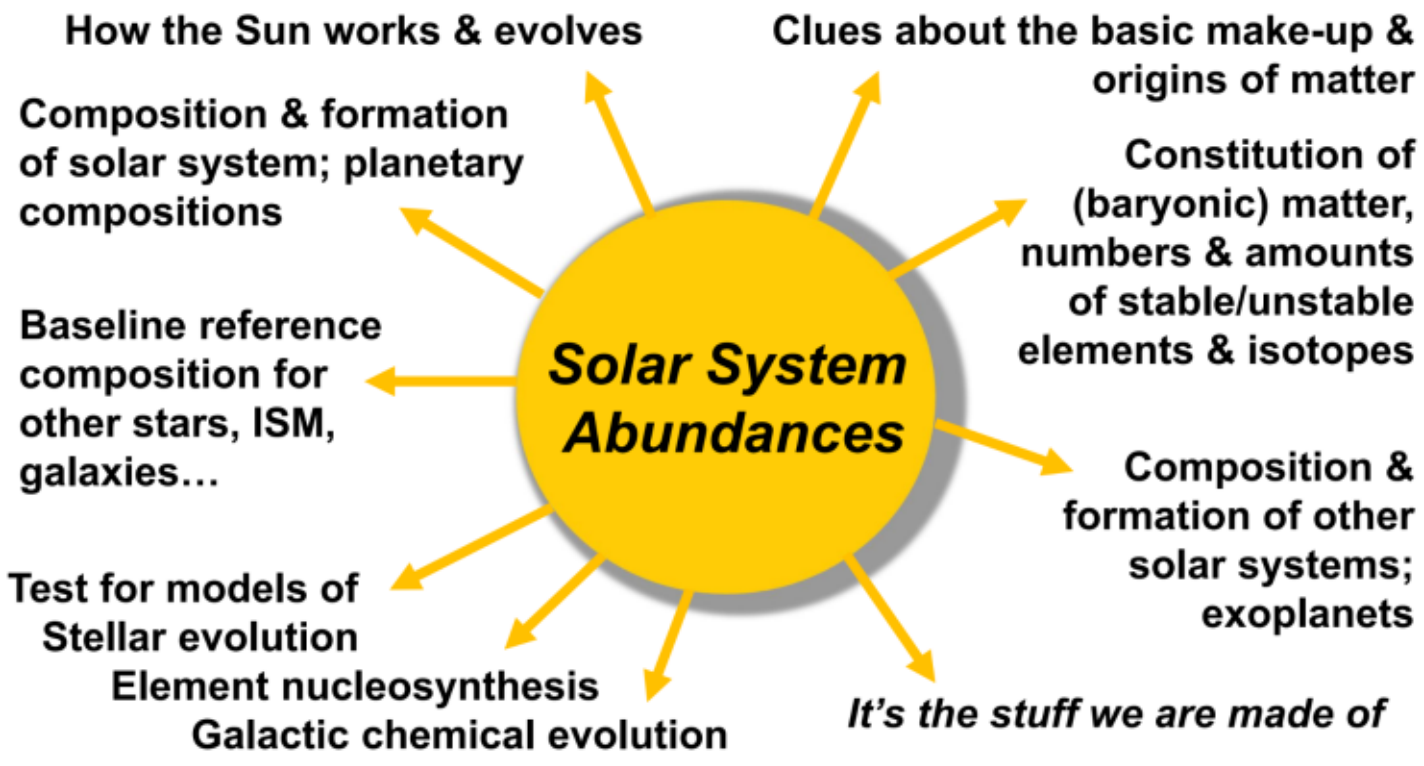

**Figure 1**. *Why solar system elemental abundances are important.*

At about the same time Victor M. Goldschmidt and others analyzed meteorites to derive the composition of average solar system matter. Many compilations followed after Goldschmidt's and Russel's lists. The Suess and Urey (1956) compilation was influential for theories of nucleosynthesis (Suess et al. 1956; Burbridge et al. 1957; Cameron 1957). Later reviews employed improved meteoritic and photospheric analyses and emphasized the excellent agreement of photospheric abundances with abundances of carbonaceous chondrites of the Ivuna-type, the CI-chondrites (Cameron, 1973; Anders and Ebihara 1982, Anders and Grevesse, 1989; Palme and Beer, 1993; Lodders 2003, Asplund et al., 2009, 2021; Lodders et al. 2009; Palme et al. 2014; Lodders 2020 gives a historical review). This and the fact that CI-chondrites are the most volatile element-rich chondrite group are the major arguments for using CI-chondrites as proxy for the condensable elemental abundances including those that currently cannot be determined quantitatively in the Sun. Only by combining solar and meteoritic data (plus some theoretical inputs) a complete set of solar system abundances for all naturally occurring elements and their isotopes is obtained. Sometimes "Solar System Abundances" are called cosmic abundances because of compositional similarities among some G-type stars like the Sun and other dwarf stars (e.g., Valenti and Fischer 2005, Nissen 2015, Bedell et al. 2018).

Table 1

Table 1 lists sources and limitations for abundance determinations. Meteorite abundances can be determined more precisely than photospheric abundances and are therefore often preferred, however, improvements in spectroscopy and more realistic physical models for the solar atmosphere give increasingly accurate and precise photospheric abundances.

"Solar abundances" means abundances primarily obtained from spectral analyses of the solar photosphere, sunspots, and the corona. There are also in-situ measurements of the solar corpuscular radiation (solar wind, SW) by spacecraft (Gloeckler et al. 1998) and laboratory analyses of solar wind implanted in lunar surface materials and returned samples collected by the Genesis mission (e.g., Reisenfeld et al. 2007, Burnett et al. 2011, Pilleri et al. 2015, Heber et al. 2021 and references therein).

"Meteoritic" or "chondritic" abundances are independent sources for bulk solar system abundances from laboratory chemical analyses of CI-chondrites. Their elemental abundances compare most closely to the composition of the solar photosphere for condensable elements, except for ultra-volatile elements H, C, N, O and noble gases which are not fully retained in meteorites and Li which is destroyed in the Sun.

The photospheric spectrum gives the composition of the present-day solar convective envelope (CE). This is not the proto-solar composition at the birth of the solar system (see Lodders 2020 for a review). Over the Sun's lifetime, elements heavier than H diffused ("settled") from the CE into the solar interior. Protosolar abundances are described below.

Currently, noble gas abundances cannot be determined from the photosphere but may be derived from the solar wind. Photospheric C, N, and O abundances are uncertain (see below). Other sources for these abundances are nearby B-type stars. Future entry probes may give atmospheric noble gas data for Saturn, Uranus, and Neptune as the Galileo probe did for Jupiter, but deviations from solar abundances are to be expected. Interpolations using the elemental (or isotopic) abundance curve as a function of atomic number (or mass number) remain useful (e.g., Goldschmidt 1937; Suess 1947a,b), and with modern nucleosynthetic systematics and Galactic chemical evolution models such interpolations give abundance estimates of Kr and Xe. The solar wind measurements from the Genesis mission yielded relative noble gas abundances and data for some other major elements (e.g., Heber et al. 2009, 2012, 2021; Vogel et al. 2011; Pepin et al. 2012; Meshik et al. 2014). The determination of H in the Genesis samples was necessary to compare solar wind to photospheric abundances relative to H (Huss et al. 2020, Heber et al. 2021). Genesis data can give the composition of the solar convection zone (with photosphere on top), but it requires full understanding of elemental fractionations between the photosphere and coronal solar wind sources. Atoms with low ionization energies (FIP < 10 eV) are more abundant in the corona and in solar energetic particles (SEP) than in the photosphere. The solar wind is ultimately derived from coronal sources, and the "FIP-bias" also applies to abundances determined from the Genesis samples.

At least two element fractionation processes must be considered to get the proto-solar or solar system elemental abundances from solar wind analyses: (1) fractionations during settling from the CE and (2) ionization potential and/or first ionization time-driven fractionations between the photosphere and the solar corona. Compositional differences exist between SEPs and the slow and fast solar winds and require model-dependent corrections for fractionations during acceleration of solar wind into different solar wind regimes in order to back-track solar wind to compositions of the CE.

## Solar Photospheric Analyses

The analysis of the solar photosphere cannot be performed directly. No in-situ experimental probes can reach the photosphere through the extremely hot corona and chromosphere. The only sources for quantitative analysis are solar spectra. These data are accessible with ground-based facilities, such as the Kitt Peak National Observatory (KPNO) Fourier Transform Spectrometer (FTS; Kurucz et al. 1984), FTS of the Institut Astrophysik Göttingen (IAG; Reiners et al. 2016), and the Swedish Solar Telescope (SST; Pietrow et al. 2023a), and from space-based facilities, such as HINODE (Caffau et al. 2015). The positions of absorption lines can be derived from experimental or theoretical atomic and molecular data. For extracting elemental abundances, the depths and detailed shapes (so-called "profiles") of lines are measured for each chemical element. The solar optical spectrum harbors hundreds of thousands of absorption lines (the exact number is unknown) of neutral and singly ionized atoms, and lines of diverse molecules.

The lines are caused by absorption of the outgoing radiation field by the photospheric plasma. The depths and shapes of lines are defined by the complex physical structure of the photosphere, specifically by distributions of gas temperature, density, and gas velocities to a depth of about ~2000 km. The shapes of spectral lines also depend on how photons interact with gas, i.e., how much true absorption or scattering occurs. Proper calculations require detailed models of the solar photosphere (see Nordlund et al. 2009 and references therein). The synthetic spectra calculated from these models are then compared to the observed spectra to find the best model fit. The corresponding synthetic spectral model is then taken as the preferred one and the abundance of the element used to compute it as the representative solar photospheric abundance. A short summary of the currently used methods is given by Bergemann and Serenelli (2014).

Table 2.

Recommended photospheric abundances are listed in Table 2 with a new system flagging the solar abundances (all highly model-dependent quantities) by their accuracy and precision. Flags range from A+ (top, most reliable value, accuracy ~0.05 dex approximately 10 %) to E (lowest quality, highly unreliable, accuracy worse than 0.25 dex, approximately a factor of 2). The error parameter "sigma" is partly based on this model-dependent assessment. The sigma here is the fiducial error of the value and is not related to confidence intervals. The criteria for assigning the flags to abundances are as follows:

**Error from 0.04 to 0.06 dex:**

**A+** NLTE based on time-dependent 3D atmosphere models, comprehensive NLTE model atom (complete level system, quantum-mechanical estimates of photo-ionization cross-sections and inelastic X+H collisions), accurate log(gf) values, multiple diagnostic lines in the optical and IR solar spectrum, lack of significant blending, independent consistent estimates; sigma = 0.04 dex.

**A** NLTE based on time-dependent 3D atmosphere models, minor concerns about the NLTE model atom or atomic data (e.g., incomplete knowledge of collisional cross-sections); other criteria as in [A+], no independent validation; sigma = 0.05 dex.

**A-** NLTE based on time-dependent horizontally-averaged 3D atmosphere models; other criteria as in [1] but measurement uncertainties due to blending and/or lack of a statistically significant number of clean diagnostic lines (e.g., solar Li), and/or lack of possibility to reliably test the excitation-ionization balance (e.g., Ba, Y, Eu); sigma = 0.06 dex.

**Error from 0.07 to 0.11 dex:**

**B+** as in group A, but more problematic atomic data (e.g., only theoretical oscillator strengths), minor differences between estimates by different groups; sigma = 0.07 dex.

**B** as in group A, but significant differences reported by independent groups (either in 1D or 3D calculations), despite reliable atomic and molecular data; or a mixture of 3D LTE and 3D NLTE (e.g., C, N); sigma = 0.09 dex.

**B-** 1D NLTE spectrum synthesis or 3D LTE + 1D NLTE modeling, no validation through direct 3D NLTE calculations; sigma = 0.11 dex.

**Error from 0.12 to 0.2 dex:**

**C+** 1D LTE synthesis + 1D NLTE abundance correction, or direct 3D LTE spectrum synthesis, multiple diagnostic lines available, reliable atomic or molecular data, multiple estimates by different independent groups; sigma=0.12 dex

**C** 3D LTE spectrum synthesis, limited number of lines, uncertain atomic or molecular data; sigma=0.14 dex;

**C-** 3D LTE spectrum synthesis or 1D LTE synthesis + 3D LTE abundance correction, no independent validation; sigma = 0.16 dex.

**D** 1D LTE calculations, multiple diagnostic lines available, reliable atomic or molecular data; sigma=0.20 dex; or 1D LTE+3D LTE correction, but heavily blended 1 diagnostic line in the blue (e.g., W) or UV (e.g., Os, Au, Pb).

**E** 1D LTE calculations, very limited number of diagnostic lines and/or substantial concerns over the quality of atomic or molecular data, any other critical concern as described for individual elements below; sigma at least 0.25 dex.

## Solar Model Atmospheres

Early compilations of the solar composition derived from absorption line spectra (Anders and Grevesse 1989, Grevesse and Sauval 1998) relied on results from simplified one-dimensional (1D) model atmospheres in hydrostatic equilibrium (HE) and local thermodynamic equilibrium (LTE). One of the main drawbacks of such models - owing to the assumption of HE - is the lack of considering turbulence and convection. Convective energy transport is usually parameterized using the mixing length theory (Böhm-Vitense 1958), whereas velocity fields are represented by an ad-hoc correction to opacity (the so-called "micro-turbulence") and an artificial broadening applied to emergent monochromatic intensities ("macro-turbulence"). These corrections are used in standard 1D HE LTE models, such as MARCS (Gustafsson et al. 2008) and Kurucz (Castelli and Kurucz 2003). In the vast majority of astrophysical research, these models serve as the basis of stellar abundance calculations, for example for the studies of other Sun-like stars.

The Sun has a convective envelope (CE) occupying roughly the outermost 30% of the Sun in radius (e.g. Serenelli et al. 2009). This envelope has a major impact on the thermodynamic structure of the atmosphere, the latter being in comparison a very thin layer of only 0.1% of the Sun. Convection manifests itself observationally through granulation (see Nordlund et al. 2009 and references therein). The scales and properties of sub-surface convection are defined by the Standard Solar Models (SSM, Basu and Antia 2004, 2008, Bahcall et al. 2004, Serenelli et al. 2009). These sophisticated models describe the evolution of the Sun from the pre-main sequence to the present. The present-day interior structure of the Sun including the depth of the convective envelope and the sound speed profile can be probed precisely by several thousands of oscillation modes measured via helioseismology methods (e.g., Christensen-Dalsgaard 2002, Basu and Antia 2008). Early tests for the depth of the solar convection zone compared abundances of Li, Be, and B of the Sun to meteorites. The isotopes of Li are destroyed at temperatures > 2.5 MK whereas Be and B (unclear whether depleted or not) require higher temperatures not attained near the bottom of the convective envelope. Far more advanced testable observables are neutrino fluxes resulting from the pp-chain and CNO cycles (Appel et al., 2022, Basilico et al. 2023). Neutrino fluxes were measured as a function of neutrino energy with the Borexino experiment, including p-p, pep, $^{7}$Be, and $^{8}$B, the latter two very accurately to 3.5% and 2%, respectively, providing stringent constraints on the structure of the solar interior.

Over the past decade, much work in modeling of the outer structure of the Sun concentrated on 3-dimensional (3D) Radiation-Hydro Dynamics (RHD) models (e.g., Nordlund 1982, Spruit et al. 1990, Vögler et al. 2005, Nordlund et al. 2009, Freytag et al. 2012). These simulations involve solving self-consistently the equations of radiation transfer and time-dependent (magneto)-hydrodynamics, and eliminate the need for ad-hoc user-dependent corrections, which are inherent to 1D HE models. The new generation of models are commonly referred to as "3D models", although the main physical improvement is not in multi-D geometry. Physically exact 2D or 3D replicas of a 1D model can be made. The key difference lies in the thermo-dynamic structure, including velocity fields and temperature-density-pressure inhomogeneities (caused by sub-surface convection), and the loss of radiation at the surface modeled from first principles. The 3D RHD model atmospheres vastly improve the agreement of synthetic observables with various observations of the Sun, including the observed granulation at the solar surface, the time variability and contrast of the granules, and fits to high-resolution solar spectra across the limb.

## Non-local thermodynamic equilibrium (NLTE)

All stellar atmosphere and radiative transfer models, either 1D or 3D, rely heavily on atomic and molecular data. Choices have to be made on parameters such as wavelength, excitation potentials, transition probabilities, damping parameters, ionization and dissociation potentials and cross-sections, hyperfine splitting and electric dipole and magnetic quadrupole constants, partition functions and isotopic structure (in all solar analyses, fixed isotope ratios from studies of meteorites are assumed). In NLTE, further physical quantities play a role, such as rates of transitions in inelastic collisions with free electrons and charge exchange with H atoms (e.g., Barklem 2016, Belyaev et al. 2019), but also completeness of the representation of atomic and molecular systems through the energy structure and transitions between states (e.g., Mashonkina et al. 2011, Bergemann et al. 2012). Calculating lines of molecules in NLTE requires inclusion of photo-dissociation and photo-attachment (e.g., Heays et al. 2017, Hrodmarsson and van Dishoeck et al. 2023), as well as corresponding collisional destruction and attachment reactions. Most of these atomic and molecular datasets are theoretical and are difficult to verify experimentally, especially for collisional data (Barklem 2016). For example, for O see discussion in Bergemann et al. (2021). Atomic structure calculations rely on various assumptions about the representation of nuclear potentials, electron-electron correlations, relativistic effects, etc. (e.g., Bautista et al. 2000, 2022), and systematic effects in the solar abundance analysis are intricately tied to the quality of atomic data. For molecules the situation is worse. Currently, CH is the only molecule with detailed NLTE abundance predictions for solar atmospheric conditions (Popa et al. 2023).

For discussion of the individual elements, the following precautions apply:

- Solar photospheric abundances are not observed quantities. All methods used for the determination of solar photospheric abundances (1D LTE, 3D LTE, 3D NLTE) require theoretical modeling and depend on choices made regarding the sub-grid physics and the numerical approach. This causes differences among individual estimates of solar abundances by different groups (e.g., Asplund et al. 2009, 2021, Caffau et al. 2012, Magg et al. 2022). Only a few alternative, less model-dependent, methods to derive solar abundances exist (e.g., Ramos et al. 2022).

- The 3D NLTE is not an objective self-consistent methodology which is uniformly and unambiguously applied by different authors, see examples and references below. The term “3D NLTE” entails a sequence of approximations assumed by different authors. In some 3D NLTE calculations (e.g., Asplund et al. 2009, Klevas et al. 2016, Bergemann et al. 2019, Amarsi et al. 2019), 3D LTE model atmospheres are used to compute 3D LTE or 3D NLTE synthetic spectra, but both with background opacities in LTE. Other calculations use time- and spatially-averaged 3D model atmospheres to compute NLTE synthetic spectra with the improvement that background opacities are handled in NLTE (Magg et al. 2022). Other studies (e.g., Caffau et al. 2011 or Asplund et al. 2021) resort to 3D LTE calculations for most elements, using 1D NLTE calculations by other authors (with 1D or <3D> models, where the brackets represent spatially-averaged models) to inform the line selection or to correct 3D LTE values by 1D NLTE (see references below). Sometimes, 3D LTE spectra are computed and 1D NLTE corrections are applied during post-processing, with the latter computed using different means and techniques. These abundances are sometimes quoted as ‘3D, NLTE’, ‘3D + NLTE’, ‘3D - NLTE’ (with a comma, a plus, or a minus signs placed in between the two shortcuts), e.g., Caffau et al. 2011 (- sign, their Table 1 for K), Grevesse et al. 2015 (+), and Scott et al. 2015a,b (+ sign, their Table 1).
- The ‘3D, NLTE’ or ‘3D + NLTE’, or ‘NLTE corrected 3D’ abundances are not necessarily physically better than 1D NLTE, or even 1D LTE. No absolute test exists for the accuracy of the resulting abundances. The precision is usually defined as the statistical scatter (1-standard deviation) between abundances obtained from various spectral lines of an element. Over the years, several diagnostic tests were developed including the excitation balance, ionization balance, line-by-line scatter, and center-to-limb variation tests (Korn et al. 2003, Bergemann et al. 2012, Lind et al, 2017, Pietrow et al. 2023b). All of these are applied to various extents in different studies, but typically a comprehensive analysis of all these tests for consistency is not available.
- All 1D hydrostatic models rely on arbitrary parameters to correct physical structures of models for the absence of gas dynamics (convection and turbulent flows) and for other physical limitations. Some models, such as the semi-empirical 1D LTE Holweger-Mueller (HM) model (Holweger and Mueller 1974) used by Asplund et al. (2009, 2021), Grevesse et al. (2015a,b), Scott et al. (2015), have a much steeper temperature and pressure gradient than strictly theoretical 1D models (MARCS, MAFAGS, Kurucz). This model was constructed and tuned to achieve the best fit of solar observations in 1D LTE. NLTE effects obtained with such models can be of opposite sign and amplitude making it difficult to analyze and interpret results of different authors.

## Beryllium

The Be value is from a 3D NLTE study by Amarsi et al. (2024) which is adopted here with an increased error. The value is based on one Be II in the far-UV. This wavelength regime is very difficult to interpret in terms of abundances, because the radiation field, and hence the UV spectrum, is formed in the chromosphere (Vernazza et al. 1981). Chromosphere is not included in standard 1D or 3D models. The diagnostic line is strongly blended by an unknown feature (Figure 8 in Amarsi et al. 2024). Therefore, despite the use of 3D NLTE models, the error in the Be abundance could be underestimated because of strong blending, lack of chromosphere in the physical modelling, and lack of independent Be diagnostics that does not allow to verify the measurement. Hence, we recommend A(Be) = 1.21±0.14 with a larger error than that (±0.05) given by Amarsi et al. (2024).

## Carbon, Nitrogen, and Oxygen

The elements C, N, and O make up around 60-70 % by mass of all elements heavier than He and provide most of the opacity in the solar interior. Their abundances also determined the amount of condensable ices, in turn affecting oxidation states in the solar nebula materials and planet compositions (Krot et al. 2000). The amounts of C, O, and particularly the C/O ratio are also important for modeling AGB stars stellar evolution and nucleosynthesis, since the initial amounts assumed in models affect how quickly these evolve to become carbon stars.

Our recommended C abundance is the mean of C I based values from Caffau et al. (2010), Asplund et al. (2021), and Magg et al. (2022). We do not consider C abundances from molecular lines ($C_2$, CH, CO), because no 1D NLTE or 3D NLTE modeling is available for them and another open and poorly-understood issue is the assumption of chemical equilibrium. Magg et al. (2022) used several optical C I lines and handled self-consistently effects of blends and opacities using average 3D models. In the NLTE study by Amarsi et al. (2019), the IR and far-IR C I lines are preferred and their NLTE effects are estimated to <0.01 dex. However, Caffau et al. (2010) found the largest NLTE effects in the IR. The three optical C I lines selected by Magg et al. (2022) are neither sensitive to NLTE nor to 3D effects, and in agreement with the results of Alexeeva and Mashonkina (2015), when re-normalized to new f-values from Li et al. (2021). Caffau et al. (2010) reported significant differences between C I EWs measured by different authors. The values of Amarsi et al. (2019) and Asplund et al. (2021) are systematically lower compared to other studies, which is also discussed in Ryabchikova et al. (2022) in relation to the diagnostic $C_2$ and CN features. The differences between these analyses are due to unresolved systematic differences in the methodology (blends, continuum) and/or choice of the solar observational data.

For N, we adopt the values from Magg et al. (2022) corrected for 3D - <3D> difference based on Caffau et al. (2009, here -0.04 dex following their value for the 8683 Å line). This value is based on measurements of two least-blended atomic N I lines in the solar spectrum (8629, 8683 Å) using new theoretical f-values for these lines computed in the same study. Even these atomic N I lines, despite being more reliable, are blended by CN lines. As shown in M22, the result is only different by 0.011 dex, if the solar CO5BOLD or Stagger model is used. The N value by Amarsi et al. (2020) is not used because it relies on empirical re-scaling of the strengths of CN features in the diagnostic N I lines. The N value adopted here is in agreement with the 3D N I value of Caffau et al. (2009) within the respective errors of both values. If we use the f-values of Amarsi et al. (2020), which are taken from Tachiev and Fischer (2002), the measurements based on the two N I lines would be in excellent agreement and give A(N) = 7.94 ± 0.017 dex. We retain the larger error, as the value is based on only two blended lines and reliable NLTE modelling of CN is needed to confirm the atomic results.

Our recommended abundance of O is based on the average of 7 values from Steffen et al. (2015, O I 777 nm lines), Caffau et al. 2013, 2015 (630, 636 nm [O I] lines), Cubas Armas et al. 2020 (630 nm line), Bergemann et al. (2021), Asplund et al. (2021), and Magg et al. (2022). We avoid multiple measurements by the same group, except when the group used different spectral indicators (e.g. 777 nm vs 630 nm, as in Caffau et al. and Steffen et al.) or different codes (Turbospectrum vs MULTI3D, as in Magg et al. vs. Bergemann et al.). No molecule-based abundances are included, as NLTE effects are unknown. Ayres et al. (2013) find A(O) = 8.78 ± 0.02 dex based on CO lines. For atomic O lines it is critical to account for 3D NLTE effects. The 1D LTE value of 8.83±0.06 dex from Grevesse and Sauval (1998) was revised to lower values once 3D NLTE calculations became possible. Asplund et al. (2009, 2021) derived the solar A(O) = 8.69 ± 0.04 dex in 3D NLTE. These low abundances received much attention, however, subsequent independent studies did not confirm them (Caffau et al. 2011, Bergemann et al. 2021, Magg et al. 2022)

The solar A(O) = 8.77 ± 0.04 dex from Magg et al. (2022) relies on the new NLTE model atom of O from Bergemann et al. (2021). In the latter paper, both the forbidden [O I] line and the permitted lines of O I were modeled in full 3D NLTE. These authors also used, for the first time, 3D NLTE formation for the critical Ni I blend in the [O I] feature, finding that the LTE assumption for Ni adopted in previous studies (Allende-Prieto et al. 2001, Asplund et al. 2004, 2021) is inadequate. The solar 3D NLTE O abundance in Caffau et al. (2008) is 8.76 ± 0.07 dex, and a further 3D NLTE estimate by the same group is 8.76 ± 0.02 dex (Steffen et al. 2015) from the analysis of center-to-limb variation of the O 777 nm lines. Caffau et al. (2013) noted that the O I forbidden line at 636 yields A(O) of 8.78 ± 0.02 dex. Magg et al. (2022) pointed out that the difference in the O abundance between CO5BOLD and Stagger 3D models does not exceed 0.015 dex. An independent estimate of the solar O abundance was proposed in Socas-Navarro et al. (2015, see also Centeno and Socas-Navarro 2008), who analyzed the polarization (Stokes V) profile of the NLTE-insensitive 630 nm [O I] line. They find A(O) = 8.86 ± 0.03 dex (Centeno and Socas-Navarro 2008), O/Ni = 210 ± 24, and the improved analysis (Cubas Armas et al. 2017, 2020) yields A(O) = 8.80 ± 0.03 dex, where estimates for the granular and intergranular regions amount to 8.83 ± 0.02 and 8.76 ± 0.02 dex, respectively. This approach only weakly depends on models and it yields the same result when different model atmospheres are used.

The recommended abundances from combining measurements by independent groups are:

A(C) = 8.51 ± 0.09 dex

A(N) = 7.94 ± 0.11 dex

A(O) = 8.76 ± 0.05 dex

Here uncertainties primarily reflect limitations of theoretical models. Problems that need to be resolved include:

- Significant difference between the cross-sections for O+H charge transfer reactions by two different groups (Barklem 2018, Belyaev et al. 2019) → effect on the O abundance at the level of 0.07 dex, which is highly significant at the level required for solar abundance diagnostics
- Lack of understanding the NLTE effects in molecules (e.g., OH, $C_2$, CN, CH, CO, NH lines). Calculations suggest that molecular abundances are under-estimated owing to the overlooked effect of photo-dissociation of molecules (Popa et al. 2023). The systematic bias influences all other species with low photo-dissociation potentials, such as OH, NH, CO. Lines from low-excited states of molecules like CN may be significantly affected. These are essential for reliable modelling of N I optical lines, as these are affected by CN.
- A systematic difference between the 636 and 630 nm O I line (Caffau et al. 2013) affects the O abundance at the level of 0.05 dex
- Photo-ionization cross-sections for Ni I are needed to test the size of NLTE effects on Ni, and consequently how much of an impact this has on the [O I] diagnostics.

## Alpha-elements: Mg, Si, Ca, and S

The elements Mg, Si, Ca, and S are mainly produced by successive He nuclei capture (hence "alpha-elements") during hydrostatic (C-, O-, and Si-burning) and explosive nucleosynthesis in massive stars (e.g. Rauscher et al. 2002). For most of these elements, detailed abundance estimates including 3D and/or NLTE effects are available (Alexeeva et al. 2018, Osorio et al. 2015, Bergemann 2017a, Asplund et al. 2021, Magg et al. 2022). Abundances of Mg, Si, and Ca were not reported by Caffau et al. (2011).

Solar abundances of Mg and Si are better constrained because of more accurate atomic data and a wealth of lines of different excitation potentials can be utilized. However, contrary to Fe-group elements, no reliable constraints on the abundance can be made using single-ionized species of these elements. Calcium is comparatively accurate, perhaps even more accurate than Mg and Si, because also diagnostic lines of Ca II are available, the f-values are reliable, and Ca I lines of different excitation potentials can be used.

For Mg, we use the averages from Osorio et al. (2015), Bergemann et al. (2017), Asplund et al. (2021), and Magg et al. (2022). Here we do not distinguish between 3D NLTE and <3> NLTE because the abundances obtained with both approaches agree to better than 0.01 dex (Asplund et al. 2021, Table A1). The effects of NLTE depend on the choice of Mg I lines in the analysis (Bergemann et al. 2017), and the associated uncertainty is at least 0.05 dex. Alexeeva et al. (2018) found an imbalance between Mg I and Mg II, with Mg I lines yielding lower abundances than Mg II. The solar abundance of A(Mg) =7.66 ± 0.07 dex was reported by Osorio et al. 2015. The results of Asplund et al. (2021) disagree with other, similar studies, also their reported NLTE effects are of opposite sign compared to these studies (e.g., Bergemann et al. 2017a, Osorio et al. 2015). It is not clear whether this is due to the line selection or the NLTE model atom employed by Asplund et al. (2021). Mg I lines accessible in optical and near-IR solar high-resolution solar spectra suffer from a strong sensitivity to damping and/or blends (Bergemann et al. 2017a).

For Si, our recommended value is based on the mean of Asplund et al. (2021), Magg, et al. (2022), and Deshmukh et al. (2022), who reported 7.51 +/- 0.03 dex, 7.59 +/0.07 dex, and 7.57 +/- 0.04 dex, respectively. The choice of oscillator strengths seems to be the main source of disagreement between these three estimates. The Asplund et al. study relied on the older source of gf-values for Si I lines, whereas both latter studies used newer laboratory values from Pehlivan Rhodin (2018), and these data were later published in Pehlivan Rhodin et al. (2021). An independent, although unpublished, analysis of a benchmark metal-poor star by C. Sneden (priv. comm) supports the higher quality of the f-values determined by Pehlivan Rhodin et al. (2024). The NLTE effects in optical Si I lines are at the level of ±0.01 dex or less (Bergemann et al. 2013, Amarsi and Asplund 2017). The quality of the only diagnostic Si II line at 6371.37 Å is debated, and discrepant results are obtained by different groups (e.g., Magg et al. 2022, Asplund et al. 2021). Deshmukh et al. (2022) found that the difference between 3D MHD and 3D RHD results is roughly -0.005 dex in abundance for Si I lines and max +0.015 dex for the Si II line. Hence, the magnetic field is not a major source of uncertainty in the solar Si abundance. Line selection also influences the Si abundance. For example, the 1D NLTE value quoted by Mashonkina (2020) would be A(Si) = 7.55, dex if re-normalized to Pehlivan Rhodin et al. (2024), but even higher (A(Si) = 7.60 dex) if the line list is limited to the Si I lines used by Magg et al. (2022). Asplund et al. (2021) did not provide details for their choice of Si lines, hence it is not possible to validate their finding of an ionization imbalance.

For S, only inhomogeneous and rather inconsistent estimates are available and no 3D NLTE analysis has been carried out to date. The 1D LTE estimate of 7.33 ± 0.11 dex (Grevesse and Sauval 1998) was superseded by Scott et al. (2015a), who found only 7.06 in 1D LTE, but 7.12 ± 0.03 dex in 3D LTE with a 1D NLTE abundance correction obtained from the ATLAS models using 8 S I lines in the optical and IR. However, the model atom was chosen to produce the most consistent “3D + NLTE” abundances, given 3D LTE values and ad-hoc scaled collisional data (using a so-called Sh scaling factor of 0.4). Another 3D estimate, A(S) = 7.16 ± 0.05 dex by Caffau et al. (2011). That study also used NLTE corrections based on Korotin (2009) model and pointed out that a significantly higher, A(S) = 7.30 dex can be obtained depending on the choice of diagnostic lines. None of these are self-consistent 3D NLTE analysis of S I lines. In Asplund et al. (2021), the solar S abundance was adopted from Scott et al. (2015a). The quality of the NLTE model of S is under debate, as radiative transitions and quantum-mechanical data for S+H collisions (Belyaev and Voronov 2020) have not yet been integrated into NLTE models and no rigorous analysis of the line formation of S I in the solar atmosphere in 3D NLTE has been undertaken so far.

For Ca, our value is based on averaging three recent Ca estimates (Mashonkina et al. 2017, Asplund et al. 2021, Magg et al. 2022). The 3D NLTE estimate by Asplund et al. (2021), 6.30 ± 0.03 dex, is lower than the 3D NLTE estimate by Magg et al. 2022, A(Ca) = 6.37 ± 0.05 dex. The latter brackets other independent NLTE estimates, e.g., Mashonkina et al. (2017) with A(Ca) = 6.33 ± 0.06 dex (from Ca I lines) and 6.40 ± 0.05 dex (from Ca II lines), and Osorio et al. (2019) albeit with a larger range of line-by-line Ca abundances. Whereas the latter estimates refer to 1D MARCS models, 3D effects are strictly positive for the diagnostic lines of both ions (when compared to MARCS, Delta (3D -1D) of +0.05 dex for Ca I and +0.02 for Ca II, see Scott et al. 2015a). Thus, the solar Ca abundance is likely higher than the value by Asplund et al. (2021). Calcium results obtained with <3D> NLTE and 3D NLTE models are identical (Scott et al. 2015a, their Table 5). It is therefore expected that our value for the solar Ca abundance is reliable.

We recommend the following abundances of alpha-elements:

A(Mg) = 7.58 ± 0.05 dex

A(Si) = 7.56 ± 0.05 dex

A(S) = 7.16 ± 0.22 dex

A(Ca) = 6.35 ± 0.06 dex

The current uncertainties to be resolved include:

- Mg: Significant scatter remains between different Mg I lines in the solar spectra. The only clean feature is at 5711 Å, other features are very weak and/or blended.
- S: Full 3D NLTE calculations with a comprehensive NLTE model atom are lacking. Quantum-mechanical data for S+H collisions are available (Belyaev and Voronov 2020). These still need to be implemented into the NLTE models. Accurate photo-ionization cross-sections for S I are also missing.
- Ca: Systematic differences exist between abundances derived by different groups, even for Ca II lines that are nearly unaffected by NLTE (Scott et al. 2015a; but see Mashonkina et al. 2017). The line-by-line scatter is small, but different groups differ over 30% in the Ca abundance using the same gf-values and same models.
- Si: The disagreement of the f-values of Si I and Si II lines, specifically the values from Garz (1973) lead to systematically lower abundances of Si in Asplund et al. (2021) compared to using laboratory values from Pehlivan Rhodin et al. (2021) which are preferred in recent studies (Magg et al. 2022, Deshmukh et al. 2022).

## Low-charge odd-Z elements: Na, Al, P, K, and Sc

The solar abundances of low-charge elements, Na, Al, P, K, and Sc are still under debate. The Na abundances obtained by Zhao et al. (2016) and Asplund et al. (2021) differ by 20% for the two diagnostic Na I lines in common (at 6145, 6160 Å). The values of the latter group are 0.1 dex lower compared to Zhao et al. (2016), and the difference exceeds the line-by scatter (of 0.02) and is five times higher than the error of the atomic data for the Na I transitions. The recommended NLTE value is adopted from Zhao et al. (2016). It was derived using QM data for Na+H collisions. The total error is increased to 0.05 dex to reflect the unexplained mismatch in the estimates by the two groups. Contrary to the statement by Asplund et al. (2021), the NLTE effects in Na do not change significantly depending on the NLTE model atom. This can be seen by comparing the Zhao et al. (2016) value of the NLTE correction of -0.04 dex with the Asplund et al. (2021) value of the NLTE correction of -0.045 dex for the 6145 and 6160 Å lines in common between the two independent studies.

The solar 3D NLTE Al abundance by Nordlander and Lind (2017) used the same models and the same NLTE code as Asplund et al. (2021). This value is identical to the solar 3D NLTE value by Scott et al. (2015a) based on Al I lines, A(Al) = 6.43 ± 0.04 dex, although the uncertainty is slightly smaller. This value is adopted here. The 1D NLTE estimate of the Al abundance is only 0.03 dex lower (Scott et al. 2015a, based on MARCS). Gehren et al. (2004) computed the Al abundance in 1D NLTE, A(Al) = 6.43 dex, although they relied on a semi-empirical NLTE model atom, with a scaling factor to Al+H collisions computed using the Drawin's formula. They note the possibility of a systematic error in the van der Waals damping constants for the diagnostic Al I transitions.

The solar abundance of P remains uncertain, owing primarily to the lack of a reliable NLTE model atom and limited quality of the diagnostic lines. The earlier 1D LTE estimate by Grevesse and Sauval (1998), A(P) = 5.45 ± 0.04 dex, is higher than the 1D LTE estimate by Scott et al. (2015a). However, the solar 3D LTE P abundance of 5.46 ± 0.04 by Caffau et al. (2011) is almost identical to that of Grevesse and Sauval (1998). The value from Scott et al. (2015a) is 5.41 ± 0.03 and it relies on eight very weak P I lines in the near-IR. Similar to S I, the NLTE effects in the P I lines are likely significant and detailed NLTE modeling is needed. According to Scott et al. (2015a), 3D effects increase the P abundance by ~+0.03 dex, compared to the MARCS 1D LTE result. Our value, A(P) = 5.44 ± 0.10 dex, is based on the average of C11 and S15, and the error is set to 0.10 dex due to the lack of knowledge of NLTE effects in P I lines.

For K, several estimates are available in the literature, employing different model atmospheres, atomic data, NLTE models, and sources of solar observations. The dispersion between these values is still significant, considering uncertainties tabulated by different authors. In 1D NLTE, the estimates range from 5.02 (Asplund et al. 2021) to 5.11± 0.01dex (Reggiani et al. 2019), although both studies used the same NLTE model atom. The latter value is corroborated by an independent analysis of Zhang et al. 2006 (5.12 ± 0.03 dex), while the value by Caffau et al. (2011) falls in the middle of these, 5.06 ± 0.04 dex. In 3D LTE, the K abundances are even more discrepant, e.g., Caffau et al. (2011) estimate A(K) of 5.26 dex, whereas the 3D LTE value by Asplund et al. (2021) is 5.12 dex. Mixed (3D LTE + <3D> NLTE) values are primarily within the range of the published 1D NLTE values. The average of 3D NLTE value from Asplund et al. (2021) and 3D NLTE corrected value from Caffau et al. (2011) is adopted here, with a conservative error due to lack of knowledge on the accuracy of atomic data, and discrepant values obtained with similar 3D and 1D model atmospheres by different groups.

The solar Sc abundance derived by different groups differs despite adopting similar models and line selections. In 1D LTE, the abundance estimates range from 2.90 ± 0.09 dex (Zhang et al. 2008) for Sc I lines to 3.21 ± 0.046 dex for Sc II (Scott et al. 2015b, the average and standard deviation are based on Table 1 in their Appendix, the Holweger-Mueller HM model in LTE). In LTE, systemically lower abundances from Sc I lines were reported (Zhang et al. 2008), however Lawler et al. (2019) obtained a perfect ionization balance for Sc I and Sc II in LTE, with A(Sc) = 3.15 ± 0.06 and 3.16 ± 0.01 dex, respectively, using the HM model. Lawler et al. (2019) comment on the unresolved systematic difference with Asplund et al. (2009) and Scott et al. (2015a) using the same HM model atmosphere. All studies consistently derive a smaller error based on Sc II lines compared to Sc I. Mashonkina and Romanovskaya (2022) analyzed 17 optical Sc II lines using experimental f-values and found A(Sc) = 3.12 ± 0.04 dex in NLTE and 3.14 = 3.12 ± 0.05 dex in LTE. These values are consistent with NLTE estimates from Zhang et al. 2008 (3.07 ± 0.04 dex for Sc II lines). The values from Scott et al. (2015) were obtained by co-adding the 3D LTE abundances with 1D NLTE corrections computed using the 1D LTE MAFAGS model atmosphere (Zhang et al. 2008), but this hybrid approximation for Sc has not been validated. Renormalizing the Scott et al. (2015b) values to the new experimental log(gf) values from Lawler et al. (2019), we obtain a smaller dispersion of line-by-line abundances for Sc II. The less NLTE sensitive Sc II lines are preferred because substantially more work on the atomic data has been performed (Pehlivan Rhodin et al. 2017, Lawler et al. 2019). No direct estimates of 3D NLTE Sc abundances are published yet.

Our recommended abundances are:

A(Na) = 6.29 ± 0.05 dex

A(Al) = 6.43 ± 0.05 dex

A(P) = 5.44 ± 0.12 dex

A(K) = 5.09 ± 0.09 dex

A(Sc) = 3.13 ± 0.11 dex

## Fe-peak elements: Ti, V, Cr, Mn, Fe, Co, Ni

The iron-peak group encompasses Ti, V, Cr, Mn, Fe, Co, and Ni produced by hydrostatic Si-burning in massive stars and during explosive nucleosynthesis in SN Ia and SN II. The abundance of Fe has been extensively investigated over the past decade. Asplund et al. (2009) were among the first to determine the solar Fe abundance in 3D. Their estimate, based on 3D LTE modeling of Fe lines plus a correction for NLTE effects using a <3D> model with an unpublished model atom, is 7.50 ± 0.04 dex. Scott et al. (2015b) recommended A(Fe) = 7.47 ± 0.04 dex using a similar approach, but note a positive difference between the lines of two ionization stages, with Fe II yielding ~0.05 dex higher abundances compared to the lines of Fe I. Lind et al. (2017) used the list of Fe lines from Scott et al. (2015) and a new model atom of Fe based on the QM collisional data for Fe+H collisions from Barklem 2016. They tested the center-to-limb variation (CLV) of Fe I lines across the solar disc and found the solar Fe abundance of 7.48 ± 0.04 dex.

In Caffau et al. (2011), the Fe abundance is based on the 3D LTE analysis of 15 Fe II lines with the CO5BOLD 3D model. Their estimate is A(Fe) = 7.52 ± 0.06 dex. They found the central Fe abundance is immune to the choice of transition probabilities, but the error and line-by-line scatter is sensitive to the source of f-values. The solar Fe abundance by Asplund et al. (2021), A(Fe) = 7.46 ± 0.04 rests upon the 3D NLTE calculations with the Stagger model. Sitnova et al. (2015) and Mashonkina et al. (2019) found A(Fe) = 7.54 dex from a 1D NLTE analysis. The <3D> NLTE value by Bergemann et al. (2012), based on the analysis of over 50 Fe I and Fe II lines, is 7.46 ± 0.06 dex, fully consistent with 1D NLTE. Magg et al. (2022) employed an updated NLTE model atom of Fe, as well as average CO5BOLD and Stagger models atmospheres and obtained A(Fe)= 7.50 ± 0.06 dex. 3D effects are not significant for the solar Fe abundance. Different groups arrive at different conclusions using very similar solar atmospheric models and line formation methods. The value adopted here is based on the average of Mashonkina et al. (2011), Caffau et al. (2011), Lind et al. (2017), Asplund et al. (2021), and Magg et al. (2022) The error primarily reflects the line-by-line scatter, and the residual uncertainties of the gf-values, damping, unresolved blends, and problems with continuum normalization of the data.

Among other Fe-group elements, 3D NLTE estimates are only available for Mn (Bergemann et al. 2019) and in <3D> for Ni (Magg et al. 2022). We do not consider the Mn and Co values from Asplund et al. (2021), because they used the outdated Mn and Co model atoms from Bergemann and Gehren (2007), which rely on incomplete collisional and radiative data. Updated models of Mn and Co were presented in Bergemann et al. (2019) and in Yakovleva et al. (2020). For Co and Cr, we use the 1D NLTE estimates from Bergemann et al. (2010) and Bergemann and Cescutti (2010), respectively. We increase the total uncertainty to 0.11 dex for Cr and Co to account for the absence of full 3D NLTE calculations.

Our recommended values for Fe-group elements are:

A(Ti) = 4.97 ± 0.11 dex

A(V) = 3.89 ± 0.16 dex

A(Cr) = 5.74 ± 0.11 dex

A(Mn) = 5.52 ± 0.05 dex

A(Fe) = 7.51 ± 0.05 dex

A(Co) = 4.95 ± 0.11 dex

A(Ni) = 6.24 ± 0.06 dex

## Copper and Zinc

Neither of the two elements were previously considered in full 3D NLTE. For Cu, we recommend the 1D NLTE value from Shi et al. (2014), which was obtained using new radiative transition probabilities computed using the method by Liu et al. (2011). The NLTE value should still be taken with caution as no detailed quantum-mechanical data for Cu+H collisions are integrated into the model atom. The solar Cu abundance from Asplund et al. (2021) was adopted from Grevesse et al. (2015) and it relies on old f-values for Cu I lines from Kock and Richter (1968). As demonstrated in Shi et al. (2014), these f-values lead to a significantly lower Cu abundance, an excitation imbalance, and a larger line-to-line scatter. A comprehensive discussion of the problems of the Kock and Richter (1968) atomic data is given in Shi et al. (2014).

For Zn, the careful NLTE analysis by Sitnova et al. (2022) is preferred here over the value by Grevesse et al. (2015). The improvements in Sitnova et al. (2022) include the high quality of atomic data, NLTE models including quantum-mechanical collisional data for Zn+H and Zn+e data, and the use of f-values from measurements of Roederer and Lawler (2012). The value A(Zn) = 4.56 ± 0.05 dex in Grevesse et al. (2015) applies a NLTE correction that was computed using a model atom lacking realistic cross-sections for collisional and radiative transitions. We did not find a new estimate of the solar Zn abundance in Asplund et al. (2021). The 1D LTE estimates of both elements are systematically lower in Grevesse et al. (2015) compared to other studies mentioned above. We adopt the NLTE value from Sitnova et al. (2022) but increase the uncertainty to 0.11 dex owing to the lack of 3D NLTE calculations.

A(Cu) = 4.24 ± 0.11 dex

A(Zn) = 4.55 ± 0.11 dex

## Neutron-capture (trans-Fe) elements

Limited progress exists in improving precision and accuracy in abundance determinations for elements beyond the Fe-peak. For several elements, new f-values were determined via laboratory experiments and/or theoretical calculations. The 3D NLTE calculations are available for a few elements (Ba, Y, and Eu).

For the majority of neutron-capture elements only 3D LTE, 1D LTE and 1D NLTE calculations have been performed. The solar abundance of Rb analyzed by Korotin (2020) is 2.47 ± 0.05 dex in 1D LTE, based on both Rb I lines, which is similar to the LTE value of Grevesse et al. (2015). However, the NLTE abundance of Rb is 2.35 ± 0.05 dex, over 0.1 dex lower compared to LTE (Korotin 2020). 3D LTE estimates range from 2.47 ± 0.07 dex in Grevesse et al. (2015) to 2.44 ± 0.08 dex in Asplund et al. (2021). Mixed 3D LTE and 1D NLTE estimates are somewhat divergent, and show a large uncertainty, exceeding that of 1D NLTE values. Our recommended 1D NLTE value is from Korotin (2020), as the effect of 3D is very small (~0.005 dex). We raised the total uncertainty to 0.11 dex to account for the sparse number of diagnostic Rb I lines and the lack of full 3D NLTE estimates.

For Sr, we use the estimate from Bergemann et al. (2012c), who employed NLTE calculations with a realistic model atom of Sr to provide the solar Sr abundance based on Sr I and Sr II lines. Abundances derived from lines of both ionization potentials are consistent, with the internal precision error of 0.04 dex. The NLTE effects are consistent with those obtained by Mashonkina and Gehren (2001). The value recommended by Grevesse et al. (2015) although referred to as "3D + NLTE" is based on LTE modeling with the averaged <3D> model atmosphere and co-added with a 1D NLTE correction. Due to a significant difference of 0.05 dex between their estimates for both ionization stages, their abundance is not used here. No 3D NLTE calculations for Sr have been performed yet.

For Y and Eu, we adopt the new full 3D NLTE estimates from Storm et al. (2024) who present calculations with novel quantum-mechanical atomic data for Y+H and Eu+H collisional processes. The 3D NLTE values for Y are higher than previously available 3D LTE estimates from Grevesse et al. 2015 (2.21±0.05 dex). The latter value relies on oscillator strengths from Hannaford et al. (1982), whereas the value from Storm et al. (2024) is based on new laboratory lifetime and branching fraction measurements from Palmeri et al. (2017), which is an update of Biemont et al. (2012). In LTE, the 3D - 1D differences for the diagnostic Y II lines are positive ranging from +0.03 dex to +0.07 dex, depending on the line, supporting the results from Grevesse et al. (2015) (their 3D - MARCS differences). For Eu II, our results are larger than the estimate in Grevesse et al. (2015). This is due to the full 3D NLTE radiative transfer (positive NLTE effects in Eu II, consistent with Mashonkina and Gehren 2000) as well as self-consistent 3D treatment of blending features (Si and Cr). In LTE, the 3D - 1D estimate corrections for Eu are fully consistent with the results of Mucciarelli et al. (2008).

The abundance of Zr is the NLTE-corrected average of the 3D LTE estimates from Grevesse et al. (2015) and Caffau et al. (2011). The large ionization imbalance for Zr I and Zr II in Grevesse et al. (2015) is likely due to neglect of NLTE effects. Velichko et al. (2010) found in NLTE A(Zr) = 2.63 ± 0.07 dex and consistent Zr abundances based on both ionization stages. We use their NLTE correction of +0.078 dex (Velichko et al. 2010, their Table 3, kh=0.1 recommended) for the Zr II lines representative of the selection in Grevesse et al. (2015). We find that Grevesse et al. unjustly criticize the selection of lines by Caffau et al. (2011). The recommended Zr results of Grevesse et al. (2015) are fully based on highly blended and strong Zr II lines in the UV and near-UV (see Table 1 in the Appendix of Grevesse et al. (2015) for the list of Zr II lines, and discussion in Velichko et al. 2010). We increased the error to 0.11 dex to account for the absence of self-consistent <3D> NLTE or 3D NLTE calculations for Zr lines.

For Ba, the 3D NLTE values are directly from calculations of Gallagher et al. (2020), who used the up-to-date atomic model of Ba based on quantum-mechanical Ba+H data. These values were also adopted in Asplund et al. (2021) and are higher than the 3D LTE+1D NLTE estimates from Grevesse et al. (2015).

No estimates of NLTE abundances are available for La. The La abundance (1.13 ± 0.03 dex) in Lawler et al. (2001) is based on the 1D LTE analysis of 14 near-UV and optical La II lines. We adopt the 1D value from Lawler et al. (2001) and correct it for the 3D - 1D (HM) differences (-0.03 dex) based on the estimates from Grevesse et al. (2015). We increased the error to 0.16 dex to account for the lack of 1D NLTE and 3D NLTE calculations.

A(Rb) = 2.35 ± 0.11 dex

A(Sr) = 2.93 ± 0.11 dex

A(Y) = 2.30 ± 0.06 dex

A(Zr) = 2.68 ± 0.11 dex

A(Ba) = 2.27 ± 0.06 dex

A(La) = 1.10 ± 0.16 dex

A(Eu) = 0.57 ± 0.06 dex

No new measurements are available for the solar abundances of Ga, Ge, Ce, Pr, Nd, Sm, Gd, Tb, Dy, Lu, and other n-capture elements. Of this group, the only elements with NLTE estimates of abundances in 1D available are Sr, Y, Zr, Ba, Pr, Eu, Gd, and Nd.

## Solar Helium Abundance and Present-day Mass Fractions of H (X), He (Y), and the Heavy Elements Li-U (Z)

The mass fractions of the elements in the photosphere or present-day solar system matter (from meteoritic and solar data) requires the knowledge of the He abundance. As described in Basu and Antia (2004, 2008) the H and He mass fractions (called X and Y, respectively) are constrained by helioseismology. The analysis depends on the ratio Z/X, which is the mass fraction Z of heavy elements from Li to U, relative to the mass fraction of H.

Basu and Antia (2004) used two model compositions for calibration with two different Z/X ratios (Z/X(mix1) = 0.0171, and Z/X(mix2) = 0.0218) and found Y is more dependent on Z/X than X. Their analyses of data from the Global Oscillation Network Group (GONG) gave X(GONG, mix1) = 0.7392 ±0.0034 and from MDI data X(MDI,mix1) = (0.7385 ±0.0034) averaging to X(mix1) = 0.7389 ±0.0048. Similarly, for mix2 the average is X(mix2) = 0.7390 ±0.0048 (assuming that for mix2, the individual uncertainties on X for GONG and MDI are the same as given for mix1, ±0.0034).

Both calibration mixtures give approximately the same value of X = 0.7389 ±0.0068. Thus, the H mass fraction from helioseismology is relatively insensitive to the mass fraction of heavy elements and is essentially constant for Z/X from 0.0171 to 0.0218 in the calibration models (Basu and Antia 2004). We adopted X = 0.7389 ±0.0068 for the present-day solar convective envelope. However, the mass fractions for He by Basu and Antia (2004) were slightly different: the averages from GONG and MDI gave Y(mix1) = 0.2485 ± 0.048 and similarly, Y(mix2) = 0.2449 ± 0.048.

The Z/X can be computed without knowledge of the He abundance from the atomic abundances by multiplying the atomic elemental abundances for elements heavier than He (atomic number z>2; here lower-case z is used to distinguish it from mass fraction of He usually written as capitalized Z) with the appropriate atomic masses:

$$\frac{Z}{X} = \frac{\sum_{z=Li-U} \varepsilon_z atwt_z}{atwt(H)}$$

Here ε is the abundance relative to H; $\varepsilon_{z>2} = N_{z>2}/N_H$. Atomic weights (atwt) of the elements are computed using the masses of the isotopes (Wang et al. 2021; appendix Table A13) weighted by the relative isotopic composition for each element because the atomic weights can be different for solar material than for terrestrial matter usually tabulated (e.g., H, O, Ar, see Lodders 2020, 2021). We obtain $Z/X_{photo}$ = 0.0217(±8%) for the photosphere, and $Z/X_{ss\ present}$ = 0.0216(±8%) for the present-day "solar system". The small difference comes from the combination of meteoritic and photospheric data in the latter dataset.

The He mass fraction is from the relations Y/X = 1/X – 1 – Z/X and X + Y + Z = 1. This yields $Y_{photo}$= 0.2451 ±0.0069 = $Y_{ss\ present}$ where the uncertainty is taken to be the same as for X as is in Basu and Antia (2004, 2008). The largest uncertainty (8%) is in Z/X from the combined uncertainties of the heavy elements in Z, of which O, Ne, C, N alone constitute 79%.

The mass fraction for the solar convection zone, taken as present-day solar system values, are X = 0.7389 ±0.0068 (±0.9%), Y = 0.2451 ±0.0069 (±2.8%), and Z = 0.0160 ±0.0013 (±8%); see below for protosolar mass fractions.

Using the mass fractions X and Y and corresponding atomic weights, the atomic He abundance is $\varepsilon_{He} = N_{He}/N_H$ = 0.0836 ±0.0025 (2.8%) and A(He) = 12+log $\varepsilon_{He}$ = 10.922±0.012 (2.8%).

The photospheric and present-day solar system Z/X ratios are essentially those of mix2 in the calibration models by Basu and Antia (2004), hence the mix2 calibration and results are more relevant. The approach adopting a constant X instead of Y in the mass balance equations gives the same Y (0.2450) as the average of GONG & MDI for mix2 (Y= 0.2449) and maintains proper mass balance when estimating the He abundance if Z/X is between the calibration values. A priori we did not know how the Z/X would come out. The frequently used Y = 0.2485 from the mix1 calibration cannot be used with our current Z/X, as this leads to X = 0.7355, clearly different from X in any of the calibration mixtures (mix1 and mix2) in Basu and Antia (2004, 2008).

## Solar Noble Gases: Ne, Ar, Kr, Xe

Noble gas abundances are summarized in Table 3. Helium is discussed in the previous section. The Ne and Ar are derived from elemental abundance ratios with other elements from photospheric and meteoritic data. Krypton and Xe from isotope systematics.

The neon abundance, A(Ne) = 8.15±0.12 dex is calculated as in Lodders (2020) and Magg et al. (2022) using Ne/O = 0.244±0.05 from Young (2018) for the solar quiet transition region and the recommended photospheric O abundance. For Ar, we adopt 6.50±0.10 dex proposed in Lodders (2008). This Ar abundance is somewhat low when compared to A(Ar) = 6.56 dex from an interpolated estimate for Ar made with the semi-statistical equilibrium method described by Cameron (1973) and our recommended Si and Ca solar system abundances both from averaged scaled meteoritic and photospheric values. The A(Ar) = 6.5 dex adopted here is, however, high in comparison to other values around 6.4 dex based on solar wind values (see below). The photospheric values used here are also often higher than in Asplund et al. (2021), involving higher O and Ca values in the ratios for estimating Ne (from Ne/O) and Ar (from Ca/Ar) also yield higher Ne and Ar than theirs. The higher metallicity, also seen from different scale coupling factors (SCF, see below) also leads to higher absolute isotopic solar system abundances, and noble gas abundances estimates are higher when nuclear systematics involving other elements are used to estimate them.

Krypton and Xe are from interpolation of s-process nuclide abundances of neighboring elements using the s-process model and Galactic chemical evolution yields from Prantzos et al. (2020) which are within 3-4% of the exact match; see also section on isotopes and Figures 15,16 below why the interpolation result is preferred. Here Kr/Xe =10.2 is consistent with the elemental Kr/Xe =10.45 on Jupiter (Mahaffy et al. 2000), which should represent the solar and proto-solar ratio, and should be compared to the lower solar wind Kr/Xe of about five.

Huss et al. (2020) derived Ne (8.060± 0.033 dex) and Ar (6.38±0.12 dex) abundances from Ne/He and Ar/He correlations with the respective He/H ratios in the four solar wind regimes as captured in the targets of the Genesis mission. Their values were essentially adopted by Asplund et al. (2021). For neon from the Ne/O ratio by Young (2018), the assumed lower O abundance from Asplund et al. (2021) also results in A(Ne) = 8.06 dex, which we believe is too low.

The Ne and Ar abundances by Huss et al. (2020) are around 25% lower than our values. For their fits, they used a solar H/He of 11.90±0.17 (Basu and Antia 2008), somewhat lower than the H/He = 11.976 used here. Huss et al. note that their "approach to estimating the Ne and Ar abundances in the Sun is model independent", but it depends on the adopted photospheric He abundance (H/He ratio) which in turn is sensitive to the amounts of heavy elements, including the more abundant O and Ne. The He/H ratio used for fitting the Ne/H thus depends on the O and Ne abundances used in the He determination. Our He abundance (10.9217 dex) is for a heavy element mass fraction that includes Ne as A(Ne) = 8.15 dex; the He abundance would slightly increase to A(He) = 10.9225 dex if A(Ne) = 8.05 dex were adopted.

The fit method by Huss et al. (2020) is intriguing but seems to have some problems. Among the four regimes (bulk solar wind (their B/C), interstream wind (L), coronal hole wind (H), and coronal mass ejections (CME, their E) used for fits in their Figure 19, the coronal hole wind (H) should be the least fractionated material from the photosphere (Huss et al. 2020). Coronal hole, interstream, and bulk solar wind plot relatively close together compared to the CME, which is the most fractionated component among the wind regimes and plots as lowest Ne/He, Ar/He, and H/He. If the coronal hole wind is the least fractionated, then photospheric values should fall near that end of the correlation, however, their derived photospheric Ne and Ar values fall at the opposite end beyond the CME values. This seems counterintuitive to expectations from solar wind fractionations of photospheric source compositions.

The Kr and Xe abundances by Meshik et al. (2020) and Asplund et al. (2021) are lower than our recommended values. Their proposed solar Kr only accounts for about 70% of the expected pure s-process nuclides $^{80}$Kr and $^{82}$Kr, and 84% of pure s-process $^{128}$Xe and $^{130}$Xe from stellar models by Prantzos et al. (2020), which is used here to obtain Kr and Xe abundances by interpolation (see isotope section below). Asplund et al. (2021) derived Xe as done in Lodders (2003) using the s-process nuclide cross-sections for Xe measured by Reifarth et al. (2002) and scaling to their $^{150}$Sm abundance to obtain A(Xe) = 2.22 ± 0.10 dex. Using this procedure with the higher $^{150}$Sm here gives A(Xe) = 2.297 dex, mainly because of the higher solar metallicity and larger scale-coupling factor (1.551) here than 1.51 in Asplund et al. (2021).

**Table 3**.

Meshik et al. (2020) used the "σN-curve" approach for estimating photospheric abundances of Kr and Xe (Table 3). This classical model approximation remains useful to estimate abundances in mass regions with about constant σNs between magic neutron numbers, but also has limitations (see below).

## Meteorites and the Significance of CI-Chondrites

There are two types of meteorites, differentiated and undifferentiated ones. Differentiated meteorites are derived from once melted planetesimals, while undifferentiated meteorites, such as chondrites, never were heated to melting temperatures. They represent aggregates of primary solar system material. Their first-order uniform composition approximates the average composition of the Solar System, except for ultra-volatile elements. However, small variations in their elemental compositions divide them into different chondrite sub-groups (e.g., Krot et al., 2014; Scott and Krot, 2014). These compositions reflect processing in the solar nebula prior to accretion, such as incomplete condensation, evaporation, preferred accumulation or separation of metal by magnetic forces, differential movement of fine vs. coarse grained material, etc.

## Cosmochemical and Geochemical Classification of the Elements

The geochemical classification of the elements is based on the chemical affinity to silicate and oxides (lithophile elements), sulfides (chalcophile elements and metal alloys (siderophile elements). The cosmochemical classification is based on the relative volatility of the elements during condensation and evaporation. Condensation temperatures are calculated assuming thermodynamic equilibrium between condensed solid and nebular gas at a given total pressure. They are measures of the relative volatility of the elements. Major elements condense as minerals whereas minor and trace elements often condense in solid solution with major minerals and/or melts (but see (1) below). The temperature where 50% of an element is condensed (or evaporated) is called the 50% condensation temperature (Lodders 2003, Fegley and Schaefer 2010, Lodders et al. 2025a, these proceedings). Condensation temperatures only apply to H- and He-rich solar-like elemental compositions and may be very different under more oxidizing conditions than in the canonical solar nebula. Condensation temperatures calculated for solar composition should not be applied to evaporation processes in the absence of abundant H. Five components can account for the variations in the elemental abundances in primitive meteorites. In addition, their oxidation state controls abundance variations during condensation and evaporation, adding to the complexity of chondrites.

(1) *Refractory component:* The first major phases to condense from a cooling gas of solar composition are Ca, Al-oxides and minor silicates associated with a large number of refractory lithophile elements (RLE) including Al, Ti, Ca, Zr, Hf, Sc, Y and the REE (Rare Earth Elements). Trace elements condense into solid solution with each other (ultra-refractory phases) or with host phases made of more abundant elements. Refractory siderophile elements (RSE) comprise all metals with lower vapor pressures than those of Fe and Ni. They include W, Os, Re, and Ir and condense as refractory metal alloys, e.g. Palme et al. 1994.

Constant ratios among refractory elements in most chondritic meteorites allow the determination of representative abundances in CI-chondrites where parent body processes may have caused heterogenous re-distribution of some refractory mobile elements, such as for example Ca or U (see below). A notable exception of constant refractory element ratios is REE in CV-chondrites. The REE pattern of bulk Allende is fractionated relative to CI-chondrites (e.g., Stracke et al. 2012).

(2) *Mg-silicates and iron-alloy:* The major fraction of condensable matter is associated with the three most abundant elements heavier than O: Si, Mg and Fe. Iron first condenses as metal alloy (FeNi), whereas Mg and Si form forsterite ($Mg_2SiO_4$) which converts to enstatite ($MgSiO_3$) at lower temperatures by reaction with SiO(gas). Below 0.1 mbar, FeNi-metal condenses at lower temperatures than forsterite, at higher total pressures, FeNi-metal condenses at higher temperatures than forsterite.

(3) Moderately volatile elements have condensation temperatures between those of Mg-silicates and FeS (troilite). This includes Mn and Na. The most abundant moderately volatile element is sulfur which starts condensing at 704 K (independent of total pressure). Half of all S is condensed at 664 K (see Palme et al., 1988, Lodders 2003).

(4) Highly volatile elements have condensation temperatures below that of FeS (704 K) and above water ice (e.g. Cd, Bi, Pb).

(5) Ultra-volatile elements have condensation temperatures at and below that of water ice. This group includes H, C N, O, and the noble gases. About 20-25% of oxygen can be removed by silicate formation at higher temperature, but because the 50% condensation temperature of O (as water ice) <200 K, O is regarded as an ultra-volatile element.

## Carbonaceous Chondrites of the Ivuna-Type (CI)

The three major types of chondritic meteorites are carbonaceous chondrites (CC), ordinary chondrites (OC), and enstatite chondrites (EC). Figure 2 illustrates compositional differences between these types. The elements in Figure 2 represent elements in various cosmochemical groups, e.g., Al represents refractory elements such as Ca and Ti; Si the elements of intermediate volatility, and Mn and S represent moderately volatile elements. The concentration ratios in Figure 2 are relative to solar photospheric abundances and are further normalized to Mg. Only CI-chondrites match solar abundances closely (i.e., element ratios for CI-chondrites plot at or close to unity). All other types of chondrites diverge in some way from CI-chondrite abundances and solar abundances. The close correspondence of solar abundances with CI-chondrites is the major argument why CI-chondrites are combined with solar data for the solar system composition (e.g., Anders 1971, Holweger 2001).

The major differences among chondrites are the depletions of moderately volatile elements (exemplified by S and Mn in Figure 2) and variations in the level of refractory elemental abundances. The Al/Mg ratio (squares in Figure 2) for CI-chondrites is similar to the Al/Mg ratio in the Sun, but Al/Mg ratios are higher in other carbonaceous chondrites and lower in other chondrite groups. Ratios of the moderately volatile elements Mn/Mg and S/Mg in CI-chondrites, and possibly EH-chondrites, closely match the solar ratios. The EH-chondrites come close to solar and CI-chondritic ratios for Al/Mg, S/Mg, and Mn/Mg, but their Si/Mg is much higher and their element/Mg ratios for elements more volatile than S are lower than in CI-chondrites, making them less suitable for solar system proxies.

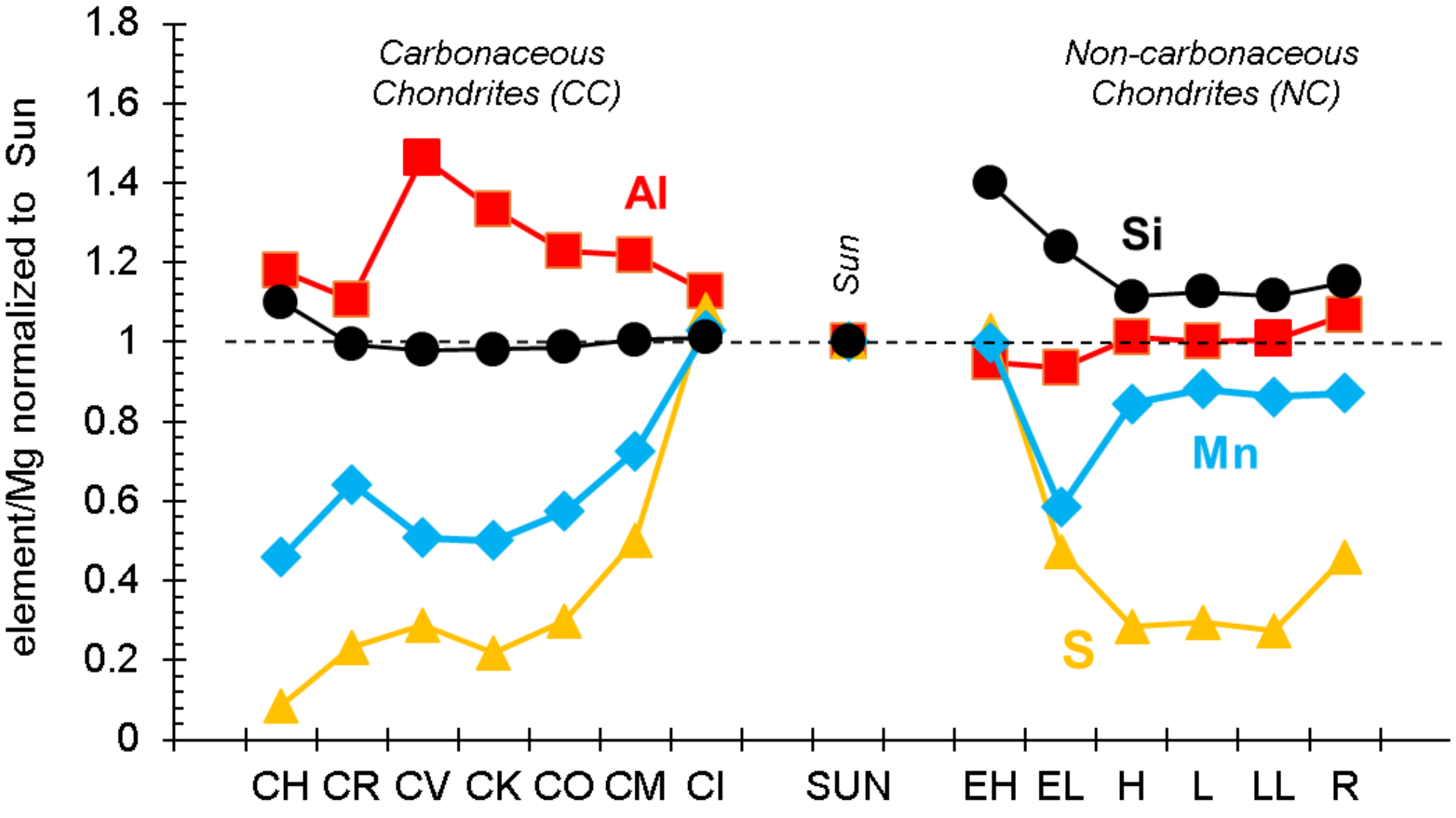

**Figure 2.** *Si/Mg, Al/Mg, Mn/Mg and S/Mg ratios in various chondrite groups normalized to solar abundance ratios presented in this paper. The CI-chondrites show the best overall match to solar photospheric abundances. The EH chondrites also fit solar element ratios for moderately volatile elements (represented by S/Mg and Mn/Mg), but they miss the solar Si/Mg ratio. Solar and CI-chondritic data are from this study, other chondrite data are from Lodders (2021).*

The five CI-chondrites Orgueil, Ivuna, Alais, Tonk, and Revelstoke are observed falls of this rare meteorite group. CI-chondrites are fragile and easily break up during atmospheric entry. The most mass is preserved from Orgueil, and most chemical and isotopic analyses were done on this meteorite. Much less material is preserved from the other CI-chondrites. Problems with representative sampling may arise when only small fragments or samples (< 50 mg) can be analyzed. Over the years sensitivity and precision of analytical methods have improved significantly: 50-gram size samples were used for analyses of Orgueil in 1864 after its fall, later published analyses of Orgueil in the 1950s used 1 to 2 g samples with wet chemical gravimetric, colorimetric and spectrographic methods. Results were largely limited to major elements. Neutron activation analyses since the 1960s expanded analyses to trace elements and sample sizes dropped to the range of about 100 mg and below. Since the 1990s, methods using inductively coupled plasma (ICP), mainly with mass spectrometry (ICP-MS) considerably improved the precision of trace element analyses and allowed the analysis of very small samples (1 mg and below). The most accurate method is isotope dilution (ID) which requires no standard (see Stracke et al. 2014). The method is very labor intensive and has, so far, been applied to some lithophile and siderophile elements (see below). While the advance in analyzing small samples seems to eliminate the need for using up precious sample material, it comes at the price of non-representative sampling.

Representative sampling is required because CI-chondrites have some chemical and mineralogical heterogeneities (Alfing et al. 2019, Greshake et al. 1998, King et al. 2020, Morlok et al. 2006). CI-chondrites contain about 10-20% bound and absorbed water affecting the distribution of aqueously mobile elements, either on the parent asteroid, or on Earth when exposed to humidity. Rare variations of up to 30% may stem from accessory phases (such as carbonates, phosphates, sulfides, sulfates) which can concentrate major and trace elements. Here trace elements are elements which do not form their own minerals but occupy positions in crystal lattices normally populated by major elements, e.g., Sr can substitute for Ca in carbonates and phosphates.

The returned materials from asteroids Ryugu (Japanese Hayabusa 2 mission, Nakamura et al. 2022, Ito et al. 2022, Yokoyama et al. 2023, 2024) and Bennu (US OSIRIS-REx mission, Lauretta et al. 2024) are similar to CI-chondrites in bulk chemical and isotopic composition. These pristine samples are not discussed here due to space limitations.

## Elemental Abundances in CI-Chondrites

A discussion of all elements is beyond the scope of this paper and will be published elsewhere (Lodders et al. 2025b). References to the data and reference codes (first letter of first author's last name and 2-digit year) are listed in the electronic appendix.

Recommended element concentrations by mass are given in μg/g (parts per million, ppm) for CI-chondrites are in Table 4. In column (2) absolute 1σ uncertainties are listed, which are converted to % uncertainties (SD%) in column (3). The quality index in column (4) gives an estimate for the reliability of the CI-concentrations with A = highest quality. It is based on 1-sigma standard deviation, element variability and mobility, and issues with analytical methods. Column (6) shows the percent deviation to data in Palme et al. (2014). Major deviations are described below. Other mass concentration units used here are weight-percent (wt%) for major elements and parts per billion, ppb, for trace elements (where 1 ppm = 1 g/ton = 1 microgram/gram (μg/g) = 0.0001 wt% = 1000 ppb).

**Table 4**

## Averaging Method

The chemical and mineralogical compositions of CI-chondrites are broadly similar but subtle differences exist. For obtaining average CI-compositions we used procedures similar to Lodders (2003) and Palme et al. (2014). The compiled chemical data for each CI-chondrite are screened for outliers and straight averages are calculated for each. The average CI-chondritic composition is obtained by taking the weighted average of the individual meteorite compositions where the statistical weight is given by the number of selected analyses per meteorite. This procedure makes Orgueil dominant as most of the analyses are done for it.

Orgueil analyses were used to evaluate results obtained by various analytical methods. Computing the distributions around the mean and 1-sigma standard deviation for each method provides a check for consistency among analytical methods and shows which analytical method(s) are most suitable for a given element. For some elements, outliers can be associated with particular analytical methods often involving extensive wet chemical processing or separations. Instrumental neutron activation analysis (INAA) does not require chemical processing and is superior in this regard. However, for CI-compositions INAA is limited to about 15 elements with uncertainties below 5 %, including Al, Na, Mn, Cr, Sc, Fe, Co, Sm, and Ir (see Palme and Zipfel 2022).

Histograms aid to spot outliers in the distributions; and the lowest and/or highest values outside the two-sigma standard deviations (SD) range are removed when Gaussian distributions and single mode distributions are indicated. Two examples, Cr and Fe, illustrate the approach. The variations in most samples deemed representative are usually to within 10% and variations among elements are less than those between different chondrite groups (e.g., smaller than differences between CI- and CM-chondrites).

Some elements vary more than expected from the quality of analytical methods used. Most of these elements are aqueous mobile elements (see below). This includes Na, K, and Ni analyzed by INAA, where variations due to chemical processing during analysis can be excluded. The variations of some elements are often correlated, suggesting the presence of minor phases enriched in trace elements such as CAIs, carbonates, sulfates, and phosphates. For these elements, sampling and sample sizes become relevant.

## Chromium (Cr)

As an example, we discuss Cr concentrations. Chromium has similar concentrations in all CI-chondrites. In analyzing Cr in Orgueil five analytical methods were used for 47 samples as shown in Figure 3. The first row shows histograms of literature data for Orgueil organized by analytical methods, the averages, 1-σ standard deviations and number of samples (N). The curves are calculated normal distributions functions (PDF) about the mean. Neutron activation analyses (mainly INAA) contribute around half of all data, followed by ICP methods (10; mainly ICP-AES; inductively coupled plasma-atomic emission spectrometry), and five XRF (x-ray fluorescence) analyses. These three methods give similar averages once outliers are removed (bottom row of Figure 3). The five older spectrographic results match tightly among themselves but yield systematically lower values whereas prompt gamma ray analyses (only two) give the highest values. The latter two methods introduced bias and are excluded in the Orgueil average of 2647 ±88 ppm from 36 measurements.

Studies analyzing more than one Orgueil sample allow inter-laboratory comparisons and allow to gauge intrinsic variability. Three different sets of NAA gave: 2663 ±95 ppm (N = 6, Gooding 1979), 2645 ±85 ppm (N = 4, Kallemeyn and Wasson 1981); and 2632 ±66 ppm (N = 9, Palme and Zipfel 2022). Agreements among different groups are excellent and similar standard deviations suggest homogeneous distribution of Cr in Orgueil. Similar conclusions follow from 2618 ±43 ppm (N =6) by ICP-AES from Barrat et al. (2012) and 2630 ±53 ppm (N = 4) by XRF from Wolf and Palme (2001). Further, concentrations have no apparent dependence on sample size (see Palme and Zipfel, 2022).

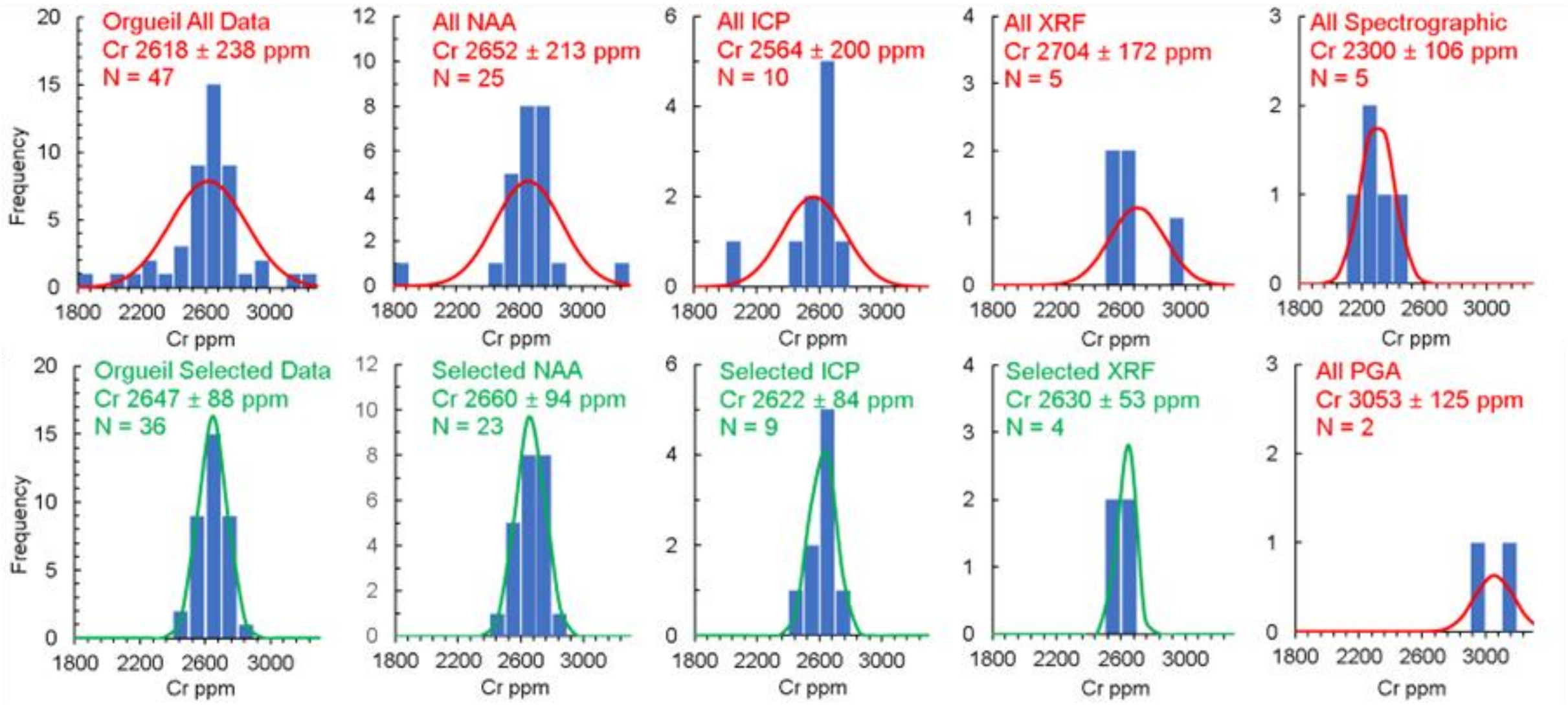

***Figure 3.*** *Published Cr concentrations for Orgueil analyzed by five methods. Neutron activation analysis, ICP methods (mainly ICP-AES), and XRF give similar Cr concentrations.*

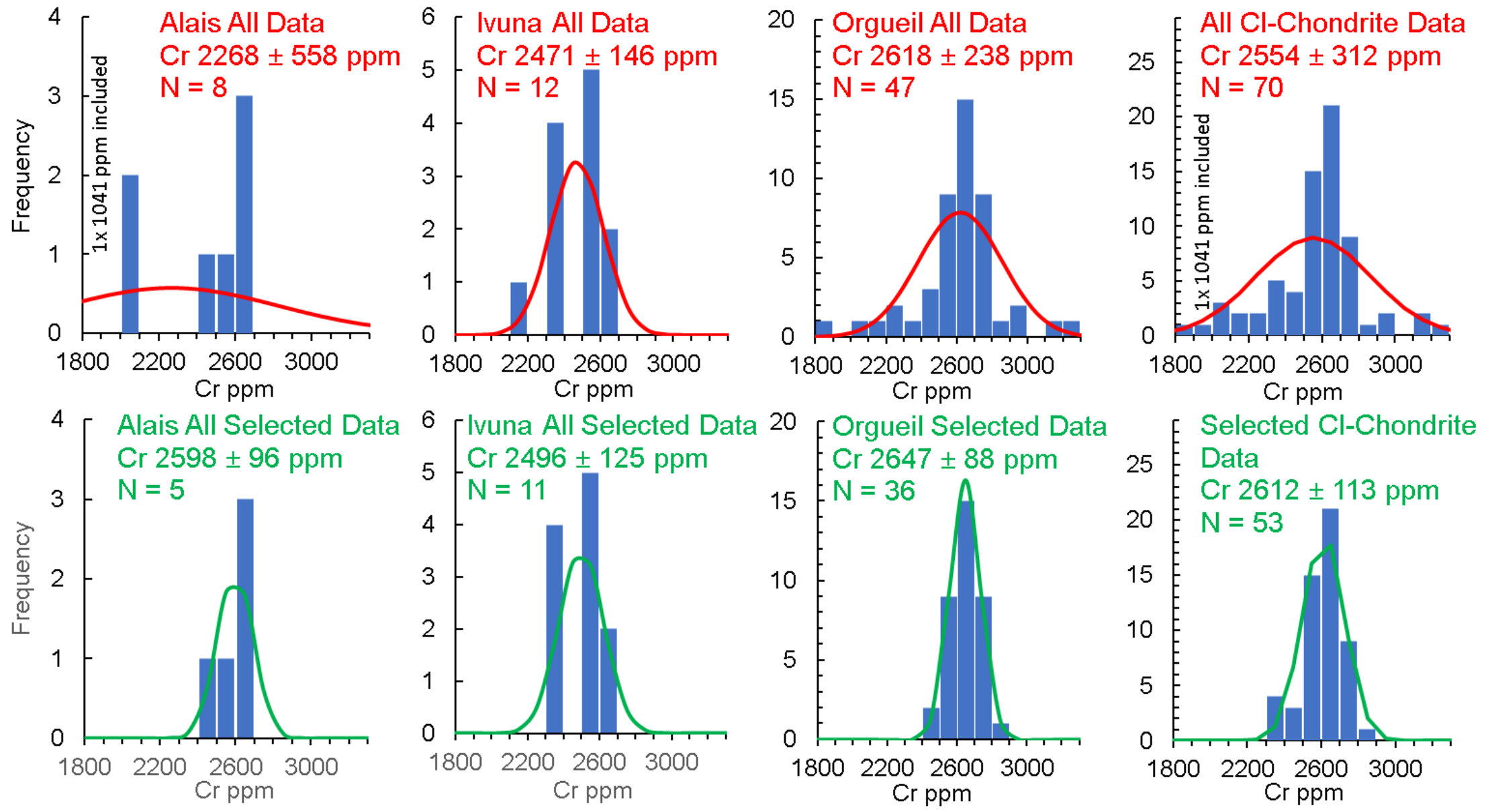

***Figure 4**. Chromium analyses for all CI-chondrites from the literature. Only methods using ICP, NAA, and XRF are included (see Figure 3). Ivuna and Orgueil agree within uncertainties.*

Figure 4 shows all literature data for Cr for the CI-chondrites Alais, Ivuna and Orgueil; the plot "All CI-chondrite data" includes Tonk (1x) and Revelstoke (2x). Only Tonk was retained for the group mean because the XRF Cr values for Revelstoke (2000 and 3200 ppm, Folinsbee et al. 1967) are at the extreme concentration limits of the overall observed range. As for Orgueil, no spectrographic or PGA measurements were used, which removes one measurement each for Ivuna and Alais. The Ivuna average of Cr = 2496 ±125 ppm is about 150 ppm smaller than in Orgueil but agrees within uncertainties. Alais has similar Cr = 2596 ± 95 ppm after removing two samples (1041 and 2010 ppm) which are also unusual in several other elements. The Cr content of Alais is essentially the same as Orgueil within uncertainties. Overall, three different analytical procedures give the same Cr concentrations among CI-chondrites within uncertainties and Cr is fairly homogeneously distributed in them. The combined data give a CI-chondrite Cr-content of 2612 ± 113 ppm or ± 4.3%.

## Iron (Fe)

The second example of the averaging method is Fe. Figure 5 shows the raw (56) and selected (39) data of Orgueil by methods, Figure 6 shows the raw and selected values for the individual CI-chondrites.

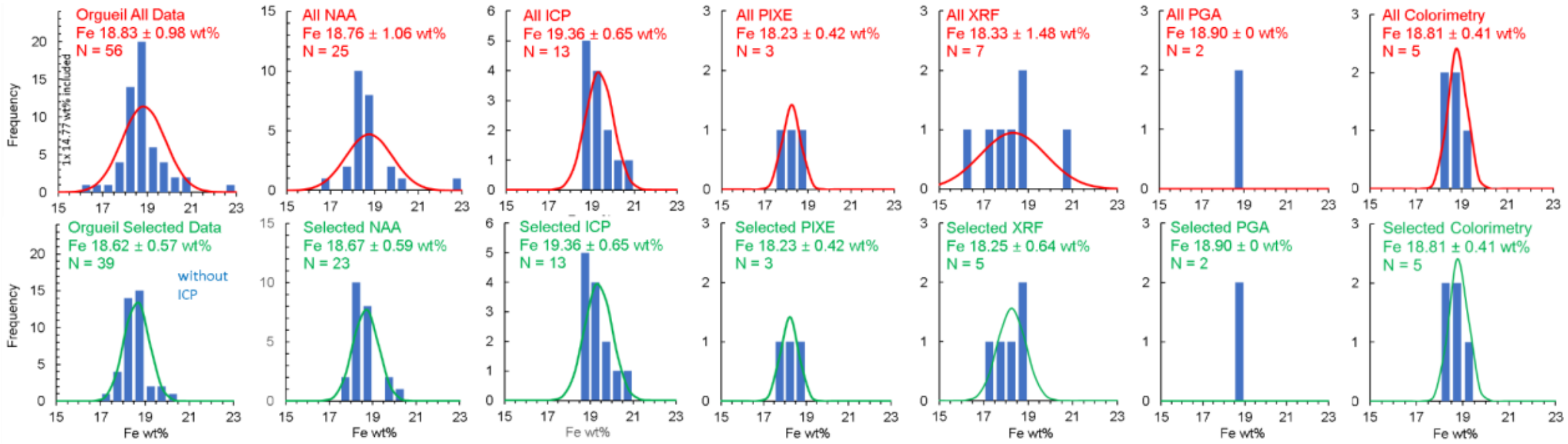

**Figure 5.** *Results of Fe analyses for the CI-chondrite Orgueil. Four methods give similar results. The ICP-AES/OES data systematically overestimate the Fe-contents, are right-skewed, and therefore were excluded from the average.*

Most of the 56 Orgueil analyses were done by NAA, followed by ICP-AES/OES, XRF, colorimetry, and PIXE. Like Cr, Fe is an ideal element for NAA because of the long half lives of the radioactive nuclides produced during irradiation and excellent counting statistics. Averaging the Fe-contents of all 25 samples measured by NAA after removing the highest outlier and the lowest value gives 18.67 ± 0.59 wt% (3.2 %SD). The good interlaboratory agreement also suggests homogeneous distribution of Fe in Orgueil.

Among the different analytical methods, results from ICP (inductively coupled plasma combined with AES/ OES (Atomic Emission spectroscopy or Optical Emission spectroscopy) are systematically higher with a right-skewed distribution than results from NAA, PIXE (particle induced x-ray emission) and XRF although they agree within combined uncertainties. Seven XRF-analyses from different labs have a comparatively large spread. The reason for the higher ICP results is unknown and might be related to matrix effects; this disagreement with other methods is not addressed in the original papers and needs to be re-visited by analysts. We excluded the ICP-data in the grand average because the unusual distribution and the higher average indicate a larger analytical issue for the ICP-AES/OES values.

Figure 6 shows iron averages for Alais, Ivuna and Orgueil. The lower Fe-content in Alais (17.7 ± 0.4 wt.%) and its differences in other major elements compared to other CI-chondrites was found by several groups, e.g., Palme and Zipfel (2022). The Fe content of Ivuna is intermediate between Alais and Orgueil. The CI-chondrite group average for iron (including Alais, but excluding ICP-data, is 18.50% ± 0.64 or ± 3.5%.

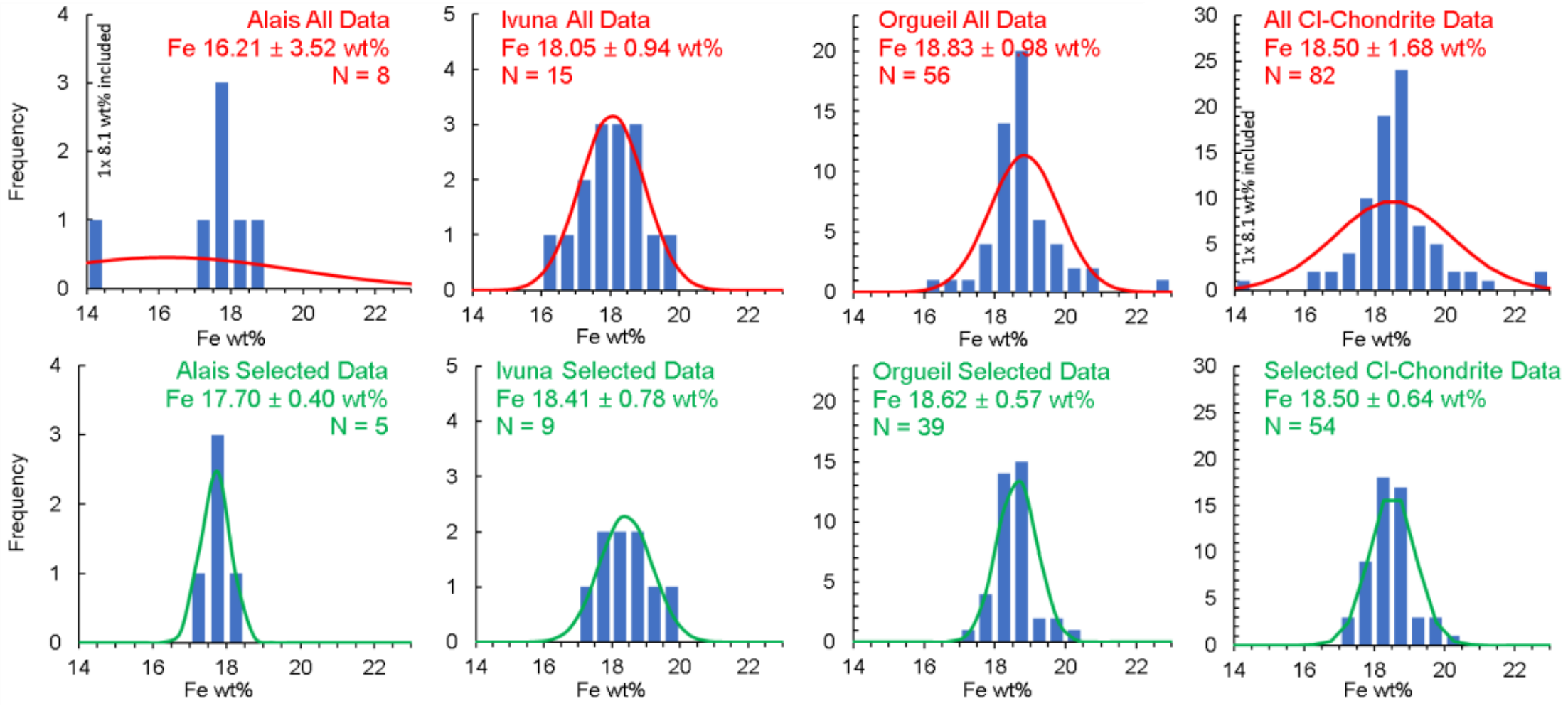

**Figure 6**. *Iron in CI-chondrites. Averages of Ivuna and Orgueil agree but Alais is somewhat lower. The five Alais and the few Tonk and Revelstoke analyses only have a small influence on the overall CI-chondrite average.*

## Carbon (C)

The 57 C concentrations (sources in Table A2) in Figure 7 are bimodally distributed in CI-chondrites, this is also indicated in Orgueil samples only. The high mode is dominated by Alais samples, which include several recent analyses.

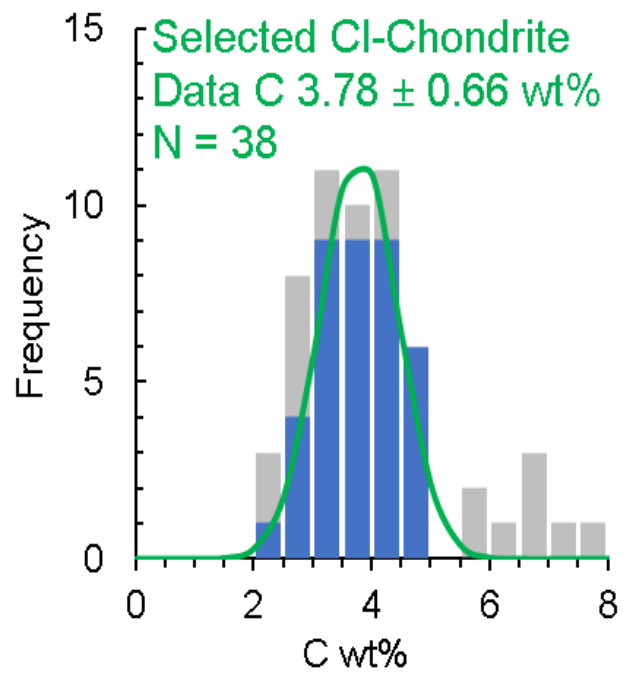

**Figure 7.** *Distribution of C concentrations in all CI-chondrites using various methods. The gray bars are for all data, the blue ones for selected data from combustion analyses. Stepwise-heating results are excluded, as are several high concentration results (mainly for Alais).*

The two major C reservoirs are assorted organics, and carbonates (dolomite, breunnerite, calcite). During stepwise heating/pyrolysis under oxidizing conditions of CI-chondrites (e.g., K70, G72, G91, W85; see appendix) the low temperature (200-400°C) C is sourced from carbonates and low-mass molecular organics, whereas C released at temperatures up to around 1350-1400°C is associated with refractory organics. Classical chemical combustion analyses pointed to 0.9 wt% C as carbonate in Alais (B834) and for Orgueil to around 0.15 wt% (C864c), as confirmed by G88 (0.16 wt%).

Several C determinations by stepwise heating were done on acid residues of CI-chondrites where carbonate C and acid-soluble organics were lost and can amount to about one percent of total C (e.g., S70, W85). This is apparent when results for Orgueil (after obvious outliers are removed) from different methods are compared: Classical combustion (C = 3.84 ± 0.47 wt%, N=8), and "element analyzer" combustion (C = 3.70 ± 0.62 wt% N=13) average to about half a percent higher than stepwise heating results (C = 3.15 ± 0.83 wt%, N=9 with a bimodal distribution). However, within the larger standard deviations, results are consistent among these methods.

The measurements by Pearson et al. (2005; P05) for Alais (5.40 wt%) yield higher C concentrations than for Orgueil (4.88 wt%), and generally somewhat higher than previous results. Alais has six out of ten C analyses with about 7 wt% (P05, W86, B834) not included in the grand mean. Pearson et al. (2005) found two data clusters for small aliquots from two Alais specimens, indicating heterogeneity. Among 33 Orgueil analyses, only three are close to 6 wt% or higher (C864a, W86, P05). Five Ivuna measurements by combustion compare well with corresponding Orgueil data, and one Tonk (C14) and one Revelstoke (F67) analysis are in the group mean.

Carbon concentrations show an inverse trend with analyzed sample mass (Figure 8). As the mass scale in the figure is logarithmic, the use of approximate masses (as reported in the literature) seems reasonable to gauge possible dependencies. Larger Orgueil samples show lower C concentrations (Figure 8). Alais appears similar but the old results on its two highest-mass samples (M806, T806) should not be over-interpreted. The two data clusters for Alais (P05) noted above are well resolved. Leaving the possible bias from analytical methods aside, data in Figure 8 suggest that samples below 50 mg are prone to higher C concentrations up to twice the recommended group average of 3.78±0.66 wt%. The recommended value excludes Alais and Orgueil values >5.8 wt% based on the data distribution in Figure 7.

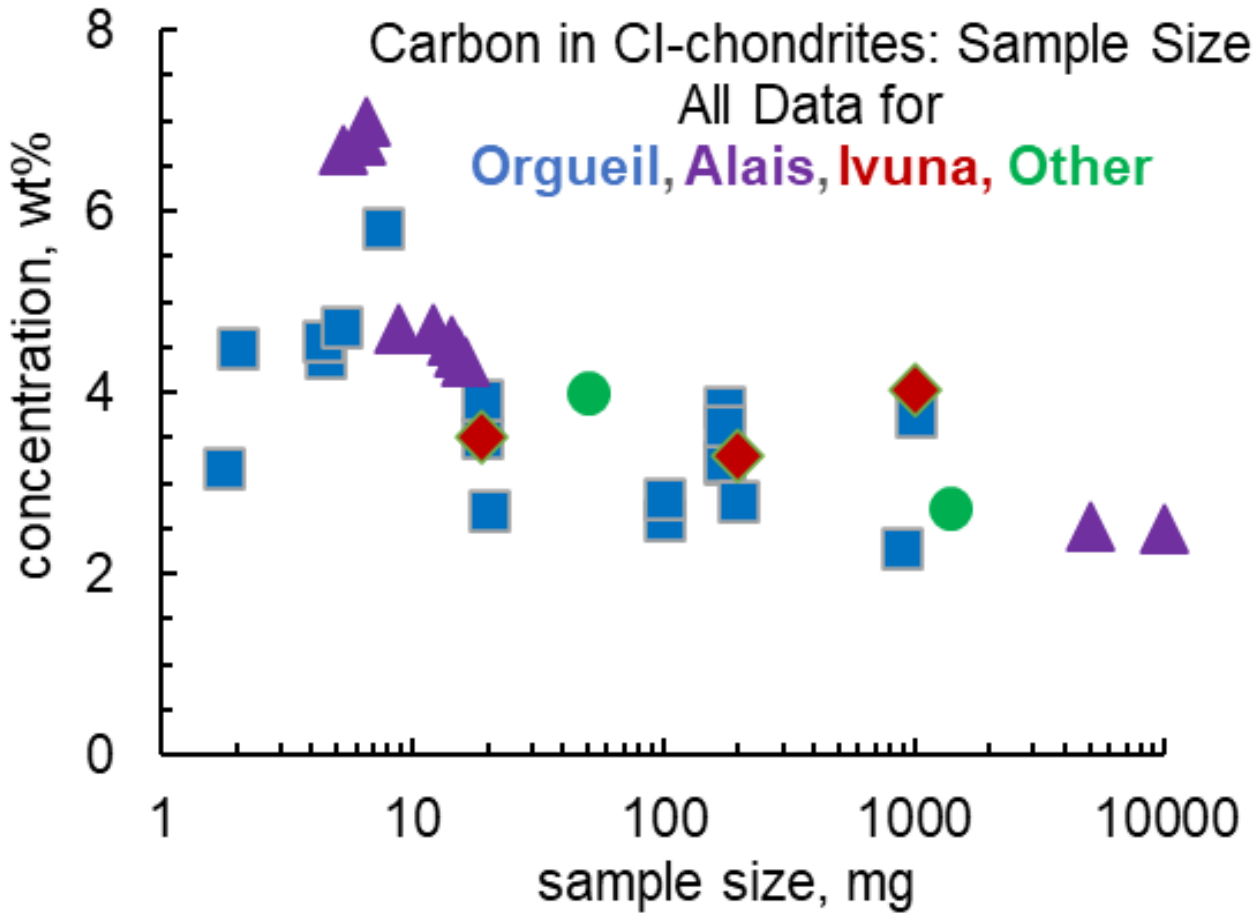

**Figure 8.** *Carbon concentrations as a function of logarithmic sample mass where (approximate) masses analyzed were reported. Orgueil = squares, Alais = triangles; Ivuna (B54, S70) = diamonds; Tonk (C14) and Revelstoke (F67) =circles.*

**Nitrogen (N)**

Measured N contents of Orgueil vary from 800 to 8200 ppm in 20 analyses by different authors (Table A2). Thirteen measurements for Orgueil give 1934 ± 458 ppm, excluding the lowest value, two extreme results from P05, and all high chromatographic values from G71. Seven Alais measurements yield N = 2021 ± 521 ppm; two for Ivuna give N = 1962 ± 151 ppm. All selected values provide the grand average N = 1965 ± 447 ppm for CI-chondrites.

**Oxygen (O)**

Total O contents determined by fast neutron activation analysis exist for Orgueil (2x Palme and Zipfel 2022, 1x Wing 1964) and one for Ivuna (Palme and Zipfel 2022). The Orgueil value by Wing (1964) is about 1 wt% lower than the average of two measurements reported by Palme and Zipfel (2022) but without further measurements, Wing's (1964) value is retained in the mean. For Alais, an estimate by difference gives 47.05% oxygen. The recommended CI-chondrite value is O = 46.57 ± 0.8 wt%.

**Silicon (Si)**

The average Si concentration from all CI-chondrites, Si = 10.66 ± 0.43 wt%, is from 26 analyses by fast-neutron activation analysis (FNAA) and XRF. The scatter in the distribution is caused by Alais which has lower Si than Ivuna and Orgueil. Silicon is traditionally used to anchor the atomic cosmochemical abundance scale to one million Si atoms, and element/Si ratios are frequently reported in the literature. However, Si is not routinely measured, and it is not included in several modern ICP-MS measurements. In this procedure samples are dissolved in HF and some Si may be removed. Thus, one has to rely on the adopted average concentration of Si for each CI-chondrite to derive atomic abundances for the Si-based abundance scale. Given the natural variations, one ideally should always normalize to the actual Si concentration of the sample at hand, which is not possible for most existing analyses. We recommend normalizing abundances to Mg instead of Si. Silicon is more volatile than Mg, and Si/Mg are fractionated in different meteorites and planetary objects (including bulk silicate Earth).

**Rare Earth Elements (REE)**

The CI-chondritic averages for the REE are based on ICP-MS and ID-MS measurements (e.g., B12, B16; B84, D15, N74, M06, P12). Outlying results, often with high concentrations, from samples analyzed by B18, E78, K73, and R93 are excluded. No NAA and PGA values (K81, P22, I12) are used because these are largely limited to La, Sm, Eu, and Lu, and would introduce heterogeneity in the weighted averages of all REE.

The REE were redistributed within CI-chondrites into carbonates, phosphates, and sulfates and sampling and sample sizes are important. In addition, analytical uncertainties remain as e.g., Nd and Lu averages from different laboratories differ. Many individual sample patterns show variations within 5-10%, depending on which samples were selected for use as normalizing values (e.g., see Barrat et al. 2012, Pourmand et al. 2012).

As quality control we used cross-correlations among the REE. Correlations of REE with their nearest neighbors can identify outlier samples with possible analytical issues (such as standards);

often this is associated with the mono-isotopic REE. Another check is the consistency of the elemental data with isotope systematics of Sm-Nd for which we calculated $^{147}Sm/^{144}Nd$ = 0.1967 which compares to 0.1966 (Bouvier et al. 2008) and 0.1967 (Jacobsen and Wasserburg 1984), see also Table A11.

**Mobile Elements**

The CI-chondrites are not a completely homogenous group in terms of chemical composition. It was long suspected (e.g., Wiik, 1969; Schmitt et al., 1972; Kallemeyn and Wasson, 1981, Ebihara et al. 1982) that Alais is lower in Al, Au, Br, Ca, Cr, Fe, and Si, but higher in Bi, In, Sb, and Se than Orgueil and Ivuna. We find similar differences ($\geq$5%, rarely up to 10%) for individual CI-chondrite averages, but cannot confirm lower Au and Cr or higher in Bi and Se for Alais. In addition, Alais is lower (5-10%) in Li, Rb, Re, Ru, V, Zr, and higher in Nb, Sn, W, Y, and REE than Orgueil. For some elements (e.g., Nb, Ru, V, W, Y, Zr) the differences are probably due to paucity of and difficulties in the chemical analyses, and not always real compositional differences. Ivuna has the highest concentrations of Na and Br. Variations beyond analytical uncertainties were found among individual meteorites and within a given CI-chondrite, for Na, K, and Ca (Palme and Zipfel 2022, Gooding 1979; Barrat et al. 2012).

Variable elements in CI-chondrites include the same elements accumulated abundantly in terrestrial ocean water: e.g., Na, K, Mg, Ca, Sr, B, C, S halogens, also Au and Os. Aqueous (re)distribution of mobile elements into host phases that are concentrating them account for some heterogeneity in bulk samples if carbonates, sulfates, and phosphates are sampled in varying proportions. Redistribution of sulfates during terrestrial storage at varying humidity is seen by appearance of efflorescence on CI-chondrite samples. If redistributions are within a closed system (isochemical) mobile element net loss does not occur but they influence sampling and representative sample sizes. For example, if efflorescence or veined material is avoided, the elements making such materials might be lower in bulk analyses.

The “chemical homogeneity” of CI-chondrites depends on the sample size and is different for each element because of different host phases. Most elements do not show any discernible correlation of concentration with sample sizes larger than about 20-50 mg (but see e.g., example for C above). This comparison is not completely unbiased because different methods also can affect the results.

Mobile elements have concentration histograms with either widely spread values or bi- to tri-modal distributions as shown in Figure 9 for e.g., K, Rb, Ca, Os, Ni, and S. In Figure 9, visual divisions were done into low (red), main (blue), and high (purple) modes for which the averages and PDF curves are shown. These divisions are non-unique but were guided by criteria to define bimodal or higher mode distributions. There is no practical analytical division into modes, which among other things depends on the chosen bin sizes. We apply Sturge’s rule (Scott 2009) and Scott’s rule (Scott 1979) for bin size optimization. For the mode division we use the rule that the means of the modes should be separated by more than two standard deviations and the standard deviations for each mode must be (about) the same absolute values. The recognition and division of the distributions into modes is potentially biased by analytical methods. In the ideal cases, enough reliable analyses by one or more proven methods with similar analytical uncertainties are available, but this may not be the case for each element considered. In all multi-modal cases (e.g., diagrams like in Figure 9) the most populated mode is taken as representative average. This “main” mode (in blue) is typically the central mode in trimodal distributions.

Ca-bearing phases (carbonates, sulfates, phosphates) are important carriers of trace elements such as Sr, Ba, Sc, Zr, Hf, U, Th, and REE. Samples enriched in these elements could be explained by carbonate and phosphate accumulations. Sodium, K, and Ni enhancements often indicate deposited sulfates, depletions point to samples poor in sulfates and other salts. Element associations cannot be checked easily using trace element correlations because only a few measurements exist where major and trace elements were measured for the same sample.

**Alkalis (Na, K, Rb, Cs)**

Sodium, K, and Rb enhancements suggest deposited sulfates, whereas depletions point to samples poor in sulfates and other salts, as observed in matrix samples (see e.g., Morlok et al. 2006, McSween and Richardson, 1977). Water soluble host phases (mainly sulfates, halides, possibly phosphates) facilitate alkali mobility, and essentially all Na and K are extractable by water from CI-chondrites (F88). Bimodality (possibly trimodality, Figure 9) and well-defined correlations of K and Rb are known (Goles 1971). However, Cs is unimodal with a well-defined average, and no correlations of Cs with other alkalis are observed suggesting different host phases for Cs. Due to its larger ionic radius, Cs might be stronger intercalated in layer silicates than the other alkalis, minimizing redistribution during aqueous processing.

**Calcium (Ca)**

In CI-chondrites, Ca is a major element in carbonates (breunnerite, calcite, dolomite), phosphates, and in (rare) sulfates (e.g., gypsum). Calcium was analyzed by NAA, ICPAES, PGA, XRF and classical wet chemical analytical methods in all CI-chondrites. Following the division into modes as described above, the nominal main mode average for Orgueil is 0.893 ±0.088 wt%, and for all CI-chondrites 0.897 ±.080 wt%. However, the criterion requiring similar standard deviations for the modes is not fulfilled, and the division into modes is not trivial for Ca. The distribution might be better described by a continuous log-normal distribution as expected from dispersal of several different Ca-rich phases (in contrast to other elements where the variations are influenced by one main carrier phase). For a representative Ca abundance, we use element ratios as described in Palme et al. (2014) and below.

Large concentration variations become apparent when small sample volumes or areas are analyzed. Morlok et al. (2006) described different lithologies in CI-chondrites with variable chemical compositions on a scale of 50 to 100 μm, thus the estimated analyzed volume is about $10^{-6}$cm$^3$ which is orders of magnitudes below the volume of the analyzed bulk samples (about 10-2000 mg for Ca). Their Ca contents vary by a factor of 60, from 0.02 to 1.5 % Ca in 115 different fragments. Greshake et al. (1998) found large K, Ca, S, and Al variations among 35 Orgueil matrix particles of about 100 μm in size. During aqueous redistribution Ca is enriched into phosphates, sulfates, and carbonates. This causes the micrometer-scale variability. The Ca-content of bulk samples on the scale of around $10^{-1}$cm$^3$ is about constant, but varies on the scale of 100 μm.

Matrix samples were sought out by Greshake et al. (1998) and Ca-bearing phases were rarely among their "matrix" samples. Such phases were studied by Morlok et al. (2006). In both studies, samples with Ca > 1 wt% often have high and correlating S and/or P contents. Tiny high-Ca samples can include minute phosphates, sulfates, and carbonates which can explain some trace element variations in CI-chondrite samples.

**Nickel (Ni)**

Nickel is typically not considered a “mobile element”, however, variable Ni concentrations in CI-chondrites are not just due to potential analytical difficulties. Nickel analyses for Orgueil range from about 0.9 to 1.4 wt% and indicate bimodal distributions in NAA and ICP datasets. Similarly large variations appear in Alais and Ivuna, and are best seen in the combined dataset (Figure 9). A critical number of analyses has to be available before different modes are clearly discernible. All data (without any outlier removal) are distributed into 2-3 modes. The criteria for the modes are fulfilled, and the main mode average gives 1.118±0.033 wt% Ni for CI-chondrites.

Tonk and Revelstoke data do not contribute to the main mode of the Ni data. The two Ni measurements (0.8, 1.3 wt%) of Revelstoke from Folinsbee et al. (1967) both fall outside the two-sigma range of the main mode mean. Their analytical method for Ni was optical emission spectrography with a stated uncertainty of about 16%. The single Tonk value by INAA (Palme and Zipfel 2022) is rather low (0.80 wt%) and compares to the lowest value for Revelstoke by Folinsbee et al. (1967).

One interpretation of the modes is soluble epsomite with isomorphous Ni substitution for Mg. Whether this sulfate is indigenous to CI-chondrites or formed in the terrestrial environment is inconclusive and related to the sulfur distribution (e.g., Gounelle and Zolensky 2001, Lodders and Fegley 2011). The sulfate is highly soluble, and Ni was found in water extracts from CI-chondrites (e.g., Thenard 1806, Berzelius 1834, Cloez 1864a,b, Fredriksson and Kerridge 1988). The low mode may represent samples that lost sulfates, and the high mode samples with sulfate deposits Samples very high in Ni (yellow bars in Figure 9) likely stem from sulfate-veined lithologies of CI-chondrites. Nickel concentrations in sulfates reach up to 10 wt% (Fredriksson and Kerridge 1988). Addition of 1 mass% of such sulfates to samples with normal Ni concentrations of 1.12wt% increases the Ni content to 1.21 wt%, just what is found for the high mode average.

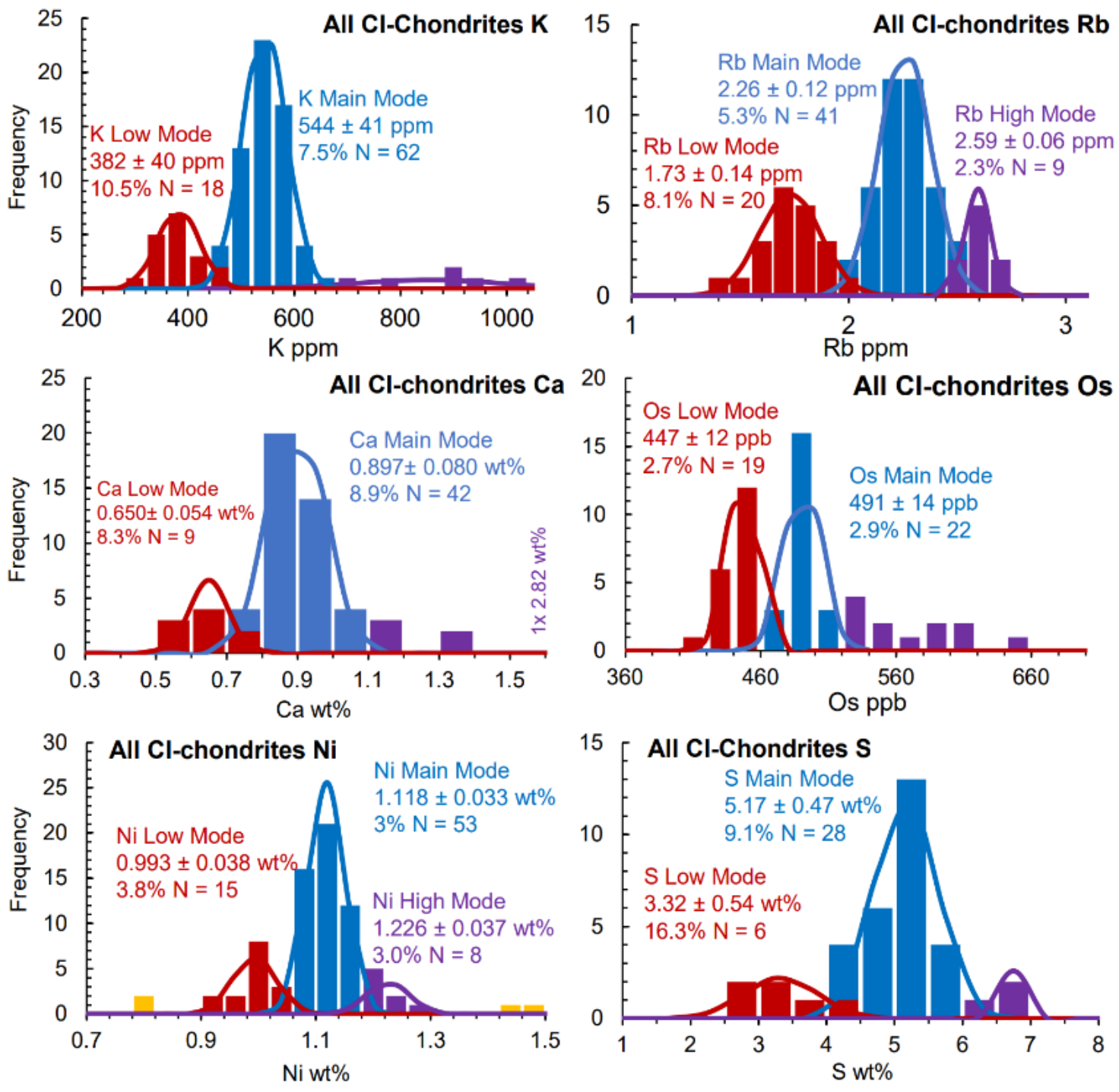

**Figure 9**. *Examples of mobile element concentrations in CI-chondrites.*

## Phosphorus (P)

The recommended concentration of P = 989 ± 63 ppm from 24 analyses of all CI-chondrites excludes wet chemical values deemed less reliable. Barrat et al. (2012) measured six Orgueil samples both with ICP-AES and ICP-MS-SF and obtained reproducible results for a given sample, and found a range of 940-1080 ppm P among their six samples. Seven XRF measurements (A89, M69, mainly W01) for Orgueil range from 800-1200 ppm. The two values from A69 and M69 seem systematically too low whereas J14 is the highest value of the XRD set. The intra-laboratory spread among samples compares to inter-laboratory results by the same methods making it difficult to uncover data with underlying analytical issues.

Phosphates are likely involved for elemental distributions with "trailing" values, e.g., for Sc, Zr, Hf, Sr, Ba, Th, U and some REE. Some other elements (Nb, Ta, B, As, Sb, Na, and Mg) hint at such distributions. However, fewer trailing values for P are observed than expected for isolated phosphate enrichments but there are also fewer P analyses and analytical difficulties for P.

## Sulfur (S)

The histogram for S for all CI-chondrite data shows a trimodal distribution, consistent with redistribution of S in CI-chondrites and associated changes in oxidation states of S as sulfides, elemental sulfur, hyposulfites, and sulfates. The main mode from all CI-chondrites is taken as 4.25 < S (wt%) < 6.25, which includes 29 analyses giving an average CI-chondritic S concentration of 5.18 ±0.48 wt%.

Only three S measurements for Alais are in the literature. Two older analyses gave relatively low sulfur (3.5%, Thenard 1806, 1.6% Berzelius 1834) which are nominally grouped into the low mode, and one analyses with high S (6.7%) by Wiik (reported in Mason 1962) is among the high-mode samples. The paucity of measurements and divergence of data for Alais do not provide a representative individual S value for Alais.

**Other elements with changes above 3% from Palme et al. (2014)**

**Antimony (Sb)**

Compared to P14, Sb increased by 8.3%. Out of 65 Sb values for all CI-chondrites, 51 are obtained by RNAA and INAA, which vary widely (110 – 220 ppb). The Sb distribution for the NAA results is unusually right-skewed and maybe low chemical yields are to blame, but the cause is difficult to track. The RNAA by Krähenbühl et al. (1973) and Takahashi et al. (1978) are often lower than those by Ebihara et al. (1982) for aliquots of the same samples, and the Sb results remain puzzling (see Ebihara et al. 1982). The average excludes K73 and T78 and the highest (220 ppb, C73) and lowest (110 ppm, X92) NAA values. The 7 ICP-MS results give 170 ±12 ppb for Orgueil, here the low 124 ppb by W05 is an outlier. Among the 3 ID-SMSS one unusually high value was not used (250 ppb, R93). All selected values yield the recommended Sb 157 ±25 ppb (N=51) from a right-skewed PDF. If none of the NAA data are kept, the average is 161 ±22 ppb with a more gaussian-shaped histogram. Overall, Sb remains ill-determined, and we assign an overall rating of "C".

**Mercury (Hg)**

Mercury (Hg) is difficult to analyze and is a frequent contaminant in CI-chondrites, especially in Orgueil (e.g., Palme and Beer 1993, Palme et al. 2014). Alais and Ivuna show more plausible values. The selected value from three analyses is 0.283 ± 0.106 ppm (36%). However, this selected concentration is only 85% of the expected s-process systematics, which implies 0.35 ppm instead. Another test for the Hg abundance is the curve of the product of the neutron-cross section times the isotopic abundances of s-process elements ($\sigma N_S$) as a function of mass number (shown in isotope section below). This downward stepping curve has regions of constant $\sigma N_S$ and pure s-process $^{198}$Hg should follow that trend. However, its $\sigma N_S$ plots slightly above other heavy s-process only isotopes. This comparison is more complicated as nearby abundances ($^{192}$Pt, $^{186}$Os) are also somewhat problematic.

Shirai et al. (2024) obtained Hg with NAA in four Ryugu samples averaging to 0.806 ±0.083 ppm. Ryugu shows many compositional similarities to CI-chondrites and the carefully curated Ryugu samples are unlikely to be contaminated with Hg. However, the more than twice higher Hg concentration in Ryugu than in CI-chondrites is puzzling. In the odd-numbered nuclear abundance systematics, the $^{199}$Hg abundance would then be similar to $^{197}$Au, and the right-sided curvature of the prominent r-process peak would take an asymmetric form. Gold and Tl abundances are unlikely to be grossly underestimated in the existing samples, so Hg remains the exception here. Similarly, $^{198}$Hg in the "pure" s-process systematics would be far too high (see Figures below). Even allowing for somewhat generally higher overall concentrations in Ryugu because of lower water contents than in CI-chondrites, the Hg in Ryugu is too high, possibly suggesting localized enrichment of Hg in the small samples analyzed. The concentrations of Hg need to be re-determined in larger (500-1000 mg), more representative samples.

Given the analytical uncertainties and high mobility of Hg, the best source for the solar system abundances would be the Sun itself but photospheric Hg determinations are not available and have their own challenges.

### Halogens

The halogen (F, Cl, Br, I) abundances in CI-chondrites, the Sun, and other astronomical environs were reviewed in Lodders and Fegley (2023) and are adopted here. The problems related to halogen abundances are also discussed by Palme and Zipfel (2022). Reliable measurements on 0.5-1 gram well characterized samples are highly recommended.

### Noble Gases in CI-chondrites

The noble gas data are the same as in Lodders (2003) and Palme et al. (2014), which were computed from the principal isotope data in Anders and Grevesse (1989) and date back to Mazor et al. (1970). Although the measurements themselves are precise, we added 10% uncertainties because the actual assignment of indigenous C-chondritic noble gases is difficult amid the considerable noble wind and cosmic ray exposure contributions.

## Element Ratio or Element Correlation Method for Refractory Elements

In some cases, we used well determined element ratios in CI-chondrites and other carbonaceous chondrites to increase the accuracy of CI-abundances. Of particular importance are element ratios used for dating, such as Os/Re, Lu/Hf, Sm/Nd, or Rb/Sr. A ratio such as the $^{176}$Lu/$^{177}$Hf ratio can be precisely determined to within one percent, and such a ratio is readily converted to an element ratio. Absolute concentrations of Lu and Hf require additional analytical efforts and are often not given in papers focusing on age determination. Thus, one has to rely on one element usually determined with less accuracy than the ratio and calculate the other element using the precise element ratio and error propagation.

The method is particularly useful when ratios are constant in various types of chondrites. In the following we only consider refractory element ratios from carbonaceous chondrites. The isotope ratio of $^{176}$Lu/$^{177}$Hf is a good example (see below). Both elements, Lu and Hf, are refractory elements and the calculated Lu/Hf ratio has the same uncertainty as the $^{176}$Lu/$^{177}$Hf ratio. The CI-chondrites contain up to 15% water, more than any other chondrite group, and mobile elements may have been redistributed (see above). By using pairs of refractory elements analyzed in other carbonaceous chondrites with minimal or no aqueous alterations, and with one element being more mobile than the other it may be possible to improve the CI-abundance of the more mobile element if analyses were done on small samples, and/or sampling avoided or mistakenly incorporated phases that concentrate mobile elements.

The most accurate method for determining the concentrations of trace elements with more than one isotope is isotope dilution (ID) combined with thermal ionization mass spectrometry (TIMS) or with inductively coupled plasma mass spectrometry (ICP-MS). A calibrated tracer with an isotope ratio different from the natural isotope ratio is mixed with the dissolved sample. The isotope ratio of an element in the dissolved sample plus the ratio in the spike are then measured and concentration of the trace element can be calculated. The mass of the mixed solution is irrelevant, and no rock standards are required (e.g., Stracke et al. 2014). The method is labor intensive and is primarily used to determine the concentration for pairs of elements used for dating purposes, such as Rb/Sr, Lu/Hf etc. Most other methods require well analyzed standard rocks as reference. The ID method cannot be used for determining the abundances of monoisotopic elements, e.g., Sc, Nb, and Rh but is good for most REE and other refractory lithophile elements, such as Zr, Hf, and Ta and for siderophile refractory trace elements, like Re, Os, Ru, Ir, W, Mo. Abundances of trace elements analyzed with ID can be determined with an accuracy of one percent or less, whereas data obtained with ICP-MS methods involving dissolution of fragments or homogenized powers are less accurate, in the best cases about 3 to 5%.

Other methods are less accurate for most of the refractory elements. An exception is the monatomic Sc, which can be determined with high accuracy by neutron activation analysis. Aluminum, Ca, and Ti are typical elements for high accuracy analysis with XRF. Another problem in analyzing bulk samples of CI-chondrites is sample heterogeneity. The variability of analytical results of some major and trace elements obtained with ID is primarily the result of inhomogeneous distribution of elements within the meteorite. The contribution of analytical uncertainties is often minor. Analyzing large enough samples, e.g., a total of 1-2 grams of homogenized powder is often recommended (e.g., Morlok et al. 2006, Stracke et al. 2012). However, analyzing aliquots of homogeneous powders from larger masses is not necessarily a guarantee to avoid the nugget effect.

In the following sections we use values obtained from correlations only, if the uncertainty of an element calculated from correlations is below the uncertainty of the grand mean.

## Refractory Lithophile Elements

### Ca/Al

Relying on published data and using the averaging procedure described above gives CI-chondritic grand average values of Ca = 0.897 ± 0.08 wt.% and Al = 0.847 ± 0.048 wt.%, with a Ca/Al ratio of 1.059. The uncertainty of 8.9% in the Ca abundance is higher than the uncertainty of Al (5.7 %). Calcium is more mobile than Al and concentrates into minor phases such as carbonates (see above) which affects the Ca/Al ratio in heterogenous, small samples.

If it can be shown that other carbonaceous chondrites with fewer hydrous phases have constant Ca/Al ratios, one may assume that CI-chondrites have the same ratio. Then the Ca abundance can be calculated from the Al content in CI-chondrites and the Ca/Al ratio of the other carbonaceous chondrites. Table A5 (appendix) gives Ca/Al ratios of carbonaceous chondrites with less water than CI-chondrites. Ahrens et al. (1969, 1970) used XRF on chondritic sample masses of around one gram. Data for twelve carbonaceous chondrites reported by Ahrens (1970) give very constant Ca/Al ratios with a mean of 1.077±0.02. Earlier Ahrens et al. (1969) summarized four Orgueil datasets with highly variable Ca/Al ratios ranging from 1.0 to 2.05, confirming the high mobility of Ca in CI-chondrites.

A broader survey of the literature yields a mean Ca/Al ratio of 1.080 ± 0.023 for carbonaceous chondrites. In calculating this ratio 126 samples of carbonaceous chondrites were considered, and 17 samples were excluded because their Ca/Al ratios were outside a 2σ limit (Ahrens 1970; Kallemeyn and Wasson 1981; Jarosewich 1990; Wolf and Palme 2001; Patzer et al. 2010; Stracke et al. 2012). A summary is given in the appendix (Table A5).

Assuming that all carbonaceous chondrite groups have the same Ca/Al ratio and using an Al content of 0.847% for CI-chondrites (Table 4) a Ca content of 0.847*1.08 = 0.915 wt% is calculated for CI-chondrites. This is somewhat higher, but within the uncertainty of the grand mean Ca concentration. The combined uncertainties of the Al content and the Ca/Al ratio give an uncertainty of 6.1% for Ca, which is less than the uncertainty for Ca of the grand average (8.9%). Table 4 thus gives the Ca content from the Ca/Al correlation. The difference between the grand average and the correlation method is small (2-3%) but it serves to demonstrate the procedure.

**Lu/Hf**

The refractory lithophile element ratio Lu/Hf connects the abundances of the refractory trivalent REE with tetravalent Zr and Hf. The Lu/Hf ratio in geochronology requires precise knowledge of the abundances of $^{176}$Lu (half-life 37.1 Ga) and $^{176}$Hf. From the precisely determined $^{176}$Lu/$^{177}$Hf ratios, Lu/Hf mass ratios can be calculated. Using only isotope dilution data an average Lu/Hf of 0.2384 ± 1% is obtained (Table A6, appendix).

This Lu/Hf ratio and the absolute Lu concentration of 25.5 ± 1.5 ppb from the grand average (Tables 4 & A4) are used to calculate Hf = 107 ± 6 ppb, which is close to the grand average of Hf of 108 ± 8 ppb. We adopt the value with lower uncertainty (Table A4). The uncertainty in this Hf determination is only slightly higher than for Lu because the uncertainty in the Lu/Hf ratio is only around 1%. The grand average of Lu = 25.5 ppb is well determined. Two large, representative Orgueil samples analyzed with ID (Beer et al. 1984) gave 25.3 and 25.4 ppb Lu (Table 4), and six recent ID-analyses of Lu in Orgueil average to 25.5 ppb (Münker et al. 2025).

**Zr/Hf**

Both elements are the first lithophile elements to condense from a hot nebular gas. Their condensation temperatures as pure oxides or in solid solution are above those of the REE and Al. Hence the Zr/Hf ratio should be constant in carbonaceous chondrites (Table A7). The grand average CI-chondrite concentration for Zr is 3.73 ± 0.30 ppm. No differences in Zr/Hf ratios between CI-chondrites and other carbonaceous chondrites are apparent (e.g., Münker et al. 2003, 2025). An average carbonaceous chondritic Zr/Hf ratio of 34.1 ± 1 is obtained from isotope dilution data (see appendix A7). This ratio is used in further calculations. With a Hf content of 107 ± 6 ppb a Zr value of 3.65 ± 0.21 ppm is calculated. The uncertainty of this value is lower than that of the grand average (± 0.30 ppm). We therefore adopt the Hf and Zr values obtained from the correlations. This example demonstrates the correlation method. The Zr/Hf of all carbonaceous chondrites has an uncertainty of 3%; using only ID data it reduces to about 1%. The combined uncertainties of the Zr/Hf ratio and the Hf content were calculated by taking the square root of the sum of the squared uncertainties in all cases. The minimum uncertainty is the uncertainty of 5.7% from Lu which enters into the Hf abundance. Thus, using the correlation method requires a single reference element with known absolute concentration.

**Zr/Nb**

Both ratios, Lu/Hf and Zr/Hf, are identical, within uncertainties, in CI, CM, CV and CO chondrites (e.g., Münker et al. 2003, 2025; Bouvier et al. 2008). All three elements have similar high condensation temperatures: Lu (1659 K), Hf (1703 K), Zr (1764 K). But Zr and Nb (1559 K) differ by nearly 200 K in condensation temperature, encompassing the REE condensing between 1578

K and 1659 K (e.g., Lodders et al., 2009). The refractory element and REE patterns in the Allende CV chondrite are fractionated with a deficit of early condensing refractory REE (Stracke et al. 2012 and references therein), similar effects could be expected for the Zr/Nb ratio, with enhancements of the more volatile Nb. Allende (Zr/Nb = 12.56, Münker et al. 2003) has indeed a slightly lower Zr/Nb ratio than Orgueil (14.07 in Münker et al. 2003, and 14.3 in Münker et al. 2025). Stracke et al. (2012) showed that the Zr/Nb ratio is variable in 500 mg-size samples of Allende. A range of Zr/Nb ratios from 10 to 13 is observed in their data (Table A8), and these authors concluded that homogenization of 60 grams of Allende is required for determining its average Nb abundance. The variability of Nb is probably caused by the inhomogeneous distribution of group II inclusions in Allende (Mason and Taylor, 1982; Stracke et al. 2012). The abundances of group II CAIs in other carbonaceous chondrite groups are unclear.

To avoid problems with the potential addition of group II inclusions we restrict element ratios involving Nb to CI- and CM-chondrites. The average Zr/Nb ratio is 13.47 ±0.78 (Table A9, appendix), leading to Nb = 0.271 ± 0.022 ppm, not far from the grand average value for Nb of 0.281±0.028 ppm. The grand average Zr/Nb ratio is 13.28 ± 1.7.

**Nb/Ta**

Niobium is monoisotopic and cannot be analyzed with ID. This lowers the accuracy of the Nb/Ta ratio. Tantalum shows the same tendency as perhaps Nb, higher Zr/Ta ratios in CI- and CM-chondrites than in CV-chondrites. The trend is stronger for Ta than for Nb: Münker et al. (2003) have six CI and CM chondrites with a mean Nb/Ta ratio of 20.40 ± 0.77, whereas seven CV- and CK-chondrites give a mean ratio of 17.73 ± 2.15. However, data by Barrat et. al (2012) and Göpel et al. (2015) do not show this effect. Eight CI-chondrite measurements give a mean Nb/Ta of 19.6 and three CM-chondrites a mean Nb/Ta ratio of 19.3. Lu et al. (2007) did not find significant differences between CI-, CM- and CV-chondrites. More high-quality data of Nb and Ta in chondrites are needed. Adding the results of Münker (2025) to the average Nb/Ta ratio of carbonaceous chondrites (Table A9) and using only CI- and CM-chondrites gives an average Nb/Ta =19.40 ±0.76, compared to the grand average Nb/Ta of 18.79. The Ta concentration of 0.0140 ppm (±9.2%) from correlations is somewhat lower than the grand average of 0.0149 ppm (± 4.7%), but with lower uncertainties. The latter value is used for CI-chondrites in Table 4.

**Zr/Y**

The Y concentration is calculated from Zr/Y ratios (Table A10) but only a few data of this ratio are in the literature. The REE datasets often include Y, but not Zr, which is analyzed together with other HFSE elements (High Field Strength Elements). The average Zr/Y ratio from analyses by Jochum et al. (2000), Stracke et al. (2012), Barrat el. (2012) and Göpel et al. (2015) is 2.4 ± 0.16. Insufficient data exist to address whether the Zr/Y ratio differs between CI-chondrites and other types of carbonaceous chondrites. From the CI-chondrite abundance of Zr 3.65 ± 0.21 ppm and the Zr/Y ratio we calculate an Y abundance of 1.52 ± 0.13 ppm, close to the grand mean of 1.51 ± 0.14 ppm. We use the value from the correlation, Y = 1.52 ± 0.13 ppm.

**Sm/Nd**

The Sm/Nd ratio is probably the most accurately determined element ratio in chondrites because the Sm isotopes $^{146}$Sm and $^{147}$Sm are used for dating. The large dataset shows no significant variations among carbonaceous chondrite groups (Table A11) and averaging over all members of the CC groups is justified. Isotope dilution results and LA-ICP-MS data give a Sm/Nd elemental ratio of 0.3268 with an uncertainty <1 %. The CI grand mean ratio is 0.3231 ± 0.0237 (7.3%), only about 1% below the average carbonaceous chondritic ratio from correlations.

**Th/U**

Neither the absolute concentrations of U and Th, nor the Th/U ratio in chondrites are well known. We do not find a well-defined Th-U correlation for CI-chondrites. Uranium is sensitive to aqueous mobilization (Rocholl and Jochum 1993). Variations in Th/U ratios exist in all carbonaceous chondrites. It is unclear if CI- and CV-chondrites have the same Th/U ratio. The Th and U histograms display trailing distributions like e.g., Ba, suggesting that U and Th are in phosphates and possibly carbonates. The mean Th/U of carbonaceous chondrites (Table A12) is 3.75 ± 0.38 (10%), compared to 3.75 ± 0.54 (14%) for CI-chondrites from the grand average. The agreement of the ratios could be coincidental because wide variations in the Th/U ratio exist among chondrites.

**Summary of Refractory Lithophile Elements**

Figure 10 summarizes using the ratio method for elemental abundances in CI-chondrites. The main thrust is to use the information in other carbonaceous chondrites to improve the quality of CI-chondrite values.

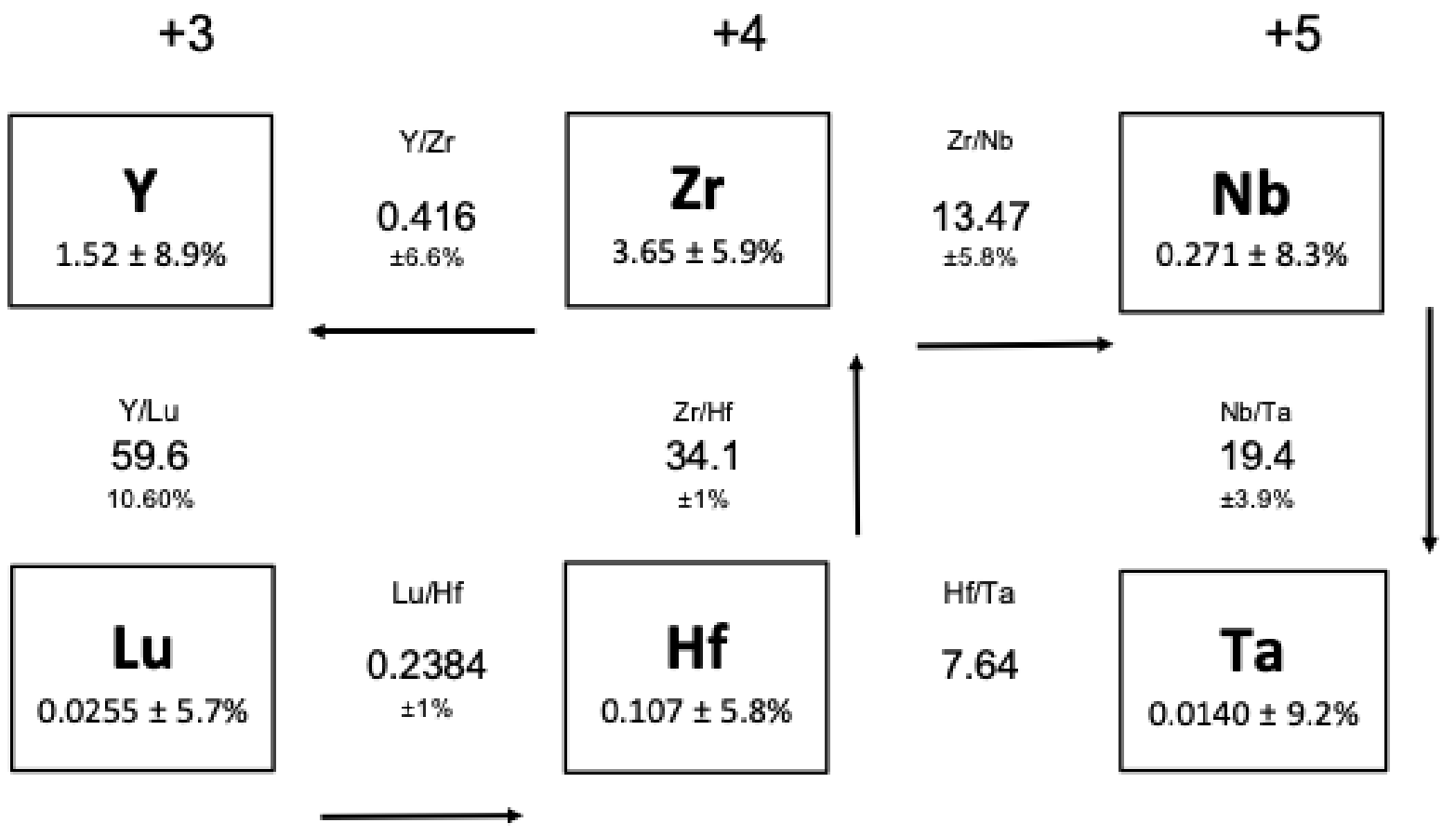

**Figure 10.** *Relationships among incompatible refractory lithophile elements. Starting with the absolute Lu abundance all other abundances of elements in the Figure are derived from correlations involving all types of carbonaceous chondrites, except those including Nb, where only CI and CM ratios are used. Improvement of the accuracy of Lu will improve the accuracy of all other elements in the Figure. Element ratios are listed in Tables A4 to A12.*

Starting with the grand average CI-chondritic Lu content of 0.0255 ppm and using the well-defined Lu/Hf ratio leads to a Hf content of 0.107 ppm with a combined uncertainty of 5.8%. By continuing with this procedure (Figure 10), CI-abundances for Zr, Nb, Ta, and Y can be calculated. The difference to the directly measured grand average values of these elements is in all cases within error bars. However, the uncertainties of the elements derived from correlations are often lower than those calculated from the grand averages.

The premise of constant element ratios in all groups of carbonaceous chondrites is the case for Lu/Hf and Zr/Hf ratios, but Zr/Nb and Nb/Ta ratios may be slightly different in CI- and CM-compared to CV-chondrites, as explained before. Except for ratios involving Nb we used all types of carbonaceous chondrites for averaging element ratios.

## Refractory Siderophile Elements

### Hf/W

The Hf/W ratio connects refractory lithophile (Hf) elements with refractory siderophile (W) elements. Under the reducing conditions of the solar nebula, W condenses as metal in an alloy with other refractory metals, Re, Os, Ir etc. (Palme and Wlotzka 1976). At higher oxygen fugacities enhanced formation of gaseous $WO_3$ lowers condensation temperatures of W (Fegley and Palme 1985). Thus, concentrations of W in bulk chondrites and Ca,Al-rich inclusions can be variable.

The application of the Hf-W clock, using the decay of $^{182}$Hf ($T_{1/2}$ = 9 Ma) into $^{182}$W, led to precise measurements of $^{180}$Hf/$^{184}$W ratios, which can be converted to Hf/W ratios. Kleine et al. (2004) analyzed 18 carbonaceous chondrites and found a mean Hf/W ratio of 1.08 ± 11%, indicating inhomogeneous distribution of W and Hf in chondrites. Analytical uncertainties are much lower (around 1%) because both $^{180}$Hf and $^{184}$W were analyzed by isotope dilution. Yin et al. (2002) obtained, with the same method, 1.14 for the Hf/W ratio in Allende. Braukmüller et al. (2018) reported a mean Hf/W ratio of 1.10 ± 10% for 23 carbonaceous chondrites, whereas Barrat et al. (2012) determined a mean Hf/W ratio of 0.92 ± 30% for 11 carbonaceous chondrites, significantly lower than other analysts. Using an average Hf/W ratio of 1.08 and a Hf content of 107 ppb leads to a CI-chondritic W content of 99 ppb with an estimated uncertainty of ± 12.3%, compared to 101 ± 11.9% ppb determined by the grand average. Münker et al. (2025) measured a Hf/W ratio for six Orgueil samples of 1.16 ± 0.08 and an average Hf/W CI-chondritic ratio of 1.12 ± 0.15. Kleine et al. (2004) reported data on two Orgueil samples with Hf/W ratios of 0.95 and 1.00. In view of these discrepancies, we use the grand average W-content of 101 ± 12 ppb.

The variability in W may reflect its strong sensitivity to aqueous alteration but it is also unclear if the CI-chondritic Hf/W ratio differs from those of other carbonaceous chondrites. More high-quality data for W and Hf in CI-chondrites are necessary to resolve this question (see Hellmann et al. 2024).

**Table 5.**

### Refractory metals other than W

The other refractory metals (Ru, Re, Pt, Rh) can be scaled to Ir and Os (Table 5) using the grand means for Ir = 456 ± 27 ppb and Os = 491 ± 14 ppb. The nearly identical Os/Re ratios of Horan et al. (2003) and Fischer-Gödde et al. (2010) provide a reliable Re content (39.9 ± 1.6 ppb). The grand mean Re of 37.1 ± 3.4 ppb is lower than that from the Os/Re ratio and has significantly larger uncertainty. The Ru and Pt values are also derived from correlations. The grand mean for Rh is used for Table 4 because it is more accurate than the correlation derived Rh value.

## Present-Day and Proto-Solar System Abundances

A scale-coupling factor is needed to combine the meteoritic and solar abundance sets to obtain a complete set of a solar system elemental abundances (combining the sets is different than renormalizing the scales to Si=$10^6$ atoms). The meteoritic set is relative to $10^6$ Si atoms. The solar set is normalized to $\log_{10}$ N(H) = 12. The scale-coupling factor to combine both sets is from elements that (1) have a top "A" rating for the goodness of analysis (Table 2), and (2) are more abundant non-highly volatile elements from 3D NLTE measurements in the solar photosphere: Na, Mg, Al, Si, Ca, Mn, Fe, and Ni. The neutron capture elements Y, Ba, and Eu with “A-" ratings are not included because they have only a few clean diagnostic lines and a reliable test of the excitation-ionization balance is not possible. The scale-coupling factor (a constant in log space) from the average of the eight elements is log SCF = 1.551±0.020. It is calculated from the meteoritic mass concentration C(E) for element E normalized to Si = $10^6$ atoms, and converted to logarithmic values, viz.,

$\log_{10}\mathrm{SCF(E)} = \mathrm{A(E)}_{\mathrm{solar}} - \log_{10}(\mathrm{N(E)/(Si)}_{\mathrm{CI}})$ where E = Na, Mg, Al, Si, Ca, Mn, Fe, and Ni, and

$\log_{10}(\mathrm{N(E)/(Si)}_{\mathrm{CI}}) = \log_{10}\{(\mathrm{C(E)}\ [\mathrm{in\ g/g}] / \mathrm{atwt(E)}\ [\mathrm{in\ g/mol}]) / (\mathrm{Si}\ [\mathrm{in\ g/g}]/ \mathrm{atwt(Si)}\ [\mathrm{in\ g/mol}] + 6\}$.

Elemental photospheric abundances are then convertible to the meteoritic abundance scale via:

$\mathrm{N(E/Si)}_{\mathrm{solar}} = 10^{(\mathrm{A(E)solar}\ -\ 1.551)}$

or CI-chondritic values to the photospheric abundance scale via:

$\mathrm{A(E)}_{\mathrm{CI}} = \log \mathrm{N(E/Si)}_{\mathrm{CI}} + 1.551$

The solar and meteoritic abundances are then easily compared (see Table 4). The meteoritic/photospheric abundance ratios are shown in Figure 11 as a function of atomic number.

Table 6 summarizes the recommended solar system abundances; the isotope abundances are given in the appendix. Both present-day (column 3) and proto-solar (column 4) abundances in Table 6 are a combination of solar photospheric and meteoritic values as indicated in the note column. The CI-chondritic values in column (1) are from Table 4 (strictly normalized to Si = $10^6$ atoms) and are used when no or uncertain photospheric values exist. Photospheric values in column (2) are obtained via $\mathrm{N(E)}=10^{\mathrm{A(E)}-1.551}$ from Table 2 and are used for H, C, N, and O; the noble gases are from Table 3. Solar and meteoritic abundances (weighted by uncertainties) were only averaged for elements (including Si) that were used to link photospheric and meteoritic abundance sets. The solar system values in column (3) were re-normalized to Si = $10^6$ because the average of solar and meteoritic abundances is also used for Si (see note column) and does not exactly equal $10^6$. This explains the small differences in columns (1) and (3) when meteoritic values are used or differences in columns (2) and (3) when photospheric values are used.

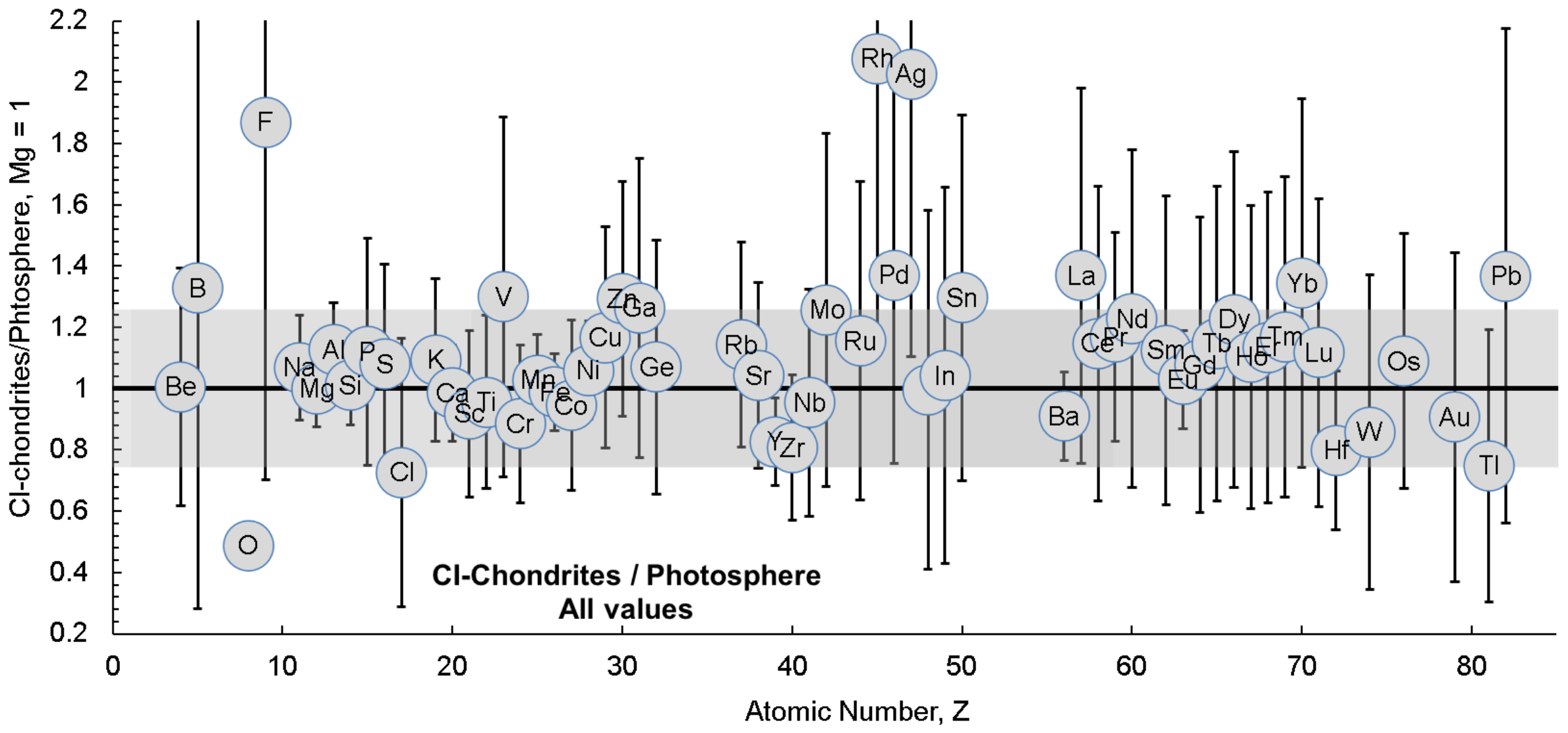

**Figure 11:** *Concentration ratios of CI-chondrites/photosphere for all elements where the comparison can be made as a function of atomic number. Solar F, Cl, In, and Tl are from sunspot spectra. The gray band indicates agreement within 26% (0.1 dex). The most deviant elements shown are Ag, Rh, Pd, F, B, B, V, La, and Yb. The reason(s) are long-standing analytical issues for photospheric but also CI-chondritic abundance determinations. Ratios for H, C, N and noble gases plot <<0.2. These and O are low because ultra-volatile elements are not fully retained on meteorite parent bodies. Lithium is destroyed in the Sun, and is not shown; the ratio is 188 ± 50.*

**Table 6**

## Derivation of Proto-Solar System Abundances and Mass Fractions (X,Y, and Z)

Heavy element settling towards the solar interior during the Sun' lifetime depleted the convection zone and thus the photosphere in heavy elements (Li to U) relative to H. Atomic diffusion is element specific and may not be a smooth function of atomic mass, and is somewhat counteracted by radiative levitation which depends on the radiative properties of the atomic species, specifically the ionization stage (e.g., Turcotte et al. 1998, Turcotte and Wimmer-Schweingruber 2002). Within the limits of analytical precisions, depletions are the same for all elements heavier than Li. Currently there is no discernable *systematic* gradual depletion suggesting "non-uniform" gravitational settling of the elements with increasing atomic mass between meteoritic and photospheric abundances; aside of this, how would such a baseline for gravitationally undisturbed or non-levitated abundances be defined to begin with? The comparison of meteoritic and photospheric abundances in Figure 11 might suggest some mass-dependent fractionation because several of the heaviest elements are higher in the photosphere (or lower in CI-chondrites). However, the analytical uncertainties are still too large for solar and/or meteoritic abundances involved in defining the "apparent fractionations".

The present-day photospheric abundances relative to H are lower than at the time the Sun formed. Settling corrections assume that all elements heavier than Li were reduced in the convective envelope (CE) by the same factor over time. Lithium is depleted because of pre-main sequence Li destruction and ongoing settling combined with the destruction of the fragile Li nuclei at the hot bottom of the solar convection zone. Beryllium and B could be affected by this type of mixing (e.g., Boesgaard et al. 2005, 2016), but the CI-chondritic/photospheric ratios for Be = 1.5 ± 0.6 and B = 1.3 ± 1.0 suggest no Be and B depletions in the solar photosphere.

To obtain proto-solar (= solar system) abundances, settling correction factors, SF, are applied. The protosolar mass fractions (indicated by subscript "0") and present-day ratios are related as

$Y_0/X_0 = (Y/X) \times SF(He/H)$, and $Z_0/X_0 = (Z/X) \times SF(Li/H\text{-}U/H)$.

The SF for Li-U was taken to be somewhat smaller than that for He (e.g., Lodders 2003, Piersanti et al. 2007, Asplund et al. 2009) but it might be the other way around (see Lodders 2020). Here we use SF(He/H) = 1.175 (17.5% change in He/H, $\log_{10}$SF(He/H) = 0.070 dex) and SF(Li/H-U/H) = 1.225 (22.5% change in heavier element/H ratios, $\log_{10}$SF(Li/H-U/H) = 0.0882 dex) and are based on models by Yang (2019). For more details and discussion see Lodders (2020). The SF here are similar to those found by Piersanti et al. (2007) for different solar compositions, e.g., from data in their Table 3, SF(He/H) = 1.181-1.197 and a smaller SF(Li/H-U/H) = 1.164-1.177.

Using $Y_0/X_0$ and $Z_0/X_0$, and the mass-balance relation, $X_0 = 1/(1 + Y_0/X_0 + Z_0/X_0)$, all proto-solar mass fractions in Table 7 are derived.

**Table 7**

The atomic weights cancel in the relation of present-day and protosolar ratios and the log of settling factors can be directly applied to obtain individual protosolar abundances on the log atomic scale relative to hydrogen. The settling correction increases the present-day abundances on the scale normalized to H by a factor "SF(E/H)," meaning a constant $\log_{10}$SF(E/H) is added to the logarithmic scale with A(H) = 12:

$$A(E)_{\text{proto-solar}} = 12+\log_{10} \varepsilon_{E,\text{proto-solar}} = A(E/H)_{\text{present}}+\log_{10}SF(E/H) = 12+\log_{10}\varepsilon_{E,\text{present}}+ \log_{10}SF(E/H)$$

This equation can be used to compute other protosolar abundances if other settling factors are preferred.

The settling corrections are applied to the element/H ratios and not to the absolute mass fraction of an element, or X, Y, and Z (this is an important difference when comparing depletions in abundance ratios and absolute mass fractions). *Absolute* mass fractions drop by $(Y_0\text{-}Y)/Y_0$ = 11% for He, and $(Z_0\text{-}Z)/Z_0$ = 14.5% for heavy elements, which are directly comparable quantities to the "depletion efficiency δ" listed by Piersanti et al. (2007). They find around 10% for each element for several different solar compositions.

The presence of long-lived (above the lifetime of the current solar system age) radioactive isotopes requires another adjustment to protosolar abundances from present-day values. The abundances of radioactive parent isotopes were adjusted for decay loss over time and the stable daughter isotopes for gain.

## Comparison to BOREXINO C+N abundances

A test for the adopted settling factors is the comparison of the C and N abundances from spectroscopy with those in the solar core from the BOREXINO particle physics experiment, for which Basilico et al. (2023) estimated (C+N)/H $(5.81(+1.22,-0.94)10^{-4}$. They found the best agreement to the photospheric (C+N)/H by Magg et al. (2022), although still only within the lower margin uncertainties. Here the photospheric (C+N)/H = $4.11(\pm 0.79)10^{-4}$ agrees within error limits.

However, this comparison is flawed because the 23% loss of heavy elements (including C and N) from the CE (hence photosphere) must be considered. The solar interior essentially remains at the protosolar values plus the settling gains from the CE. The latter are very small in absolute terms because the CE is only about 2% of solar mass. The protosolar (C+N)/H = $5.03(\pm 0.97)10^{-4}$ is closer to the BOREXINO value (within 13%), and indicates that the larger settling corrections applied here are plausible. To match the BOREXINO (C+N)/H, a reduction of heavy elements from the CE requires doubling the settling factor to around 42% (log SF(Li-U) = 0.151). Whether such high settling losses are consistent with standard solar models needs to be investigated. Such high settling corrections would increase the protosolar metallicity to $Z_0 = 0.0216$, or $[Fe/H]_{protosolar} = 0.13$, still consistent with the metallicity dispersions of B stars (e.g., Mashonkina et al. 2020) and independent measurements in the local ISM (e.g., Ritchey et al. 2023).

## Comparison of Solar and Meteoritic Abundances: No Trend with Condensation Temperatures

By the late 1960s, CI-chondrites were firmly identified as the best proxy for solar system abundances. Anders (1971), Holweger (2001), and others found no apparent differences in relative element abundances except for ultra-volatile elements. However, some doubts occasionally remain (e.g., Grevesse 2019). González (2010, 2014) concluded that there is a significant trend of CI-chondritic/photospheric abundance ratios with condensation temperatures, similar claims were made by Melendez et al. (2014), Asplund et al. (2021), and Jurewicz et al. (2024).

Figure 12 shows CI-chondritic and solar abundance ratios as a function of 50% condensation temperatures for all elements where the comparison can be made. Within uncertainties solar and meteoritic abundances for most elements agree within 26% (grey bar in Figure 12). A trend with 50% condensation temperatures could be suggested by lower ratios for refractory Y, Hf, Zr, and W, and higher volatiles (e.g., Pb, B, Ga) but considerable uncertainties remain for solar and CI-chondritic Zr, Hf, W, Pb, Ga, and B abundances (see above).

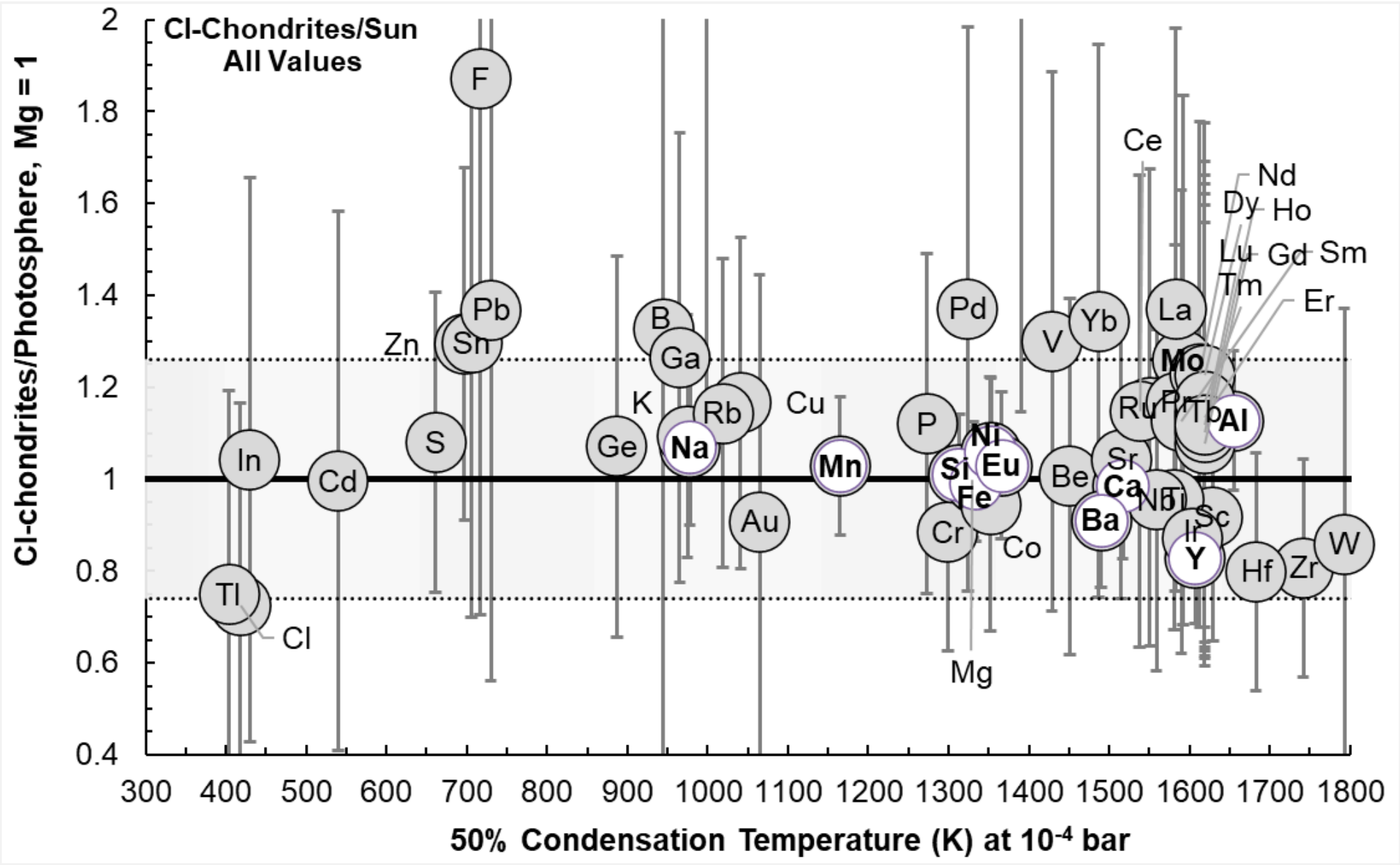

**Figure 12.** *CI-chondrite/solar abundance ratios as a function of 50% condensation temperatures. White symbols with bold labels are for elements with low uncertainties in the Sun. Gray shaded bar indicates agreement within 26%. Abundances are normalized to Mg.*

Restricting the comparison to elements with high quality solar analyses (Table 2), few elements remain as shown in Figure 13a. A "trend" could be interpreted into these data. However, the refractory Al is out of trend, and the meteoritic Y and Ca concentrations are from element ratios here, and not strictly from CI-chondrites. Given the uncertainties, the conclusion of a volatility trend for solar photospheric and CI-chondritic data is premature and strongly depends on the elements and their uncertainties selected for comparison. However, using photospheric data from Asplund et a. (2021), Jurewicz et al. (2024) found a trend of CI-chondritic/solar abundance ratios with condensation temperatures. They also argue for a similar trend for CI-chondrite/solar wind abundance ratios but did not compare solar wind/photospheric abundance ratios as a function of condensation temperatures. All these abundances ratios as functions of condensation temperatures are compared in Figure 13b, and 13c. Figure 13b may suggest depletions of refractory Ca and Al and enrichments of more volatile Na and K in CI-chondrites relative to solar wind, which seems to support the notion that CI-chondritic abundances are fractionated from the "solar" composition if solar is taken from either photosphere or solar wind. However, using the same arguments, the data in Figure 13c would then suggest that the solar wind is enriched in refractories over photospheric values, or that alkalis are enriched in the photosphere over those in the solar wind. A comparison of solar wind/photospheric ratios with condensation temperature is not discussed in Jurewicz et al. (2024), and FIP corrections would not resolve this issue. The correlations in Figures 13b and 13c appear stronger than in 13a, and assuming the arguments by Jurewicz et al. (2024) hold, the conclusion could be that solar wind abundances are volatility fractionated from the photosphere (Figure 13c). This seems implausible in the solar plasma.

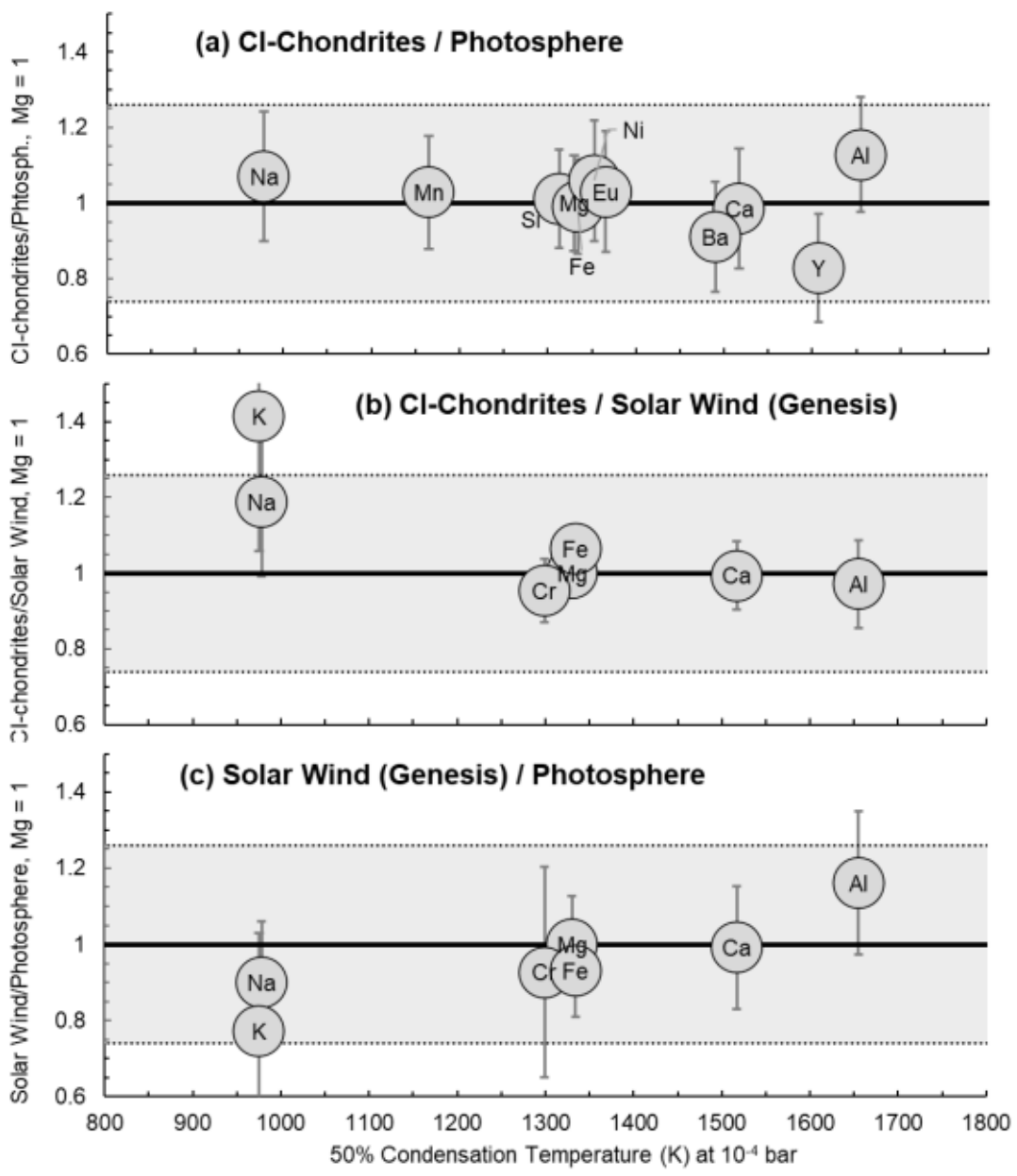

**Figure** ***13a-c***. *Abundance ratios versus 50% condensation temperatures at 0.1 mbar for (a) CI-chondrites/photosphere for top-rated solar analyses (Table 2; also Figure 12), (b) CI-chondrites/solar wind, and (c) solar wind/photospheric abundances. Solar wind data are from Jurewicz et al. (2024); photospheric and meteoritic abundances from this study. Gray shading indicates agreement within 26% (0.1 dex).*

## Isotope Systematics as Test for Abundances

One type of empirical abundance scaling and testing is to use isotopes mainly created by the slow neutron capture process (s-process) and to plot the product of their neutron cross-sections (σ) with their isotopic abundances ($\sigma N_S$) versus mass number (see comprehensive summaries in Käppler et al. 1989, Reifarth et al. 2014).

In the classical s-process model approach, the "$\sigma N_S$" of the s-process only nuclides give a smooth stepping curve with intervals of more or less constant $\sigma N_S$ which drop at mass numbers for nuclides with magic neutron numbers N = 50, 82, and 126. This is shown in Figure 14 for our data (Table A13).

The $\sigma N_S$ are relatively constant along the plateau for isotopes only formed by the s-process. This approach employs Maxwellian averaged neutron capture cross sections (MACS) values (e.g., from the Kadonis database) for a single thermal energy of $k_BT$ = 30 keV for a scenario of constant temperature and neutron density. Then, elemental abundances for e.g., Kr and Xe can be estimated by fitting desired isotopic abundances (e.g., s-only isotopes for Kr or Xe) to the $\sigma N_S$-curve and then scaling to elemental abundances using the isotopic composition.

The high mass range beyond the Ba region shows more scatter, and Os, and possibly Pt, seem somewhat low, however, these isotopic abundances are affected by radioactive decay or gain. The two s-process isotopes $^{176}$Lu and $^{176}$Hf are linked through decay of $^{176}$Lu, and their sum follows the linear trend going from $^{160}$Dy to $^{170}$Yb to $^{204}$Pb. In this region, the pure s-isotopes follow a little valley seated below the trend of isotopes with less than 70% s-process contributions (gray symbols). This and the log-scale for $\sigma N_S$ complicates abundance estimates and comparisons in that mass region. Furthermore, most of the “pure” s-process isotopes have small p-process contributions which are not subtracted from the overall observed isotopic abundances to obtain the “truly pure” s-process concentrations for in Figure 14.

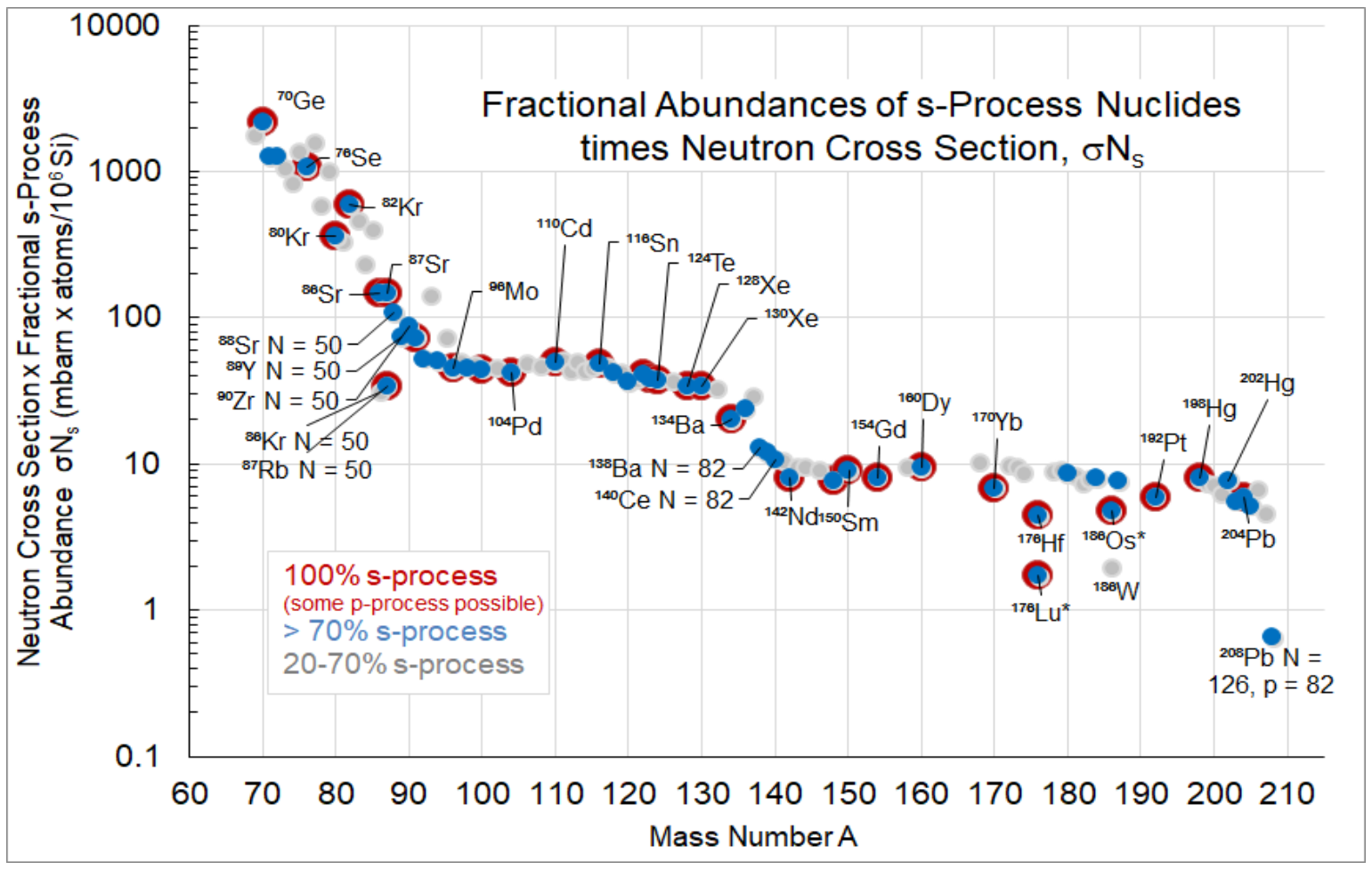

**Figure 14.** *The product of the neutron capture cross-sections with the isotopic abundances ($\sigma N_S$) versus mass number. “Pure” s-process produced isotopes are shown with red circles. Blue symbols without rim are for isotopes with more than 70% s-process contributions, grey ones for less than 70%. Similar diagrams are discussed in Anders and Grevesse (1989) and Palme and Beer (1993). See text.*

Reifarth et al. (2014) describe how this classical model is superseded by more sophisticated stellar models considering different conditions (temperatures, neutron fluences and densities in stars of different mass and metallicities). The classical model is useful for quickly estimating abundances in mass regions with about constant $\sigma N_s$ between magic neutron numbers (see Käppler et al. 1989, Reifarth et al. 2014).

Another test is the comparison of pure s-process isotopic abundances to those expected from independent models using nucleosynthetic network computations and galactic chemical evolution models (e.g., Arlandini et al. 1999, Reifarth et al. 2014). Here abundances of “pure” s-process elements are compared to results from Prantzos et al. (2020). In Figure 15 the horizontal line at unity is for perfect agreement, the dotted lines indicate agreement within 5%. The agreement for most isotopes is within 4%, except for Hg. Many of the isotopes show small excesses (1-4%) from the pure s-process abundances predicted by the model, and possibly lighter isotopes are slightly more enriched than heavy ones. These observed “excesses” are likely due to p-process contributions (Käppeler et al. 1989).

Figure 15 is also used to obtain Kr and Xe elemental abundances (Table 3) by interpolation of their “pure” s-process nuclide abundance ratios between those of their neighboring s-only nuclides. The abundances derived from the interpolated values are within 4% of the exact match to the model values. They are also following the curvature in the $\sigma N_S$ diagram (Figure 14). The lower Kr and Xe abundances by other authors (Table 3) only reach 70 or 84% of the model abundances.

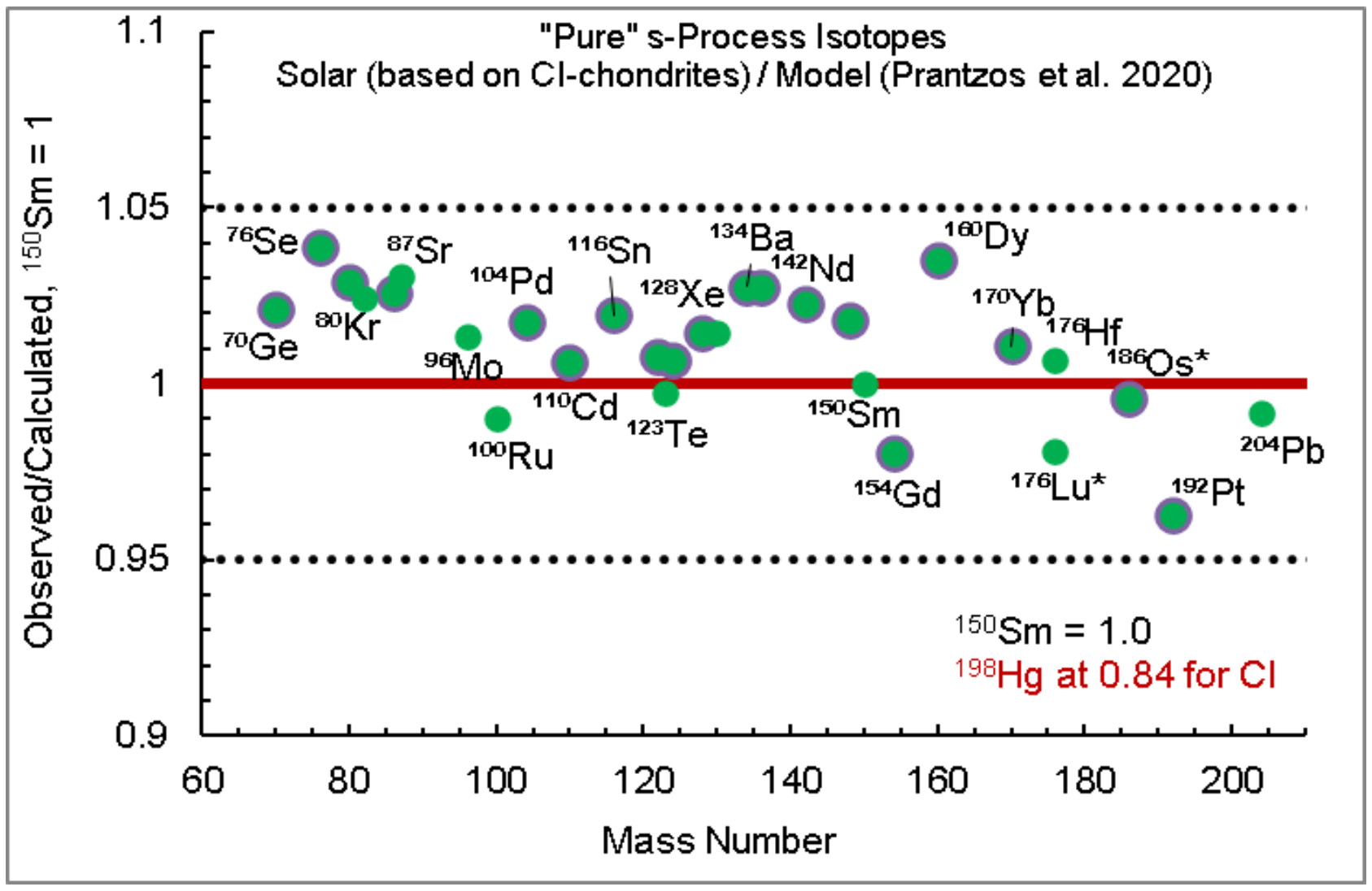

**Figure 15.** *Comparison of observed abundances of "pure" s-process isotopes abundances with model predictions by Prantzos et al. (2020). Isotopes with purple rims can have a few percent of p-process contributions. The abundances of Kr and Xe are fitted into the trends given by neighboring isotopes.*

In Figure 15, most elements (other than Kr, Xe) show an excess (ratio > 1) which is why we interpolated the Kr and Xe abundances to neighboring "pure" s-process isotopes instead of fixing the Kr and Xe abundances to the model values. For example, about 15% p-process contribution was estimated for $^{80}$Kr by Käppeler et al. (1989) and about 9% p-process contribution to $^{128}$Xe; here 3% and 4%, resp., are needed. The declining excesses for "s-process only" nuclide abundances over the model values with increasing mass numbers (Figure 15) mimic the declining p-process nuclide abundances in Figure 16. Figure 16 also shows the nominal fractional contribution from the p-process to the "pure" s-isotopes as defined from the excesses to the s-process model abundances by Prantzos et al. (2020) that are apparent in Figure 15.

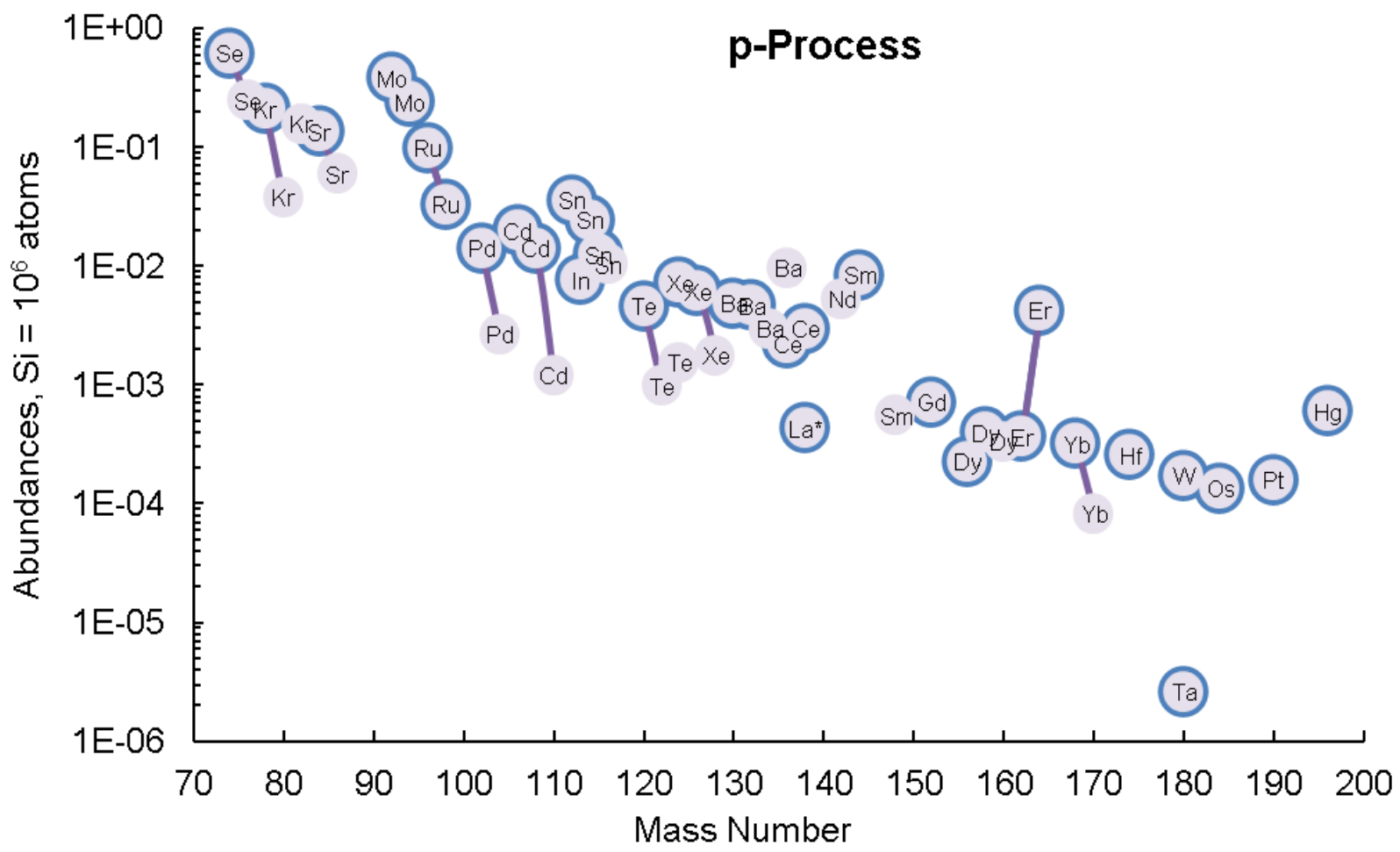

**Figure 16.** *Abundances of isotopes mainly made by the p-process. Symbols with solid rims indicate (almost) pure p-process isotopes. For other isotopes only the fractional isotopic abundances from*

*the observed excesses to the model abundances of "pure" s-process isotopes (seen in Figure 15) are plotted.*

## Conclusions and Outlook

We report on progress in establishing the single chemical composition, parental to the Sun and the planets in our solar system. This composition is established from two sources: (1) The composition of the Sun, (2) the composition of CI-chondrites. The increasing agreement of solar and meteoritic abundances with nucleosynthetic calculations demonstrate the unprocessed nature (in terms of elemental composition) of this material.

Solar photospheric abundance data substantially improved over the last two decades. Solar and meteoritic abundances agree within 25% for many condensable elements. Much work remains to be done to obtain full 3D NLTE results for many elements. Eleven elements have full 3D NLTE results and excellent agreement with CI-chondrites exists for the non-volatile elements among these. Until more laboratory and theoretical work bring improvements, uncertainties in solar abundances remain relatively large for many elements and it is too early to debate a possibly abnormal composition of the Sun. The solar metallicity, especially abundances of C, N, and O, is slightly higher than previously thought with the more conservative approach chosen here. This is consistent with other independent measurements, such as helioseismology and constraints from the solar neutrino experiments.

The use of solar wind data as potential abundance standards is currently limited to the noble gases and a few major elements and depends heavily on modeling how elemental abundances become fractionated when passing through the corona to become the solar wind. Suggested elemental fractionations between photospheric, solar wind, and meteoritic abundances with condensation temperatures are just apparent at this stage and should not be over-interpreted in light of existing uncertainties with the abundances in each regime.

The abundances of the refractory lithophile elements (Y, Zr, Hf, Nb, Ta, U, Th) are much improved. However, the apparent heterogeneous distribution of some of them (e.g., U, W) in CI-meteorites needs to be better understood. All siderophile element abundances in CI-chondrites would benefit from refined analyses, especially As, Sb, Mo, W, noble metals including Au. More isotope dilution analyses of multi-isotope elements would increase the accuracy of CI-elemental abundances.

New reliable halogen analyses are needed for more than one specimen of each CI-chondrite because the halogens are among the most aqueously mobile elements. Many elements with tri-bimodal and/or trailing concentration distributions are mobile elements and the mechanism and timing of their redistribution (parent body versus museum) needs to be better understood. The dissolving and reprecipitating phases are likely carbonates and phosphates; sulfates are mainly secondary precipitates after sulfide decomposition. The trace element distributions into accessory phases require further study to assess the scale of homogeneity for mobile elements for better estimates of required sample sizes for analyses. New analytical data of samples from the asteroids Ryugu and Bennu confirm that there is much more material of CI-composition than earlier thought. The analyses of these pristine asteroidal materials will help to better understand redistribution of aqueously mobile elements in CI-chondrite parent bodies.

## Acknowledgements

We thank ISSI for the invitations to an interesting workshop and for hosting the meeting. Work by KL is supported by NSF AST 2108172, NASA 80NSSC24K1284, and the McDonnell Center for the Space Sciences. MB is supported through the Lise Meitner grant from the Max Planck Society. This project has received funding from the European Research Council (ERC) under the European Union's Horizon 2020 research and innovation programme (Grant agreement No. 949173). We acknowledge support from the DAAD grant Project No.: 57654415. We thank the reviewers Andy Davis and Rainer Wieler and the editor Klaus Mezger for detailed comments that helped to improve the manuscript.

## Declarations

**Competing Interests:** The authors declare that they have no conflict of interest.

## References

Ahrens, L.H. (1970). The composition of stony meteorites (IX) abundance trends of the refractory elements in chondrites, basaltic achondrites and Apollo 11 fines. *Earth and Planetary Science Letters* 10,1-6.

Ahrens, L.H., Von Michaelis, H. and Fesq, H. (1969). The composition of the stony meteorites (IV) some analytical data on Orgueil, Nogoya, Ornans and Ngawi. *Earth and Planetary Science Letters* 6, 285-288.

Alexeeva, S. A. and Mashonkina, L. I. (2015). Carbon abundances of reference late-type stars from 1D analysis of atomic C I and molecular CH lines. *Monthly Notices of the Royal Astronomical Society* 453, 1619–1631.

Alexeeva, S., Ryabchikova, T., Mashonkina, L. and Hu, S. (2018). NLTE line formation for Mg I and Mg II in the atmospheres of B-A-F-G-K stars. *Astrophysical Journal* 866, 153.

Alfing, J., Patzek, M. and Bischoff, A. (2019). Modal abundances of coarse-grained (> 5 µm) components within CI-chondrites and their individual clasts–Mixing of various lithologies on the CI parent body(ies). *Geochemistry* 79, 125532.

Allende Prieto, C., Lambert, D.L. and Asplund, M. (2001). The forbidden abundance of oxygen in the Sun. *Astrophysical Journa*l 556, L63-L66.

Amarsi, A. M., Barklem, P. S., Collet, R., Grevesse, N. and Asplund, M. (2019). 3D non-LTE line formation of neutral carbon in the Sun. *Astronomy and Astrophysics* 624, A111.

Amarsi, A.M. and Asplund, M. (2017). The solar silicon abundance based on 3D non-LTE calculations. *Monthly Notices of the Royal Astronomical Society* 464, 264-273.

Amarsi, A.M., Grevesse, N., Grumer, J., Asplund, M., Barklem, P.S. and Collet, R. (2020). The 3D non-LTE solar nitrogen abundance from atomic lines. *Astronomy and Astrophysics* 636, A120.

Amarsi, A.M., Ogneva, D., Buldgen, G., Grevesse, N., Zhou, Y. and Barklem, P.S. (2024). The solar beryllium abundance revisited with 3D non-LTE models. *Astronomy and Astrophysics* 690, A128.

Anders E. (1971). How well do we know "Cosmic" abundances? *Geochimica et Cosmochimica Acta* 35, 516-522.

Anders E. and Ebihara M. (1982). Solar system abundances of the elements. *Geochimica et Cosmochimica Acta* 46, 2363–2380.

Anders E. and Grevesse N. (1989). Abundances of the elements— Meteoritic and solar. *Geochimica et Cosmochimica Acta* 53,197– 214.

Appel, S., Bagdasarian, Z., Basilico, D., Bellini, G., Benziger, J., Biondi, R., Caccianiga, B., Calaprice, F., Caminata, A., Cavalcante, P., and Chepurnov, A. (Borexino Collaboration) (2022). Improved measurement of solar neutrinos from the carbon-nitrogen-oxygen cycle by Borexino and its implications for the standard solar model. *Physical Review Letters* 129, 252701.

Asplund M., Amarsi, A. M. and Grevesse, N. (2021). The chemical make-up of the Sun: A 2020 vision. *Astronomy and Astrophysics* 653, A141.

Asplund M., Grevesse, N., Sauval, A. J. and Scott, P. (2009). The chemical composition of the Sun. *Annual Review of Astronomy and Astrophysics* 47, 481–522.

Ayres, T.R., Lyons, J.R., Ludwig, H.G., Caffau, E. and Wedemeyer-Böhm, S. (2013). Is the sun lighter than the earth? Isotopic CO in the photosphere, viewed through the lens of three-dimensional spectrum synthesis. *Astrophysical Journal* 765, A46.

Bahcall, J.N., Serenelli, A.M. and Pinsonneault, M. (2004). How accurately can we calculate the depth of the solar convective zone? *Astrophysical Journal* 614, 464-471.

Barklem, P.S. (2016). Accurate abundance analysis of late-type stars: advances in atomic physics. *Astronomy and Astrophysics Review* 24,1-54.

Barrat, J.A., Zanda B., Moynier F., Bollinger C., Lior Zou C. and Bayon G. (2012) Geochemistry of CI chondrites: Major and trace elements, and Cu and Zn Isotopes. *Geochimica et Cosmochimica Acta* 83, 79-92.

Basilico, D., Bellini, G., Benziger, J., Biondi, R., Caccianiga, B., Calaprice, F., Caminata, A., Chepurnov, A., D'Angelo, D., Derbin, A. and Di Giacinto, A. (Borexino Collaboration) (2023). Final results of Borexino on CNO solar neutrinos. *Physical Review* D 108, 102005.

Basu, S. and Antia, H. M. (2004). Constraining solar abundances using helioseismology. *Astrophysical Journal* 606, L85–L88.

Basu, S. and Antia, H. M. (2008). Helioseismology and Solar Abundances. *Physics Reports* 457, 217–283.

Bautista, M. (2000). Theoretical Calculations of Atomic Data for Spectroscopy. Atomic Data Needs for X-ray Astronomy. Editors: Bautista, Kallmann, Pradhan. NASA CP-2000-209968, p.25

Bautista, M.A., Bergemann, M., Gallego, H.C., Gamrath, S., Palmeri, P. and Quinet, P. (2022). Atomic radiative data for oxygen and nitrogen for solar photospheric studies. *Astronomy and Astrophysics* 665, A18.

Bedell, M., Bean, J. L., Meléndez, J., Spina, L., Ramírez, I., Asplund, M., Alves-Brito, A., dos Santos, L., Dreizler, S., Yong, D., Monroe, T. and Casagrande, L. (2018), The Chemical Homogeneity of Sun-like Stars in the Solar Neighborhood. *Astrophysical Journal* 865, id68.

Beer, H., Walter, G., Macklin, R. L. and Patchett, P. J. (1984). Neutron Capture Cross Sections and Solar System Abundances of 160,161Dy, 170, 171 Yb, 175,176 Lu and 176,177 Hf for the S-Process Analysis of the Radionuclide 176Lu. *Physical Review C* 30-2, 464-478.

Belyaev, A.K. and Voronov, Y. V. (2020). Inelastic Processes in Low-energy Sulfur-Hydrogen Collisions. *Astrophysical Journal* 893, 59.

Belyaev, A.K., Voronov, Y.V., Mitrushchenkov, A., Guitou, M. and Feautrier, N. (2019). Inelastic processes in oxygen–hydrogen collisions. *Monthly Notices of the Royal Astronomical Society* 487, 5097-5105.

Bergemann, M. and Cescutti, G. (2010). Chromium: NLTE abundances in metal-poor stars and nucleosynthesis in the Galaxy. *Astronomy and Astrophysics* 522, A9.

Bergemann, M. and Gehren, T. (2007). Formation of Mn I lines in the solar atmosphere. *Astronomy and Astrophysics* 473, 291-302.

Bergemann, M., Collet, R., Amarsi, A. M., Kovalev, M., Ruchti, G. and Magic, Z. (2017). Non-local thermodynamic equilibrium stellar spectroscopy with 1D and <3D> models. I. Methods and application to magnesium abundances in standard stars. *Astrophysical Journal* 847, A15.

Bergemann, M., Gallagher, A. J., Eitner, P., Bautista, M., Collet, R., Yakovleva, S. A., Mayriedl, A., Plez, B., Carlsson, M., Leenaarts, J., Belyaev, A. K. and Hansen, C. (2019). Observational constraints on the origin of the elements. I. 3D NLTE formation of Mn lines in late-type stars. *Astronomy and Astrophysics* 631, A80.

Bergemann, M., Hoppe, R., Semenova, E., Carlsson, M., Yakovleva, S.A., Voronov, Y.V., Bautista, M., Nemer, A., Belyaev, A.K., Leenaarts, J. and Mashonkina, L. (2021). Solar oxygen abundance. *Monthly Notices of the Royal Astronomical Society* 508, 2236-2253.

Bergemann, M., Kudritzki, R.-P., Würl, M., Plez, B., Davies, B. and Gazak, Z. (2013) Red Supergiant Stars as Cosmic Abundance Probes. II. NLTE Effects in J-band Silicon Lines. *Astrophysical Journal* 764, id115.

Bergemann, M., Lind, K., Collet, R., Magic, Z. and Asplund, M. (2012). Non-LTE line formation of Fe in late-type stars—I. Standard stars with 1D and< 3D> model atmospheres. *Monthly Notices of the Royal Astronomical Society* 427, 27-49.

Bergemann, M., Pickering, J. C. and Gehren, T. (2010) NLTE analysis of CoI/CoII lines in spectra of cool stars with new laboratory hyperfine splitting constants. *Monthly Notices of the Royal Astronomical Society* 401, 1334-1346.

Berzelius, J.J. (1834). Om Meteorstenar. *Handl. Svenska Vetenskaps-Akad.* 115–183. Also in *Ann. Physik (Poggendorff's)* 33, 1–32, and 113–148.

Boesgaard, A.M., Deliyannis, C.P. and Steinhauer, A. (2005). Boron depletion in F and G dwarf stars and the beryllium-boron correlation. *Astrophysical Journal* 621, 991-998.

Boesgaard, A.M., Lum, M.G., Deliyannis, C.P., King, J.R., Pinsonneault, M.H. and Somers, G. (2016). Boron abundances across the "Li-Be dip" in the Hyades cluster. *Astrophysical Journal* 830, 49.

Böhm-Vitense, E. (1958). Über die Wasserstoffkonvektionszone in Sternen verschiedener Effektivtemperaturen und Leuchtkräfte. *Zeitschrift für Astrophysik* 46, *108*-143.

Bouvier A., Vervoort J.D. and Patchett P.J. (2008). The Lu–Hf and Sm–Nd isotopic composition of CHUR: constraints from unequilibrated chondrites and implications for the bulk composition of terrestrial planets. Earth and Planetary Science Letters 273, 48–57.

Braukmüller N., Wombacher F., Hezel D.C., Escoube R. and Münker C. (2018). The chemical composition of carbonaceous chondrites: Implications for volatile element depletion, complementarity and alteration. *Geochimica et Cosmochimica Acta* 239, 17-48.

Burbidge, E.M., Burbidge, G.R., Fowler, W.A. and Hoyle, F. (1957). Synthesis of the elements in stars. *Rev. Mod. Phys.* 29, 547-650.

Burnett, D.S. and Genesis Science Team (2011). Solar composition from the Genesis Discovery Mission. *Proc. Natl. Acad. Sci.* 108, 19147-19151.

Caffau, E., Bonifacio, P., François, P., Spite, M., Spite, F., Zaggia, S., Ludwig, H.G., Steffen, M., Mashonkina, L., Monaco, L. and Sbordone, L. (2012). A primordial star in the heart of the Lion. *Astronomy and Astrophysics* 542, A51.

Caffau, E., Ludwig, H. G., Steffen, M., Freytag, B. and Bonifacio, P. (2011). Solar chemical abundances determined with a CO5BOLD 3D model atmosphere. *Solar Physics* 268, 255–269.

Caffau, E., Ludwig, H.-G., Bonifacio, P., Faraggiana, R., Steffen, M., Freytag, B., Kamp, I. and Ayres, T. R. (2010). The solar photospheric abundance of carbon. Analysis of atomic carbon lines with the CO5BOLD solar model. *Astronomy and Astrophysics* 514, A92.

Caffau, E., Ludwig, H.-G., Malherbe, J.-M., Bonifacio, P., Steffen, M. and Monaco, L. (2013). The photospheric solar oxygen project. II. Non-concordance of the oxygen abundance derived from two forbidden lines. *Astronomy and Astrophysics* 554, A126.

Caffau, E., Ludwig, H.G., Steffen, M., Ayres, T.R., Bonifacio, P., Cayrel, R., Freytag, B. and Plez, B. (2008). *Astronomy and Astrophysics* 488, 1031-1046.

Caffau, E., Ludwig, H.-G., Steffen, M., Livingston, W., Bonifacio, P., Malherbe, J.-M., Doerr, H.P. and Schmidt, W. (2015). The photospheric solar oxygen project. III. Investigation of the centre-to-limb variation of the 630 nm [O I]-Ni I blend. *Astronomy and Astrophysics* 579, A88.

Cameron, A. G. W. (1957). Nuclear reactions in stars and nucleogenesis. *Publications of the Astronomical Society of the Pacific* 69, 201–222.

Cameron, A.G.W. (1973). Abundances of the elements in the solar system. *Space Science Reviews* 15, 121–146.

Castelli, F. and Kurucz, R.L. (2003). Modelling of Stellar Atmospheres. IAU Symp. 210, ed. N. E. Piskunov, W. W. Weiss and D. F. Gray, PASP Conf. Ser. A20.

Centeno, R. and Socas-Navarro, H. (2008). A new approach to the solar oxygen abundance problem. *Astrophysical Journal* 682, L61.

Christensen-Dalsgaard, J. (2002). Helioseismology. *Reviews of Modern Physics* 74, 1073-1129.

Cloez, S. (1864a). Analyse chimique de la pierre météorique d'Orgueil. *Comptes Rendus de l'Académie des Sciences Paris* 59, 37-40.

Cloez, S. (1864b). Note sur la composition chimique de la pierre météorique d'Orgueil. *Comptes Rendus de l'Académie des Sciences Paris* 59, 986-988.

Cloez, S. (1864c). Dosage de l'acide carbonique contenue dans la météorite d'Orgueil. *Comptes Rendus de l'Académie des Sciences Paris* 59, 830–831.

Cubas Armas, M., Asensio Ramos, A. and Socas-Navarro, H. (2017). Uncertainties in the solar photospheric oxygen abundance. *Astronomy and Astrophysics* 600, A45.

Cubas Armas, M., Asensio Ramos, A. and Socas-Navarro, H. (2020). Spatially resolved measurements of the solar photospheric oxygen abundance. *Astronomy and Astrophysics* 643, A142.

Cunha, K. and Smith, V.V. (1999). A determination of the solar photospheric boron abundance. *Astrophysical Journal* 512, 1006-1013.

Deshmukh, S.A., Ludwig, H.G., Kučinskas, A., Steffen, M., Barklem, P.S., Caffau, E., Dobrovolskas, V. and Bonifacio, P. (2022). The solar photospheric silicon abundance according to CO5BOLD-Investigating line broadening, magnetic fields, and model effects. *Astronomy and Astrophysics* 668, A48.

Ebihara, M., Shinotsuka, K., Ozaki, H. and Oura, Y. (2000). Critical Evaluation of CI Chondrites as the Solar System Standard of Elemental Abundances. In Origin of elements in the solar system: implications of post-1957 observations. Edited by O. Manuel, Kluwer Academic/Plenum Publishers, pp. 289-300.

Ebihara, M., Wolf, R. and Anders, E. (1982). Are C1 chondrites chemically fractionated? A Trace Element Study. *Geochimica et Cosmochimica Acta* 46, 1849-1861.

Fegley, B. and Palme, H. (1985). Evidence for oxidizing conditions in the solar nebula from Mo and W depletions in refractory inclusions in carbonaceous chondrites. *Earth and Planetary Science Letters* 72 ,311-326.

Fegley, B. and Schaefer, L. (2010). Cosmochemistry. in Principles and Perspectives in Cosmochemistry 16, 347. doi:10.1007/978-3-642-10352-0_7

Fischer, D.A. and Valenti, J. (2005). The planet-metallicity correlation. *Astrophysical Journal* 622, 1102-1117.

Fischer-Gödde, M., Becker, H. and Wombacher, F. (2010). Rhodium, gold and other highly siderophile element abundances in chondritic meteorites. *Geochimica et Cosmochimica Acta* 74, 356–379.

Folinsbee, R.E., Douglas, J.A.V. and Maxwell, J.A. (1967). Revelstoke, a New Type I Carbonaceous Chondrite. *Geochimica et Cosmochimica Acta* 31, 1625-1635.

Fredriksson, K. and Kerridge, J.F. (1988). Carbonates and sulfates in CI chondrites: Formation by aqueous activity on the parent body. *Meteoritics* 23, 35-44.

Freytag, B., Steffen, M., Ludwig, H.G., Wedemeyer-Böhm, S., Schaffenberger, W. and Steiner, O. (2012). Simulations of stellar convection with CO5BOLD. *Journal of Computational Physics* 231, 919-959.

Gallagher, A.J., Bergemann, M., Collet, R., Plez, B., Leenaarts, J., Carlsson, M., Yakovleva, S.A. and Belyaev, A.K. (2020). Observational constraints on the origin of the elements-II. 3D non-LTE formation of Ba II lines in the solar atmosphere. *Astronomy and Astrophysics* 634, A55.

Garz, T. (1973). Absolute Oscillator Strengths of SI I Lines between 2500 A and 9000 A. *Astronomy and Astrophysics* 26, 47-477.

Gehren, T., Liang, Y.C., Shi, J.R., Zhang, H.W. and Zhao, G. (2004). Abundances of Na, Mg and Al in nearby metal-poor stars. *Astronomy and Astrophysics* 413, 1045-1063.

Gloeckler G., Cain, J., Ipavich, F.M., Tums, E.O., Bedini, P., Fisk, L.A., Zurbuchen, T.H., Bochsler P., Fischer, J., Wimmer-Schweingruber, R.F., Geiss, J. and Kallenbach, R. (1998). Investigation of the composition of solar and interstellar matter using solar wind and pickup ion measurements with SWICS and SWIMS on the ACE spacecraft. *Space Science Reviews* 86, 497-539.

Goldschmidt V.M. (1937). Geochemische Verteilungsgesetze der Elemente, IX. Die Mengenverhältnisse der Elemente und der Atom-Arten. *Skrifter Norske Videnskaps-Akad. Akademi I, Oslo,* Mathematisk-Naturvidenskapelig klasse No 4, 99-101.

Goles, G. (1970) Potassium. In: Handbook of elemental abundances in meteorites (Mason, B. ed.) Gordon and Breach Science Publ. NY, 555 pp.

Gooding, J.L. (1979). Petrogenetic properties of chondrules in unequilibrated H-, L-, and LL-group chondritic meteorites. Ph. D. Thesis, Univ. of New Mexico, Albuquerque, 392pp.

Göpel, C., Birck, J. L., Galy, A., Barrat, J. A. and Zanda, B. (2015). Mn–Cr systematics in primitive meteorites: Insights from mineral separation and partial dissolution. *Geochimica et Cosmochimica Acta* 156, 1-24.

Gounelle, M. and Zolensky, M.E. (2001). A terrestrial origin for sulfate veins in CI1 chondrites. *Meteoritics and Planetary Science* 36, 1321-1329.

Greshake A., Kloeck W., Arndt P., Maetz M., Flynn G.J., Bajt S. and Bischoff A. (1998) Heating experiments simulating atmospheric entry heating of micrometeorites: Clues to their parent body sources. *Meteoritics and Planetary Science* 33, 267-290.

Grevesse, N. and Sauval, A.J. (1998). Standard solar composition. *Space Science Reviews* 85, 161-174.

Grevesse, N., Scott, P., Asplund, M., Sauval, A.J. (2015) The elemental composition of the Sun. III. The heavy elements Cu to Th. *Astronomy and Astrophysics* 573, A27.

Gustafsson, B., Edvardsson, B., Eriksson, K., Jørgensen, U.G., Nordlund, Å. and Plez, B. (2008). A grid of MARCS model atmospheres for late-type stars. I. Methods and general properties. *Astronomy and Astrophysics* 486, A951.

Heays, A.N., Bosman, A.V. and Van Dishoeck, E.F. (2017). Photodissociation and photoionisation of atoms and molecules of astrophysical interest. *Astronomy and Astrophysics* 602, A105.

Heber V.S., Wieler, R., Baur, H., Olinger, C., Friedmann, T. A. and Burnett, D. S. (2009). Noble gas composition of the solar wind as collected by the Genesis mission. *Geochimica et Cosmochimica Acta* 73, 7414–7432.

Heber, V.S., Baur, H., Bochsler, P., McKeegan, K.D., Neugebauer, M., Reisenfeld, D.B., Wieler, R. and Wiens, R.C. (2012). Isotopic mass fractionation of solar wind: evidence from fast and slow solar wind collected by the Genesis mission. *Astrophysical Journal* 759, id121.

Heber, V.S., McKeegan, K.D., Steele, R.C., Jurewicz, A.J., Rieck, K.D., Guan, Y., Wieler, R. and Burnett, D.S. (2021). Elemental abundances of major elements in the solar wind as measured in Genesis targets and implications on solar wind fractionation. *Astrophysical Journal* 907, id15.

Hellmann, J.L., Van Orman, J.A. and Kleine, T. (2024). Hf-W isotope systematics of enstatite chondrites: Parent body chronology and origin of Hf-W fractionations among chondritic meteorites. *Earth and Planetary Science Letters* 626, 118518.

Holweger H. (2001). Photospheric abundances, problems, updates, implications. *AIP Conference Proceedings* 598, 23–30. College Park, MD: American Institute of Physics.

Holweger, H. (1977). The solar Na/Ca and S/Ca ratios: A close comparison with carbonaceous chondrites. *Earth and Planetary Science Letters* 34, 152-154.

Holweger, H. and Müller, E.A. (1974). The photospheric barium spectrum: solar abundance and collision broadening of Ba II lines by hydrogen. *Solar Physics* 39, 19-30.

Horan, M.F., Walker, R.J., Morgan, J.W., Grossman, J.N. and Rubin, A.E. (2003). Highly siderophile elements in chondrites. *Chemical Geology* 196, 5–20.

Hrodmarsson, H.R. and Van Dishoeck, E.F. (2023). Photodissociation and photoionization of molecules of astronomical interest-Updates to the Leiden photodissociation and photoionization cross section database. *Astronomy and Astrophysics* 675, A25.

Huss, G.R., Koeman-Shields, E., Jurewicz, A.J., Burnett, D.S., Nagashima, K., Ogliore, R. and Olinger, C.T. (2020). Hydrogen fluence in Genesis collectors: Implications for acceleration of solar wind and for solar metallicity. *Meteoritics and Planetary Science* 55, 326-351.

Ito, M., Tomioka, N., Uesugi, M., Yamaguchi, A., Shirai, N., Ohigashi, T., Liu, M.C., Greenwood, R.C., Kimura, M., Imae, N. and Uesugi, K. (2022). A pristine record of outer Solar System materials from asteroid Ryugu's returned sample. *Nature Astronomy* 6, 1163-1171.,

Jacobsen, S.B. and Wasserburg, G.J. (1984). Sm-Nd isotopic evolution of chondrites and achondrites, II. *Earth and Planetary Science Letters* 67, 137-150.

Jarosewich, E. (1990). Chemical analyses of meteorites: A compilation of stony and iron meteorite analyses. *Meteoritics* 25, 323-337.

Jurewicz, A.G.J., Amarsi, A.M., Burnett, D.S. and Grevesse, N. (2024). Differences in elemental abundances between CI chondrites and the solar photosphere. *Meteoritics and Planetary Science* 59, 3193-3214.

Kallemeyn G.W. and Wasson J.T. (1981) The compositional classification of chondrites: I. The carbonaceous chondrite groups. *Geochimica et Cosmochimica Acta* 45, 1217-1230.

Käppler, F., Beer, H. and Wisshak, K. (1989). S-process nucleosynthesis-nuclear physics and the classical model. *Reports on Progress in Physics* 52, 945.

King, A.J., Phillips, K.J.H., Strekopytov, S., Vita-Finzi, C. and Russell, S.S. (2020). Terrestrial modification of the Ivuna meteorite and a reassessment of the chemical composition of the CI type specimen. *Geochimica et Cosmochimica Acta* 268, 73-89.

Kleine, T., Mezger, K., Palme, H. and Münker, C. (2004). The W isotope evolution of the bulk silicate Earth: constraints on the timing and mechanisms of core formation and accretion. *Earth and Planetary Science Letters* 228, 109-123.

Klevas, J., Kučinskas, A., Steffen, M., Caffau, E. and Ludwig, H.G. (2016). Lithium spectral line formation in stellar atmospheres - The impact of convection and NLTE effects. *Astronomy and Astrophysics* 586, A156.

Kock, M. and Richter, J. (1968). Experimentelle Übergangswahrscheinlichkeiten und die solare Häufigkeit des Kupfers. *Zeitschrift für Astrophysik* 69, 180.

Korn, A. J., Shi, J. and Gehren, T. (2003). Kinetic equilibrium of iron in the atmospheres of cool stars. III. The ionization equilibrium of selected reference stars. *Astronomy and Astrophysics* 407, 691-703.

Korotin, S.A. (2009). The effects of deviations from LTE in sulphur lines for late-type stars. *Astronomy Reports* 53, 651-659.

Korotin, S.A. (2020). Non-LTE Effects in Rubidium Lines in Cool Stars. *Astronomy Letters 46(8)*, 541-549.

Krot, A.N., Keil, K., Scott, E.R.D., Goodrich, C.A. and Weisberg, M. K. (2014). Classification of meteorites and their genetic relationships. *Treatise on geochemistry. Meteorites and cosmochemical processes*, *1*, 1-63.

Krot, A.N., Fegley, B., Palme, H. and Lodders, K. (2000). Meteoritical and astrophysical constraints on the oxidation state of the solar nebula. In Protostars and Planets IV (eds. Mannings, V., Boss, A.P., Russell, S.S.) Univ. of Arizona Press, 1019-1054.

Kurucz, R. L., Furenlid, I., Brault, J. and Testerman, L. (1984). National Solar Observatory Atlas No. 1.: Solar Flux Atlas from 296 to 1300 nm. Tucson: NOAO.

Lauretta, D.S., Connolly Jr, H.C., Aebersold, J.E., Alexander, C.M.O.D., Ballouz, R.L., Barnes, J.J., Bates, H.C., Bennett, C.A., Blanche, L., Blumenfeld, E.H. and Clemett, S.J. (2024). Asteroid (101955) Bennu in the laboratory: Properties of the sample collected by OSIRIS-REx. *Meteoritics and Planetary Science* 59, 2453-2486.

Lawler, J.E., Sneden, C., Nave, G., Wood, M.P. and Cowan, J.J. (2019). Transition probabilities of Sc I and Sc II and Scandium abundances in the Sun. Arcturus, and HD 84937. *Astrophysical Journal Supplement Series*, 241, id21.

Lind, K., Bergemann, M. and Asplund, M. (2012). Non-LTE line formation of Fe in late-type stars—II. 1D spectroscopic stellar parameters. *Monthly Notices of the Royal Astronomical Society* 427, 50-60.

Liu, Y. P., Gao, C., Zeng, J. L. and Shi, J. R. 2011, Atomic data of Zn I for the investigation of element abundances. *Astronomy and Astrophysics* 536, A51.

Lodders, K. (2003). Solar system abundances and condensation temperatures of the elements. *Astrophysical Journal* 591, 1220–1247.

Lodders, K. (2020). Solar elemental abundances. In: Online Oxford Research Encyclopedia of Planetary Science. Oxford University Press, https://doi.org/10.1093/acrefore/9780190647926.013.145.

Lodders, K. (2021). Relative atomic solar system abundances, mass fractions, and atomic masses of the elements and their isotopes, composition of the solar photosphere, and compositions of the major chondritic meteorite groups. *Space Science Reviews* 217, A44.

Lodders, K. et al. (2025b).The Composition of CI-chondrites. in preparation.

Lodders, K. and Fegley, B. (2011). *Chemistry of the solar system*. Royal Society of Chemistry, (Cambridge, UK), pp. 476

Lodders, K. and Fegley, B. (2023). Solar system abundances and condensation temperatures of the halogens fluorine, chlorine, bromine, and iodine. *Geochemistry 83*(1), 125957.

Lodders, K., Fegley, B., Mezger, K. and Ebel, D.S. (2025a). Condensation and the "Volatility Trend" of the Earth. These proceedings.

Lodders, K., Palme H. and Gail, H.P. (2009). Abundances of the elements in the solar system. In Landolt-Börnstein, New Series, Vol. VI/4B, Chapter 4.4, J.E. Trümper ed., Berlin, Heidelberg, New York: Springer-Verlag, p. 560-630.

Lu, Y., Makishima, A. and Nakamura, E. (2007). Coprecipitation of Ti, Mo, Sn and Sb with fluorides and application to determination of B, Ti, Zr, Nb, Mo, Sn, Sb, Hf and Ta by ICP-MS. *Chemical Geology* 236(1-2), 13-26.

Maas, Z. G., Pilachowski, C. A. and Hinkle, K. (2016). Chlorine abundances in cool stars. *Astrophysical Journal* 152, id196.

Magg, E., Bergemann, M., Serenelli, A., Bautista, M., Plez, B., Heiter, U., Gerber, J.M., Ludwig, H.G., Basu, S., Ferguson, J.W. and Gallego, H.C., 2022. Observational constraints on the origin of the elements-IV. Standard composition of the Sun. *Astronomy and Astrophysics* 661, A140.

Mahaffy, P.R., Niemann, H.B., Alpert, A., Atreya, S.K., Demick, J., Donahue, T.M., Harpold, D.N. and Owen, T.C. (2000). Noble gas abundance and isotope ratios in the atmosphere of Jupiter from the Galileo Probe Mass Spectrometer. *Journal of Geophysical Research: Planets 105*(E6), 15061-15071.

Mashonkina, L. and Gehren, T., 2001. Heavy element abundances in cool dwarf stars: An implication for the evolution of the Galaxy. *Astronomy and Astrophysics* 376, 232-247.

Mashonkina, L., Gehren, T., Shi, J.R., Korn, A.J. and Grupp, F. (2011). A non-LTE study of neutral and singly-ionized iron line spectra in 1D models of the Sun and selected late-type stars. *Astronomy and Astrophysics* 528, A87.

Mashonkina, L., Ryabchikova, T., Alexeeva, S., Sitnova, T. and Zatsarinny, O. (2020). Chemical diversity among A-B stars with low rotational velocities: non-LTE abundance analysis. *Monthly Notices of the Royal Astronomical Society* 499, 3706–3719.

Mashonkina, L.I. and Romanovskaya, A.M. (2022). Scandium Abundance in F–G–K Stars in a Wide Metallicity Range. *Astronomy Letters* 48, 455-468.

Mason B. (1962) The carbonaceous chondrites. *Space Science Reviews* 1, 621-646.

Mason, B. and Taylor, S.R. (1982). Inclusions in the Allende meteorite. *Smithson. Contrib. Earth Science* 25, 1–30.

Mazor, E., Heymann, D. and Anders, E., (1970). Noble gases in carbonaceous chondrites. *Geochimica et Cosmochimica Acta* 34, 781-824.

McSween Jr, H.Y. and Richardson, S.M. (1977). The composition of carbonaceous chondrite matrix. *Geochimica et Cosmochimica Acta* 41, 1145-1161.

Meshik, A., Hohenberg, C., Pravdivtseva, O. and Burnett, D. (2014). Heavy noble gases in solar wind delivered by Genesis mission. *Geochimica et Cosmochimica Acta* 127, 326-347.

Meshik, A., Pravdivtseva, O. and Burnett, D. (2020). Refined composition of Solar Wind xenon delivered by Genesis NASA mission: Comparison with xenon captured by extraterrestrial regolith soils. *Geochimica et Cosmochimica Acta* 276, 289-298.

Morlok, A., Bischoff, A., Stephan, T., Floss, C., Zinner, E. and Jessberger, E.K. (2006). Brecciation and chemical heterogeneities of CI chondrites. *Geochimica et Cosmochimica Acta* 70, 5371-5394.

Münker C. et al. (2025). High field strength elements in chondrites and their components. *Geochemical Perspectives Letters*, submitted.

Münker, C., Pfänder, J. A., Weyer, S., Buchl, A., Kleine, T. and Mezger, K. (2003). Evolution of planetary cores and the Earth-Moon system from Nb/Ta systematics. *Science* 301, 84-87.

Nakamura, E., Kobayashi, K., Tanaka, R., Kunihiro, T., Kitagawa, H., Potiszil, C., Ota, T., et al. (2022). On the Origin and Evolution of the Asteroid Ryugu: A Comprehensive Geochemical Perspective. Proceedings of the Japanese Academy. Series B: Physical and Biological Sciences 98, 227–282.

Nissen, P.E. (2015). High-precision abundances of elements in solar twin stars-Trends with stellar age and elemental condensation temperature. *Astronomy and Astrophysics* 579, 52.

Nordlander, T. and Lind, K. (2017). Non-LTE aluminium abundances in late-type stars. *Astronomy and Astrophysics* 607, A75.

Nordlund, A. (1982). Numerical simulations of the solar granulation. I-Basic equations and methods. *Astronomy and Astrophysics* 107, 1-10.

Nordlund, Å., Stein, R.F. and Asplund, M. (2009). Solar surface convection. *Living Reviews in Solar Physics* 6, 1-117.

Osorio, Y., Barklem, P.S., Lind, K., Belyaev, A.K., Spielfiedel, A., Guitou, M. and Feautrier, N. (2015). Mg line formation in late-type stellar atmospheres-I. The model atom. *Astronomy and Astrophysics* 579, A53.

Palme, H. Larimer, J.W. and Lipschutz, M.E. (1988). Moderately volatile elements. In Meteorites and the Early Solar System (eds. J. F. Kerridge and M. S. Matthews), pp. 436-461.

Palme H., Hutcheon, I.D. and Spettel, B. (1994). Composition and origin of refractory-metal-rich assemblages in a Ca–Al-rich Allende inclusion. *Geochimica Cosmochimica Acta* 58, 495–513.

Palme H., Lodders K. and Jones A. (2014). Solar system abundances of the elements. In Planets, Asteroids, Comets and the Solar System; Treatise on Geochemistry (ed. A. M. Davis), pp. 15-36.

Palme, H. and Beer, H. (1993). Abundances of the elements in the solar system. In H. H. Voigt (Ed.), Landolt-Bornstein, Group VI (pp. 196–221). Astronomy and Astrophysics Vol. 3, Extension and Supplement to Vol. 2, Subvolume A. Berlin, Germany: Springer Verlag.

Palme, H. and Wlotzka, F. (1976). A metal particle from a Ca, Al-rich inclusion from the meteorite Allende, and the condensation of refractory siderophile elements. *Earth and Planetary Science Letters 33*, 45-60.

Palme, H. and Zipfel, J. (2022). The composition of CI chondrites and their contents of chlorine and bromine; Results from instrumental neutron activation analysis. *Meteoritics and Planetary Science* 57, 317–333.

Patzer, A., Pack, A. and Gerdes, A. (2010). Zirconium and hafnium in meteorites. *Meteoritics and Planetary Science 45*, 1136-1151.

Pearson, V.K., Sephton, M.A., Franchi, I.A., Gibson, J.M. and Gilmour, I., 2006. Carbon and nitrogen in carbonaceous chondrites: Elemental abundances and stable isotopic compositions. *Meteoritics and Planetary Science* 41, 1899-1918.

Pehlivan Rhodin, A. (2018). Experimental and computational atomic spectroscopy for astrophysics : Oscillator strengths and lifetimes for Mg I, Si I, Si II, Sc I, and Sc II, Ph.D. Thesis, Lund University, Sweden.

Pehlivan Rhodin, A., Hartman, H., Nilsson, H. and Jönsson, P. (2024). Accurate and experimentally validated transition data for Si I and Si II. *Astronomy and Astrophysics* 682, A184.

Pepin, R. O., Schlutter, D. J., Becker, R. H. and Reisenfeld, D. B. (2012). Helium, neon, and argon composition of the solar wind as recorded in gold and other Genesis collector materials. *Geochimica et Cosmochimica Acta* 89, 62–80.

Piersanti, L., Straniero, O. and Cristallo, S. (2007). A method to derive the absolute composition of the Sun, the solar system, and the stars. *Astronomy and Astrophysics* 462, 1051-1062.

Pietrow, A. G. M., Kiselman, D., Andriienko, O., Petit dit de la Roche, D. J. M., Díaz Baso, C. J., & Calvo, F. (2023a), Center-to-limb variation of spectral lines and continua observed with SST/CRISP and SST/CHROMIS, Astronomy and Astrophysics, 671, A130.

Pietrow, A.G., Hoppe, R., Bergemann, M. and Calvo, F. (2023b). Solar oxygen abundance using SST/CRISP center-to-limb observations of the O I 7772 Å line. *Astronomy and Astrophysics* 672, L6.

Pilleri, P., Reisenfeld, D.B., Zurbuchen, T.H., Lepri, S.T., Shearer, P., Gilbert, J.A., von Steiger, R. and Wiens, R.C. (2015). Variations in solar wind fractionation as seen by ACE/SWICS and the implications for GENESIS mission results. *Astrophysical Journal* 812, id.1.

Popa, S.A., Hoppe, R., Bergemann, M., Hansen, C.J., Plez, B. and Beers, T.C. (2023). Non-local thermodynamic equilibrium analysis of the methylidyne radical molecular lines in metal-poor stellar atmospheres. *Astronomy and Astrophysics* 670, A25.

Pourmand A., Dauphas N. and Ireland T. J. (2012). A novel extraction chromatography and MC-ICP-MS technique for rapid analysis of REE, Sc and Y: revising CI-chondrite and Post-Archean Australian Shale (PAAS, abundances. *Chemical Geology* 291, 38–54.

Prantzos, N., Abia, C., Cristallo, S., Limongi, M. and Chieffi, A. (2020). Chemical evolution with rotating massive star yields II. A new assessment of the solar s-and r-process components. *Monthly Notices of the Royal Astronomical Society* 491, 1832-1850.

Rauscher, T., Heger, A., Hoffman, R.D. and Woosley, S.E. (2002). Nucleosynthesis in massive stars with improved nuclear and stellar physics. *Astrophysical Journal* 576(1), 323-348.

Reggiani, H., Amarsi, A.M., Lind, K., Barklem, P.S., Zatsarinny, O., Bartschat, K., Fursa, D.V., Bray, I., Spina, L. and Meléndez, J., 2019. Non-LTE analysis of KI in late-type stars. *Astronomy and Astrophysics* 627, A177.

Reifarth, R., Heil, M., Käppeler, F., Voss, F., Wisshak, K., Becvář, F., Krticka, M., Gallino, R. and Nagai, Y. (2002). Stellar neutron capture cross sections of 128, 129, 130 Xe. *Physical Review C 66*, p.064603.

Reifarth, R., Lederer, C. and Käppeler, F. (2014). Neutron reactions in astrophysics. *Journal of Physics G: Nuclear and Particle Physics 41*, p.053101.

Reiners, A., Mrotzek, N., Lemke, U., Hinrichs, J., and Reinsch, K. (2016). The IAG solar flux atlas: Accurate wavelengths and absolute convective blueshift in standard solar spectra. *Astronomy and Astrophysics* 587, A65.

Reisenfeld, D.B., Burnett, D.S., Becker, R.H., Grimberg, A. G., Heber, V.S., Hohenberg, C. M., Jurewicz, A.J. G., Meshik, A., Pepin, R. O., Raines, J. M., Schlutter, D.J., Wieler R., Wiens, R.C. and Zurbuchen, T. H. (2007). Elemental abundances of the bulk solar wind: Analyses from Genesis and ACE. *Space Science Reviews* 130, 79–86.

Ritchey, A.M., Jenkins, E.B., Shull, J.M., Savage, B.D., Federman, S.R. and Lambert, D.L. (2023). The Distribution of Metallicities in the Local Galactic Interstellar Medium. *Astrophysical Journal* 952, id57.

Rocholl, A. and Jochum, K.P. (1993). Th, U and other trace elements in carbonaceous chondrites: implications for the terrestrial and solar-system Th/U ratios. *Earth Planetary Science Letters* 117, 265–278.

Roederer, I.U. and Lawler, J.E. (2012). Detection of elements at all three r-process peaks in the metal-poor star HD 160617. *Astrophysical Journal 750*(1), id76.

Russell H.N. (1929). On the composition of the sun's atmosphere. *Astrophysical Journal* 70, 11-28.

Ryabchikova, T., Piskunov, N. and Pakhomov, Y. (2022). Using Molecular Lines to Determine Carbon and Nitrogen Abundances in the Atmospheres of Cool Stars. *Atoms 10*, p.103.

Schmitt R.A., Goles G. G., Smith R.H. and Osborn T. W. (1972). Elemental abundances in stone meteorites. *Meteoritics* 7, 131-214.

Scott, D.W. (1979). On optimal and data-based histograms; *Biometrika* 66, 605-610.

Scott, D.W. (2009). Sturges' rule. WIREs Wiley's Interdisciplinary Reviews Comp Stat, 1, 303-306. https://doi.org/10.1002/wics.35.

Scott, E.R.D. and Krot, A.N. (2014). Chondrites and their components. *Treatise on geochemistry. Meteorites and cosmochemical processes*. Vol. *1*, pp. 65-137.

Scott, P., Asplund, M., Grevesse, N., Bergemann, M. and Sauval, A.J. (2015a). The elemental composition of the Sun. II. The iron group elements Sc to Ni. *Astronomy and Astrophysics* 573, A26.

Scott, P., Grevesse, N., Asplund, M., Sauval, A.J., Lind, K., Takeda, Y., Collet, R., Trampedach, R. and Hayek, W. (2015b). The elemental composition of the Sun. I. The intermediate mass elements Na to Ca. *Astronomy and Astrophysics* 573, A25.

Serenelli, A.M., Basu, S., Ferguson, J.W. and Asplund, M. (2009). New solar composition: the problem with solar models revisited. *Astrophysical Journal 705*, p.L123.

Shi, J.R., Gehren, T., Zeng, J.L., Mashonkina, L. and Zhao, G. (2014). Statistical equilibrium of copper in the solar atmosphere. *Astrophysical Journal 782*, id80.

Shirai, N., Ito, M., Yamaguchi, A., Tomioka, N., Uesugi, M., Imae, N., Kimura, M., Greenwood, R., Liu, M.C., Ohigashi, T. and Sekimoto, S. (2024). Mercury (Hg) in Ryugu particles and implications for the origin of volatile elements in early Earth. Preprint https://doi.org/10.21203/rs.3.rs-4002901/v1

Sitnova, T., Zhao, G., Mashonkina, L., Chen, Y., Liu, F., Pakhomov, Y., Tan, K., Bolte, M., Alexeeva, S., Grupp, F. and Shi, J.R. (2015). Systematic non-LTE study of the $-2.6 \leq [Fe/H] \leq 0.2$ F and G dwarfs in the solar neighborhood. I. Stellar atmosphere parameters. *Astrophysical Journal* 808(2), 148.

Socas-Navarro, H. (2015). The solar oxygen abundance from an empirical three-dimensional model. *Astronomy and Astrophysics* 577, id25.

Spruit, H.C., Nordlund, A. and Title, A.M. (1990). Solar convection. *Annual Review of Astronomy and Astrophysics* 28, 263–301. doi:10.1146/annurev.aa.28.090190.001403.

Steffen, M., Prakapavičius, D., Caffau, E., Ludwig, H.G., Bonifacio, P., Cayrel, R., Kučinskas, A. and Livingston, W.C. (2015). The photospheric solar oxygen project-IV. 3D-NLTE investigation of the 777 nm triplet lines. *Astronomy and Astrophysics 583*, A57.

Storm, N., Barklem, P.S., Yakovleva, S.A., Belyaev, A.K., Palmeri, P., Quinet, P., Lodders, K., Bergemann, M. and Hoppe, R. (2024). 3D NLTE modelling of Y and Eu-Centre-to-limb variation and solar abundances. *Astronomy and Astrophysics 683*, A200.

Stracke, A., Palme, H., Gellissen, M., Münker, C., Kleine, T., Birbaum, K., Günther, D., Bourdon, B. and Zipfel, J. 2012. Refractory element fractionation in the Allende meteorite: Implications for solar nebula condensation and the chondritic composition of planetary bodies. *Geochimica et Cosmochimica Acta*, *85*, 114-141.

Stracke, A., Scherer, E.E. and Reynolds, B.C. (2014). Application of Isotope Dilution in Geochemistry. Treatise on Geochemistry: Second Edition. 15. 71-86. 10.1016/B978-0-08-095975-7.01404-2.

Suess, H. E. (1947a). Über kosmische Kernhäufigkeiten. I. Mitteilung, Einige Häufigkeitsregeln und ihre Anwendung bei der Abschätzung der Häufigkeitswerte für die mittelschweren und schweren Elemente. *Zeitschrift für Naturforschung* 2a, 311–321.

Suess, H. E. (1947b). Über kosmische Kernhäufigkeiten II. Mitteilung, Einzelheiten in der Häufigkeitsverteilung der mittelschweren und schweren Kerne. *Zeitschrift für Naturforschung* 2a, 604–608.

Suess, H. E. and Urey H. C. (1956). Abundances of the elements. *Reviews in Modern Physics* 28, 53–74.

Tachiev, G.I. and Fischer, C.F. (2002). Breit-Pauli energy levels and transition rates for nitrogen-like and oxygen-like sequences. *Astronomy and Astrophysics* 385(2), 716-723.

Thenard, L.J. (1806). Analyse d'un aérolithe tombée dans l'arrondissement d'Alais, le 15 mars 1806. *Ann. Chim. et Phys.* 59, 103–110.

Turcotte, S., Richter, J., Michaud, G., Iglesia, C. A. and Rogers, F. J. (1998). Consistent solar evolution model including diffusion and radiative acceleration effects. *Astrophysical Journal* 504, 539–558.

Turcotte, S. and Wimmer-Schweingruber, R. F. (2002). Possible in situ tests of the evolution of elemental and isotopic abundances in the solar convection zone. *Journal of Geophysical Research* 107(A12), 1442.

Vernazza, J. E., Avrett, E. H. and Loeser, R. 1981, Structure of the solar chromosphere. III. Models of the EUV brightness components of the quiet sun. *Astrophysical Journal Supplement Series* 45, 635.

Vogel, N., Heber, Veronika, V.S., Baur, H., Burnett, D. S. and Wieler, R. (2011). Argon, krypton, and xenon in the bulk solar wind as collected by the Genesis mission. Geochimica et Cosmochimica Acta 75, 3057–3071.

Vögler, A., Shelyag, S., Schüssler, M., Cattaneo, F., Emonet, T. and Linde, T. (2005). Simulations of magneto-convection in the solar photosphere-Equations, methods, and results of the MURaM code. *Astronomy and Astrophysics 429*, 335-351.

Wang, M., Huang, W. J., Kondev, F. G., Audi, G., and Naimi, S. (2021). The AME 2020 atomic mass evaluation (II). Tables, graphs and references. *Chinese Physics* C, 45(3) #030003 (512 pp).

Wiik, H. B. (1969). On the regular discontinuities in the composition of meteorites. *Commentationes physico-mathematicae societatis scientiarum Fennicae* 34,135–145.

Wing, J. (1964) Simultaneous Determination of Oxygen and Silicon in Meteorites and Rocks by Nondestructive Activation Analysis with Fast Neutrons. *Anal. Chemistry* 36, 559-564.

Wolf, D. and Palme, H. (2001, The solar system abundances of phosphorus and titanium and the nebular volatility of phosphorus. *Meteoritics and Planetary Science* 36, 559-571.

Yakovleva, S. A., Belyaev, A. K. and Bergemann, M. (2020). Cobalt-Hydrogen Atomic and Ionic Collisional Data. *Atoms* 8, 34.

Yang, W. (2019). Rotating solar models with low metal abundances as good as those with high metal abundances. *Astrophysical Journal* 873, id18.

Yin, Q., Jacobsen, S. B., Yamashita, K., Blichert-Toft, J., Télouk, P. and Albarede, F. (2002). A short timescale for terrestrial planet formation from Hf–W chronometry of meteorites. *Nature*, *418*, 949-952.

Yokoyama, T., Dauphas, N., Fukai, R., Usui, T., Tachibana, S., Schönbächler, M., Busemann, H., Abe, M. and Yada, T. (2024). The Chemical Composition of Ryugu: Prospects as a Reference Material for Solar System Composition. *arXiv preprint arXiv:2405.04500*.https://doi.org/10.1126/science. abn7850.

Yokoyama, T., Nagashima, K., Hakai, I., Young, E. D., Abe, Y., Aleon, J., Alexander, C. M. O'D., et al. (2023). Samples Returned from the Asteroid Ryugu are Similar to Ivuna-Type Carbonaceous Chondrites. *Science* 379: eabn7850.

Young, P. R. (2018). Element abundance ratios in the quiet Sun transition region. *Astrophysical Journal* 855, id15.

Zhang, H.W., Gehren, T. and Zhao, G. (2008). A non-local thermodynamic equilibrium study of scandium in the Sun. *Astronomy and Astrophysics 481*, 489-497.

Zhang, H.W., Gehren, T., Butler, K., Shi, J.R. and Zhao, G. (2006). Potassium abundances in nearby metal-poor stars. *Astronomy and Astrophysics 457*, 645-650.

Zhao, G., Mashonkina, L., Yan, H.L., Alexeeva, S., Kobayashi, C., Pakhomov, Y., Shi, J.R., Sitnova, T., Tan, K.F., Zhang, H.W. and Zhang, J.B. (2016). Systematic non-LTE study of the $-2.6 \leq [Fe/H] \leq 0.2$ F and G dwarfs in the solar neighborhood. II. Abundance patterns from Li to Eu. *Astrophysical Journal* 833, 225.

# Tables

## Table 1

**Table 1. Sources for Solar System Abundances**

| Abundances from: | Limitations: |
|---|---|
| Sun: Photosphere & Sunspots<br><br>representative of present-day solar convection envelope (CE) | 68 of 83 natural occurring elements were analyzed, about 10-20 elements with nominal uncertainties <10%. Limitations are line accessibilities, atomic parameters and transition probabilities. Model atmosphere (1D vs 3D) and choices between local thermodynamic equilibrium (LTE) vs. non-LTE (NLTE) are required. The best solar system source for abundances of C, N, O are currently debated. No direct method for noble gases is available. To obtain proto-solar values from the present-day convective envelope (CE), corrections for atomic diffusion (gravitational settling and radiative acceleration) corrections are needed. |
| Sun: Corona/solar wind (SW) from GENESIS<br><br>representative of FIP/FIT biased photospheric values | Genesis provides direct measurements of all noble gases, but other) data are limited to abundant elements. To derive photospheric values first ionization potential (FIP) & first ionization time (FIT) corrections from fast and slow SW, and solar energetic particles are required. Coronal sources corrected for FIP/FIT bias from photosphere need further settling and diffusion corrections to obtain proto-solar values (see Lodders 2020 for a review). |
| CI-Chondrites<br><br>representative of proto-solar condensable fraction | All elements are measurable. Ultra volatile elements (H,C,N,O, noble gases) are strongly depleted in CI-chondrites. Elements are usually determined with 3-10% relative uncertainties. This requires representative sampling. Limited available material may be a problem. |
| Other Sources<br><br>indirect and/or model-dependent | He abundance can be calculated from helioseismology, Ne from O/Ne of solar wind and B stars, Ar, Kr, Xe from nucleosynthesis systematics and abundance curve interpolations. Additional data are provided by B stars, the ISM (interstellar medium), and gas-giant planets in the solar system. |

# Table 2

**Table 2. Solar photospheric abundances**

| | This Work | | | | | Asplund et al. (2021) | | |
|---|---|---|---|---|---|---|---|---|
| Element | 12+log N(E/H) dex | ±1σ dex | 1σ% | Quality Index | Notes * | 12+log N(E/H) dex | ±1σ dex | Difference (1) - (6) |
| E | (1) | (2) | (3) | (4) | (5) | (6) | (7) | (8) |
| H | 12 | 0.004 | 0.9 | NA | … | 12.00 | 0.00 | 0.00 |
| He | 10.922 | 0.012 | 2.8 | NA | see text | 10.91 | 0.013 | 0.01 |
| Li | 1.04 | 0.09 | 23 | B | see text | 0.96 | 0.06 | 0.08 |
| Be | 1.21 | 0.14 | 38 | C | A24, U | 1.38 | 0.09 | 0.00 |
| B | 2.7 | 0.25 | 78 | E | CS99, U | 2.70 | 0.20 | 0.00 |
| C | 8.51 | 0.09 | 23 | B | see text | 8.46 | 0.04 | 0.05 |
| N | 7.98 | 0.11 | 29 | B- | see text?? | 7.83 | 0.07 | 0.15 |
| O | 8.76 | 0.05 | 12 | A | see text | 8.69 | 0.04 | 0.07 |
| F | 4.4 | 0.2 | 58 | D | sunspot; s. text | 4.40 | 0.25 | 0.00 |
| Ne | 8.15 | 0.12 | 32 | D | see text | 8.06 | 0.05 | 0.09 |
| Na | 6.29 | 0.05 | 12 | A | see text | 6.22 | 0.03 | 0.07 |
| Mg | 7.58 | 0.05 | 12 | A | see text | 7.55 | 0.03 | 0.03 |
| Al | 6.43 | 0.05 | 12 | A | see text | 6.43 | 0.03 | 0.00 |
| Si | 7.56 | 0.05 | 12 | A | see text | 7.51 | 0.03 | 0.05 |
| P | 5.44 | 0.12 | 32 | C+ | see text | 5.41 | 0.03 | 0.03 |
| S | 7.16 | 0.11 | 29 | B- | see text | 7.12 | 0.03 | 0.04 |
| Cl | 5.43 | 0.2 | 58 | D | sunspot; s. text | 5.31 | 0.20 | 0.12 |
| Ar | 6.5 | 0.12 | 32 | D | see text | 6.38 | 0.10 | 0.12 |
| K | 5.09 | 0.09 | 23 | B | see text | 5.07 | 0.03 | 0.02 |
| Ca | 6.35 | 0.06 | 15 | A- | see text | 6.30 | 0.03 | 0.05 |
| Sc | 3.13 | 0.11 | 29 | B- | see text | 3.14 | 0.04 | -0.01 |
| Ti | 4.97 | 0.11 | 29 | B- | see text | 4.97 | 0.05 | 0.00 |
| V | 3.89 | 0.16 | 45 | C- | see text | 3.90 | 0.08 | -0.01 |
| Cr | 5.74 | 0.11 | 29 | B- | see text | 5.62 | 0.04 | 0.12 |
| Mn | 5.52 | 0.05 | 12 | A | see text | 5.42 | 0.06 | 0.10 |
| Fe | 7.51 | 0.05 | 12 | A | see text | 7.46 | 0.04 | 0.05 |
| Co | 4.95 | 0.11 | 29 | B- | see text | 4.94 | 0.05 | 0.01 |
| Ni | 6.24 | 0.06 | 15 | A- | see text | 6.20 | 0.04 | 0.04 |
| Cu | 4.24 | 0.11 | 29 | B- | see text | 4.18 | 0.05 | 0.06 |
| Zn | 4.55 | 0.11 | 29 | B- | see text | 4.56 | 0.05 | -0.01 |
| Ga | 3.02 | 0.14 | 38 | C | A21 | 3.02 | 0.05 | 0.00 |
| Ge | 3.62 | 0.14 | 38 | C | A21 | 3.62 | 0.10 | 0.00 |
| As | … | ... | | … | see G15 | … | … | … |
| Se | … | ... | | … | … | … | … | … |
| Br | … | ... | | … | … | … | … | … |
| Kr | 3.31 | 0.12 | 3table 42 | D | see text | 3.12 | 0.1 | 0.19 |
| Rb | 2.35 | 0.11 | 29 | B- | see text | 2.32 | 0.08 | 0.03 |
| Sr | 2.93 | 0.11 | 29 | B- | see text | 2.83 | 0.06 | 0.10 |
| Y | 2.3 | 0.06 | 15 | A- | see text | 2.21 | 0.05 | 0.09 |
| Zr | 2.68 | 0.11 | 29 | B- | see text | 2.59 | 0.04 | 0.09 |
| Nb | 1.47 | 0.14 | 38 | C | A21, U | 1.47 | 0.06 | 0.00 |
| Mo | 1.88 | 0.16 | 45 | C- | A21, U | 1.88 | 0.09 | 0.00 |
| Ru | 1.75 | 0.16 | 45 | C- | A21, U | 1.75 | 0.08 | 0.00 |
| Rh | 0.78 | 0.16 | 45 | C- | A21, U | 0.78 | 0.11 | 0.00 |
| Pd | 1.57 | 0.16 | 45 | C- | A21, U | 1.57 | 0.1 | 0.00 |
| Ag | 0.96 | 0.16 | 45 | C- | A21, U | 0.96 | 0.1 | 0.00 |
| Cd | 1.77 | 0.2 | 58 | D | G15, U, s. text | … | … | … |
| In | 0.8 | 0.2 | 58 | D | Sunspot, s. text | 0.80 | 0.2 | 0.00 |
| Sn | 2.02 | 0.16 | 45 | C- | A21, U | 2.02 | 0.1 | 0.00 |
| Sb | … | ... | | … | … | … | … | … |
| Te | … | ... | | … | … | … | … | … |
| I | … | ... | | … | … | … | … | … |
| Xe | 2.3 | 0.12 | 32 | D | see text | 2.22 | 0.05 | 0.08 |
| Cs | … | ... | | … | … | … | … | … |
| Ba | 2.27 | 0.06 | 15 | A- | see text | 2.27 | 0.05 | 0.00 |
| La | 1.1 | 0.16 | 45 | C- | see text | 1.11 | 0.04 | -0.01 |
| Ce | 1.58 | 0.16 | 45 | C- | G15, U | 1.58 | 0.04 | 0.00 |
| Pr | 0.75 | 0.11 | 29 | B- | see text | 0.75 | 0.05 | 0.00 |
| Nd | 1.42 | 0.16 | 45 | C- | G15, U | 1.42 | 0.04 | 0.00 |
| Sm | 0.95 | 0.16 | 45 | C- | G15, U | 0.95 | 0.04 | 0.00 |
| Eu | 0.57 | 0.06 | 15 | A- | see text | 0.52 | 0.04 | 0.05 |
| Gd | 1.08 | 0.16 | 45 | C- | G15, U | 1.08 | 0.04 | 0.00 |
| Tb | 0.31 | 0.16 | 45 | C- | G15, U | 0.31 | 0.1 | 0.00 |
| Dy | 1.1 | 0.16 | 45 | C- | G15, U | 1.10 | 0.04 | 0.00 |
| Ho | 0.48 | 0.16 | 45 | C- | G15, U | 0.48 | 0.11 | 0.00 |
| Er | 0.93 | 0.16 | 45 | C- | G15, U | 0.93 | 0.05 | 0.00 |
| Tm | 0.11 | 0.16 | 45 | C- | G15, U | 0.11 | 0.04 | 0.00 |
| Yb | 0.85 | 0.16 | 45 | C- | G15, U | 0.85 | 0.11 | 0.00 |
| Lu | 0.1 | 0.16 | 45 | C- | G15, U | 0.10 | 0.09 | 0.00 |
| Hf | 0.86 | 0.12 | 32 | C+ | see text | 0.85 | 0.05 | 0.01 |
| Ta | … | ... | | … | … | … | … | … |
| W | 0.79 | 0.2 | 58 | D | A21, U | 0.79 | 0.11 | 0.00 |
| Re | … | ... | | … | … | … | … | … |
| Os | 1.36 | 0.14 | 38 | C | see text | 1.35 | 0.12 | 0.01 |
| Ir | 1.42 | 0.2 | 58 | E | G15, U, s. text | … | … | … |
| Pt | … | ... | | … | … | … | … | … |
| Au | 0.91 | 0.2 | 58 | D | A21, U | 0.91 | 0.12 | 0.00 |
| Hg | … | ... | | … | … | … | … | … |
| Tl | 0.95 | 0.2 | 58 | D | Sunspot, s. text | 0.92 | 0.17 | 0.03 |
| Pb | 1.95 | 0.2 | 58 | D | A21, U | 1.95 | 0.08 | 0.00 |
| Bi | … | ... | | … | … | … | … | … |
| Th | [0.08] | [0.03] | | ** | See footnote | 0.03 | 0.1 | … |
| U | ... | ... | | … | … | … | … | … |

* Notes: A21: Asplund et al. 2021. A24: Amarsi et al. 2024. CS99: Cunha and Smith 1999. G15: Grevesse et al. 2015. U: uncertainty changed from value given in A21, A24, CS99, or G15. ** Th from Caffau et al. (2011). This 3D value is based on a weak and blended line at 401.9 nm, and is only listed for reference here.

## Table 3

**Table 3. Noble Gas Abundances**

| | Solar A(X), dex Recommended | Solar A(X), dex other authors | Reference | Bulk Solar Wind, A(X) dex | Reference |
|---|---|---|---|---|---|
| **He** | 10.922 ± 0.012 | 10.914 ± 0.013<br>10.924<br>10.914 ± 0.05 | Asplund et al. 2021<br>Huss et al. 2020<br>Meshik et al. 2020 | 10.706 ± 0.02 | Heber et al. 2021<br>Huss et al. 2020 |
| **Ne** | 8.15 ± 0.12 | 8.06 ± 0.05<br>8.060 ± 0.033<br>8.06 ± 0.05<br>8.15 ± 0.11<br>8.15 ± 010 | Asplund et al. 2021<br>Huss et al. 2020<br>Meshik et al. 2020<br>Magg et al. 2022<br>Young 2018 using O from Caffau et al. 2011 | 7.920 ± 0.004<br>7.92 ± 0.02 | Heber et al. 2021<br>Meshik et al. 2020 |
| **Ar** | 6.50 ± 0.12 | 6.38 ± 0.10<br>6.38 ± 0.12<br>6.38 ± 0.08 | Asplund et al. 2021<br>Huss et al. 2020<br>Meshik et al. 2020 | 6.337 ± 0.004<br>6.34 ± 0.02 | Heber et al. 2021<br>Meshik et al. 2020 |
| **Kr** | 3.31 ± 0.12 | 3.12 ± 0.10<br>3.24 ± 0.06 | Asplund et al. 2021<br>Meshik et al. 2020 | 3.132 ± 0.009<br>3.12 ± 0.02 | Heber et al. 2021<br>Meshik et al. 2020 |
| **Xe** | 2.30 ± 0.12 | 2.22 ± 0.05<br>2.26 ± 0.06 | Asplund et al. 2021<br>Meshik et al. 2020 | 2.474 ± 0.025<br>2.42 ± 0.04 | Heber et al. 2021<br>Meshik et al. 2020 |

## Table 4

**Table 4. CI-Chondrite Composition and Comparison to Solar Abundances [a]**

| E | Concentration by mass ± 1σ (SD) | | %SD | n | SE | %SE | 95%C.I. | N(E) ± 1σ Si = $10^6$ atoms | | Quality Index & Notes | | Change from P14, % | $A(E)_{CI}$ ± 1σ [b] dex | | $A(E)_{Sun}$ [c] dex | CI/Sun |
|---|---|---|---|---|---|---|---|---|---|---|---|---|---|---|---|---|
| H | 18598 | 1719 | 9.2 | 16 | 430 | 2.3 | 916 | 4.86E+6 | 4.5E+5 | B | | -5.6 | 8.24 | 0.04 | 12.00 | 1.7E-4 |
| He | 9.17E-3 | 9.2E-4 | 10 | | | | | 0.604 | 0.060 | B | see text | 0 | 1.33 | 0.04 | 10.92 | 3E-10 |
| Li | 1.48 | 0.07 | 4.7 | 20 | 0.02 | 1.1 | 0.03 | 56.2 | 2.7 | A | | 2.1 | 3.30 | 0.02 | 1.04 | 182 |
| Be | 0.0225 | 0.0013 | 5.8 | 13 | 0.0004 | 1.6 | 0.0008 | 0.658 | 0.038 | B | | 2.8 | 1.37 | 0.02 | 1.21 | 1.44 |
| B | 0.744 | 0.095 | 13 | 5 | 0.042 | 5.7 | 0.118 | 18.1 | 2.3 | C | | -4.0 | 2.81 | 0.05 | 2.70 | 1.29 |
| C | 37813 | 6600 | 17 | 38 | 1071 | 2.8 | 2169 | 8.29E+5 | 1.45E+5 | D | see text | 8.7 | 7.47 | 0.07 | 8.51 | 0.09 |
| N | 1965 | 970 | 49 | 22 | 206 | 10.5 | 428 | 37000 | 18200 | D | | -33 | 6.12 | 0.17 | 7.94 | 0.02 |
| O | 465700 | 8000 | 1.7 | 4 | 4000 | 0.9 | 12730 | 7.7E+6 | 1.3E+3 | A | see text | 1.5 | 8.44 | 0.01 | 8.76 | 0.47 |
| F | 92 | 20.0 | 22 | 6 | 8.2 | 8.8 | 21.0 | 1280 | 277 | D | see LF23 | 59 | 4.66 | 0.09 | 4.40 | 1.81 |
| Ne | 1.80E-4 | 1.8E-5 | 10 | | | | | 2.36E-3 | 2.4E-4 | B | see text | 0 | -1.08 | 0.04 | 8.15 | 6E-10 |
| Na | 4960 | 512 | 10 | 59 | 67 | 1.3 | 133 | 56800 | 5900 | B | | -0.05 | 6.31 | 0.04 | 6.29 | 1.04 |
| Mg | 95600 | 3000 | 3.1 | 51 | 420 | 0.4 | 844 | 1.0E+6 | 32.5E+3 | A | | 0.2 | 7.57 | 0.01 | 7.58 | 0.97 |
| Al | 8470 | 480 | 5.7 | 48 | 69 | 0.8 | 139 | 82700 | 4700 | B | | 0.8 | 6.47 | 0.02 | 6.43 | 1.09 |
| Si | 106600 | 4000 | 4.1 | 26 | 863 | 0.8 | 1777 | 1.0E+6 | 41.3E+3 | A | | -0.4 | 7.55 | 0.02 | 7.56 | 0.98 |
| P | 989 | 89 | 9.1 | 32 | 16 | 1.6 | 32 | 8410 | 760 | B | | 0.4 | 5.48 | 0.04 | 5.44 | 1.09 |
| S | 51800 | 4600 | 8.9 | 29 | 854 | 1.6 | 1750 | 4.26E+5 | 3.8E+4 | B | | -3.2 | 7.18 | 0.04 | 7.16 | 1.05 |
| Cl | 717 | 110 | 15 | 16 | 28 | 3.8 | 59 | 5330 | 820 | C | see LF23 | 2.7 | 5.28 | 0.06 | 5.43 | 0.70 |
| Ar | 1.33E-3 | 1.3E-4 | 10 | | | | | 9.66E-3 | 9.7E-4 | B | see text | 0 | -0.46 | 0.04 | 6.50 | 1.1E-7 |
| K | 544 | 41 | 7.5 | 62 | 5.21 | 1.0 | 10 | 3670 | 280 | B | | -0.33 | 5.12 | 0.03 | 5.09 | 1.06 |
| Ca | 9148 | 554 | 6.1 | R | | | | 60100 | 3600 | B | Ca/Al Tab. A5 | 0.41 | 6.33 | 0.03 | 6.35 | 0.96 |
| Sc | 5.76 | 0.37 | 6.4 | 60 | 0.05 | 0.8 | 0.10 | 33.8 | 2.2 | B | | -0.81 | 3.08 | 0.03 | 3.13 | 0.89 |
| Ti | 442 | 30 | 6.8 | 43 | 5 | 1.0 | 9 | 2430 | 170 | B | | -1.1 | 4.94 | 0.03 | 4.97 | 0.93 |
| V | 53.1 | 3.9 | 7.3 | 42 | 0.6 | 1.1 | 1.2 | 275 | 20 | B | | -2.7 | 3.99 | 0.03 | 3.89 | 1.26 |
| Cr | 2616 | 116 | 4.4 | 54 | 16 | 0.6 | 32 | 13300 | 600 | A | see text | -0.3 | 5.67 | 0.02 | 5.74 | 0.86 |
| Mn | 1936 | 153 | 7.9 | 72 | 18 | 0.9 | 36 | 9280 | 730 | B | | 1.0 | 5.52 | 0.03 | 5.52 | 1.00 |
| Fe | 185000 | 6000 | 3.5 | 54 | 871 | 0.5 | 1747 | 872.8E+3 | 30.2E+3 | A | see text | -0.9 | 7.49 | 0.01 | 7.51 | 0.96 |
| Co | 514 | 26 | 5.1 | 73 | 3 | 0.6 | 6 | 2300 | 120 | B | | 0.16 | 4.91 | 0.02 | 4.95 | 0.92 |
| Ni | 11180 | 300 | 3.0 | 53 | 45 | 0.4 | 91 | 50200 | 1500 | A | | 2.5 | 6.25 | 0.01 | 6.24 | 1.03 |
| Cu | 133 | 15 | 11 | 40 | 2 | 1.8 | 5 | 552 | 62 | C | | 0.17 | 4.29 | 0.05 | 4.24 | 1.13 |
| Zn | 310 | 22 | 7.1 | 81 | 2 | 0.8 | 5 | 1250 | 90 | B | | 0.5 | 4.65 | 0.03 | 4.55 | 1.25 |
| Ga | 9.54 | 0.68 | 7.1 | 59 | 0.09 | 0.9 | 0.18 | 36.1 | 2.6 | B | | -0.8 | 3.11 | 0.03 | 3.02 | 1.22 |
| Ge | 33.5 | 2.3 | 6.9 | 33 | 0.4 | 1.2 | 0.8 | 122 | 8 | B | | 2.9 | 3.64 | 0.03 | 3.62 | 1.04 |
| As | 1.75 | 0.17 | 9.7 | 40 | 0.03 | 1.5 | 0.05 | 6.16 | 0.60 | B | | 0.7 | 2.34 | 0.04 | ... | ... |
| Se | 21.5 | 0.70 | 3.3 | 32 | 0.1 | 0.6 | 0.3 | 71.7 | 2.3 | A | | 5.8 | 3.41 | 0.01 | ... | ... |
| Br | 3.77 | 0.90 | 24 | 15 | 0.2 | 6.2 | 0.5 | 12.4 | 3.0 | D | see LF23 | 16 | 2.65 | 0.09 | ... | ... |
| Kr | 5.22E-5 | 5.2E-6 | 10 | | | | | 1.64E-4 | 1.6E-5 | B | see text | 0 | -2.23 | 0.04 | 3.31 | 2.9E-6 |
| Rb | 2.26 | 0.12 | 5.3 | 41 | 0.02 | 0.8 | 0.04 | 6.98 | 0.37 | B | | -2.4 | 2.39 | 0.02 | 2.35 | 1.11 |
| Sr | 8.04 | 0.27 | 3.4 | 22 | 0.06 | 0.7 | 0.12 | 24.2 | 0.8 | A | | 3.3 | 2.93 | 0.01 | 2.93 | 1.01 |
| Y | 1.52 | 0.13 | 8.8 | R | | | | 4.50 | 0.40 | C | Zr/Y, Tab. A10 | 4.0 | 2.20 | 0.04 | 2.30 | 0.80 |
| Zr | 3.65 | 0.21 | 5.8 | R | | | | 10.5 | 0.6 | C | Zr/Hf, Tab. A7 | 0.4 | 2.57 | 0.02 | 2.68 | 0.78 |
| Nb | 0.271 | 0.022 | 8.2 | R | | | | 0.767 | 0.063 | C | Zr/Nb, Tab. A8 | -4.4 | 1.44 | 0.03 | 1.47 | 0.92 |
| Mo | 0.947 | 0.100 | 11 | 23 | 0.021 | 2.2 | 0.043 | 2.60 | 0.27 | C | | -1.5 | 1.97 | 0.04 | 1.88 | 1.22 |
| Tc | ... | ... | ... | ... | ... | ... | ... | ... | ... | ... | | ... | ... | ... | ... | ... |
| Ru | 0.680 | 0.042 | 6.2 | R | | | | 1.77 | 0.11 | B | Os/Ru, Tab. 5 | -1.5 | 1.80 | 0.03 | 1.75 | 1.12 |
| Rh | 0.133 | 0.006 | 4.5 | 11 | 0.002 | 1.4 | 0.004 | 0.341 | 0.015 | A | | 1.0 | 1.08 | 0.02 | 0.78 | 2.01 |
| Pd | 0.561 | 0.026 | 4.6 | 36 | 0.004 | 0.8 | 0.009 | 1.39 | 0.06 | A | | 0.13 | 1.69 | 0.02 | 1.57 | 1.33 |
| Ag | 0.206 | 0.019 | 9.2 | 30 | 0.003 | 1.7 | 0.007 | 0.504 | 0.046 | B | | 2.6 | 1.25 | 0.04 | 0.96 | 1.96 |
| Cd | 0.682 | 0.047 | 6.9 | 45 | 0.007 | 1.0 | 0.014 | 1.60 | 0.11 | B | | 1.1 | 1.75 | 0.03 | 1.77 | 0.96 |
| In | 0.0781 | 0.0055 | 7.0 | 38 | 0.001 | 1.1 | 0.002 | 0.179 | 0.013 | B | | 0.4 | 0.80 | 0.03 | 0.80 | 1.01 |
| Sn | 1.67 | 0.20 | 12 | 39 | 0.03 | 1.9 | 0.06 | 3.70 | 0.44 | C | | 2.2 | 2.12 | 0.05 | 2.02 | 1.26 |
| Sb | 0.157 | 0.026 | 17 | 51 | 0.004 | 2.3 | 0.007 | 0.340 | 0.056 | D | | 8.3 | 1.08 | 0.07 | ... | ... |
| Te | 2.30 | 0.09 | 3.9 | 26 | 0.02 | 0.8 | 0.04 | 4.75 | 0.19 | A | | 1.0 | 2.23 | 0.02 | ... | ... |
| I | 0.772 | 0.310 | 40 | 5 | 0.139 | 18 | 0.385 | 1.60 | 0.64 | D | see LF23 | 46 | 1.76 | 0.15 | ... | ... |
| Xe | 1.74E-4 | 1.7E-5 | 10 | | | | | 3.49E-4 | 3.49E-5 | B | see text | 0 | -1.91 | 0.04 | 2.30 | 6.2E-5 |
| Cs | 0.186 | 0.012 | 6.5 | 58 | 0.002 | 0.8 | 0.003 | 0.368 | 0.024 | B | | -1.3 | 1.12 | 0.03 | ... | ... |
| Ba | 2.41 | 0.14 | 5.8 | 36 | 0.02 | 1.0 | 0.05 | 4.62 | 0.27 | B | | -0.5 | 2.22 | 0.02 | 2.27 | 0.88 |
| La | 0.248 | 0.010 | 4.0 | 28 | 0.002 | 0.8 | 0.004 | 0.470 | 0.019 | A | | 2.6 | 1.22 | 0.02 | 1.10 | 1.33 |
| Ce | 0.633 | 0.026 | 4.2 | 28 | 0.005 | 0.8 | 0.010 | 1.19 | 0.05 | A | | 2.2 | 1.63 | 0.02 | 1.58 | 1.11 |
| Pr | 0.0957 | 0.0042 | 4.4 | 28 | 0.0008 | 0.8 | 0.0016 | 0.179 | 0.008 | A | | 1.9 | 0.80 | 0.02 | 0.75 | 1.13 |
| Nd | 0.482 | 0.025 | 5.2 | 29 | 0.005 | 1.0 | 0.010 | 0.880 | 0.046 | B | | 1.8 | 1.50 | 0.02 | 1.42 | 1.19 |
| Pm | ... | ... | ... | ... | ... | ... | ... | ... | ... | | | | ... | ... | ... | ... |
| Sm | 0.156 | 0.008 | 5.2 | 29 | 0.002 | 1.0 | 0.003 | 0.273 | 0.014 | B | | 1.6 | 0.99 | 0.02 | 0.95 | 1.09 |
| Eu | 0.0601 | 0.0027 | 4.5 | 28 | 0.0005 | 0.8 | 0.0010 | 0.104 | 0.005 | A | | 2.2 | 0.57 | 0.02 | 0.57 | 1.00 |
| Gd | 0.211 | 0.010 | 4.9 | 29 | 0.002 | 0.9 | 0.004 | 0.353 | 0.017 | A | | 1.8 | 1.10 | 0.02 | 1.08 | 1.04 |
| Tb | 0.0385 | 0.0017 | 4.3 | 24 | 0.0003 | 0.9 | 0.0007 | 0.0638 | 0.003 | A | | 1.4 | 0.36 | 0.02 | 0.31 | 1.11 |
| Dy | 0.259 | 0.011 | 4.2 | 28 | 0.002 | 0.8 | 0.004 | 0.421 | 0.018 | A | | 1.4 | 1.17 | 0.02 | 1.10 | 1.19 |
| Ho | 0.0568 | 0.0023 | 4.0 | 24 | 0.0005 | 0.8 | 0.0010 | 0.0908 | 0.004 | A | | 0.7 | 0.51 | 0.02 | 0.48 | 1.07 |
| Er | 0.167 | 0.007 | 4.3 | 28 | 0.001 | 0.8 | 0.003 | 0.263 | 0.011 | A | | 0.9 | 0.97 | 0.02 | 0.93 | 1.10 |
| Tm | 0.0263 | 0.0012 | 4.6 | 24 | 0.0002 | 0.9 | 0.0005 | 0.0410 | 0.002 | A | | 0.9 | 0.16 | 0.02 | 0.11 | 1.13 |
| Yb | 0.170 | 0.007 | 4.0 | 28 | 0.001 | 0.8 | 0.003 | 0.259 | 0.010 | A | | 1.0 | 0.96 | 0.02 | 0.85 | 1.30 |
| Lu | 0.0255 | 0.0015 | 5.7 | 31 | 0.0003 | 1.0 | 0.0005 | 0.0384 | 0.002 | B | | 1.8 | 0.13 | 0.02 | 0.10 | 1.08 |
| Hf | 0.107 | 0.006 | 5.8 | R | | | | 0.158 | 0.009 | C | Lu/Hf, Tab. A6 | 0.4 | 0.75 | 0.02 | 0.86 | 0.77 |
| Ta | 0.0149 | 0.0007 | 4.7 | R | | | | 0.0218 | 0.0010 | D | Nb/Ta, Tab. A9 | -0.3 | -0.11 | 0.02 | ... | ... |
| W | 0.101 | 0.012 | 12 | R | | | | 0.144 | 0.017 | D | Hf/W, s. text | 4.9 | 0.71 | 0.05 | 0.79 | 0.83 |
| Re | 0.0399 | 0.0017 | 4.1 | R | | | | 0.0564 | 0.0023 | A | Os/Re, Tab. 5 | -0.3 | 0.30 | 0.02 | ... | ... |
| Os | 0.491 | 0.014 | 2.8 | 28 | 0.003 | 0.5 | 0.005 | 0.680 | 0.019 | A | | -0.7 | 1.38 | 0.01 | 1.36 | 1.06 |
| Ir | 0.456 | 0.027 | 5.9 | 70 | 0.003 | 0.7 | 0.006 | 0.625 | 0.037 | B | | -2.8 | 1.35 | 0.02 | 1.42 | 0.84 |
| Pt | 0.904 | 0.070 | 7.8 | R | | | | 1.22 | 0.09 | B | Pt/Ir, Tab. 5 | -2.3 | 1.64 | 0.03 | ... | ... |
| Au | 0.150 | 0.013 | 8.8 | 50 | 0.002 | 1.2 | 0.004 | 0.201 | 0.018 | B | | 1.6 | 0.85 | 0.04 | 0.91 | 0.88 |
| Hg | 0.293 | 0.106 | 36 | 3 | 0.061 | 21 | 0.263 | 0.385 | 0.139 | D | | -16 | 1.14 | 0.13 | ... | ... |
| Tl | 0.141 | 0.014 | 9.9 | 31 | 0.003 | 1.8 | 0.005 | 0.182 | 0.018 | B | | 0.8 | 0.81 | 0.04 | 0.95 | 0.73 |
| Pb | 2.61 | 0.19 | 7.3 | 26 | 0.04 | 1.4 | 0.08 | 3.32 | 0.24 | B | | -0.2 | 2.07 | 0.03 | 1.95 | 1.33 |
| Bi | 0.113 | 0.007 | 6.2 | 25 | 0.001 | 1.2 | 0.003 | 0.142 | 0.009 | B | | 2.5 | 0.70 | 0.03 | ... | ... |
| Th | 0.0304 | 0.0033 | 11 | 37 | 0.0005 | 1.8 | 0.0011 | 0.0345 | 0.0037 | C | | 1.3 | 0.09 | 0.04 | ... | ... |
| U | 8.10E-3 | 7.70E-4 | 9.5 | 44 | 1.2E-4 | 1.4 | 2.3E-4 | 8.96E-3 | 8.52E-4 | D | | -0.1 | -0.50 | 0.04 | ... | ... |

(a) Concentrations mainly from CI-chondrites, except when concentrations were calculated from element ratios, as noted (R). Quality index (A highest quality) is based on 1-sigma standard error (SD in %) and element variability and mobility, and issues with analytical methods. n = number of analyses included in grand mean. SD: standard deviation. SE: standard error. 95%C.I.: 95% confidence interval. LF23 = Lodders & Fegley 2023. P14 = Palme et al. 2014.
(b) A(E) = 1.551+ log N(E). (c) from Table 2.

# Table 5

**Table 5. Concentration Ratios and Concentrations of Refractory Metals in Carbonaceous Chondrites**

| Reference | N [a] | Os/Ir | Os/Re | Os/Ru | Pt/Ir | Pt/Rh |
|---|---|---|---|---|---|---|
| | | (1) | (2) | (3) | (4) | (5) |
| Horan et al. 2003 | 20 | 1.06 ±2.3% | 12.37±3.1% | 0.727 ±6.2% | 1.91±7.2% | … |
| Fischer-Gödde et al. 2010 | 13 | 1.06 ±3.6% | 12.23 ±2.7% | 0.716 ±3.9% | 2.05±3.6% | 7.24 ±7.5% |
| Selected samples: | 33 | 1.06 ±2.9% | 12.32 ±3.0% | 0.723 ±5.3% | 1.982 ±5.0% | 7.24 ±7.5% |
| Abundances of refractory metals in CI-chondrites (ppb) | | | | | | |
| | Ir | Os | Re | Ru | Pt | Rh |
| Via ratios: | N/A | 483 ±6.6% | **39.9 ±4.1%** | **680 ±6.0%** | **904 ±7.7%** | 126 ±8.0% |
| Via grand mean: | **456 ±5.9%** | **491 ±2.8%** | 37.1 9.2% | 648 ±10.2% | 926 ±11.2% | **133±4.5%** |
| Recommended: (Table 4) | 456 ±5.9% | 491 ±2.8% | 39.9 ±4.1 | 680 ±6.0% | 904 ±7.7% | 133 ±4.5% |

Selected values are in bold-face. Excluded from average: (1) EET 9204; (2) Kainsaz, Karoonda, one Orgueil sample; (3) Ornans; (4) Kainsaz, Ornans, Lance (5) Ninqiang. (a) N = number of samples

# Table 6

**Table 6. Recommended Atomic Solar System Abundances: Present and Proto-Solar 4.567 Ga Ago**

| Z | E | CI-Chondrites (from Table 4) | | Sun Convection Zone (mainly photosphere) | | Solar System * Present | | Solar System * Proto-Solar †,‡ | | | Solar System * Present | | Solar System* Proto-Solar †,‡ | |
|---|---|---|---|---|---|---|---|---|---|---|---|---|---|---|
| | . | N(E), $10^6$ Si | ±σ | N(E)=$10^{A(E)+1.551}$ | ±σ | N(E), $10^6$ Si | ±σ | N(E), $10^6$ Si | ±σ | Note | log (E/H)+12 | ±σ | log (E/H)+12 | ±σ |
| 1 | H | 4.86E+06 | 4.49E+05 | 2.81E+10 | 2.57E+08 | 2.81E+10 | 2.57E+08 | 2.29E+10 | 2.1E+08 | s | 12.000 | 0.004 | 12.000 | 0.004 |
| 2 | He | 0.604 | 0.060 | 2.35E+09 | 7E+07 | 2.34E+09 | 7E+07 | 2.25E+09 | 6E+07 | s,t | 10.922 | 0.012 | 10.992 | 0.012 |
| 3 | Li | 56.2 | 2.7 | 0.308 | 0.071 | 56.1 | 2.7 | 56.1 | 2.7 | m | 3.30 | 0.02 | 3.39 | 0.02 |
| 4 | Be | 0.658 | 0.038 | 0.456 | 0.174 | 0.657 | 0.038 | 0.657 | 0.038 | m | 1.37 | 0.02 | 1.46 | 0.02 |
| 5 | B | 18.1 | 2.3 | 14.1 | 11.0 | 18.1 | 2.3 | 18.1 | 2.3 | m | 2.81 | 0.05 | 2.90 | 0.05 |
| 6 | C | 8.29E+05 | 1.45E+05 | 9.10E+06 | 2.10E+06 | 9.08E+06 | 2.10E+06 | 9.08E+06 | 2.10E+06 | s | 8.51 | 0.09 | 8.60 | 0.09 |
| 7 | N | 3.70E+04 | 1.82E+04 | 2.45E+06 | 7.1E+05 | 2.44E+06 | 7E+05 | 2.44E+06 | 7.1E+05 | s | 7.94 | 0.11 | 8.03 | 0.11 |
| 8 | O | 7.67E+06 | 1.3E+05 | 1.62E+07 | 2.0E+06 | 1.62E+07 | 2.0E+06 | 1.62E+07 | 2.0E+06 | s | 8.76 | 0.05 | 8.85 | 0.05 |
| 9 | F | 1280 | 277 | 706 | 413 | 1278 | 277 | 1278 | 277 | m | 4.66 | 0.09 | 4.75 | 0.09 |
| 10 | Ne | 2.36E-03 | 2.4E-04 | 3.97E+06 | 1.26E+06 | 3.96E+06 | 1.26E+06 | 3.96E+06 | 1.26E+06 | s,t | 8.15 | 0.12 | 8.24 | 0.12 |
| 11 | Na | 56838 | 5868 | 54838 | 6691 | 55852 | 1414 | 55852 | 1414 | a | 6.30 | 0.01 | 6.39 | 0.01 |
| 12 | Mg | 1.04E+06 | 3E+04 | 1.07E+06 | 1.3E+05 | 1.04E+06 | 2E+04 | 1.04E+06 | 2E+04 | a | 7.57 | 0.01 | 7.66 | 0.01 |
| 13 | Al | 82707 | 4687 | 75698 | 9237 | 81102 | 4956 | 81102 | 4956 | a | 6.46 | 0.03 | 6.55 | 0.03 |
| 14 | Si | 1.00E+06 | 4E+04 | 1.02E+06 | 1.2E+05 | 1.00E+06 | 1E+04 | 1.00E+06 | 1E+04 | a | 7.55 | 0.01 | 7.64 | 0.01 |
| 15 | P | 8413 | 757 | 7746 | 2465 | 8395 | 757 | 8395 | 757 | m | 5.48 | 0.04 | 5.56 | 0.04 |
| 16 | S | 425635 | 37798 | 406521 | 117180 | 424747 | 37798 | 424747 | 37798 | m | 7.18 | 0.04 | 7.27 | 0.04 |
| 17 | Cl | 5326 | 817 | 7570 | 4428 | 5315 | 817 | 5315 | 817 | m | 5.28 | 0.06 | 5.37 | 0.06 |
| 18 | Ar | 0.0097 | 0.0010 | 88937 | 28305 | 88752 | 28305 | 88751 | 28305 | s,t | 6.50 | 0.12 | 6.59 | 0.12 |
| 19 | K | 3667 | 276 | 3460 | 797 | 3659 | 276 | 3664 | 277 | m | 5.12 | 0.03 | 5.20 | 0.03 |
| 20 | Ca | 60135 | 3641 | 62963 | 9328 | 60382 | 2000 | 60378 | 2000 | a | 6.33 | 0.01 | 6.42 | 0.01 |
| 21 | Sc | 33.8 | 2.2 | 38 | 11 | 33.7 | 2.2 | 33.7 | 2.2 | m | 3.08 | 0.03 | 3.17 | 0.03 |
| 22 | Ti | 2433 | 165 | 2625 | 757 | 2428 | 165 | 2428 | 165 | m | 4.94 | 0.03 | 5.03 | 0.03 |
| 23 | V | 275 | 20 | 218 | 97 | 274 | 20 | 274 | 20 | m | 3.99 | 0.03 | 4.08 | 0.03 |
| 24 | Cr | 13255 | 588 | 15456 | 4455 | 13227 | 588 | 13227 | 588 | m | 5.67 | 0.02 | 5.76 | 0.02 |
| 25 | Mn | 9282 | 734 | 9313 | 1136 | 9272 | 22 | 9272 | 22 | a | 5.52 | 0.00 | 5.61 | 0.00 |
| 26 | Fe | 8.73E+05 | 3.0E+04 | 9.10E+05 | 1.11E+05 | 8.74E+05 | 2.6E+04 | 8.74E+05 | 2.6E+04 | a | 7.49 | 0.01 | 7.58 | 0.01 |
| 27 | Co | 2297 | 116 | 2507 | 723 | 2292 | 116 | 2292 | 116 | m | 4.91 | 0.02 | 5.00 | 0.02 |
| 28 | Ni | 50184 | 1481 | 48875 | 7241 | 50026 | 926 | 50026 | 926 | a | 6.25 | 0.01 | 6.34 | 0.01 |
| 29 | Cu | 552 | 62 | 489 | 141 | 551 | 62 | 551 | 62 | m | 4.29 | 0.05 | 4.38 | 0.05 |
| 30 | Zn | 1251 | 89 | 998 | 288 | 1248 | 89 | 1248 | 89 | m | 4.65 | 0.03 | 4.74 | 0.03 |
| 31 | Ga | 36 | 3 | 29.4 | 11.2 | 36 | 3 | 36.0 | 2.6 | m | 3.11 | 0.03 | 3.20 | 0.03 |
| 32 | Ge | 122 | 8 | 117 | 45 | 121 | 8 | 121 | 8 | m | 3.64 | 0.03 | 3.72 | 0.03 |
| 33 | As | 6.16 | 0.60 | ... | ... | 6.15 | 0.60 | 6.15 | 0.60 | m | 2.34 | 0.04 | 2.43 | 0.04 |
| 34 | Se | 71.7 | 2.3 | ... | ... | 71.5 | 2.3 | 71.5 | 2.3 | m | 3.41 | 0.01 | 3.49 | 0.01 |
| 35 | Br | 12.4 | 3.0 | ... | ... | 12.4 | 3.0 | 12.4 | 3.0 | m | 2.65 | 0.09 | 2.73 | 0.09 |
| 36 | Kr | 0.000164 | 0.000016 | 57.4 | 18.3 | 57.3 | 18.3 | 57.3 | 18.3 | t | 3.31 | 0.12 | 3.40 | 0.12 |
| 37 | Rb | 6.98 | 0.37 | 6.30 | 1.81 | 6.96 | 0.37 | 7.09 | 0.38 | m | 2.39 | 0.02 | 2.49 | 0.02 |
| 38 | Sr | 24.2 | 0.8 | 23.9 | 6.9 | 24.1 | 0.8 | 24.0 | 0.8 | m | 2.93 | 0.01 | 3.02 | 0.01 |
| 39 | Y | 4.50 | 0.40 | 5.61 | 0.83 | 4.49 | 0.40 | 4.49 | 0.40 | m | 2.20 | 0.04 | 2.29 | 0.04 |
| 40 | Zr | 10.5 | 0.6 | 13.5 | 3.9 | 10.5 | 0.6 | 10.5 | 0.6 | m | 2.57 | 0.02 | 2.66 | 0.02 |
| 41 | Nb | 0.77 | 0.06 | 0.830 | 0.316 | 0.77 | 0.06 | 0.77 | 0.06 | m | 1.44 | 0.03 | 1.52 | 0.03 |
| 42 | Mo | 2.60 | 0.27 | 2.13 | 0.95 | 2.60 | 0.27 | 2.60 | 0.27 | m | 1.97 | 0.04 | 2.05 | 0.04 |
| 43 | Tc | | ... | ... | ... | | | ... | ... | ... | | ... | ... | ... |
| 44 | Ru | 1.77 | 0.11 | 1.58 | 0.70 | 1.77 | 0.11 | 1.77 | 0.11 | m | 1.80 | 0.03 | 1.89 | 0.03 |
| 45 | Rh | 0.341 | 0.015 | 0.169 | 0.075 | 0.341 | 0.015 | 0.341 | 0.015 | m | 1.08 | 0.02 | 1.17 | 0.02 |
| 46 | Pd | 1.39 | 0.06 | 1.04 | 0.47 | 1.39 | 0.06 | 1.39 | 0.06 | m | 1.69 | 0.02 | 1.78 | 0.02 |
| 47 | Ag | 0.504 | 0.046 | 0.256 | 0.114 | 0.503 | 0.046 | 0.503 | 0.046 | m | 1.25 | 0.04 | 1.34 | 0.04 |
| 48 | Cd | 1.60 | 0.11 | 1.66 | 0.97 | 1.59 | 0.11 | 1.59 | 0.11 | m | 1.75 | 0.03 | 1.84 | 0.03 |
| 49 | In | 0.179 | 0.013 | 0.177 | 0.104 | 0.179 | 0.013 | 0.179 | 0.013 | m | 0.80 | 0.03 | 0.89 | 0.03 |
| 50 | Sn | 3.70 | 0.44 | 2.94 | 1.31 | 3.69 | 0.44 | 3.69 | 0.44 | m | 2.12 | 0.05 | 2.21 | 0.05 |
| 51 | Sb | 0.340 | 0.056 | ... | ... | 0.339 | 0.056 | 0.339 | 0.056 | m | 1.08 | 0.07 | 1.17 | 0.07 |
| 52 | Te | 4.75 | 0.19 | ... | ... | 4.74 | 0.19 | 4.74 | 0.19 | m | 2.23 | 0.02 | 2.32 | 0.02 |
| 53 | I | 1.60 | 0.64 | ... | ... | 1.60 | 0.64 | 1.60 | 0.64 | m | 1.76 | 0.15 | 1.84 | 0.15 |
| 54 | Xe | 3.49E-04 | 3.5E-05 | 5.61 | 1.79 | 5.60 | 1.79 | 5.60 | 1.79 | t | 2.30 | 0.12 | 2.39 | 0.12 |
| 55 | Cs | 0.368 | 0.024 | ... | ... | 0.367 | 0.024 | 0.367 | 0.024 | m | 1.12 | 0.03 | 1.20 | 0.03 |
| 56 | Ba | 4.62 | 0.27 | 5.24 | 0.78 | 4.61 | 0.27 | 4.61 | 0.27 | m | 2.22 | 0.02 | 2.30 | 0.02 |
| 57 | La | 0.470 | 0.019 | 0.354 | 0.158 | 0.469 | 0.019 | 0.469 | 0.019 | m | 1.22 | 0.02 | 1.31 | 0.02 |
| 58 | Ce | 1.190 | 0.049 | 1.07 | 0.48 | 1.187 | 0.049 | 1.187 | 0.049 | m | 1.63 | 0.02 | 1.71 | 0.02 |
| 59 | Pr | 0.179 | 0.008 | 0.158 | 0.046 | 0.179 | 0.008 | 0.179 | 0.008 | m | 0.80 | 0.02 | 0.89 | 0.02 |
| 60 | Nd | 0.880 | 0.046 | 0.740 | 0.330 | 0.879 | 0.046 | 0.877 | 0.046 | m | 1.50 | 0.02 | 1.58 | 0.02 |
| 61 | Pm | 0.000 | ... | ... | ... | ... | ... | ... | ... | ... | ... | ... | ... | ... |
| 62 | Sm | 0.273 | 0.014 | 0.251 | 0.112 | 0.273 | 0.014 | 0.274 | 0.014 | m | 0.99 | 0.02 | 1.08 | 0.02 |
| 63 | Eu | 0.104 | 0.005 | 0.104 | 0.015 | 0.104 | 0.005 | 0.104 | 0.005 | m | 0.57 | 0.02 | 0.66 | 0.02 |
| 64 | Gd | 0.353 | 0.017 | 0.338 | 0.151 | 0.352 | 0.017 | 0.352 | 0.017 | m | 1.10 | 0.02 | 1.19 | 0.02 |
| 65 | Tb | 0.0638 | 0.0028 | 0.0574 | 0.0256 | 0.0637 | 0.0028 | 0.0637 | 0.0028 | m | 0.36 | 0.02 | 0.44 | 0.02 |
| 66 | Dy | 0.421 | 0.018 | 0.354 | 0.158 | 0.420 | 0.018 | 0.420 | 0.018 | m | 1.17 | 0.02 | 1.26 | 0.02 |
| 67 | Ho | 0.0908 | 0.0037 | 0.0849 | 0.0378 | 0.0906 | 0.0037 | 0.0906 | 0.0037 | m | 0.51 | 0.02 | 0.60 | 0.02 |
| 68 | Er | 0.263 | 0.011 | 0.239 | 0.107 | 0.263 | 0.011 | 0.263 | 0.011 | m | 0.97 | 0.02 | 1.06 | 0.02 |
| 69 | Tm | 0.0410 | 0.0019 | 0.0362 | 0.0161 | 0.0410 | 0.0019 | 0.0410 | 0.0019 | m | 0.16 | 0.02 | 0.25 | 0.02 |
| 70 | Yb | 0.259 | 0.010 | 0.199 | 0.089 | 0.259 | 0.010 | 0.259 | 0.010 | m | 0.96 | 0.02 | 1.05 | 0.02 |
| 71 | Lu | 0.0384 | 0.0022 | 0.0354 | 0.0158 | 0.0383 | 0.0022 | 0.0384 | 0.0022 | m | 0.13 | 0.02 | 0.22 | 0.02 |
| 72 | Hf | 0.158 | 0.009 | 0.204 | 0.065 | 0.157 | 0.009 | 0.157 | 0.009 | m | 0.75 | 0.02 | 0.84 | 0.02 |
| 73 | Ta | 0.0218 | 0.0010 | ... | ... | 0.0217 | 0.0010 | 0.0217 | 0.0010 | m | -0.11 | 0.02 | -0.02 | 0.02 |
| 74 | W | 0.144 | 0.017 | 0.173 | 0.101 | 0.144 | 0.017 | 0.144 | 0.017 | m | 0.71 | 0.05 | 0.80 | 0.05 |
| 75 | Re | 0.0564 | 0.0023 | ... | ... | 0.0563 | 0.0023 | 0.0591 | 0.0024 | m | 0.30 | 0.02 | 0.41 | 0.02 |
| 76 | Os | 0.680 | 0.019 | 0.644 | 0.245 | 0.679 | 0.019 | 0.676 | 0.019 | m | 1.38 | 0.01 | 1.47 | 0.01 |
| 77 | Ir | 0.625 | 0.037 | 0.740 | 0.433 | 0.624 | 0.037 | 0.624 | 0.037 | m | 1.35 | 0.03 | 1.44 | 0.03 |
| 78 | Pt | 1.221 | 0.095 | ... | ... | 1.218 | 0.095 | 1.218 | 0.095 | m | 1.64 | 0.03 | 1.73 | 0.03 |
| 79 | Au | 0.201 | 0.018 | 0.229 | 0.134 | 0.201 | 0.018 | 0.201 | 0.018 | m | 0.85 | 0.04 | 0.94 | 0.04 |
| 80 | Hg | 0.385 | 0.139 | ... | ... | 0.384 | 0.139 | 0.384 | 0.139 | m | 1.14 | 0.13 | 1.22 | 0.13 |
| 81 | Tl | 0.182 | 0.018 | 0.251 | 0.147 | 0.181 | 0.018 | 0.181 | 0.018 | m | 0.81 | 0.04 | 0.90 | 0.04 |
| 82 | Pb | 3.32 | 0.24 | 2.51 | 1.47 | 3.32 | 0.24 | 3.29 | 0.24 | m | 2.07 | 0.03 | 2.16 | 0.03 |
| 83 | Bi | 0.142 | 0.009 | ... | ... | 0.142 | 0.009 | 0.142 | 0.009 | m | 0.70 | 0.03 | 0.79 | 0.03 |
| 90 | Th | 0.0345 | 0.0037 | ... | ... | 0.0344 | 0.0037 | 0.0431 | 0.0047 | m | 0.09 | 0.04 | 0.27 | 0.04 |
| 92 | U | 0.00896 | 0.00085 | ... | ... | 0.00894 | 0.00085 | 0.02382 | 0.00227 | m | -0.50 | 0.04 | 0.02 | 0.04 |

* CI-chondritic values are strictly normalized to Si = $10^6$ atoms. Photospheric values via $N(E)=10^{A(E)-1.551}$. **Solar system values are from the combined solar and meteoritic datasets in columns (1) and (2) and were re-normalized to Si = $10^6$ because the average of solar and meteoritic values is used for Si. Note column: m = meteoritic, s = from sun, o = by other means †corrected for radioactive decay. ‡present day photospheric values corrected for element settling from the convection zone to obtain protosolar values.

# Table 7

Table 7. Mass Fractions for Solar System Composition*

| Mass Fraction | Present-Day | Protosolar |
|---|---|---|
| X | 0.7389 ± 0.0068 (±0.9%) | 0.7060 ± 0.0065 (±0.9%) |
| Y | 0.2452 ± 0.0069 (±2.8%) | 0.2753 ± 0.0077 (±2.8%) |
| Z | 0.0160 ± 0.0013 (±8%) | 0.0187 ± 0.0015 (±8%) |
| Z/X | 0.0216 ± 0.0017 (±8%) | 0.0265 ± 0.0021 (±8%) |

* Composition derived from photospheric and CI-chondritic data. Mass fraction X is for H, Y for He, and Z is for the sum of Li to U. See text.

# Supplementary Information *Space Science Reviews* 2025

## Solar System Elemental Abundances from the Solar Photosphere and CI-Chondrites

**Katharina Lodders, Maria Bergemann, Herbert Palme**

This supplementary information contains the following Tables and documentation:

**Table A1. References for CI-Chondrite Concentration Data [a]**

| Reference | Code | Method | Number of samples and meteorites [b] | Elements Analyzed |
|---|---|---|---|---|
| Ahrens et al. 1969 | A69 | XRF | 1 Org | Al, Ca, [Fe], K, [Mg], Mn, P, [Si], Ti |
| Akaiwa 1966 | A66 | RNAA | 2 Org | [In], [Se], [Te] |
| Arden & Cressey 1984 | A84 | IDMS | 1 Org | Pb, Tl, U |
| Babechuk et al. 2010 | B10 | ICPMS | 1 Org | Ba, [Hf], U, W La, Sm |
| Baedecker et al. 1973 | B73 | RNAA | 1 Org | Au, Ge, Ir, [Ni] |
| Baker et al. 2010 | B10 | ICPMS-ID | 1 Org | Pb,[Tl] |
| Barrat et al. 2012 | B12 | ICPMS-SF, ICP-AES | 1 Alais, 1 Ivuna, 6 Org | Ba, Be, Co Cs*, Cu, Ga, Hf, K*,Li, Ni*, Mn, P, Pb, Rb*, Sc, Sr*, Ta, U*, V, W*, Y, Zn, Zr. All REE*. ICP-AES: [Al], [Ca], Co, Cr, [Fe], Mg, Mn, Na, P, Ni, Ti |
| Barrat et al. 2016 | B16 | ICPMS-SF | 1 Alais, 1 Ivuna, 5 Org | All REE |
| Beer et al. 1984 | B84 | IDMS, TIMS | 2 Org | Ba, Cs, Hf, K, Rb, [Sr]. La, Ce, Nd, Sm, Eu, Gd, Dy, Er, Yb, Lu. |
| Bendel 2013 | B13 | ICPMS-MC | 1 Alais 2 Ivuna, 2 Org | Hf, Nb*, Ta, W, Zr. Nd, Sm, Lu |
| Bermingham et al. 2016 | B16 | ICPMS | 2 Org | Ba |
| Blichert-Toft & Albarede 1997 | B97 | ID-MS | 1 Org | [Hf], Lu |
| Bouvier et al. 2008 | B08 | ICPMS-ID-MC | 1 Ivuna, 1 Org | Hf, Lu |
| Braukmüller et al. 2018 | B18 | ICPMS-SF | 1 Ivuna, 2 Org | Ag, Al, As,[Ba],Ca, Cd, [Cs],Co, Cr, Cu, Ga, [Fe], [Hf], In, Ir, [K], Mg, Mn, Na, Nb, Ni, P, Pt, [Rb], Re, Rh, Sn, [Te], [Ti], Tl, U*, V, W, Zn, [Zr] All REE: La* |
| Braukmüller et al. 2020 | B20 | ICPMS-ID-Q | 1 Org | Ag, Cd, Cu, Ga, In, Sn,Tl, Zn |
| Briggs & Mamikunian 1963 | B63 | Wet chem.(?) | 1 Org | [Al], [Ca], [Co[, [Cr], Fe, [K], Mg, [Mn], [Na], [Ni], [P], Si, [Ti] |
| Burkhardt et al. 2012 | B12 | ICPMS | 1 Org | Mo, [W] |
| Burnett et al. 1989 | B89 | PIXE | 3 Ivuna, 3 Org | As*, [Cu], Fe, Ga, Ge, Mo, Ni, Pb*, Rb*, Ru*, Se, Sr*, [Y], Zn*, Zr |
| Case et al. 1973 | C73 | RNAA | 2 Ivuna, 2 Org | As, Au*, Co*, Ga, [Mo], [Re],Sb*, [Se], [Te], Zn |
| Chou et al. 1976 | C76 | RNAA | 4 Org | Au, Cd, Ge*, In, Ir, Ni*,Zn |
| Curtis & Gladney 1985 | C85 | PGA (NAA for Na) | 1 Ivuna, 2 Org | B, Na*, S, Si |
| Curtis et al. 1980 | C80 | PGA | [Ivuna], 6 Org | [B] |
| Craddock & Dauphas 2010 | C10 | ICPMS | 1 Ivuna | [Fe] |
| Crocket et al. 1967 | C67 | RNAA | 1 Alais, 2 Ivuna, 3 Org | Au, Ir*, Os*, Pd*, Pt*, Ru |
| Dauphas & Pourmand 2011 | D11 | IDMS | 1 Ivuna | Hf, U |
| Dauphas & Poumand 2015 | D15 | ICPMS-MC | 3 Org | All REE |
| David et al. 2000 | D00 | ICPMS-ID-MC | 1 Org | [Hf], [Zr] |
| De Laeter et al. 1974 | D74 | IDMS | 1 Org | Sn |
| De Laeter & Hosie 1978 | D78 | IDMS | 1 Org | Ba |
| De Laeter et al. 1998 | D98 | IDMS | 1 Ivuna, 4 Org | Ba |
| Easton, Lovering 1964 | E64 | Flame photometry | 1 Org | [K] |

**Table A1. References for CI-Chondrite Concentration Data [a]**

| Reference | Code | Method | Number of samples and meteorites [b] | Elements Analyzed |
|---|---|---|---|---|
| Ebihara et al. 1982 | E82 | RNAA | 1 Alais,1 Ivuna, 4 Org | Ag, Au, Bi, Cd, Cs, Ge, In, Ir, Ni*, Os*, Pd*, Pt, Rb*, Re, Sb, Se, Sn, [Te], Tl, Zn. [Ce], [Nd], [Eu], [Tb], [Yb], [Lu] |
| Edwards & Urey 1956 | E55 | Flame photo. | 2 Ivuna, 2 Org | K, [Na] |
| Ehmann & Chyi 1974 | E74 | RNAA | Ivuna, 4 Org | Hf, [Zr] |
| Ehmann & Gillum 1972 | E72a | RNAA | 1 Ivuna, 2 Org | Au, Pt |
| Ehmann & Rebagay 1970 | E72b | INAA/RNAA | 1Ivuna, 1 Org | [Zr,Hf] Do not use at all |
| Ehmann et al. 1970a | E70 | INAA/RNAA | 1 Ivuna, 2 Org | Au, Ir* |
| Evensen et al. 1978 | E78 | IDMS | 1 Ivuna, 2 Or | [La], [Ce], [Nd], [Sm], [Eu], [Gd],[Dy], [Er], [Yb] |
| Fehr et al. 2005 | F05 | ICPMS | 2 Org | [Sn], Te |
| Fehr et al. 2018 | F18 | ICPMS | 4 Org | Te |
| Fischer-Gödde et al. 2010 | F10 | ICPMS-ID | Ivuna, 3 Org | Au, Ir, Os, Pd, Pt, Re, Rh, Ru |
| Fisher 1972 | F72 | Fission track | 2 Org | [U] |
| Folinsbee et al. 1967 | F67a | XRF | 2 Revelstoke | Co,[Cu], [Fe], [K], [Ni], Ti* |
| Fouche & Smales 1967a | F67b, | RNAA | 1 Ivuna, 1 Org | As, Au, Re, Pd, Sb |
| Fouche & Smales 1967b | F67c | RNAA | 1 Iv Ivuna, 1 Org | Au, Ga, Ge, In |
| Fredriksson et al. 1997 | F97 | Pellet EMP | Org | [Al], [Ca], [Fe], [Mg], [Na], [Ni], [Si] |
| Friedrich et al. 2002 | F02 | ICPMS-Q | Org | As, Ba, Co, Cs, Cu, Ga,[Hf], Ir, [Li], [Nb], [Mo], Rb, Re, Mn, Pd, Pt, [Ru],Sb, Sc, Se, Sn, [Sr], Te, Ti, U, V, [W], [Y], Zn, [Zr]. All REE. |
| Funk 2015 | F15 | ICPMS-ID | 1 Ivuna,2 Org | Se, Te |
| Ganapathy et al. 1976 | G76-2 | RNAA | 2 Org | [Hf], [Zr] |
| Gooding 1979 | G79 | NAA | 7 Org | Al, [Au], Ca*,Co*, Cr, Fe*, Ir*, Mg, Mn, Na, Ni*,Sc*, [Ta], [Ti], V, Zn<br>[La], [Sm], [Eu],[Yb], [Lu] |
| Graham & Mason 1972 | G72 | SMSS | 1 Ivuna | Nb |
| Greenland 1967 | G67 | RNAA | 1 Ivuna, 1 Org | [Ag], [Cd], [Pd], [Se], [Te] ,[Zn] |
| Greenland & Goles 1965 | G65a | RNAA | 1 Ivuna, 3 Org | Cu*, [Zn] |
| Greenland & Lovering 1965 | G65b | spectrometric | Org | Emission spectrographic: [Ba], [Cr], [Cu], [P], [Sc], [Sr], [V]<br>Colorimetric/ spectrophotom.: Co, Fe, [Ga] Ge, [Mn], [Ni], Ti, Zn |
| Grossman et al. 1985 | G85 | NAA | 1 Org | Al, As, Au, Ca, Co, Cr, Cs, [Fe], Ga, Hf, Ir, K, [Mg], Mn, Na, [Ni], Os, Ru, Sb, Se, [Te], V, Zn, [La], [Sm], [Eu], [Yb],[Lu] |
| Grossman & Ganapathy 1975 | G75 | NAA | Alais, 2 Org | Mn*, Na |
| Grossman & Ganapathy 1976 | G76 | NAA | [1 Al],1 Org | Au, Co, Cr, Fe, Ir*, Os, Sc, [La], [Dy], [Sm], [Eu], |
| Hamaguchi et al. 1969 | H69 | RNAA | 3 Org | As, Sb, Sn |
| Hellmann et al. 2020 | H20 | ICPMS | 1 Ivuna, 1 Org | Te |
| Hermann & Wichtl 1974 | H74 | NAA | 1 Org | [As],Au, [Cs], Ir, [Mo], [Pd], Re, [Rb], Ru, Sb,[Sc],[Se],Sn, [Zn] |
| Hidaka & Yoneda, 2011 | H11 | ICPMS | 1 Org | Ba, Cs |
| Hintenberger et al. 1973 | H73 | IDSSMS | 1 Org | [Au], [Ir], [Pb],[Pd], [Pt],[Tl], [U], [W] |
| Horan et al. 2003 | H03 | IDMS; N-TIMS: Re, Os | 2 Ivuna, 2 Org | Ir, Os, Pd, Pt, Re, Ru |
| Hu et al. 2023a | H23 | ICPMS | 1 Org | Al, [K],[Li], Na, Pb, Ti, U, V, All REE: [Lu] |

**Table A1. References for CI-Chondrite Concentration Data [a]**

| Reference | Code | Method | Number of samples and meteorites [b] | Elements Analyzed |
|---|---|---|---|---|
| Humayun & Clayton 1995 | H95 | IDMS | 1 Org | [K] |
| Islam et al. 2012 | I12 | PGA | 1 Alais,1 Ivuna, 2 Org | Ca, Co, [Cr], Fe, K, [Mg], Mn, Na, Ni, [Si], [Ti], [Sm], [Gd] |
| James & Palmer 2000 | J00 | ID-MS | 1 Org | Li |
| Jarosewich 1972 | J72 | Flame photometry | 1 Org | [Al], Ca, Co, [Cr], Fe, K, Mg, [Mn], [Na], [Ni], P, Si, Ti |
| Jenniskens et al. 2014 | J14 | ICPMS-Q | 1 Org | Ba, Be, Ca, Cd, Co, Cr, Cs, Cu, Ga, [Fe], [Hf], K, Li, Mg, Mn, Na, Nb, Ni, P, Pb, Rb, Sb, Sc,[Sr], [Ta], Ti, Tl, U, [V], Y, Zn, [Zr] REE |
| Jochum et al. 1986 | J86 | ID-SSMS | 1 Ivuna, 2 Org | Nb, Sc, U, Y. NAA for [Ta] |
| Jochum et al. 1993 | J93 | ID-SSMS | 1 Ivuna, 1 Org | Sn |
| Jochum et al. 1996 | J96 | ID-SSMS | 1 Ivuna, 2 Org | Au, Ir, Mo, Os, Pt, Re, Rh, Ru |
| Jochum et al. 2000 | J00 | ID-SSMS | 2 Org | Nb, Y, Zr |
| Jochum & Seufert 1995 | J95 | ID-SSMS | 1 Org | Nb, Rh, Y, Zr |
| Kallemeyn & Wasson 1981 | K81 | RNAA/INAA | 2 Alais, 4 Org | Al, As, Au, Ca, Cd, Co, Cr, Fe, Ga, Ge, In, Ir, K, Mg*, Mn, Na, Ni*, Os, Ru, Sb, Sc, Se, V, Zn. [La], [Sm], [Eu], [Yb], [Lu] |
| Kaushal & Wetherill 1970 | K70 | IDMS | 1 Ivuna, 3 Org | K*, Rb*, Sr* |
| Kiesl 1979 | K79 | NAA | 1 Org | Al, Ca, [Cr], Fe, K, Mg, [Mn], Na, Ni, [P], Si, [Ti] |
| King et al. 2020 | K20 | ICP-OES,MS | 2 Ivuna | Al, [Ba], Be, Ca, Cd, Co, Cs, Cu, [Fe], [Hf], K*, Na*, Ni, Mo, Rb, Sn, P, Sc, Sr*,Ti, [U], V, Y, Zn, Zr*, Dy |
| Kleine et al. 2004 | K06 | IXPMS | 2 Org | Hf, [W] |
| Knab & Hintenberger 1978 | K78 | IDMS | 1 Org | Ba, Cu, Ga, Ge, Hf, Pb, Pt, [Re], Sb, Sn, Sr, [Te], [W], Zr, Dy |
| Koefoed et al. 2023 | K23 | ICPMS | 1 Alais, 1 Ivuna, 5 Org | K* |
| Kolesov 1974 | K74 | RNAA | 1 Org | [subset of REE] |
| Krähenbühl et al. 1973 | K73 | RNAA | 1 Alais, 3 Ivuna,4 Org | [Ag], Au, Bi, [Br], Cd*, Cs, Ge*, I, In, [Ir], Rb*, Re, [Sb], Se, [Te], Tl, U*, Zn |
| Krankowsky & Müller 1964 | K64 | ID-MS | 1 Org | [Li] |
| Labidi et al. 2016 | L16 | ICPMS | 2 Alais, 2 Org | Se |
| Laul et al. 1970a | L70a | RNAA | Ivuna,1 Org | Bi |
| Laul et al. 1970b | L70b | RNAA | Ivuna,1 Org | Tl |
| Loss et al. 1984 | L84 | IDMS | 5 Org | Ag, Cd*, Pd, Te |
| Loss et al. 1989 | L89 | IDMS | 4 Org | Sn |
| Loveland et al. 1969 | L69 | NAA | 1 Alais, 2 Ivuna, 1 Org | Al |
| Lu et al. 2007 | L07 | ICPMS-ID | 1 Ivuna, 4 Org | Hf*,Nb*. Mo, Sb, Sn ,Ta,*Ti, Zr* |
| Luck et al. 2005 | L05 | ICPMS-ID | 1 Ivuna, 2 Org | Cu, Zn |
| Makishima & Nakamura 2006 | M06 | ICPMS-ID | 2 Org | Al, Ba, Be, Bi, Ca, Co, Cs, Cu, Ga, [Fe], In, K, Li, Mn, Na, [Ni], P, Pb, Rb, Sc, Sr*, Tl, U, V, Y, Zn, All REE. |
| Mermelengas et al. 1979 | M79 | IDMS | 1 Org | Pd |
| Mittlefehldt 2002 | M02 | NAA | 1 Org | Al, As, Au, Ca, Co, Cr, Cs, Fe, [Hf], Ir, K, Mg, Mn, [Na], [Ni], Sb, Sc, Se, V, Zn. [La], [Eu], [Yb], [Lu] |
| Mittlefehldt & Wetherill 1979 | M79b | IDMS | 1 Ivuna, 3 Org | K, Rb, Sr* |
| Morgan & Lovering 1967 | M64 | RNAA | 2 Org | Os |

**Table A1. References for CI-Chondrite Concentration Data [a]**

| Reference | Code | Method | Number of samples and meteorites [b] | Elements Analyzed |
|---|---|---|---|---|
| Morgan & Lovering 1967 | M67 | RNAA | 2 Org | Os, Re |
| Morgan & Lovering 1968 | M68 | RNAA | 2 Alais, 6 Ivuna, 8 Org, 2 Tonk | Th*, U* |
| Morgan & Walker 1988 | M88 | IDMS RIMS | 1 Org | Os, Re |
| Münker et al. 2003 | M03 | ICPMS-ID-MC | 1 Org | [Hf], [Nb], [Ta], [Zr] |
| Murty et al. 1983 | M83 | RNAA | 1 Org | [Li] |
| Murthy & Compston 1965 | M65 | IDMS | 1 Org | K*, Rb*,Sr* |
| Nakamura et al. 2022 | N22 | ICPMS | 1 Org | Al, As, [B], Ba, [Be], [Bi], [Ca], Cd, [Cr], [Co], Cs, [Fe], Ga, Ge, Hf, Ir, K,[Li], [Mg], Mn, Na, [Nb], [Ni], Os, P, Pb, Pd, Pt, Rb, Re, Ru, Sb, Se, Sn, [Sr], Te, Th, Ti, [Tl], V, W, Y , Zn, Zr, All REE: [La] |
| Nakamura 1974 | N74a | AES | 2 Org | Ba*, K, Fe, Mg, Na, Pd, Sc, U, La,Ce, Nd, Dy, Sm, Eu, Gd, Dy, Er, Yb, Lu. |
| [Nichiporuk & Bingham. 1970] | N70b | AES | 1 Org | [Cu], [V] |
| Nichiporuk & Moore 1970 | N70 | AAS | 1 Org | [Li] |
| Nichiporuk & Moore 1974 | N74b | AAS | 1 Ivuna, 1 Org | [Fe], K, Li, Na |
| Nichiporuk et al. 1967 | N67 | XRF | 1 Org | Ca,[Co], [Cr], [Fe], Mn, Ni |
| Nie et al. 2021 | N21 | ICPMS | 1 Ivuna, 1 Org | K*, Rb* |
| Palme & Zipfel 2021 | P21 | INAA | 2 Alais, 3 Ivuna, 9 Org, 1 Tonk | Al, As*, Au, Ca*, Co*, Cs*, Cr, Cu, Fe, Ga, Ge, Hf*, Ir, K*, Mg, Mn, Na*, Nb, Ni*, [Mo], O, Os*, [Pd], [Pt], Re*, Ru*, Sb, Sc, Se*, Si, [Ti], V, [W], Zn, [La], [Ce], [Nd], [Sm], [Eu], [Gd],[[Tb], Dy], [Ho], [Er], [Tm], [Yb], [Lu] |
| Pinson et al. 1953 | P53 | AAS/ES | 1 Org | [Sc] |
| Pogge von Strandmann et al. 2011 | P11 | ICPMS | 1 Org | Li* |
| Pourmand et al. 2012 | P12 | ICPMS-MC | 1 Alais,3 Ivuna, 7 Org | Sc, Y, All REE |
| Pringle & Moynier 2017 | P17 | ICPMS | 1 Org | Rb |
| Rambaldi et al. 1978 | R78 | RNAA | 1 Org | Au, Ir, Ni, Os, Pt, Re |
| Rammensee & Palme 1982 | R82 | RNAA | 2 Org | As, Au, Co, Cu, Ir, Mo, Ni, Os, Pt, Sb, W |
| Reed et al. 1960 | R60 | RNAA | 2 Org | Ba, Pb*, Tl, U |
| Reed & Allen1966 | R66 | RNAA | 2 Ivuna,2 Org | [Te], U |
| Rocholl & Jochum 1993 | R93 | ID-SSMS | 1 Ivuna, 5 Org | Ba*,Cs, Hf*, Nb*, Rb, Pb, Sb*, [Sr], Th*, U*, Y, Zr*, La*, Ce, Pr, Nd, Sm, Eu*, Gd, Tb, Dy, Ho, Er, Yb |
| Rosman & De Laeter 1974 | R74 | IDMS | 1 Org | Cd |
| Rosman & De Laeter 1986 | R86 | IDMS | 1 Org | Cd |
| Schmitt et al. 1963 | S63 | INAA/RNAA | 1 Ivuna, 1 Org | [Cd] |
| Schmitt et al. 1964 | S64-1 | RNAA | 1 Ivuna, 1 Org | Sc*, Y. [All REE except Gd] |
| Schmitt et al. 1972 | S72 | NAA | 1 Alais, 2 Ivuna, 1 Org | Al, Co*, Cr, Cu, Fe*, In, Mn, Na, Sc |
| Schönbächler et al. 2005 | S05 | ICPMS | 1 Org | Ag, [Nd], Pd, Zr |
| Seitz et al. 2007 | S07 | ICPMS | 1 Org | Li |
| Sephton et al. 2013 | S13 | ICPMS | 1 Org | Li |
| Shima 1979 | S79 | IDMS | 1 Org | Hf, Ti, Zr |

**Table A1. References for CI-Chondrite Concentration Data [a]**

| Reference | Code | Method | Number of samples and meteorites [b] | Elements Analyzed |
|---|---|---|---|---|
| Smales et al. 1964 | S64-2 | IDMS, RNAA | 4 Iv | Cs, Rb |
| Smales et al. 1971 | S71 | IDMS | 1 Iv | Cs, Rb |
| Smith et al. 1977 | S77 | IDMS | 2 Org | Te |
| Takahashi et al. 1978 | T78 | RNAA | 1 Org | Ag, Au, Cd, Cs, [Ge], In, Ir, Ni, Os, Pd, Rb, Re, [Sb], Se, [Te], Tl, U, Zn |
| Tanner & Ehmann 1967 | T67 | INAA/RNAA | 1 Ivuna,1 Org | Sb |
| Tatsumoto et al. 1976 | T76 | IDMS | 1 Org | Pb, Th, U |
| Vilcsek E. 1977 | V77 | AAS | 1 Org | [Be] |
| Vogt & Ehmann 1965 | V65 | NAA | 1 Org | Si |
| Vollstaedt et al. 2016 | V16 | ICPMS-ID | 1 Org | Se |
| Von Michaelis et al. 1969 | V69 | XRF | 1Org | Al, [Ca],Fe, K, Mg, Mn, P, Si, Ti |
| Walker et al. 2002 | W02 | N-TIMS | 2 Ivuna, 2 Org | Os, Re |
| Wang et al. 2013 | W13 | ICPMS-HG-ID | 1 Org | Se, Te |
| Wang et al. 2014 | W15 | ICPMS-SF-ID | 1 Ivuna,1 Org | Ag*, [Ba], [Bi], Cd, Cu, Mo, Se, Te, Tl, W, Sm. |
| Wang & Jacobsen 2016 | W16 | ICPMS | 1 Org | [K] |
| Wieser & de Laeter 2000 | W00 | IDMS | 3 Org | Mo |
| Weller et al. 1978 | W78 | Various | 1 Ivuna, 1 Org | [B] |
| Wieser, & De Laeter 2000 | W00 | IDMS | 2 Ivuna, 3 Org | Mo |
| Wiik 1956. 1969 | W56 | Flame photometry | 1 Alais, 1 Ivuna, 2 Org | [Al], Co, Cr*, [Cu],Fe, K*, Mg, [Mn], [Na], [P], Si, Ti, [V] |
| Wing 1964 | W64 | NAA | 1 Org | Si, O |
| Wolf & Palme 2001 | W02 | XRF | 1 Ivuna, 4 Org | Al, Ca, Cr, Fe, Mg, Mn, P, Si, Ti, V |
| Wolf et al. 2005 | W05 | ICPMS-Q | 1 Org | Bi, Cd, Cs, Cu, Ga, Rb, Sb, Se, Sn, Te, Tl, Zn |
| Xiao & Lipschutz 1992 | X92 | RNAA | 2 Org | Ag, Au, Bi, Cd, Co, Cs, Ga, In, Rb, Sb, Se, [Te], Zn |
| Yi & Masuda 1996 | Y96 | ICPMS-ID | 2 Org | Ir, Pd, Pt, Ru |
| Zhai & Shaw 1994 | Z94 | PGA | 2 Ivuna, 2 Org | B |

[a] For C, N, O, and S see Table 2. For halogens, see Lodders and Fegley (2023). Data in square brackets [ ] indicate that raw data were considered but not included in final meteorite and group mean. Elements marked with a star indicate that one or more samples were excluded from multiple samples reported in a given reference. No iron measurements from ICP and no REE by INAA and RNAA are included in the group mean. [b] Org =Orgueil.

**Table A2. References for Hydrogen, Carbon, Nitrogen, Sulfur in CI-Chondrites**

| Reference | Code | Method | Number of samples and meteorites [b] | Elements Analyzed |
|---|---|---|---|---|
| Alexander et al. 2012b | A12 | Combustion element analyzer | 1 Ivuna, 2 Org | C |
| Belsky & Kaplan 1970 | B70 | Stepped combustion, MS | 1 Ivuna, 1 Org | C (same as S70) |
| Berzelius 1834 | B834 | Wet. Chem/ combustion | 1 Alais | [C from carbonate & bulk], [S] |
| Boato 1954 | B54 | Stepped combustion, MS | 1 Ivuna, 1 Org | C |
| Braukmüller et al. 2018 | B18 | ICPMS | 1 Ivuna, 2 Org | S |
| Briggs & Mamikunian 1963 | B63 | Combustion | 1 Org | H, C, S |
| Burgess et al. 1991 | B91 | Stepped combustion, MS | 1 Ivuna, 2 Org* | S" |
| Christie 1914 | C14 | Wet chem., flame photom. combustion | 1 Tonk | C [as carbonate] C bulk, S |
| Cloez 1864a | C864a | Wet. Chem./ combustion | 1 Org I | C |
| Cloez 1864b | C864b | Wet. Chem./ combustion | 1 Org | C, S |
| Cloez 1864c | C864c | combustion | 1 Org | [C from carbonate only] |
| Curtis & Galdney 1985 | C85 | PGA | 2 Ivuna, 2 Org | S |
| Dreibus et al. 1993 | D93 | Combustion element analyzer | 1 Org | S |
| Dreibus et al. 2004 | D04 | Combustion element analyzer | 1 Ivuna, 1 Org | C, S |
| Filhols & Mellies 1864 | F864 | Wet. Chem./combustion | 1 Org | C, [S] |
| Folinsbee et al. 1967 | F67 | Combustion element analyzer | 2 Revelstoke | C, S |
| Fredriksson et al. 1997 | F97 | Pellet analysis | 1 Org | [C, S] |
| Funk 2015 | F15 | ICPMS | 1 Org | S |
| Gao & Thiemens 1993 | | Wet. Chem., MS | 2 Org | S* |
| Gibson et al. 1971 | G71 | Gas chromatography /spectrophotometic | 4 Org | C, N |
| Grady et al 1991 | G91 | MS, stepwise heating | 1 Alais, 1 Org | C, N |
| Grady et al. 2002 | G02 | MS, stepwise heating | 1 Org | C, N |
| Halbout et al. 1986 | H86 | MS, stepwise heating | 1 Org | C |
| Islam et al 2012 | I12 | PGA | 2 Org | S |
| Jarosewich 1972 | J72 | Combustion | 1 Org | C, S |
| Kaplan & Hulston 1966 | K66 | MS, pyrolysis | 1 Org | S |
| Kerridge 1985 | K85 | MS, pyrolysis | 1 Alais, 1 Iv | H, C, N |
| Kiesl 1979 | K79 | Unknown | 1 Org | C, S |
| Lawrence Smith 1876a | L876a | Dissolution/combustion | Orgueil 50gr, [Alais 2 grams] | C |
| Lawrence Smith 1876b | L876b | Dissolution/combustion | Orgueil 50gr, Alais 2 grams] | C* |
| Mason 1962 | M62 | Combustion | 1 Alais | C, [S] |
| Monge et al. 1864 | M864 | Wet. Chem./combustion | 1 Alais 5 gr | [C from carbonate only] |
| Monster et al. 1965 | M65-2 | | | |
| Mueller et al. 1965 | M65-3 | Combustion | 1 Org | H.C,S |
| Nakamura et al. 2022 | N22 | Combustion element analyzer | 1 Org | C, S |
| Otting & Zaehringer 1967 | O67 | Combustion | 1 Org | C |
| Pearson et al. 2005 | P05 | Combustion element analyzer, MS | 8 Alais, 4 Org | C* |
| Pisani 1864 | P864 | Wet. Chem./ combustion | 1 Orgueil | [C], S |
| Robert & Epstein 1982 | R82-2 | Combustion, stepped pyrolysis, 200-1300°C, MS | 2 Org | C |
| Roscoe 1863 | R863 | Combustion | 1 Alais | C |
| Sephton et al. 2003 | S03 | MS, stepwise heating | 1 Org | C |
| Smith & Kaplan 1970 | S70 | MS, stepwise heating | 1 Org, 1 Iv | S |

**Table A2. References for Hydrogen, Carbon, Nitrogen, Sulfur in CI-Chondrites**

| Reference | Code | Method | Number of samples and meteorites [b] | Elements Analyzed |
|---|---|---|---|---|
| Thenard 1806 | T806 | Wet. Chem. | Alais, 10 gr | C, [S] |
| Wang et al. 2014 | W15 | | 1 Ivuna, 1 Org | S |
| Wiik 1956 | W56 | Wet. Chem./combustion | 1 Ivuna, 1 Org | C, S |
| Wiik 1969 | W69 | Wet. Chem./combustion | 1 Org | C, S |
| Wright et al. 1985 | W85 | MS, stepwise heating | 1 Org | C |
| Wright et al. 1986 | W86 | MS, stepwise heating | 1 Alais, 1 Org | [C] |

**Table A3. References for Early Analyses of CI-Chondrites**

| Reference | Code | Method | Number of samples and meteorites [b] | Elements Analyzed |
|---|---|---|---|---|
| Berzelius 1834 | B834 | Wet. Chem | Alais | [Al], [Fe], [Mg], [Si] |
| Christie 1914 | C14 | Wet chem., flame photometry | Tonk | [Ca], [Fe], [P], [Si], [Ti] |
| Cloez 1864a,b | C864a,b | Wet. Chem. | Orgueil | [Ca],[Al], [Fe], [Mg], [Si] |
| Filhols & Mellies 1864 | F864 | Wet. Chem. | Orgueil | [Al], [Fe], [Mg], other? |
| Monge et al. 1864 | M864 | Wet. Chem. | Orgueil, 5 gr | [Cr],[Fe], [Si], [Mn],[Ni]- |
| Thenard 1806 | T806 | Wet. Chem. | Alais, 10 gr | |
| Pisani 1864 | P864 | Wet. Chem. | Orgueil | [Al], [Ca],[Fe], [Mg], [Si] other |

| **Table A4. Refractory Element Concentrations from Element Ratios in Carbonaceous Chondrites, and determined by averaging selected analyses [a]** | | | | | | |
|---|---|---|---|---|---|---|
| **Element** | **Concentration from** | **Source** | **Weight Ratio** | **%SD** | **Concentration ppm** | **%SD** |
| Lu | grand average | Table 4 | -- | | **0.255** | 5.7 |
| **Hf** | ratio used | Lu/Hf | 0.238 | 1 | **0.107** | **5.8** |
| Hf | grand average | | | | 0.108 | 7.4 |
| **Zr** | ratio used | Zr/Hf | 34.1 | 1 | **3.65** | **5.9** |
| Zr | grand average | | | | 3.73 | 8.0 |
| **Nb** | ratio used | Zr/Nb | 13.47 | 5.8 | **0.271** | **8.3** |
| Nb | grand average | | | | 0.29 | 10.3 |
| Ta | ratio used | Nb/Ta | 19.4 | 3.9 | 0.0140 | 9.2 |
| **Ta** | grand average | | | | **0.0149** | **4.7** |
| **Y** | ratio used | Zr/Y | 2.4 | 6.5 | **1.52** | **8.8** |
| Y | grand average | | | | 1.51 | 9.2 |
| | | | | | | |
| Th | grand average | Table 4 | | | 0.0304 | 10.9 |
| U | ratio used | Th/U | 3.75 | 11 | 0.0810 | 12.3 |
| **U** | grand average | | | | **0.0810** | **9.5** |
| | | | | | | |
| Al | grand average | Table 4 | | | 8470 | 5.7 |
| **Ca** | ratio used | Ca/Al | 1.08 | 2.1 | **9148** | **6.1** |
| Ca | grand average | | | | 8970 | 8.9 |
| | | | | | | |
| **Ir** | grand average | Table 4 | | | **0.456** | **5.9** |
| Os | ratio used | Os/Ir | 1.06 | 2.8 | 0.483 | 6.6 |
| **Os** | grand average | | | | **0.491** | **2.8** |
| **Re** | ratio used | Os/Re | 12.32 | 3.0 | **0.0399** | **4.1** |
| Re | grand average | | | | 0.0371 | 9.2 |
| **Ru** | ratio used | Os/Ru | 0.723 | 5.3 | **0.670** | **8.4** |
| Ru | grand average | | | | 0.648 | 10.2 |
| **Pt** | ratio used | Pt/Ir | 1.98 | 5.0 | **0.904** | **7.7** |
| Pt | grand average | | | | 0.926 | 11.2 |
| Rh | ratio used | Pt/Rh | 7.24 | 7.5 | 0.126 | 10.8 |
| **Rh** | grand average | | | | **0.133** | **4.5** |

[a] entries in bold are adopted final values, see Table 4. For element ratios, see tables A5-A12.

**Table A5. Ca/Al Ratios of Carbonaceous Chondrites**

| Source | Number of samples | Ca/Al | %SD | Remarks |
|---|---|---|---|---|
| Ahrens 1970 | 8 | 1.08 | 1.9 | CI, CM, CO, CV |
| Kallemeyn and Wasson 1981 | 45 | 1.07 | 8.5 | CI, CM, CO, CV |
| Wolf and Palme 2001 | 19 | 1.10 | 4.8 | one Orgueil, Axtell, Bali excluded |
| Jarosewich 2006 | 15 | 1.04 | 7.2 | Murchison, Colony, Axtell excluded |
| Patzer et al. 2010 | 7 | 1.10 | 5.6 | Axtell excluded |
| Stracke et al. 2012 | 39 | 1.10 | 8.8 | only Allende samples |
| Braukmüller et al. 2018 | 24 | 1.07 | 8.6 | Tagish Lake, EET85013 excluded |
| **Average Ratios** | | **1.080** | **6.1** | |
| from grand average method | | 1.060 | 8.9 | |

**Table A6. Lu/Hf Ratios of Carbonaceous Chondrites**

| Source | Number of samples | Lu/Hf | %SD | Remarks | Analytical Method |
|---|---|---|---|---|---|
| Beer et al. 1984 | 2 | 0.2382 | | 2 Orgueil samples | ID |
| Patchett et al. 2004 | 7 | 0.2386 | 1.4 | Karoonda & Kainsaz excluded | ID |
| Bouvier et al. 2008 | 15 | 0.2387 | 1.3 | Karoonda & Ninqiang excluded | ID |
| Dauphas & Pourmand 2011 | 13 | 0.2382 | 1.0 | | ID |
| Stracke et al. 2012 | 39 | 0.2408 | 2.7 | only Allende samples | ICP-MS |
| Barrat et al. 2012; Göpel et al. 2015 | 10 | 0.2346 | 2.9 | 2 Orgueil samples excluded | ICP-MS |
| Braukmüller et al. 2018 | 17 | 0.2478 | 2.7 | ALH85002 excluded | ICP-MS |
| **Average Ratios** | | **0.2384** | **1** | **only ID analyses** | |
| From grand average method | | 0.2355 | 9.3 | | |

Samples with excessive deviations (>2σ) from average were excluded; Barrat et al. (2012) reported data on 6 Orgueil samples, Göpel et al. (2015) included additional data analyzed by Barrat.

**Table A7. Zr/Hf Ratios of Carbonaceous Chondrites**

| Source | Number of samples | Zr/Hf | %SD | Analytical Method |
|---|---|---|---|---|
| Shima 1971 | 11 | 34.11 | 3.4 | ID |
| Weyer et al. 2002 | 5 | 34.12 | 2.1 | ID |
| Münker et al. 2003 | 14 | 34.09 | 1.3 | ID |
| Lu et al. 2007 | 8 | 33.44 | 2.8 | ICP-MS |
| Patzer et al. 2010 | 6 | 32.89 | 4.2 | ICP-MS |
| Stracke et al. 2012 | 39 | 35.99 | 2.1 | ICP-MS |
| Barrat et al.2012; Göpel et al. 2015 | 12 | 32.85 | 3.0 | ICP-Ms |
| Braukmüller et al. 2018 | 26 | 33.74 | 1.4 | ICP-MS |
| Münker et al. 2025 | 8 | 33.9 | 1.5 | ID |
| Average, all data | | 33.9 | 2.7 | |
| **Average ID analyses** | | **34.1** | **1** | |
| Grand average method | | 34.1 | 7.5 | |

**Table A8. Zr/Nb Ratios of Carbonaceous Chondrites** (only CI- and CM-Chondrites, see text)

| Source | Number of samples | Zr/Nb | %SD | Remarks |
|---|---|---|---|---|
| Schönbächler et al. 2003 | 5 | 14.11 | 10.2 | Ci, CM |
| Münker et al. 2003 | 6 | 13.58 | 9.7 | CI, CM |
| Lu et al. 2007 | 8 | 12.41 | 4.2 | CI |
| Braukmüller et al. 2018 | 16 | 12.95 | 5.0 | CI, CM |
| Münker et al. 2025 | 8 | 14.3 | 5.6 | CI, CM |
| **Average Ratios** | **6** | **13.47** | **5.8** | |
| Grand average method | | 13.27 | 13.0 | |

**Table A9. Nb/Ta Ratios of Carbonaceous Chondrites (only CI- and CM-Chondrites, see text)**

| Source | Number of samples | Nb/Ta | %SD | Remarks |
|---|---|---|---|---|
| Münker et al. 2003 | 6 | 20.40 | 3.8 | only CI and CM |
| Lu et al. 2007 | 7 | 18.67 | 6.3 | CI, CM |
| Barrat et al. 2012 Göpel et al. 2015 | 10 | 19.53 | 3.09 | CI, CM [(a)] |
| Münker et al. 2025 | 6 | 19.00 | 2.1 | CI |
| **Average Ratios** | | **19.40** | **3.6** | |
| Grand average method | | 18.94 | 11.3 | |

[a] Barrat et al. (2012) reported data for 6 Orgueil samples, Göpel et al. (2015) included additional data analyzed by Barrat. Data for heavily weathered Arch were excluded.

**Table A10. Zr/Y Ratios of Carbonaceous Chondrites**

| Source | Number of samples | Zr/Y | %SD | Remarks |
|---|---|---|---|---|
| Jochum et al. 2000 | 3 | 2.36 | 5.9 | Orgueil, Murray, Murchison |
| Stracke et al. 2012 | 39 | 2.58 | 6.7 | only Allende |
| Barrat et al. 2012; Göpel et al. 2015 | 12 | 2.27 | 8.7 | |
| **Average Ratios** | | **2.40** | **6.6** | |
| Grand average method | | 2.47 | 12.2 | |

[a] Barrat et al. (2012) reported data for 6 Orgueil samples, Göpel et al. (2015) included additional data analyzed by Barrat.

**Table A11. Sm/Nd Ratios of Carbonaceous Chondrites**

| Source | Number of samples | Sm/Nd | %SD | Remarks |
|---|---|---|---|---|
| Beer et al. 1984 | 2 | 0.3276 | | Orgueil |
| Patchett et al. 2004 | 6 | 0.3230 | 2.0 | |
| Bouvier et al. 2008 | 17 | 0.3250 | 0.9 | |
| Barrat et al. 2012; Göpel et al. 2015 | 13 | 0.3280 | 2.9 | CI, CM |
| Braukmüller et al. 2018 | 24 | 0.3306 | 2.1 | CI, CM |
| **Average Ratios** | | **0.3268** | **1** | |
| Grand average method | | 0.323 | 7.3 | |

**Table A12. Th/U Ratios of Carbonaceous Chondrites**

| Source | Number of samples | Th/U | %SD | Remarks |
|---|---|---|---|---|
| Tatsumoto et al. 1973 | 3 | 3.8 | 6 | CM, CV |
| Rocholl and Jochum 1993 | 29 | 3.08 | 24 | |
| Dauphas and Pourmand 2011 | 12 | 3.74 | 7.7 | CI, CM, CV, CO; Lance excluded |
| Makashima and Nakamura 2011 | 6 | 3.66 | 4.3 | CI, CM |
| Stracke et al. 2012 | 39 | 4.21 | 12.7 | Allende, one sample excluded |
| Barrat et al. 2012; Göpel et al. 2015 | 9 | 3.61 | 6.6 | CI, CM |
| Braukmüller et al. 2018 | 23 | 4.15 | 9.6 | Nogoya, Vigarano excluded |
| **Average** | | **3.75** | **11** | |
| Grand average method | | 3.75 | 14 | |

Table A13. Proto Solar (4.567 Ga ago) Isotopic and Elemental Compositions, Mass Fractions, and Atomic Weights *

| | Z | A | Isotope Fractions of Element atom% | Elemental Abundance N(Si)=1e6 | Isotopic Abundance Σ Si=1e6 | Elemental Mass Fractions | Isotopic Mass Fractions | Atomic Mass AMU | Mean Atomic Weight (Proto Solar) Dalton | Mean Atomic Weight (Present-Day) Dalton |
|---|---|---|---|---|---|---|---|---|---|---|
| H | 1 | 1 | 99.99803 | | 2.29E+10 | | 7.061E-01 | 1.007825032 | | |
| H (D) | 1 | 2 | 0.00197 | | 4.51E+05 | | 2.780E-05 | 2.014101778 | | |
| | | | **100** | **2.29E+10** | | **0.7061** | | | 1.007825 | 1.007845 |
| He | 2 | 3 | 0.0166 | | 3.73E+05 | | 3.443E-05 | 3.01602932 | | |
| He | 2 | 4 | 99.9834 | | 2.25E+09 | | 2.752E-01 | 4.002603254 | | |
| | | | **100** | **2.25E+09** | | **0.2752** | | | 4.002199 | 4.002439 |
| Li | 3 | 6 | 7.589 | | 4.3 | | 7.830E-10 | 6.015122885 | | |
| Li | 3 | 7 | 92.411 | | 51.8 | | 1.112E-08 | 7.016003428 | | |
| | | | **100** | **56.1** | | **1.190E-08** | | | 6.940047 | 6.940047 |
| Be | 4 | 9 | **100** | **0.657** | 0.657 | **1.810E-10** | 1.810E-10 | 9.01218291 | 9.012183 | 9.012183 |
| B | 5 | 10 | 19.83 | | 3.6 | | 1.099E-09 | 10.01293696 | | |
| B | 5 | 11 | 80.17 | | 14.5 | | 4.885E-09 | 11.00930537 | | |
| | | | **100** | **18.1** | | **5.984E-09** | | | 10.81176 | 10.81176 |
| C | 6 | 12 | 98.965 | | 8.99E+06 | | 3.299E-03 | 12 | | |
| C | 6 | 13 | 1.035 | | 94000 | | 3.738E-05 | 13.00335484 | | |
| | | | **100** | **9.08E+06** | | **3.336E-03** | | | 12.01039 | 12.01039 |
| N | 7 | 14 | 99.774 | | 2.44E+06 | | 1.044E-03 | 14.003074 | | |
| N | 7 | 15 | 0.226 | | 5520 | | 2.532E-06 | 15.0001089 | | |
| | | | **100** | **2.44E+06** | | **1.047E-03** | | | 14.00532 | 14.00532 |
| O | 8 | 16 | 99.777 | | 1.61E+07 | | 7.883E-03 | 15.99491462 | | |
| O | 8 | 17 | 0.035 | | 5700 | | 2.963E-06 | 16.99913176 | | |
| O | 8 | 18 | 0.188 | | 30400 | | 1.673E-05 | 17.99915961 | | |
| | | | **100** | **1.62E+07** | | **7.902E-03** | | | 15.99904 | 15.99904 |
| F | 9 | 19 | **100** | **1278** | 1278 | **7.424E-07** | 7.424E-07 | 18.99840317 | 18.99840 | 18.99840 |
| Ne | 10 | 20 | 93.125 | | 3.69E+06 | | 2.257E-03 | 19.99244018 | | |
| Ne | 10 | 21 | 0.224 | | 8860 | | 5.689E-06 | 20.99384668 | | |
| Ne | 10 | 22 | 6.651 | | 2.64E+05 | | 1.773E-04 | 21.99138512 | | |
| | | | **100** | **3.96E+06** | | **2.440E-03** | | | 20.12764 | 20.12764 |
| Na | 11 | 23 | **100** | **55900** | 55900 | **3.930E-05** | 3.930E-05 | 22.98976928 | 22.98977 | 22.98977 |
| Mg | 12 | 24 | 78.992 | | 8.18E+05 | | 6.003E-04 | 23.9850417 | | |
| Mg | 12 | 25 | 10.003 | | 1.04E+05 | | 7.919E-05 | 24.98583691 | | |
| Mg | 12 | 26 | 11.005 | | 1.14E+05 | | 9.060E-05 | 25.98259295 | | |
| | | | **100** | **1.04E+06** | | **7.701E-04** | | | 24.30498 | 24.30498 |
| Al | 13 | 27 | **100** | **81100** | 81100 | **6.692E-05** | 6.692E-05 | 26.98153859 | 26.98154 | 26.98154 |
| Si | 14 | 28 | 92.2297 | | 9.22E+05 | | 7.891E-04 | 27.97692653 | | |
| Si | 14 | 29 | 4.6832 | | 46800 | | 4.147E-05 | 28.97649467 | | |
| Si | 14 | 30 | 3.0872 | | 30900 | | 2.833E-05 | 29.97377017 | | |
| | | | **100** | **1.00E+06** | | **8.589E-04** | | | 28.08538 | 28.08538 |
| P | 15 | 31 | **100** | **8390** | 8390 | **7.948E-06** | 7.948E-06 | 30.973762 | 30.97376 | 30.97376 |
| S | 16 | 32 | 95.04074 | | 4.04E+05 | | 3.947E-04 | 31.97207117 | | |
| S | 16 | 33 | 0.74869 | | 3180 | | 3.207E-06 | 32.97145569 | | |
| S | 16 | 34 | 4.19599 | | 17800 | | 1.849E-05 | 33.9678669 | | |
| S | 16 | 36 | 0.01458 | | 62 | | 6.820E-08 | 35.96708076 | | |
| | | | **100** | **4.25E+05** | | **4.165E-04** | | | 32.06388 | 32.06388 |
| Cl | 17 | 35 | 75.7647 | | 4030 | | 4.310E-06 | 34.96885268 | | |
| Cl | 17 | 37 | 24.2353 | | 1290 | | 1.458E-06 | 36.96590259 | | |
| | | | **100** | **5320** | | **5.768E-06** | | | 35.45284 | 35.45284 |
| Ar | 18 | 36 | 84.596 | | 75100 | | 8.261E-05 | 35.96754511 | | |
| Ar | 18 | 38 | 15.380 | | 13600 | | 1.579E-05 | 37.96273234 | | |
| Ar* | 18 | 40 | 0.024 | | 21 | | 2.567E-08 | 39.96238312 | | |
| | | | **100** | **88800** | | **9.843E-05** | | | 36.27538 | 36.27536 |
| K | 19 | 39 | 93.132 | | 3410 | | 4.063E-06 | 38.96370649 | | |
| K* | 19 | 40 | 0.147 | | 5 | | 6.111E-09 | 39.96399848 | | |
| K | 19 | 41 | 6.721 | | 246 | | 3.082E-07 | 40.96182526 | | |
| | | | **100** | **3660.00** | | **4.378E-06** | | | 39.09830 | 39.09947 |
| Ca* | 20 | 40 | 96.941 | | 58500 | | 7.150E-05 | 39.96259086 | | |
| Ca | 20 | 42 | 0.647 | | 391 | | 5.017E-07 | 41.95861801 | | |
| Ca | 20 | 43 | 0.135 | | 82 | | 1.077E-07 | 42.95876667 | | |
| Ca | 20 | 44 | 2.086 | | 1260 | | 1.694E-06 | 43.95548173 | | |
| Ca | 20 | 46 | 0.004 | | 2 | | 2.811E-09 | 45.9536926 | | |
| Ca | 20 | 48 | 0.187 | | 113 | | 1.657E-07 | 47.9525343 | | |
| | | | **100** | **60400** | | **7.397E-05** | | | 40.07802 | 40.07802 |
| Sc | 21 | 45 | **100** | **33.7** | 3.371E+01 | **4.634E-08** | 4.634E-08 | 44.9559119 | 44.95591 | 44.95591 |
| Ti | 22 | 46 | 8.249 | | 200 | | 2.811E-07 | 45.95262889 | | |
| Ti | 22 | 47 | 7.437 | | 181 | | 2.599E-07 | 46.95176293 | | |
| Ti | 22 | 48 | 73.72 | | 1790 | | 2.625E-06 | 47.94794631 | | |
| Ti | 22 | 49 | 5.409 | | 131 | | 1.961E-07 | 48.94786998 | | |
| Ti | 22 | 50 | 5.185 | | 126 | | 1.925E-07 | 49.94479117 | | |
| | | | **100** | **2430** | | **3.554E-06** | | | 47.86688 | 47.86688 |
| V | 23 | 50 | 0.25 | | 0.7 | | 1.046E-09 | 49.9471585 | | |
| V | 23 | 51 | 99.75 | | 273.6 | | 4.263E-07 | 50.9439595 | | |
| | | | **100** | **274** | | **4.2735E-07** | | | 50.94147 | 50.94147 |
| Cr | 24 | 50 | 4.345 | | 575 | | 8.783E-07 | 49.9460442 | | |
| Cr | 24 | 52 | 83.790 | | 11100 | | 1.763E-05 | 51.94050751 | | |
| Cr | 24 | 53 | 9.501 | | 1260 | | 2.040E-06 | 52.94064943 | | |
| Cr | 24 | 54 | 2.365 | | 313 | | 5.163E-07 | 53.93888045 | | |
| | | | **100** | **13200** | | **2.107E-05** | | | 51.99612 | 51.99612 |
| Mn | 25 | 55 | **100** | **9270** | 9.270E+03 | **1.558E-05** | 1.558E-05 | 54.93804512 | 54.93804 | 54.93804 |
| Fe | 26 | 54 | 5.845 | | 51100 | | 8.430E-05 | 53.93961046 | | |
| Fe | 26 | 56 | 91.754 | | 8.02E+05 | | 1.371E-03 | 55.93493745 | | |
| Fe | 26 | 57 | 2.119 | | 18500 | | 3.221E-05 | 56.93539427 | | |
| Fe | 26 | 58 | 0.282 | | 2460 | | 4.359E-06 | 57.93327558 | | |
| | | | **100** | **873500** | | **1.492E-03** | | | 55.84514 | 55.84514 |
| Co | 27 | 59 | **100** | **2290** | 2.260E+03 | **4.073E-06** | 4.073E-06 | 58.93319506 | 58.93319 | 58.93319 |
| Ni | 28 | 58 | 68.077 | | 34100 | | 6.042E-05 | 57.9353435 | | |
| Ni | 28 | 60 | 26.223 | | 13100 | | 2.401E-05 | 59.93078635 | | |
| Ni | 28 | 61 | 1.140 | | 570 | | 1.062E-06 | 60.93105603 | | |
| Ni | 28 | 62 | 3.635 | | 1820 | | 3.447E-06 | 61.92834511 | | |
| Ni | 28 | 64 | 0.926 | | 463 | | 9.052E-07 | 63.92796594 | | |
| | | | **100** | **50030** | | **8.985E-05** | | | 58.69335 | 58.69335 |
| Cu | 29 | 63 | 69.174 | | 381 | | 7.338E-07 | 62.92959751 | | |
| Cu | 29 | 65 | 30.826 | | 170 | | 3.374E-07 | 64.92778945 | | |
| | | | **100** | **551** | | **1.071E-06** | | | 63.54556 | 63.54556 |
| Zn | 30 | 64 | 49.1704 | | 614 | | 1.200E-06 | 63.92914224 | | |
| Zn | 30 | 66 | 27.7306 | | 346 | | 6.979E-07 | 65.92603345 | | |
| Zn | 30 | 67 | 4.0401 | | 50 | | 1.032E-07 | 66.92712739 | | |
| Zn | 30 | 68 | 18.4483 | | 230 | | 4.784E-07 | 67.9248442 | | |
| Zn | 30 | 70 | 0.6106 | | 8 | | 1.630E-08 | 69.9253193 | | |
| | | | **100** | **1250** | | **2.496E-06** | | | 65.37777 | 65.37777 |
| Ga | 31 | 69 | 60.108 | | 21.6 | | 4.559E-08 | 68.9255735 | | |
| Ga | 31 | 71 | 39.892 | | 14.4 | | 3.114E-08 | 70.9247026 | | |
| | | | **100** | **36.0** | | **7.673E-08** | | | 69.72307 | 69.72307 |
| Ge | 32 | 70 | 20.526 | | 24.9 | | 5.330E-08 | 69.9242474 | | |
| Ge | 32 | 72 | 27.446 | | 33.3 | | 7.330E-08 | 71.9220758 | | |
| Ge | 32 | 73 | 7.76 | | 9.4 | | 2.101E-08 | 72.9234589 | | |
| Ge | 32 | 74 | 36.523 | | 44.3 | | 1.003E-07 | 73.92117777 | | |
| Ge | 32 | 76 | 7.745 | | 9.4 | | 2.183E-08 | 75.92140273 | | |
| | | | **100** | **121** | | **2.697E-07** | | | 72.62959 | 72.62959 |
| As | 33 | 75 | **100** | **6.15** | 6.15 | **1.410E-08** | 1.410E-08 | 74.9215965 | 74.92159 | 74.92159 |

| | | | | | | | | | |
|---|---|---|---|---|---|---|---|---|---|
| Se | 34 | 74 | 0.863 | | 0.6 | | 1.396E-09 | 73.92247594 | | |
| Se | 34 | 76 | 9.22 | | 6.6 | | 1.531E-08 | 75.91921372 | | |
| Se | 34 | 77 | 7.594 | | 5.4 | | 1.278E-08 | 76.919914 | | |
| Se | 34 | 78 | 23.685 | | 16.9 | | 4.037E-08 | 77.9173091 | | |
| Se | 34 | 80 | 49.813 | | 35.6 | | 8.708E-08 | 79.9165213 | | |
| Se^ | 34 | 82 | 8.825 | | 6.3 | | 1.581E-08 | 81.9166994 | | |
| | | | **100** | **71.5** | | **1.728E-07** | | | 78.97168 | 78.97168 |
| Br | 35 | 79 | 50.686 | | 6.29 | | 1.518E-08 | 78.9183371 | | |
| Br | 35 | 81 | 49.314 | | 6.12 | | 1.515E-08 | 80.9162906 | | |
| | | | **100** | **12.4** | | **3.033E-08** | | | 79.90361 | 79.90361 |
| Kr | 36 | 78 | 0.3652667 | | 0.21 | | 4.988E-10 | 77.92036486 | | |
| Kr | 36 | 80 | 2.3440789 | | 1.34 | | 3.283E-09 | 79.91637915 | | |
| Kr | 36 | 82 | 11.686258 | | 6.70 | | 1.678E-08 | 81.91348282 | | |
| Kr | 36 | 83 | 11.572467 | | 6.63 | | 1.682E-08 | 82.9141271 | | |
| Kr | 36 | 84 | 56.89512 | | 32.6 | | 8.367E-08 | 83.91149717 | | |
| Kr | 36 | 86 | 17.13681 | | 9.82 | | 2.580E-08 | 85.91061067 | | |
| | | | **100** | **57.3** | | **1.468E-07** | | | 83.78964 | 83.78964 |
| Rb | 37 | 85 | 70.844 | | 5.02 | | 1.305E-08 | 84.91178974 | | |
| Rb* | 37 | 87 | 29.156 | | 2.07 | | 5.497E-09 | 86.90918054 | | |
| | | | **100** | **7.09** | | **1.855E-08** | | | 85.46776 | 85.49415 |
| Sr | 38 | 84 | 0.558 | | 0.13 | | 3.440E-10 | 83.9134203 | | |
| Sr | 38 | 86 | 9.871 | | 2.37 | | 6.226E-09 | 85.9092602 | | |
| Sr | 38 | 87 | 6.898 | | 1.66 | | 4.402E-09 | 86.9088771 | | |
| Sr | 38 | 88 | 82.672 | | 19.8 | | 5.336E-08 | 87.9056122 | | |
| | | | **100** | **24.0** | | **6.433E-08** | | | 87.61369 | 87.61750 |
| Y | 39 | 89 | **100** | **4.49** | 4.49 | **1.221E-08** | 1.221E-08 | 88.9058483 | 88.90584 | 88.90584 |
| Zr | 40 | 90 | 51.49 | | 5.41 | | 1.488E-08 | 89.9047044 | | |
| Zr | 40 | 91 | 11.218 | | 1.18 | | 3.277E-09 | 90.9056458 | | |
| Zr | 40 | 92 | 17.148 | | 1.80 | | 5.064E-09 | 91.9050408 | | |
| Zr | 40 | 94 | 17.359 | | 1.82 | | 5.238E-09 | 93.9063152 | | |
| Zr^ | 40 | 96 | 2.785 | | 0.29 | | 8.583E-10 | 95.9082734 | | |
| | | | **100** | **10.5** | | **2.931E-08** | | | 91.22184 | 91.22184 |
| Nb | 41 | 93 | **100** | **0.766** | 0.766 | **2.176E-09** | 2.176E-09 | 92.9063781 | 92.90637 | 92.90637 |
| Mo | 42 | 92 | 14.649904 | | 0.380 | | 1.069E-09 | 91.90680811 | | |
| Mo | 42 | 94 | 9.1877391 | | 0.238 | | 6.847E-10 | 93.9050856 | | |
| Mo | 42 | 95 | 15.873772 | | 0.412 | | 1.196E-09 | 94.9058394 | | |
| Mo | 42 | 96 | 16.67381 | | 0.433 | | 1.269E-09 | 95.90467712 | | |
| Mo | 42 | 97 | 9.582996 | | 0.249 | | 7.370E-10 | 96.9060196 | | |
| Mo | 42 | 98 | 24.286871 | | 0.630 | | 1.887E-09 | 97.9054058 | | |
| Mo^ | 42 | 100 | 9.7449085 | | 0.253 | | 7.727E-10 | 99.9074724 | | |
| | | | **100** | **2.60** | | **7.615E-09** | | | 95.94866 | 95.94866 |
| Ru | 44 | 96 | 5.54 | | 0.098 | | 2.874E-10 | 95.9075939 | | |
| Ru | 44 | 98 | 1.87 | | 0.033 | | 9.892E-11 | 97.9052876 | | |
| Ru | 44 | 99 | 12.76 | | 0.226 | | 6.822E-10 | 98.9059393 | | |
| Ru | 44 | 100 | 12.60 | | 0.223 | | 6.805E-10 | 99.9042195 | | |
| Ru | 44 | 101 | 17.06 | | 0.302 | | 9.307E-10 | 100.9055821 | | |
| Ru | 44 | 102 | 31.55 | | 0.558 | | 1.738E-09 | 101.9043493 | | |
| Ru | 44 | 104 | 18.62 | | 0.329 | | 1.046E-09 | 103.9054326 | | |
| | | | **100** | **1.77** | | **5.464E-09** | | | 101.06498 | 101.06498 |
| Rh | 45 | 103 | **100** | **0.341** | 0.341 | **1.072E-09** | 1.072E-09 | 102.9055043 | 102.90549 | 102.90549 |
| Pd | 46 | 102 | 1.02 | | 0.014 | | 4.404E-11 | 101.9056286 | | |
| Pd | 46 | 104 | 11.14 | | 0.154 | | 4.904E-10 | 103.9040359 | | |
| Pd | 46 | 105 | 22.33 | | 0.309 | | 9.925E-10 | 104.9050847 | | |
| Pd | 46 | 106 | 27.33 | | 0.379 | | 1.226E-09 | 105.9034808 | | |
| Pd | 46 | 108 | 26.46 | | 0.367 | | 1.210E-09 | 107.9038907 | | |
| Pd | 46 | 110 | 11.72 | | 0.162 | | 5.458E-10 | 109.9051703 | | |
| | | | **100** | **1.39** | | **4.509E-09** | | | 106.41533 | 106.41533 |
| Ag | 47 | 107 | 51.8392 | | 0.261 | | 8.518E-10 | 106.9050965 | | |
| Ag | 47 | 109 | 48.1608 | | 0.242 | | 8.062E-10 | 108.9047523 | | |
| | | | **100** | **0.503** | | **1.658E-09** | | | 107.86815 | 107.86815 |
| Cd | 48 | 106 | 1.249 | | 0.020 | | 6.450E-11 | 105.9064602 | | |
| Cd | 48 | 108 | 0.89 | | 0.014 | | 4.683E-11 | 107.9041824 | | |
| Cd | 48 | 110 | 12.485 | | 0.199 | | 6.691E-10 | 109.9030035 | | |
| Cd | 48 | 111 | 12.804 | | 0.204 | | 6.924E-10 | 110.9041781 | | |
| Cd | 48 | 112 | 24.117 | | 0.385 | | 1.316E-09 | 111.9027578 | | |
| Cd^ | 48 | 113 | 12.225 | | 0.195 | | 6.731E-10 | 112.9044026 | | |
| Cd | 48 | 114 | 28.729 | | 0.458 | | 1.596E-09 | 113.9033595 | | |
| Cd^ | 48 | 116 | 7.501 | | 0.120 | | 4.239E-10 | 115.9047632 | | |
| | | | **100** | **1.59** | | **5.482E-09** | | | 112.41215 | 112.41215 |
| In | 49 | 113 | 4.281 | | 0.008 | | 2.643E-11 | 112.9040574 | | |
| In^ | 49 | 115 | 95.719 | | 0.171 | | 6.015E-10 | 114.9038788 | | |
| | | | **100** | **0.179** | | **6.279E-10** | | | 114.81827 | 114.81827 |
| Sn | 50 | 112 | 0.971 | | 0.036 | | 1.227E-10 | 111.9048218 | | |
| Sn | 50 | 114 | 0.659 | | 0.024 | | 8.474E-11 | 113.9027788 | | |
| Sn | 50 | 115 | 0.339 | | 0.013 | | 4.397E-11 | 114.9033424 | | |
| Sn | 50 | 116 | 14.536 | | 0.537 | | 1.902E-09 | 115.9017405 | | |
| Sn | 50 | 117 | 7.676 | | 0.283 | | 1.013E-09 | 116.9029516 | | |
| Sn | 50 | 118 | 24.223 | | 0.894 | | 3.224E-09 | 117.9016031 | | |
| Sn | 50 | 119 | 8.59 | | 0.317 | | 1.152E-09 | 118.9033076 | | |
| Sn | 50 | 120 | 32.593 | | 1.203 | | 4.412E-09 | 119.9022002 | | |
| Sn | 50 | 122 | 4.629 | | 0.171 | | 6.370E-10 | 121.9034391 | | |
| Sn | 50 | 124 | 5.789 | | 0.214 | | 8.098E-10 | 123.9052761 | | |
| | | | **100** | **3.69** | | **1.340E-08** | | | 118.71035 | 118.71035 |
| Sb | 51 | 121 | 57.213 | | 0.194 | | 7.175E-10 | 120.9038157 | | |
| Sb | 51 | 123 | 42.787 | | 0.145 | | 5.455E-10 | 122.904214 | | |
| | | | **100** | **0.339** | | **1.263E-09** | | | 121.75972 | 121.75972 |
| Te | 52 | 120 | 0.096 | | 0.005 | | 1.670E-11 | 119.9040452 | | |
| Te | 52 | 122 | 2.603 | | 0.124 | | 4.605E-10 | 121.9030439 | | |
| Te^ | 52 | 123 | 0.908 | | 0.043 | | 1.619E-10 | 122.9042701 | | |
| Te | 52 | 124 | 4.816 | | 0.229 | | 8.659E-10 | 123.9028176 | | |
| Te | 52 | 125 | 7.139 | | 0.339 | | 1.294E-09 | 124.9044307 | | |
| Te | 52 | 126 | 18.952 | | 0.899 | | 3.463E-09 | 125.9033117 | | |
| Te^ | 52 | 128 | 31.687 | | 1.504 | | 5.881E-09 | 127.9044621 | | |
| Te^ | 52 | 130 | 33.799 | | 1.604 | | 6.372E-09 | 129.9062228 | | |
| | | | **100** | **4.74** | | **1.851E-08** | | | 127.58559 | 127.58559 |
| I | 53 | 127 | **100** | **1.59** | 1.590E+00 | **6.171E-09** | 6.171E-09 | 126.9044728 | 126.90447 | 126.90447 |
| Xe^ | 54 | 124 | 0.129 | | 0.007 | | 2.747E-11 | 123.905893 | | |
| Xe | 54 | 126 | 0.110 | | 0.006 | | 2.365E-11 | 125.9042912 | | |
| Xe | 54 | 128 | 2.220 | | 0.124 | | 4.862E-10 | 127.9035313 | | |
| Xe | 54 | 129 | 27.428 | | 1.536 | | 6.055E-09 | 128.9047809 | | |
| Xe | 54 | 130 | 4.349 | | 0.244 | | 9.677E-10 | 129.9035094 | | |
| Xe | 54 | 131 | 21.763 | | 1.219 | | 4.879E-09 | 130.9050524 | | |
| Xe | 54 | 132 | 26.360 | | 1.476 | | 5.955E-09 | 131.9041551 | | |
| Xe | 54 | 134 | 9.730 | | 0.545 | | 2.231E-09 | 133.9053945 | | |
| Xe^ | 54 | 136 | 7.911 | | 0.443 | | 1.841E-09 | 135.9072145 | | |
| | | | **100** | **5.60** | | **2.247E-08** | | | 131.18270 | 131.18270 |
| Cs | 55 | 133 | **100** | **0.367** | 3.671E-01 | **1.492E-09** | 1.492E-09 | 132.905452 | 132.90545 | 132.90545 |
| Ba^ | 56 | 130 | 0.1058 | | 4.879E-03 | | 1.938E-11 | 129.9063215 | | |
| Ba | 56 | 132 | 0.1012 | | 4.667E-03 | | 1.883E-11 | 131.9050613 | | |
| Ba | 56 | 134 | 2.417 | | 1.115E-01 | | 4.564E-10 | 133.9045084 | | |
| Ba | 56 | 135 | 6.592 | | 3.040E-01 | | 1.254E-09 | 134.9056886 | | |
| Ba | 56 | 136 | 7.853 | | 3.621E-01 | | 1.505E-09 | 135.904576 | | |
| Ba | 56 | 137 | 11.232 | | 5.179E-01 | | 2.169E-09 | 136.9058274 | | |
| Ba | 56 | 138 | 71.699 | | 3.306E+00 | | 1.394E-08 | 137.9052473 | | |
| | | | **100** | **4.61** | | **1.937E-08** | | | 137.32692 | 137.32692 |
| La* | 57 | 138 | 0.0916 | | 4.294E-04 | | 1.811E-12 | 137.907112 | | |
| La | 57 | 139 | 99.9084 | | 4.684E-01 | | 1.990E-09 | 138.9063533 | | |
| | | | **100** | **0.469** | | **1.992E-09** | | | 138.90548 | 138.90545 |
| Ce | 58 | 136 | 0.186 | | 2.208E-03 | | 9.179E-12 | 135.9071295 | | |
| Ce^ | 58 | 138 | 0.250 | | 2.967E-03 | | 1.251E-11 | 137.905991 | | |
| Ce | 58 | 140 | 88.450 | | 1.050E+00 | | 4.493E-09 | 139.9054387 | | |
| Ce^ | 58 | 142 | 11.114 | | 1.320E-01 | | 5.727E-10 | 141.9092442 | | |
| | | | **100** | **1.19** | | **5.088E-09** | | | 140.90766 | 140.90766 |
| Pr | 59 | 141 | **100** | **0.179** | 1.786E-01 | **7.697E-10** | 7.697E-10 | 140.9076525 | 140.90766 | 140.90766 |
| Nd | 60 | 142 | 27.045 | | 0.237 | | 1.030E-09 | 141.9077233 | | |
| Nd* | 60 | 143 | 12.023 | | 0.105 | | 4.610E-10 | 142.9098143 | | |
| Nd^ | 60 | 144 | 23.729 | | 0.208 | | 9.162E-10 | 143.9100873 | | |
| Nd | 60 | 145 | 8.763 | | 0.077 | | 3.407E-10 | 144.9125736 | | |
| Nd | 60 | 146 | 17.130 | | 0.150 | | 6.706E-10 | 145.9131169 | | |
| Nd | 60 | 148 | 5.716 | | 0.050 | | 2.268E-10 | 147.9168933 | | |
| Nd^ | 60 | 150 | 5.596 | | 0.049 | | 2.251E-10 | 149.9208949 | | |
| | | | **100** | **0.877** | | **3.870E-09** | | | 144.24276 | 144.24465 |
| Sm | 62 | 144 | 3.083 | | 0.0084 | | 3.695E-11 | 143.9120046 | | |
| Sm* | 62 | 147 | 15.017 | | 0.0000 | | 0.000E+00 | 146.9148979 | | |
| Sm^ | 62 | 148 | 11.254 | | 0.0422 | | 1.907E-10 | 147.9148227 | | |
| Sm | 62 | 149 | 13.830 | | 0.0307 | | 1.397E-10 | 148.9171847 | | |
| Sm | 62 | 150 | 7.351 | | 0.0377 | | 1.729E-10 | 149.9172755 | | |
| Sm | 62 | 152 | 26.735 | | 0.0201 | | 9.333E-11 | 151.9197324 | | |
| Sm | 62 | 154 | 22.730 | | 0.0729 | | 3.433E-10 | 153.9222093 | | |
| | | | **100** | **0.0620** | | **9.770E-10** | | | 150.36500 | 150.36328 |
| Eu^ | 63 | 151 | 47.81 | | 0.0498 | | 2.296E-10 | 150.9198502 | | |
| Eu | 63 | 153 | 52.19 | | 0.0543 | | 2.540E-10 | 152.9212303 | | |
| | | | **100** | **0.1041** | 0.1041 | **4.836E-10** | | | 151.96438 | 151.96438 |
| Gd^ | 64 | 152 | 0.2029 | | 0.00071 | | 3.320E-12 | 151.9197922 | | |
| Gd | 64 | 154 | 2.1809 | | 0.00768 | | 3.616E-11 | 153.9208693 | | |
| Gd | 64 | 155 | 14.7998 | | 0.05213 | | 2.470E-10 | 154.9226276 | | |
| Gd | 64 | 156 | 20.4664 | | 0.07209 | | 3.437E-10 | 155.9221287 | | |
| Gd | 64 | 157 | 15.6518 | | 0.05513 | | 2.646E-10 | 156.9239647 | | |
| Gd | 64 | 158 | 24.8347 | | 0.08747 | | 4.225E-10 | 157.9241101 | | |
| Gd | 64 | 160 | 21.8635 | | 0.07701 | | 3.766E-10 | 159.9270585 | | |
| | | | **100** | **0.352** | | **1.694E-09** | | | 157.25205 | 157.25205 |
| Tb | 65 | 159 | **100** | **0.0637** | 6.370E-02 | **3.096E-10** | 3.096E-10 | 158.9253468 | 158.92535 | 158.92535 |
| Dy | 66 | 156 | 0.0539 | | 0.0002 | | 1.079E-12 | 155.9242829 | | |
| Dy | 66 | 158 | 0.0946 | | 0.0004 | | 1.917E-12 | 157.9244096 | | |
| Dy | 66 | 160 | 2.3288 | | 0.0098 | | 4.780E-11 | 159.9251975 | | |
| Dy | 66 | 161 | 18.8887 | | 0.0793 | | 3.901E-10 | 160.9269334 | | |
| Dy | 66 | 162 | 25.4791 | | 0.1069 | | 5.295E-10 | 161.9267984 | | |
| Dy | 66 | 163 | 24.8954 | | 0.1045 | | 5.206E-10 | 162.9287312 | | |
| Dy | 66 | 164 | 28.2596 | | 0.1186 | | 5.946E-10 | 163.9291748 | | |
| | | | **100** | **0.420** | | **2.086E-09** | | | 162.49977 | 162.49977 |
| Ho | 67 | 165 | **100** | **0.0906** | 0.0906 | **4.570E-10** | 4.570E-10 | 164.9303221 | 164.93033 | 164.93033 |
| Er | 68 | 162 | 0.139 | | 0.0004 | | 1.808E-12 | 161.9287799 | | |
| Er | 68 | 164 | 1.601 | | 0.004 | | 2.108E-11 | 163.9292065 | | |
| Er | 68 | 166 | 33.503 | | 0.088 | | 4.465E-10 | 165.9302931 | | |
| Er | 68 | 167 | 22.869 | | 0.060 | | 3.066E-10 | 166.9320482 | | |
| Er | 68 | 168 | 26.978 | | 0.071 | | 3.639E-10 | 167.9323702 | | |
| Er | 68 | 170 | 14.91 | | 0.039 | | 2.035E-10 | 169.9354643 | | |
| | | | **100** | **0.263** | | **1.343E-09** | | | 167.25908 | 167.25908 |
| Tm | 69 | 169 | **100** | **0.0410** | 0.0410 | **2.116E-10** | 2.116E-10 | 168.9342133 | 168.93422 | 168.93422 |
| Yb | 70 | 168 | 0.123 | | 0.0003 | | 1.637E-12 | 167.9338869 | | |
| Yb | 70 | 170 | 2.982 | | 0.008 | | 4.010E-11 | 169.9347618 | | |
| Yb | 70 | 171 | 14.086 | | 0.036 | | 1.905E-10 | 170.9363258 | | |
| Yb | 70 | 172 | 21.686 | | 0.056 | | 2.951E-10 | 171.9363815 | | |
| Yb | 70 | 173 | 16.103 | | 0.042 | | 2.204E-10 | 172.9382108 | | |
| Yb | 70 | 174 | 32.025 | | 0.083 | | 4.408E-10 | 173.9388621 | | |
| Yb | 70 | 176 | 12.995 | | 0.034 | | 1.809E-10 | 175.9425717 | | |
| | | | **100** | **0.259** | | **1.369E-09** | | | 173.05447 | 173.05447 |
| Lu | 71 | 175 | 97.18 | | 0.0373 | | 1.996E-10 | 174.9407712 | | |
| Lu* | 71 | 176 | 2.82 | | 0.0011 | | 5.831E-12 | 175.9426867 | | |
| | | | **100** | **0.0384** | | **2.054E-10** | | | 174.96681 | 174.96906 |
| Hf^ | 72 | 174 | 0.16 | | 0.0003 | | 1.349E-12 | 173.9400462 | | |
| Hf | 72 | 176 | 5.20 | | 0.008 | | 4.408E-11 | 175.9414091 | | |
| Hf | 72 | 177 | 18.60 | | 0.029 | | 1.585E-10 | 176.9432224 | | |
| Hf | 72 | 178 | 27.30 | | 0.043 | | 2.338E-10 | 177.9437004 | | |
| Hf | 72 | 179 | 13.63 | | 0.021 | | 1.174E-10 | 178.945817 | | |
| Hf | 72 | 180 | 35.11 | | 0.055 | | 3.041E-10 | 179.9465512 | | |
| | | | **100** | **0.157** | | **8.591E-10** | | | 178.48515 | 178.48658 |
| Ta* | 73 | 180 | 0.01201 | | 0.000003 | | 1.436E-14 | 179.9474648 | | |
| Ta | 73 | 181 | 99.98799 | | 0.0217 | | 1.202E-10 | 180.9479958 | | |
| | | | **100** | **0.0217** | | **1.202E-10** | | | 180.94788 | 180.94788 |
| W^ | 74 | 180 | 0.1198 | | 0.0002 | | 9.494E-13 | 179.9467091 | | |
| W | 74 | 182 | 26.4985 | | 0.038 | | 2.123E-10 | 181.9482042 | | |
| W | 74 | 183 | 14.3136 | | 0.021 | | 1.153E-10 | 182.9502223 | | |
| W | 74 | 184 | 30.6422 | | 0.044 | | 2.482E-10 | 183.9509312 | | |
| W | 74 | 186 | 28.4259 | | 0.041 | | 2.328E-10 | 185.9543641 | | |
| | | | **100** | **0.144** | | **8.096E-10** | | | 183.84170 | 183.84170 |
| Re | 75 | 185 | 35.6616 | | 0.0211 | | 1.191E-10 | 184.9529549 | | |
| Re* | 75 | 187 | 64.3384 | | 0.0380 | | 2.172E-10 | 186.9557531 | | |
| | | | **100** | **0.0591** | | **3.363E-10** | | | 186.20675 | 186.24152 |
| Os^ | 76 | 184 | 0.0198 | | 0.0001 | | 7.546E-13 | 183.9524891 | | |
| Os^ | 76 | 186 | 1.5973 | | 0.011 | | 6.143E-11 | 185.9538382 | | |
| Os | 76 | 187 | 1.2817 | | 0.009 | | 4.956E-11 | 186.9557505 | | |

| | | | | | | | | | |
|---|---|---|---|---|---|---|---|---|---|
| Os | 76 | 188 | 13.3269 | | 0.090 | | 5.181E-10 | 187.9558382 | | |
| Os | 76 | 189 | 16.2549 | | 0.110 | | 6.353E-10 | 188.9581475 | | |
| Os | 76 | 190 | 26.4368 | | 0.179 | | 1.039E-09 | 189.9584471 | | |
| Os | 76 | 192 | 41.0827 | | 0.278 | | 1.631E-09 | 191.9614807 | | |
| | | | **100** | **0.676** | | **3.935E-09** | | | 190.23494 | 190.24822 |
| Ir | 77 | 191 | 37.272 | | 0.232 | | 1.358E-09 | 190.9605941 | | |
| Ir | 77 | 193 | 62.728 | | 0.391 | | 2.309E-09 | 192.9629264 | | |
| | | | **100** | **0.624** | | **3.667E-09** | | | 192.21661 | 192.21661 |
| Pt* | 78 | 190 | 0.0130 | | 0.0002 | | 9.183E-13 | 189.9599321 | | |
| Pt | 78 | 192 | 0.7938 | | 0.010 | | 5.676E-11 | 191.961038 | | |
| Pt | 78 | 194 | 32.8078 | | 0.400 | | 2.371E-09 | 193.962679 | | |
| Pt | 78 | 195 | 33.7871 | | 0.412 | | 2.454E-09 | 194.9647901 | | |
| Pt | 78 | 196 | 25.2902 | | 0.308 | | 1.846E-09 | 195.9649515 | | |
| Pt | 78 | 198 | 7.3083 | | 0.089 | | 5.390E-10 | 197.967891 | | |
| | | | **100** | **1.218** | | **7.267E-09** | | | 195.08395 | 195.08395 |
| Au | 79 | 197 | **100** | **0.201** | 0.201 | **1.209E-09** | 1.209E-09 | 196.9665687 | 196.96657 | 196.96657 |
| Hg | 80 | 196 | 0.16 | | 0.001 | | 3.572E-12 | 195.9658326 | | |
| Hg | 80 | 198 | 10.04 | | 0.039 | | 2.337E-10 | 197.9667689 | | |
| Hg | 80 | 199 | 16.94 | | 0.065 | | 3.963E-10 | 198.9682804 | | |
| Hg | 80 | 200 | 23.14 | | 0.089 | | 5.440E-10 | 199.968326 | | |
| Hg | 80 | 201 | 13.17 | | 0.051 | | 3.112E-10 | 200.9703022 | | |
| Hg | 80 | 202 | 29.74 | | 0.114 | | 7.064E-10 | 201.970643 | | |
| Hg | 80 | 204 | 6.82 | | 0.026 | | 1.635E-10 | 203.9734941 | | |
| | | | **100** | **0.384** | | **2.359E-09** | | | 200.59240 | 200.59240 |
| Tl | 81 | 203 | 29.524 | | 0.054 | | 3.326E-10 | 202.9723442 | | |
| Tl | 81 | 205 | 70.476 | | 0.128 | | 8.017E-10 | 204.9744275 | | |
| | | | **100** | **0.181** | | **1.134E-09** | | | 204.38333 | 204.38333 |
| Pb^ | 82 | 204 | 1.9968 | | 0.066 | | 4.101E-10 | 203.9730436 | | |
| Pb | 82 | 206 | 18.5823 | | 0.612 | | 3.854E-09 | 205.9744653 | | |
| Pb | 82 | 207 | 20.5631 | | 0.677 | | 4.286E-09 | 206.9758969 | | |
| Pb | 82 | 208 | 58.8578 | | 1.938 | | 1.233E-08 | 207.976652 | | |
| | | | **100** | **3.293** | | **2.088E-08** | | | 207.31630 | 207.31887 |
| Bi^ | 83 | 209 | **100** | **0.142** | 0.142 | **9.064E-10** | 9.064E-10 | 208.9803987 | 208.98040 | 208.98040 |
| Th* | 90 | 232 | **100** | **0.0431** | 0.0431 | **3.062E-10** | 3.062E-10 | 232.0380553 | 232.03806 | 232.03806 |
| U* | 92 | 234 | 0.0042 | | 9.9E-07 | | 7.093E-15 | 234.0409521 | | |
| U* | 92 | 235 | 24.3016 | | 0.0058 | | 4.161E-11 | 235.0439299 | | |
| U* | 92 | 238 | 75.6942 | | 0.0180 | | 1.313E-10 | 238.0507882 | | |
| | | | **100** | **0.0238** | | **1.729E-10** | | | 238.02891 | 237.31991 |
| Sum | | | | | | 1.000E+00 | 1.000000 | | | |

* Table modified from Lodders (2020, 2021) where more references and details can be found. Abundances are for 4.567 Ga ago. Atomic Masses are from Wang et al. 2012. Elements marked with * involve long-lived radioactive nuclides with half-lives up to 10^12 years. Isotopes with half-lives above 10^12 years (marked with ^) can be considered as stable compared to the age of the solar system and some of them are of interest of studies of double-beta decay. Isotopic compositions mainly adopted from Meija et al. (2016); except for H, He, C, N, O, Ne, Ar, Kr, and Xe isotopic compositions; see Lodders 2020 for details and references.

# References

*references only relevant for Tables A4-A12 are marked with one preceding star, references in Table A13 with 2 stars.

*Ahrens, L. H. 1970. The composition of stony meteorites (IX) abundance trends of the refractory elements in chondrites, basaltic achondrites and Apollo 11 fines. *Earth and Planetary Science Letters*, *10*(1), 1-6.

Ahrens, L. H., Von Michaelis, H., Fesq, H. W. 1969. The composition of the stony meteorites (IV) some analytical data on Orgueil, Nogoya, Ornans and Ngawi. *Earth and Planetary Science Letters* 6, 285-288.

Akaiwa, H. 1966. Abundances of selenium, tellurium and indium in meteorites. *Journal of Geophysical Research* 71, 1919-1923.

Alexander C. M. O'D., Bowden R., Fogel M. L., Howard K. T., Herd C. D. K., Nittler L. R. 2012b. The provenances of asteroids, and their contributions to the volatile inventories of the terrestrial planets. *Science* 337, 721–723.

Arden, J.W., Cressey, G. 1984. Thallium and lead in the Allende C3V carbonaceous chondrite. A study of the matrix phase. *Geochimica et Cosmochimica Acta* 48, 1899-1912.

Babechuk, M.G., Kamber, B.S., Greig, A., Canil, D., Kidolanyi, J. 2010. The behaviour of tungsten during mantle melting revisited with implications for planetary differentiation time scales. *Geochimica et Cosmochimica Acta* 74, 1448-1470

Baedecker, P. A., Chou, C.-L., Grudewicz, E. B., Wasson, J.T. 1973. Volatile and Trace Elements in Apollo 15 Samples: Geochemical Implications and Characterization of the Long-lived and Short-lived Extralunar Materials. *Proc. Lunar. Sci. Conf.* 4th, 1177-1195

Baker, R.G.A., Schönbächler, M., Rehkämper, M.Williams, H.D., Halliday, A.N. 2010. The thallium isotope composition of carbonaceous chondrites — New evidence for live 205Pb in the early solar system. *Earth and Planetary Science Letters* 291, 39-47

Barrat J. A., Zanda B., Moynier F., Bollinger C., Liorzou C., Bayon G. 2012. Geochemistry of CI chondrites: Major and trace elements, and Cu and Zn Isotopes. *Geochimica et Cosmochimica Acta* 83, 79-92.

Barrat, J.A.; Dauphas, N., Gillet, P., Bollinger, C., Etoubleau, J.,Bischoff, A., Yamaguchi, A. 2016. Evidence from Tm anomalies for non-CI refractory lithophile element proportions in terrestrial planets and achondrites. *Geochimica et Cosmochimica Acta* 176, 1-17.

Beer, H., Walter, G., Macklin. R. L., Patchett, P.J. 1984. Neutron Capture Cross Sections and Solar System Abundances of 160,161Dy, 170, 171 Yb, 175,176 Lu and 176,177 Hf for the S-Process Analysis of the Radionuclide 176Lu. *Physical Review C*, 30-2, 464-478

Belsky, T., Kaplan, I.R. 1970. Light hydrocarbon gases, C 13, and origin of organic matter in carbonaceous chondrites. *Geochimica et Cosmochimica Acta* 34, 257-278.

Bendel, V. 2013. Volatilitätskontrollierte Fraktionierung refraktär-lithophiler Elemente in Meteoriten und der Erde. Ph. D. Dissertation Georg-August-Universität Göttingen, 149pp.

Bermingham, K.R., Mezger, K., Scherer, E.E., Horan, M.F., Carlson, R.W., Upadhyay, D., Magna, T., Pack, A., 2016. Barium isotope abundances in meteorites and their implications for early Solar System evolution. *Geochimica et Cosmochimica Acta* 175, 282-298.

Berzelius, J., 1834. Om Meteorstenar. 4. Meteorsten fran Alais. Kongl. Vetensk. Acad. Handl. 1834, 144-158; German German translation: Über Meteorsteine, *Poggendorfs Annalen der Physik Chemie* 33, 115-120, 143.

Blichert-Toft, J., Albarede, F. 1997. The Lu-Hf isotope geochemistry of chondrites and the evolution of the mantle-crust system. *Earth and Planetary Science Letters* 148, 1997, 243-258

Boato G. 1954. The isotopic composition of hydrogen and carbon in the carbonaceous chondrites. *Geochimica et Cosmochimica Acta* 6, 209-220

Bouvier A., Vervoort J. D. and Patchett P. J. 2008. The Lu–Hf and Sm–Nd isotopic composition of CHUR: constraints from unequilibrated chondrites and implications for the bulk com-position of terrestrial planets. *Earth and Planetary Science Letters* 273, 48–57.

Braukmüller N., Wombacher F., Hezel D.C., Escoube R., Münker, C. 2018. The chemical composition of carbonaceous chondrites: Implications for volatile element depletion, complementarity and alteration. *Geochimica et Cosmochimica Acta* 239, 17-48.

Braukmüller, N., Wombacher, F., Bragagni, A., Münker, C. 2020. Determination of Cu, Zn, Ga, Ag, Cd, In, Sn and Tl in geological reference materials and chondrites by isotope dilution ICP-MS. *Geostandards and Geoanalytical Research* 44(4), 733-752.

Briggs, M. H., Mamikunian, G. 1963. Organic Constituents of the Carbonaceous Chondrites. *Space Science Reviews* 1, 647-682.

Burgess R., Wright I.P., Pillinger C.T. 1991. Determination oi sulphur-bearing components in CI and C2 carbonaceous chondrites by stepped combustion. *Meteoritics* 26, 55-64.

Burkhardt, C., Kleine, T., Dauphas, N., Wieler, R. 2012. Origin of isotopic heterogeneity in the solar nebula by thermal processing and mixing of nebular dust. *Earth and Planetary Science Letters* 357/358, 298–307.

Burnett, D.S., Woolum, D.S., Benjamin, T.M., Rogers, P. S. Z., Duffy, C. J., Maggiore, C. 1989. A Test of the Smoothness of the Elemental Abundances of Carbonaceous Chondrites. *Geochimica et Cosmochimica Acta* 53, 471-481.

Case D.R., Laul J. C., Pelly I. Z., Wechter M.A., Schmidt-Bleek F., Lipschutz M.E. 1973. Abundance patterns of thirteen trace elements in primitive carbonaceous and unequilibrated ordinary chondrites *Geochimica et Cosmochimica Acta* 37. 19-33.

Chou, C.-L., Baedecker, P. A., Wasson, J.T. 1976. Allende Inclusions: Volatile-Element Distribution and Evidence for Incomplete Volatilization of Presolar Solids. *Geochimica et Cosmochimica Acta* 40, 85-94

Christie W.A.K. 1914. A carbonaceous aerolite from Rajputana. Records Geol. Survey India, 44, 41-51

Cloez S. 1864a. Note sur la composition chimique de la pierre meteoritique d'Orgueil. *Comptes Rendus Hebdomadaires des Séances de l'Académie des Sciences* Paris 58, 986–988.

Cloez S. 1864b. Analyse chimique de la pierre meteorique d'Orgueil. *Comptes Rendus Hebdomadaires des Séances de l'Académie des Sciences* Paris 59, 37–40.

Cloez S. 1864c. Dosage de l'acide carbonique contenue dans la meteorite d'Orgueil. *Comptes Rendus Hebdomadaires des Séances de l'Académie des Sciences* Paris 59, 830–831.

Crocket J.H., Keays R.R., Hsieh, S. 1967. Precious metal abundances in some carbonaceous and enstatite chondrites. *Geochimica et Cosmochimica Acta* 31, 1615-1623

Curtis D.B., Gladney E., Jurney E. 1980. A revision of the meteorite based cosmic abundance of boron. *Geochimica et Cosmochimica Acta* 44, 1945-1953.

Curtis, D. C., Gladney, E. S. 1985. Boron Cosmochemistry. *Earth and Planetary Science Letters* 75, 311-320.

Dauphas N., Pourmand A. 2011. Hf-W-Th evidence for rapid growth of Mars and its status as a planetary embryo. *Nature* 473, 489-492.

Dauphas, N., Pourmand, A. 2015. Thulium anomalies and rare earth element patterns in meteorites and Earth: Nebular fractionation and the nugget effect, *Geochimica et Cosmochimica Acta* 63, 234-261.

David K., Schiano P., Allegre C. J., 2000. Assessment of the Zr/Hf fractionation in oceanic basalts and continental materials during petrogenetic processes. *Earth and Planetary Science Letters* 178, 285–301.

De Laeter J.R., Rosman K. J.R., Ly C. 1998. Meteoritical barium abundance from carbonaceous chondrites. *Meteoritics and Planetary Science* 33, A40 . See also: De Laeter J.R., Rosman K. J.R. 1999, Deficiencies in the classical model of s-process nucleosynthesis. *Meteoritics and Planetary Science* 34, 717-721.

De Laeter, J.R., Hosie D. J. 1978. The abundance of barium in stony meteorites. *Earth and Planetary Science Letters* 38, 2, 416-420.

De Laeter, J.R., McCulloch M. T., Rosman K. J.R. 1974. Mass spectrometric isotope dilution analyses of tin in stony meteorites and standard rocks. *Earth and Planetary Science Letters* 22, 3, 226-232.

Dreibus G., Friedrich J. M., Haubold R., Huisl W., Spettel B. 2004. Halogens, carbon, and sulfur in the Tagish Lake meteorite: Implications for classification and terrestrial alteration. 35th Lunar and Planetary Science Conference, abstract #1268.

Dreibus, G., Palme, H., Spettel, B., Zipfel, J, Wänke, H. 1995. Sulfur and Selenium in Chondritic Meteorites. *Meteoritics* 30, 439-445.

Easton, A.J., Lovering, J.F. 1964. Determination of small quantities of potassium and sodium in stony meteoritic materials, rocks and minerals. Anal. Chim. Acta 30, 543-548.

Ebihara, M., Wolf, R., Anders, E. 1982. Are C1 chondrites chemically fractionated ? A Trace Element Study. *Geochimica et Cosmochimica Acta* 46, 1849-1861

Edwards, G., Urey, H.C. 1956. Determination of alkali metals in meteorites by a distillation process. *Geochimica et Cosmochimica Acta*, 7, 154-168

Ehmann W. D., Baedecker P. A., McKown D. M. 1970a. Gold and iridium in meteorites and some selected rocks. *Geochimica et Cosmochimica Acta* 34, 493-507.

Ehmann W. D., Rebagay T. V. 1970. Zirconium and hafnium in meteorites by activation analysis. *Geochimica et Cosmochimica Acta* 34, 649-658.

Ehmann, W. D., Chyi, L. L. 1974. Zirconium and Hafnium in Meteorites. *Earth and Planetary Science Letters* 21, 230-234.

Ehmann, W. D., Gillum, D. E. 1972. Platinum and Gold in Chondritic Meteorites. *Chemical Geology* 9, 1-11

Evensen, N. M., Hamilton, P. J., O'Nions, R. K. 1978. Rare-earth abundances in chondritic meteorites. *Geochimica et Cosmochimica Acta* 42, 1199-1212

Fehr M.A., Rehkaemper M., Halliday A.N., Wiechert U., Hattendorf B., Guenther D., Ono S., Eigenbrode J. L., Rumble D. 2005, Tellurium isotopic composition of the early solar system - A search for effects resulting from stellar nucleosynthesis, 126Sn decay, and mass-independent fractionation. *Geochimica et Cosmochimica Acta* 69, 5099-5112.

Fehr, M. A., Hammond, S. J., Parkinson, I. J. 2018. Tellurium stable isotope fractionation in chondritic meteorites and some terrestrial samples. *Geochimica et Cosmochimica Acta* 222, 17-33.

Filhol E., Mellies J. 1864.Note sur la composition chimique de l'aerolithe du 14 Mai 1864. *Mem. Acad. Sci. Toulouse* 2, 379-382, reported 2 and 20 June 1864.

Fischer- Gödde M., Becker H., Wombacher, F., 2010. Rhodium, gold and other highly siderophile element abundances in chondritic meteorites. *Geochimica et Cosmochimica Acta* 74, 356-379.

Fisher, E.F. 1972. Uranium content and radiogenic ages of hypersthene, bronzite, amphoterite and carbonaceous meteorites, *Geochimica et Cosmochimica Acta* 36, 15-33.

Folinsbee, R. E., Douglas, J. A. V., Maxwell, J. A.1967. Revelstoke, a New Type I Carbonaceous Chondrite, *Geochimica et Cosmochimica Acta* 31, 1625-1635

Fouche, K.F., Smales, A.A. 1967a, The distribution of trace elements in chondritic meteorites. I. Gallium, germanium and indium. *Chem. Geol.* 2, 5-33

Fouche, K.F., Smales, A.A. 1967b. The distribution of trace elements in chondritic meteorites. 2. Antimony, arsenic, gold, palladium and rhenium. *Chem. Geol*. 2, 105-134.

Fredriksson, L., Brenner, P.R., Fredriksson, B.J., Olsen, E. 1997. A nondestructive analytical method for stone meteorites—and a controversial discrepancy. *Meteoritics and Planetary Science* 32, 55-60.

Friedrich J. M., Wang M.-S., Lipschutz M.E. 2002. Comparison of the trace element composition of Tagish Lake with other primitive carbonaceous chondrites. *Meteoritics and Planetary Science* 37, 677-686.

Funk , C. 2015. Abundances and distribution of chalcogen volatile elements in chondritic meteorites and their components, PhD Thesis (Dissertation), University of Cologne, 2015, 127pp.

Ganapathy, R., Papia, G. M., Grossman, L. 1976. The Abundances of Zirconium and Hafnium in the Solar System. *Earth and Planetary Science Letters* 29, 302-308.

Gao, X., Thiemens, M. H. 1993. Isotopic Composition and Concentration of Sulfur in Carbonaceous Chondrites. *Geochimica et Cosmochimica Acta* 57, 3159-3169.

Garenne A., Beck P., Montes-Hernandez G., Chiriac R., Toche F., Quirico E., Bonal L., Schmitt B. 2014. The abundance and stability of "water" in type 1 and 2 carbonaceous chondrites (CI, CM and CR). *Geochimica et Cosmochimica Acta* 137, 93-112.

Gibson E. K., Moore C. B., Lewis C. F. 1971. Total nitrogen and carbon abundances in carbonaceous chondrites. *Geochimica et Cosmochimica Acta* 35, 599-604.

*Göpel, C., Birck, J. L., Galy, A., Barrat, J. A., Zanda, B. 2015. Mn–Cr systematics in primitive meteorites: Insights from mineral separation and partial dissolution. *Geochimica et Cosmochimica Acta*, *156*, 1-24.

Gooding, J.L. 1979, Petrogenetic properties of chondrules in unequilibrated H-, L-, and LL-group chondritic meteorites. Ph. D. Thesis, Univ. of New Mexico, Albuquerque, 392pp; see also Gooding, J.L., Keil, K., Fukuoka, T., Schmitt, R.A. 1980, Elemental Abundances in Chondrules from Unequilibrated Chondrites: Evidence for Chondrule Origin by Melting of Pre-Existing Materials. *Earth and Planetary Science Letters* 50, 171-180.

Grady, M.M., Wright, I.P., Pillinger C.T. 1991. Comparisons between Antarctic and non-Antarctic meteorites based on carbon isotope geochemistry, *Geochimica et Cosmochimica Acta*, 55, 49-58

Grady, M.M., Verchovsky, A.B., Franchi, I.A., Wright, I.P., Pillinger C.T. 2002. Light element geochemistry of the Tagish Lake CI2 chondrite: Comparison with CI1 and CM2 meteorites. *Meteoritics and Planetary Science* 37, 713-735.

Graham A. L., Mason B. 1972. Niobium in meteorites. *Geochimica et Cosmochimica Acta* 36, 917-922.

Greenland, L. 1967. The abundance of selenium, tellurium, silver, palladium, cadmium, and zinc in chondritic meteorites. *Geochimica et Cosmochimica Acta* 31, 849-860.

Greenland, L., Goles, G.G. 1965. Copper and zinc abundances in chondritic meteorites. *Geochimica et Cosmochimica Acta* 29, 1285-1292

Greenland, L., Lovering J.F. 1965. Minor and trace element abundances in chondritic meteorites. *Geochimica et Cosmochimica Acta* 29, 821-858

Grossman L., Ganapathy, R. 1976. Trace elements in the Allende meteorite--I. Coarse-grained, Ca-rich inclusions. *Geochimica et Cosmochimica Acta* 40, 331-344.

Grossman L., Ganapathy, R. 1975. Volatile Elements in Allende Inclusions. Proc. Lunar. Sci. Conf. $6^{th}$ , 1729-1736

Grossman, J.N., Rubin, A.E., Rambaldi, E.R., Rajan, R.S., Wasson, J.T. 1985. Chondrules in the Qingzhen type-3 enstatite chondrite: Possible precursor components and comparison to ordinary chondrite chondrules. *Geochimica et Cosmochimica Acta* 49(8), 1781-1795.

Halbout, J., Mayeda, T.K., Calyton, R.N. 1986. Carbon isotopes and light element abundances in carbonaceous chondrites. *Earth and Planetary Science Letters* 80, 1-18.

Hamaguchi, H., Onuma, N., Hirao, Y., Yokoyama, H., Bando, S., Furukawa, M. 1969. The abundance of arsenic, tin and antimony in chondritic meteorites. *Geochimica et Cosmochimica Acta* 33, 507-518.

Hellmann, J.L., Hopp, T., Burkhardt, C., Kleine, T. 2020. Origin of volatile element depletion among carbonaceous chondrites. *Earth and Planetary Science Letters* 549,116508

Hermann, F., Wichtl, M. 1974. Neutronenaktivierungsanalytische Bestimmung von Spurenelementen in Meteoriten der Vatikanischen Sammlung. In: Analyse Extraterrestrischen Materials ed W. Kiesl, M. Malissa, Springer-Verlag Wien, 163-172

Hidaka, H., Yoneda, S. 2011. Diverse nucleosynthetic components in barium isotopes of carbonaceous chondrites: Incomplete mixing of s- and r-process isotopes and extinct 135Cs in the early solar system. Geochimica et Cosmochimica Acta, 75 (13) 3687-3697 doi:10.1016/j.gca.2011.03.043

Hintenberger, H., Jochum K. P., Seufert, M. 1973. The concentrations of the heavy metals in the four new Antarctic meteorites Yamato a, b, c, and d, and in Orgueil, Murray, Abee, Allegan, Mocs and Johnstown. *Earth and Planetary Science Letters* 20, 391-394.

Horan M. F., Walker R. J., Morgan J.W., Grossman J.N. ,Rubin A.E., 2003, Highly siderophile elements in chondrites. Chem. Geol. 196, 27-42.

Hu, Y., Moynier, F., Yang, X., 2023a. Volatile-depletion processing of the building blocks of Earth and Mars as recorded by potassium isotopes. *Earth and Planetary Science Letters* 620, 118319.

Humayun, M., Clayton, R.N. 1995. Potassium isotope cosmochemistry: Genetic implications of volatile element depletion. *Geochimica et Cosmochimica Acta* 59, 2131-2148

Injerd W. G., Kaplan I. R.1974. Nitrogen isotope distribution in meteorites. *Meteoritics* 9, 352-353 (abs.).

Islam, M.A., Ebihara, M., Toh, Y., Murakami, Y. and Harada, H. 2012. Characterization of multiple prompt gamma-ray analysis (MPGA) system at JAEA for elemental analysis of geological and cosmochemical samples. *Applied Radiation and Isotopes* 70, 1531-1535.

James, R.H., Palmer, M.R. 2000. The lithium isotope composition of international rock standards. *Chemical Geology* 166, 319–326.

Jarosewich E. 1972. Quoted in: Fredriksson, K., Kerridge, J.F. 1988. *Meteoritics*, 23, 35, see also Jarosewich E. 1990. Chemical analyses of meteorites: A compilation of stony and iron meteorite analyses. *Meteoritics* 25, 323-337.

*Jarosewich, E. 2006. Chemical analyses of meteorites at the Smithsonian Institution: An update. *Meteoritics and Planetary Science*, *41*(9), 1381-1382.

Jenniskens P., Rubin A. E., Yin Q.–Z., Sears D. W. G., Sandford S. A., Zolensky M. E., Krot A. N., Blair L., Kane D., Utas J., Verish R., Friedrich J. M., Wimpenny J., Eppich G. R., Ziegler K., Verosub K. L., Rowland D. J., Albers J., Gural P. S., Grigsby B., Fries M. D., Matson R., Johnston M., Silber E., Brown P., Yamakawa A., Sanborn M. E., Laubenstein M., Welten K. C., Nishiizumi K., Meier M. M. M., Busemann H., Clay P., Caffee M. W., Schmitt–Kopplin P., Hertkorn N., Glavin D. P., Callahan M. P., Dworkin J. P., Wu Q., Zare R. N., Grady M., Verchovsky S., Emel'yanenko V., Naroenkov S., Clark D. L., Girten B., Worden P. S. 2014. Fall, recovery and characterization of the Novato L6 Chondrite Breccia. *Meteoritics and Planetary Science* 49,1388–1425.

Jochum, K. P., Seufert, H. M., Spettel, B., Palme, H. 1986. The Solar-System Abundances of Nb, Ta, and Y, and the Relative Abundances of Refractory Lithophile Elements in Differentiated Planetary Bodies. *Geochimica et Cosmochimica Acta* 50, 1173-1183

Jochum, K.P., Hofmann, A.W., Seufert, H.M. 1993. Tin in mantle-derived rocks: constraints on Earth evolution. *Geochimica et Cosmochimica Acta* 57, 3585-3595.

Jochum, K.P., Seufert, H.M. 1995. New SSMS techniques for the determination of rhodium and other platinum-group elements in carbonaceous chondrites. Meteoritics, 30, 525

Jochum K. P. 1996. Rhodium and other platinum-group elements in carbonaceous chondrites. *Geochimica et Cosmochimica Acta* 60, 3353-3357.

Jochum. K. P., Stolz J., McOrist, G. 2000. Niobium and tantalum in carbonaceous chondrites: Constraints on the solar system and primitive mantle niobium/tantalum, zirconium/niobium, and niobium/uranium ratios. *Meteoritics and Planetary Science* 35, 229-235.

Kallemeyn, G.W., Wasson J.T. 1981. The compositional classification of chondrites: I. The carbonaceous chondrite groups. *Geochimica et Cosmochimica Acta* 45, 1217-1230.

Kaushal, S. K., Wetherill, G. W. 1970. Rubidium 87-Strontium 87 age of carbonaceous chondrites. *Journal of Geophysical Research* (1896-1977). 75, 463-468.

Kerridge J. F. 1985. Carbon, hydrogen and nitrogen in carbonaceous chondrites: Abundances and isotopic compositions in bulk samples. *Geochimica et Cosmochimica Acta* 49, 1707-1714.

Kiesl, W. 1979. Kosmochemie. Springer-Verlag, Wien und New York, xx pp.

King, A.J., Phillips, K.J.H., Strekopytov, S., Vita-Finzi, C., Russell, S.S. 2020. Terrestrial modification of the Ivuna meteorite and a reassessment of the chemical composition of the CI type specimen. *Geochimica et Cosmochimica Acta 268*, 73-89.

Kleine T., Mezger K., Munker C., Palme H., Bischoff A. 2004. $^{182}$Hf-$^{182}$W isotope systematics of chondrites, eucrites, and Martian meteorites: Chronology of core formation and early *Geochimica et Cosmochimica Acta* **68**, 2935-2946.

Knab, H. J., Hintenberger, H. 1978. Isotope dilution analyses of 20 trace elements in 9 carbonaceous chondrites by spark source mass spectrography. *Meteoritics 13, 522-527*

Koefoed, P., Barrat, J. A., Pravdivtseva, O., Alexander, C. M. D., Lodders, K., Ogliore, R., Wang, K. 2023. The potassium isotopic composition of CI chondrites and the origin of isotopic variations among primitive planetary bodies. *Geochimica et Cosmochimica Acta 358*, 49-60.

Kolesov, G. M. 1978. Abundance of rare-earth elements in the main types of stony meteorites. *Meteoritika37*, 112-128.

Krähenbühl, U., Morgan, J. M., Ganapathy, R., Anders, E. 1973. Abundance of 17 Trace Elements in Carbonaceous Chondrites. *Geochimica et Cosmochimica Acta* 37, 1353-1370

Krankowsky, D., Mueller, O. 1964. Isotopenhäufigkeit des Lithiums in Steinmeteoriten. *Geochimica et Cosmochimica Acta* 28, 1625-1630

Labidi J., König S., Kurzawa T., Yierpan, A., Schoenberg R. 2018. The selenium isotopic variations in chondrites are mass-dependent, Implications for sulfide formation in the early solar system. Earth Planetary Science Letters 481, 212-222

Laul J. C., Case D.R., Schmidt-Bleek F., Lipschutz M.E. 1970a. Bismuth contents of chondrites. *Geochimica et Cosmochimica Acta* 34, 89-103.

Laul J. C., Pelly I., Lipschutz M.E. 1970b. Thallium contents of chondrites. *Geochimica et Cosmochimica Acta* 34, 909-920.

Lawrence Smith, J. 1876a. Recherches sur les composés du carbone. Annales de Chimie et Physique 9, 259-288. (see also Recherches sur les composés du carbone dans les météorites. *Comptes Rendus Hebdomadaires des Séances de l'Académie des Sciences* 82,1042-1043)

Lawrence Smith, J. 1876b. Researches in the solid carbon compounds in meteorites. *American J. Science and Arts* (3)XI, 388-395 and 433-443

Loss R.D., Rosman K. K. R., DeLaeter J.R. 1989. The solar system abundance of tin. *Geochimica et Cosmochimica Acta* 933-935.

Loss, R.D., Rosman, K. J.R., De Laeter, J.R. 1984. Mass Spectrometric Isotope Dilution Analyses of Palladium, Silver, Cadmium and Tellurium in Carbonaceous Chondrites. *Geochimica et Cosmochimica Acta* 48, 1677-1681.

Loveland, W., Schmitt, R.A., Fisher, D.E. 1969. Aluminum abundances in stony meteorites. *Geochimica et Cosmochimica Acta* 33, 375-385.

Lu, Y., Makishima, A., Nakamura, E. 2007. Coprecipitation of Ti, Mo, Sn and Sb with fluorides and application to determination of B, Ti, Zr, Nb, Mo, Sn, Sb, Hf and Ta by ICP-MS. *Chemical Geology*, *236*(1-2), 13-26.

Luck J.-M., Othman D.B., Albarede F. 2005. Zn and Cu isotopic variations in chondrites and iron meteorites: Early solar nebula reservoirs and parent-body processes. *Geochimica et Cosmochimica Acta* **69**, 5351-5363.

Makishima A., Nakamura E. 2006. Determination of major/ minor and trace elements in silicate samples by ICP-QMS and ICP-SFMS Applying Isotope Dilution-Internal Standardisation (ID-IS), and Multi-Stage Internal Standardisation. *Geostandards and Geoanalytical Research* 30, 245–271.

Mason B. 1962-1963. The carbonaceous chondrites. *Space Science Reviews* 1, 621-646.

McCulloch M. T., De Laeter, J.R., Rosman K.J.R. 1976. The isotopic composition and elemental abundance of lutetium in meteorites and terrestrial samples and the Lu-176 cosmochronometer. *Earth and Planetary Science Letters* 28, 3, 308-322

McCulloch M.T., Rosman K., De Laeter J.R. 1977. The isotopic and elemental abundance of ytterbium in meteorites and terrestrial samples. *Geochimica et Cosmochimica Acta* 41, 1703-1707.

Meier, M.M., Cloquet, C., Marty, B. 2016. Mercury (Hg) in meteorites: Variations in abundance, thermal release profile, mass-dependent and mass-independent isotopic fractionation. *Geochimica et Cosmochimica Acta* 182, 55-72.

**Meija, J., Coplen, T.B., Berglund, M., Brand, W.A., de Bievre, P., Groning, M., Holden, N.E., Irrgher, J., Loss, R.D., Walczyk, T., Prohaska, T. 2016. Isotopic compositions of the elements 2013 (IUPAC Technical Report), Pure Appl. Chem. 88, 293-306.

Mermelengas N., De Laeter, J.R., Rosman K. J.R. 1979. New data on the abundance of palladium in meteorites. *Geochimica et Cosmochimica Acta* 43, 5, 747-753

Mittlefehldt D.W. 2002. Geochemistry of the ungrouped carbonaceous chondrite Tagish Lake, the anomalous CM chondrite Bells, and comparison with CI and CM chondrites. *Meteoritics and Planetary Science* 37, 703-712.

Mittlefehldt D.W., Wetherill G.W. 1979. Rb-Sr studies of CI and CM chondrites. *Geochimica et Cosmochimica Acta* 43, 201-206.

Monge G., Fourcroy A., and Berthollet C. 1806. Extrait du rapport du juge de paix du canton de Vezenobres, premier arrondissement du Gard, sur une pierre tombee a Valence, le 15 mars 1806, and: Extrait du procès-verbal de la séance de l’Institut nationAlais, du 23 juin1806. *Annales de Chimie* 59, 34–40.

Monster, J., Anders, E. Thode, H.G. 1965. 34S/32S ratios for the different forms of sulphur in the Orgueil meteorite and their mode of formation. *Geochimica et Cosmochimica Acta* 29, 773-779.

Morgan J.W, Lovering J.F. 1967. Rhenium and osmium abundances in chondritic meteorites. *Geochimica et Cosmochimica Acta* 31, 1893-1909.

Morgan, J. W., Lovering, J.F. 1968. Uranium and thorium abundances in chondritic meteorites. *Talanta 15*, 1079-1095.

Morgan, J.W., Walker, R.J. 1988. Rhenium and osmium isotope systematics in carbonaceous chondrites. In *Abstracts of the Lunar and Planetary Science Conference* 19, 802.

Moynier, F. et al. 2020. Chondritic mercury isotopic composition of Earth and evidence for evaporation equilibrium degassing during the formation of eucrites. *Earth and Planetary Science Letters* 551, 116544.

Mueller, G., Shaw, R.A., Ogawa, T. 1965. Interrelations between volatilization curves, elemental composition and total volatiles in carbonaceous chondrites. *Nature* 206, 23-25.

Münker, C., Pfänder, J. A., Weyer, S., Buchl, A., Kleine, T., Mezger, K. 2003. Evolution of planetary cores and the Earth-Moon system from Nb/Ta systematics. *Science 301*, 84-87.

*Münker C. et al. 2025. High field strength elements in chondrites and their components. *Geochemical Perspectives Letters,* submitted.

Murthy, V.R., Compston, W. 1965. RbSr ages of chondrites and carbonaceous chondrites. *Journal of Geophysical Research* 70, 5297-5307.

Murty, S.V.S., Shukla, P. N., Goel, P.S. 1983. Lithium in stone meteorites and stony irons. Meteoritics 18, 123-136.

Nakamura, E., et al., 2022. On the origin and evolution of the asteroid Ryugu: A comprehensive geochemical perspective. Proceedings of the Japan Academy, Series B. 98, 227-282.

Nakamura, N. 1974. Determination of REE, Ba, Fe, Mg, Na and K in Carbonaceous and Ordinary Chondrites, *Geochimica et Cosmochimica Acta* 38, 757-775

Nichiporuk W., Moore C. B. 1970. Lithium in chondritic meteorites. *Earth and Planetary Science Letters* 9, 280-286.

Nichiporuk, W., Bingham, E. 1970. Vanadium and copper in chondrites. *Meteoritics*, *5*(3), 115-130.

Nichiporuk, W., Chodos, A., Helin., E., Brown, J. 1967. Determination of iron, nickel, cobalt, calcium, chromium and manganese in stony meteorites by X-ray fluorescence. *Geochimica et Cosmochimica Acta* 31, 1911-1930

Nie, NX, Chen XY, Hopp T, Hu JY, Zhang ZJ, Teng FZ, Shahar A, Dauphas N. 2021. Imprint of chondrule formation on the K and Rb isotopic compositions of carbonaceous meteorites. *Science Advances* 7(49):eabl3929. doi: 10.1126/sciadv.abl3929.

Otting W., Zaehringer J. 1967. Total carbon content and primordial rare gases in chondrites. *Geochimica et Cosmochimica Acta* 31, 1949-1960.

Palme, H., Zipfel, J. 2022. The composition of CI chondrites and their contents of chlorine and bromine; Results from instrumental neutron activation analysis. *Meteoritics and Planetary Science* 57, 317–333.

Palme, H., Lodders, K., Jones, A. 2014. Solar system abundances of the elements, in *Treatise on Geochemistry*, vol. 2, ed. by H.D. Holland, K.K. Turekian 2nd ed. (Elsevier, Oxford), pp. 15–36

Patchett, P.J., Vervoort, J.D., Soderlund, U., Salters, V.J.M. 2004. Lu–Hf and Sm–Nd isotopic systematics in chondrites and their constraints on the Lu–Hf properties of the Earth. *Earth and Planetary Science Letters* 222, 29–41.

*Patzer, A., Pack, A., Gerdes, A. 2010. Zirconium and hafnium in meteorites. *Meteoritics and Planetary Science 45*, 1136-1151.

Pearson, V. K., Sephton, M. A., Franchi, I. A., Gibson, J. M., Gilmour, I. 2006. Carbon and nitrogen in carbonaceous chondrites: Elemental abundances and stable isotopic compositions. *Meteoritics and Planetary Science 41*, 1899-1918.

** Wang, M., Huang, W. J., Kondev, F. G., Audi, G., Naimi, S. 2021. The AME 2020 atomic mass evaluation (II). Tables, graphs and references. Chinese Physics C, 45(3) #030003 (512 pp).

Pinson, W. H., Ahrens, L. H., Franck, M. L. 1953. The abundances of Li, Se, Sr, Ba and Zr in chondrites and some ultramaflc rocks. *Geochimica et Cosmochimica Acta*, *4*(5), 251-260.

Pisani, F. 1864. Etude chimique et analyse de l'aerolithe Orgueil. Comp. Rend. Acad. Sci. Paris 59, 132-135 ; see corrigenda in CR p348; and in Phipson 1864

Pogge von Strandmann, P.A. E.; Elliott, T.; Marschall, H.R.; Coath, C., Lai, Y.-J.; Jeffcoate, A.B.; Ionov, D.A. 2011. Variations of Li and Mg isotope ratios in bulk chondrites and mantle xenoliths. *Geochimica et Cosmochimica Acta* 75, 5247-5268

Pourmand A., Dauphas N., Ireland T. J. 2012. A novel extraction chromatography and MC-ICP-MS technique for rapid analysis of REE, Sc and Y: revising CI-chondrite and Post-Archean Australian Shale (PAAS, abundances. *Chemical Geology* 291, 38–54.

Pringle, E.A., Moynier, F., Beck, P., Paniello, R., Hezel, D.C. 2017. The origin of volatile element depletion in early solar system material: Clues from Zn isotopes in chondrules. Earth Planetary Science Letters 468,62-71.

Rambaldi, E. R., Cendales, M., Thacker, R. 1978. Trace element distribution between magnetic and non-magnetic portions of ordinary chondrites. *Earth and Planetary Science Letters*, *40*(2), 175-186.

Rammensee, W., Palme, H. 1982. Metal-silicate extraction technique for the analysis of geological and meteoritical samples. J. Radioanal. Chem. 71, 401-418.

Reed, G.W., Kigoshi, K, Turekevitch, A. 1960. Determinations of concentrations of heavy elements in meteorites by activation analysis. *Geochimica et Cosmochimica Acta* 20, 122–140

Robert, F., Epstein, S. 1982. The concentration and isotopic composition of hydrogen, carbon and nitrogen in carbonaceous meteorites. *Geochimica et Cosmochimica Acta* 46, 81–95.

Rocholl, A., Jochum, K. P. 1993. Th, U and other trace elements in carbonaceous chondrites: Implications for the terrestrial and solar-system Th/U ratios. *Earth and Planetary Science Letters* 117, 265-278.

Roscoe, H.E. 1863. On the existence of a crystalline carbon compound and free sulphur in Alais meteorite, *Philosophical Magazine* (4th Series) 25, 319–320.

Rosman, K.J.R., De Laeter, J.R. 1974. The abundance of cadmium and zinc in meteorites. *Geochimica et Cosmochimica Acta* 38, 1665-1677.

Rosman, K.J.R., DeLaeter, J.R. 1986. Isotopic fractionation in meteoritic cadmium revisited. *Meteoritics* 21, 495.

Schmitt R.A., Goles G. G., Smith R. H., Osborn T. W. 1972. Elemental abundances in stone meteorites. *Meteoritics* 7, 131-214.

Schmitt, R. A., Smith, R. H., Lasch, J. E., Mosen, A. W., Olehy, D. A., Vasilevskis, J. 1963. Abundances of the fourteen rare-earth elements, scandium, and yttrium in meteoritic and terrestrial matter. *Geochimica et Cosmochimica Acta 27*, 577-622.

Schmitt, R.A., Smith, R. H., Olehy, D. A. 1964. Rare-Earth, Yttrium and Scandium Abundances in Meteoritic an Terrestrial Matter – II. *Geochimica et Cosmochimica Acta* 28, 67-68

*Schönbächler, M., Lee, D.C., Rehkämper, M., Halliday, A.N. Fehr, M.A., Hattendorf, B., Günther, D 2003. Zirconium isotope evidence for incomplete admixing of r-process components in the solar nebula. *Earth and Planetary Science Letters* 216, 467-481.

Schönbächler M., Rehkämper M., Fehr M.A., Halliday A.N., Hattendorf B., Günther, D. 2005. Nucleosynthetic zirconium isotope anomalies in acid leachates of carbonaceous chondrites. *Geochimica et Cosmochimica Acta* 69, 5113-5122.

Seitz, H.A., Brey, G.P, Zipfel, J., Ott, U., Weyer, S., Durali, S., Weinbruch, S. 2007. Lithium isotope composition of ordinary and carbonaceous chondrites, and differentiated planetary bodies: Bulk solar system and solar reservoirs. *Earth and Planetary Science Letters* 260, 582-596

Sephton, M.A., James, R.H., Fehr, M.A., Bland, P.A., Gounelle, M. 2013.Lithium isotopes as indcators of meteorite parent body alteration, MAPS 48, 872-878.

Sephton, M.A., Verchovsky A. B., Bland P. A., Gilmour I., Grady, M.M., Wright I.P. 2003. Investigating the variations in carbon and nitrogen in carbonaceous chondrites. *Geochimica et Cosmochimica Acta* 67, 2093-2108.

Shima, M. 1979. The abundances of titanium, zirconium and hafnium in stony meteorites. *Geochimica et Cosmochimica Acta* 43, 353-362.

Smales, A.A. 1971. in Activation analysis in geochemistry and cosmochemistry, ed. A.O Brunfelt and E. Steinnes, Proc. Nato Advanced Study Institute, Kjeller, Norway 7-12 Sept 1970, pp. 17-24, Universitietsforlaget, Oslo.

Smales, A.A., Hughes, T.C., Mapper, D., McInnes, C.A.J., Webster, R.K. 1964. The determination of rubidium and caesium in stony meteorites by neutron activation analysis and by mass spectrometry. *Geochimica et Cosmochimica Acta* 28, 209-233.

*Stracke, A., Palme, H., Gellissen, M., Münker, C., Kleine, T., Birbaum, K., Günther, D., Bourdon, B., Zipfel, J. 2012. Refractory element fractionation in the Allende meteorite: Implications for solar nebula condensation and the chondritic composition of planetary bodies. *Geochimica et Cosmochimica Acta 85*, 114-141.

Takahashi, H., Janssens, M.-J., Morgan, J.W., Anders, E. 1978. Further Studies of Trace Elements in C3 Chondrites. *Geochimica et Cosmochimica Acta* 42, 97-106.

Tanner J.T., Ehmann, W.D. 1967. The abundance of antimony in meteorites, tektites and rocks by neutron activation analysis. *Geochimica et Cosmochimica Acta* 31, 2007-2026.

Tatsumoto, M., Unruh, D. M., Desborough, G. A. 1976. U-Th-Pb and Rb-Sr systematics of Allende and U-Th-Pb systematics of Orgueil. *Geochimica et Cosmochimica Acta 40*(6), 617-634.

Thénard, L.J. 1806. Analyse d'un aerolithe tombee dans l'arrondissement d'Alais, le 15 Mars 1806. *Ann. Chim. Phys*, 59, 103-110.

Vilcsek, E. 1977. Beryllium in Meteorites. Meteoritics, 12, 373.

Vogt, J.R., Ehmann, W.D. 1965. Silicon abundances in stony meteorites by fast neutron activation analysis. *Geochimica et Cosmochimica Acta* 29, 373-383.

Vollstaedt, H., Mezger, K., Leya, I. 2016. The isotope composition of selenium in chondrites constrains the depletion mechanism of volatile elements in solar system materials, Earth Planetary Science Letters 450, 372–380

Von Michaelis, H., Ahrens L. H., Willis J. P. 1969, The composition of stony meteorites II. The analytical data and an assessment of their quality. *Earth and Planetary Science Letters* 5, 387-394.

Walker R. J., Horan M. F., Morgan J.W., Becker H., Grossman J.N. and Rubin A.E., 2002. Comparative 187Re-187Os systematics in chondrites: Implications regarding early solar system processes. *Geochimica et Cosmochimica Acta* 66, 4187-4201.

Wang, K., Jacobsen, S. B. 2016. Potassium isotopic evidence for a high-energy giant impact origin of the Moon. Nature. 538, 487-490.

Wang, Z. Becker, H., Gawronski, T. 2013. Partial re-equilibration of highly siderophile elements and the chalcogens in the mantle: A case study on the Baldissero and Balmuccia peridotite massifs (Ivrea Zone, Italian Alps). *Geochimica et Cosmochimica Acta* 108, 21-44.

Wang, Z., Becker, H., Wombacher, F. 2014. Mass Fractions of S, Cu, Se, Mo, Ag, Cd, In, Te, Ba, Sm, W, Tl and Bi in Geological Reference Materials and Selected Carbonaceous Chondrites Determined by Isotope Dilution ICP-MS. *Geostandards and Geoanalytical Research* 39, 185-208.

Weller, M.R., Furst, M., Tombrello, T.A., Burnett, D.S. 1978. Boron concentration in carbonaceous chondrites. *Geochimica et Cosmochimica Acta* 42, 999-1009.

*Weyer, S., Münker, C., Rehkämper, M., Mezger, K. 2002. Determination of ultra-low Nb, Ta, Zr and Hf concentrations and the chondritic Zr/Hf and Nb/Ta ratios by isotope dilution analyses with multiple collector ICP-MS. *Chemical Geology*, *187*(3-4), 295-313.

Wieser M.E. and De Laeter, J.R. 2000. Stable isotope dilution analyses of molybdenum in meteorites. *Fresenius' Journal of Analytical Chemistry* 368, 303–306.

Wiik H.B. 1956. The chemical composition of some stony meteorites. *Geochimica et Cosmochimica Acta* 9, 279–289.

Wiik H.B. 1969. On the regular discontinuities in the composition of meteorites. *Commentationes physico-matematicae societatis scientiarum Fennicae* 34, 135–145 (see also: Mason B. (1962-1963) The carbonaceous chondrites. *Space Science Reviews* 1, 621-646)

Wing, J. 1964. Simultaneous Determination of Oxygen and Silicon in Meteorites and Rocks by Nondestructive Activation Analysis with Fast Neutrons. *Analytical Chemistry 36*(3), 559-564.

Wolf S.F., Unger D.L. Friedrich J.M. 2005. Determination of cosmochemically volatile trace elements in chondritic meteorites by inductively coupled plasma-mass spectrometry. *Analytica Chimica Acta* 528, 121–128.

Wolf, D., Palme H. 2001. The solar system abundances of phosphorus and titanium and the nebular volatility of phosphorus. *Meteoritics and Planetary Science* 36, 559-571.

Wright P. I., McGarvie W. D., Grady M. M., Gibson K. E., Pillinger T. C. 1986. An investigation of 13C-rich material in C1 and C2 meteorites. *Proceedings of the Eleventh Symposium on Antarctic Meteorites* 137–139.

Wright, P.I., Grady, M.M., Carr, R.H., Pillinger, C.T. 1985. Carbon and nitrogen isotope measurements of chemical and physical separates from the Murchison and Orgueil carbonaceous chondrites, Planetary Science. Unit Intern. Rep. 1, 38pp (cited in Wright et al. 1990).

Xiao X., Lipschutz M.E. 1992. Labile trace elements in carbonaceous chondrites: A survey. *Journal of Geophysical Research* 97 E6, 10199-10211.

Yi, Y.V., Masuda, A. 1996, Simultaneous Determination of Ruthenium, Palladium, Iridium, and Platinum at Ultratrace Levels by Isotope Dilution Inductively Coupled Plasma Mass Spectrometry in Geological Samples. *Analytical Chem*. 68, 1444-1450.

Zhai M., Shaw D.M. 1994. Boron cosmochemistry. Part I: Boron in meteorites. *Meteoritics* 29, 607-615.